\documentclass[aps,prd,twocolumn,showpacs,floatfix]{revtex4}

\usepackage{graphicx}
\usepackage{color}

\newenvironment{myitemize}{\begin{itemize}
                \setlength{\itemsep}{-1 pt}}{\end{itemize}}

\newenvironment{myenumerate}{\begin{enumerate}
                \setlength{\itemsep}{-1 pt}}{\end{enumerate}}

\newcommand{\met}       {\mbox{$\not\!\!E_T$}}
\newcommand{\rar}       {\rightarrow}
\newcommand{\rargap}    {\mbox{ $\rightarrow$ }}
\newcommand{\ttbar}     {\mbox{$t\bar{t}$}}
\newcommand{\ppbar}     {\mbox{$p\bar{p}$}}
\newcommand{\bbbar}     {\mbox{$b\bar{b}$}}
\newcommand{\ccbar}     {\mbox{$c\bar{c}$}}
\newcommand{\comphep}   {\sc{comphep}}
\newcommand{\singletop} {\sc{singletop}}
\newcommand{\pythia}    {\sc{pythia}}
\newcommand{\herwig}    {\sc{herwig}}
\newcommand{\alpgen}    {\sc{alpgen}}
\newcommand{\madgraph}  {\sc{madgraph}}
\newcommand{\helas}     {\sc{helas}}
\newcommand{\tauola}    {\sc{tauola}}
\newcommand{\evtgen}    {\sc{evtgen}}
\newcommand{\geant}     {\sc{geant}}
\newcommand{\qtimesz}   {\mbox{$Q(\ell)\times \hat{z}(\ell)$}}
\newcommand{\totalderiv}[2]{\frac{d #1}{d #2}}
\newcommand{\lepjets}   {\mbox{${\ttbar}{\rar}\ell$+jets}}
\newcommand{\elepjets}  {\mbox{${\ttbar}{\rar}e$+jets}}

\newcommand{\vtb}       {\mbox{$|V_{tb}|$}}
\newcommand{\vtbsq}     {\mbox{$|V_{tb}|^2$}}
\newcommand{\deta}      {\mbox{\ensuremath{\eta^{\rm det}}}}
\newcommand{\meta}      {\mbox{\ensuremath{|\eta|}}}
\newcommand{\mdeta}     {\mbox{\ensuremath{|\eta^{\rm det}|}}}
\newcommand{\pt}        {\mbox{$p_T$}}
\def \mathbi#1{\text{\em #1}}

\begin{document}
\hspace{5.2in} \mbox{FERMILAB-PUB-08/056-E}
\title{Evidence for production of single top quarks}

% LIST_OF_AUTHORS_R2.TEX               2/19/08              
%
\author{V.M.~Abazov$^{36}$}
\author{B.~Abbott$^{75}$}
\author{M.~Abolins$^{65}$}
\author{B.S.~Acharya$^{29}$}
\author{M.~Adams$^{51}$}
\author{T.~Adams$^{49}$}
\author{E.~Aguilo$^{6}$}
\author{S.H.~Ahn$^{31}$}
\author{M.~Ahsan$^{59}$}
\author{G.D.~Alexeev$^{36}$}
% ...

\affiliation{\vspace{0.1 in}(The D\O\ Collaboration)\vspace{0.1 in}}
% ...
\affiliation{$^{6}$University of Alberta, Edmonton, Alberta, Canada,
                Simon Fraser University, Burnaby, British Columbia, Canada,
                York University, Toronto, Ontario, Canada, and
                McGill University, Montreal, Quebec, Canada}
% ...
\affiliation{$^{29}$Tata Institute of Fundamental Research, Mumbai, India}
% ...
\affiliation{$^{31}$Korea Detector Laboratory, Korea University, Seoul, Korea}
% ...
\affiliation{$^{36}$Joint Institute for Nuclear Research, Dubna, Russia}
% ...
\affiliation{$^{49}$Florida State University, Tallahassee, Florida 32306, USA}
% ...
\affiliation{$^{51}$University of Illinois at Chicago,
                Chicago, Illinois 60607, USA}
% ...
\affiliation{$^{59}$Kansas State University, Manhattan, Kansas 66506, USA}
% ...
\affiliation{$^{65}$Michigan State University,
                East Lansing, Michigan 48824, USA}
% ...
\affiliation{$^{75}$University of Oklahoma, Norman, Oklahoma 73019, USA}
% ...

\date{April 23, 2008}

\begin{abstract}
We present first evidence for the production of single top quarks in
the D0 detector at the Fermilab Tevatron {\ppbar} collider. The
standard model predicts that the electroweak interaction can produce a
top quark together with an antibottom quark or light quark, without
the antiparticle top quark partner that is always produced from strong
coupling processes. Top quarks were first observed in pair production
in 1995, and since then, single top quark production has been searched
for in ever larger datasets. In this analysis, we select events from a
0.9~fb$^{-1}$ dataset that have an electron or muon and missing
transverse energy from the decay of a $W$~boson from the top quark
decay, and two, three, or four jets, with one or two of the jets
identified as originating from a $b$~hadron decay. The selected events
are mostly backgrounds such as $W$+jets and {\ttbar} events, which we
separate from the expected signals using three multivariate analysis
techniques: boosted decision trees, Bayesian neural networks, and
matrix element calculations. A binned likelihood fit of the signal
cross section plus background to the data from the combination of the
results from the three analysis methods gives a cross section for
single top quark production of $\sigma({\ppbar}{\rargap}tb+X,~tqb+X) =
4.7 \pm 1.3$~pb. The probability to measure a cross section at this
value or higher in the absence of signal is $0.014\%$, corresponding
to a 3.6~standard deviation significance. The measured cross section
value is compatible at the 10\% level with the standard model
prediction for electroweak top quark production. We use the cross
section measurement to directly determine the
Cabibbo-Kobayashi-Maskawa quark mixing matrix element that describes
the $Wtb$ coupling and find $|V_{tb}f_1^L|=1.31^{+0.25}_{-0.21}$,
where $f_1^L$ is a generic vector coupling. This model-independent
measurement translates into $0.68 < |V_{tb}| \le 1$ at the
$95\%$~C.L. in the standard model.
\end{abstract}

\pacs{\vspace{-0.2in} 14.65.Ha; 12.15.Ji; 13.85.Qk }

\maketitle 

%---------------------------------------------------------------------
%---------------------------------------------------------------------

\tableofcontents
\vfill
\clearpage

%\input{introduction}
%\input{detector}
%\input{triggers_data}
%\input{event_reconstruction}
%\input{monte_carlo}
%\input{event_selection}
%\input{background_model}
%\input{signal_acceptances}
%\input{event_yields}
%\input{systematic_uncertainties}
%\input{multivariate_intro}
%\input{dt_analysis}
%\input{bnn_analysis}
%\input{me_analysis}
%\input{multivariate_outputs}
%\input{ensembles}
%\input{cross_checks}
%\input{cs_measurement}
%\input{results}
%\input{combination}
%\input{vtb_measurement}
%\input{summary}
%\input{acknowledgements}
%\input{references}

%---------------------------------------------------------------------
%---------------------------------------------------------------------
\section{Introduction}
\label{introduction}

\subsection{Single Top Quarks}
\label{single-top-quarks}

Top quarks were first observed in top~quark -- top~antiquark pair
production via the strong interaction in
1995~\cite{top-obs-1995-cdf,top-obs-1995-d0}. The standard model also
predicts that the electroweak interaction can produce a top quark
together with a bottom antiquark or a light quark, without the
antiparticle top quark partner that is always produced in
strong-coupling processes. This electroweak process is generally
referred to as single top quark production. Since 1995, the D0 and CDF
collaborations have been searching ever larger datasets for signs of
single top quark production.

We present here the results of a search for top quarks produced singly
via the electroweak interaction from the decay of an off-shell
$W$~boson or fusion of a virtual $W$~boson with a $b$~quark.
Previously measured top quarks have been produced in pairs from highly
energetic virtual gluons via the strong interaction. The cross section
for {\ttbar} production at the Fermilab Tevatron proton-antiproton
collider (center-of-mass energy = 1.96~TeV) is
$6.77\pm0.42$~pb~\cite{ttbar-xsec-kidonakis} at next-to-leading order
(NLO) plus higher-order soft-gluon corrections, for a top quark of
mass $m_{\rm top} = 175$~GeV~\cite{top-mass-footnote}. The standard
model predicts three processes for production of a top quark without
its antiparticle partner . These are as follows: (i) the s-channel
process ${\ppbar}{\rar}t\bar{b} + X, \bar{t}b +
X$~\cite{s-channel-cortese,s-channel-stelzer,singletop-heinson}, with
a cross section of $0.88 \pm 0.14$~pb~\cite{singletop-xsec-sullivan}
at NLO for $m_{\rm top} = 175$~GeV; (ii) the t-channel process
${\ppbar}{\rar}tq\bar{b} + X,\bar{t}\bar{q}b +
X$~\cite{t-channel-willenbrock,t-channel-yuan,t-channel-ellis,singletop-heinson},
with a cross section of $1.98 \pm
0.30$~pb~\cite{singletop-xsec-sullivan} at the same order in
perturbation theory and top quark mass; and (iii) the $tW$ process
${\ppbar}{\rar}tW^{-} + X,\bar{t}W^{+} +
X$~\cite{singletop-heinson,tW-channel-tait}, where the cross section
at the Tevatron energy is small, $0.08 \pm
0.02$~pb~\cite{tW-channel-tait} at LO.

\begin{figure}[!h!btp]
\vspace{0.1in}
\includegraphics[width=0.45\textwidth]{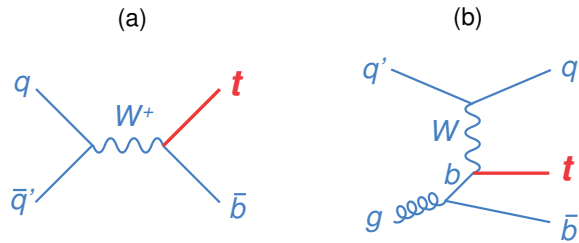}
\vspace{-0.1in}
\caption[feynman]{Main tree-level Feynman diagrams for (a) s-channel
single top quark production, and (b) t-channel production.}
\label{feynman-diagrams}
\end{figure}

The main tree-level Feynman diagrams for the dominant single top quark
production processes are illustrated in
Fig.~\ref{feynman-diagrams}. For brevity, in this paper we will use
the notation ``$tb$'' to mean the sum of $t\bar{b}$ and $\bar{t}b$,
and ``$tqb$'' to mean the sum of $tq\bar{b}$ and
$\bar{t}\bar{q}b$. The analysis reported in this paper searches only
for the s-channel process $tb$ and the t-channel process $tqb$, and
does not include a search for the $tW$ process because of its small
production rate at the Tevatron.

Top quarks are interesting particles to study since in the standard
model their high mass implies a Yukawa coupling to the Higgs boson
with a value near unity, unlike any other known particle. They also
decay before they hadronize, allowing the properties of a bare quark
such as spin to be transferred to its decay products and thus be
measured and compared to the standard model predictions. Events with
single top quarks can also be used to study the $Wtb$
coupling~\cite{Wtb-coupling-yuan1,Wtb-coupling-yuan2,singletop-heinson},
and to measure directly the absolute value of the quark mixing matrix
(the Cabibbo-Kobayashi-Maskawa (CKM)
matrix~\cite{ckm-matrix-c,ckm-matrix-km}) element $|V_{tb}|$ without
assuming there are only three generations of
quarks~\cite{singletop-vtb-jikia,singletop-vtb-stelzer}. A measured
value for $|V_{tb}|$ significantly different from unity could imply
the existence of a fourth quark family or other effects from beyond
the standard model~\cite{Vtb-not-one-alwall}.

\subsection{Search History}
\label{search-history}

The D0 collaboration has published three searches for single top quark
production using smaller datasets. We analyzed 90~pb$^{-1}$ of data
from Tevatron Run~I (1992--1996 at a center-of-mass energy of 1.8~TeV)
which resulted in the first upper limits on single top quark
production~\cite{run1-d0-prd} and we performed a more refined search
using neural networks that achieved greater
sensitivity~\cite{run1-d0-plb}. In Run~II, we used 230~pb$^{-1}$ of
data collected from 2002 to 2004 to set more stringent upper
limits~\cite{run2-d0-plb,run2-d0-prd}. Our best published 95\%
C.L. upper limits are 6.4~pb in the s-channel ($tb$ production) and
5.0~pb in the t-channel ($tqb$ production). Students in the D0
collaboration have completed ten Ph.D.\ dissertations on the single
top quark search~\cite{dissertations-d0}. Our most recent
publication~\cite{run2-d0-prl-evidence} presents first evidence for
single top quark production using a 0.9~fb$^{-1}$ dataset. We provide
a more detailed description of that result here, and also include
several improvements to the analysis methods that lead to a final
result on the same dataset with slightly higher significance.

The CDF collaboration has published two results from analyzing
106~pb$^{-1}$ of Run~I data~\cite{run1-cdf-1,run1-cdf-2}, and one that
uses 162~pb$^{-1}$ of Run~II data~\cite{run2-cdf-162}. Their best
95\% C.L. upper limits are 14~pb in the s-channel, 10~pb in the
t-channel, and 18~pb in the s-channel and t-channel combined. Students
in the CDF collaboration have completed seven Ph.D.\ dissertations on
the single top quark search~\cite{dissertations-cdf}.

\subsection{Search Method Overview}
\label{search-method-overview}

The experimental signal for single top quark events consists of one
isolated high transverse momentum (\pt), central pseudorapidity
($\eta$~\cite{eta}) charged lepton and missing transverse energy
({\met}) from the decay of a $W$~boson from the top quark decay,
accompanied by a $b$~jet from the top quark decay. There is always a
second jet, which originates from a $b$~quark produced with the top
quark in the s-channel, or which comes from a forward-traveling up- or
down-type quark in t-channel events. Some t-channel events have a
detectable $b$~jet from the gluon splitting to $b\bar{b}$. Since there
may be significant initial-state or final-state radiation, we include
in our search events with two, three, or four jets. We use data
collected with triggers that include an electron or a muon, and a jet.
In the electron channel, multijet events can fake signal ones when a
jet is misidentified as an electron, and we have stringent
identification criteria for electrons to reduce this type of
background. In the muon channel, {\bbbar}+jets events can fake signal
ones when one of the $b$'s decays to a muon. We reject much of this
background by requiring the muon to be isolated from all jets in the
event. Finally, we apply a set of simple selection criteria to retain
regions of phase space that single top quark events tend to populate.

We divide the selected events into 12 nonoverlapping samples, referred
to as analysis channels, depending on the flavor of the lepton ($e$ or
$\mu$), the number of jets (2, 3, 4), and the number of jets
identified as originating from $b$~quarks (number of ``tagged'' jets =
1, 2), because the signal-to-background ratios and fractions of
expected signal in each channel differ significantly. The dominant
background in most of these channels is $W$+jets events. We model this
background using events simulated with Monte Carlo (MC) techniques and
normalized to data before $b$~tagging. We also use an MC model to
simulate the background from {\ttbar} events. Finally, we use data
events with poorly identified leptons to model the multijet background
where a jet is misidentified as an electron, or a muon in a jet from
$c$ or $b$~decay is misidentified as a muon from a $W$~boson decay. We
apply a neural-network-based $b$-identification algorithm to each jet
in data and keep events with one or two jets that are identified as
$b$~jets. We model this $b$~tagging in the MC event samples by
weighting each event by the probability that one or more jets is
tagged.

After event selection, we calculate multivariate discriminants in each
analysis channel to separate as much as possible the expected signal
from the background. We then perform a binned likelihood fit of the
background model plus possible signal to the data in the discriminant
output distributions and combine the results from all channels that
improve the expected sensitivity. Finally, we calculate the
probability that our data are compatible with background only, use the
excess of data over background in each bin to measure the signal cross
section, and calculate the probability that the data contains both
background and signal produced with at least the measured cross
section value.

For each potential analysis channel, the relevant details are the
signal acceptance and the signal-to-background ratio.
Table~\ref{acceptance-percentages} shows the percentage of the total
signal acceptance for each jet multiplicity and number of $b$-tagged
jets, and the associated signal-to-background ratios. We used this
information to determine that the most sensitive channels have two,
three, or four jets, and one or two $b$ tags. In the future, it could
be beneficial to extend the analysis to include events with only one
jet, $b$~tagged, since the signal-to-background ratios are not bad,
and to study the untagged events with two or three jets where there is
significant signal acceptance.

\begin{table}[!h!tbp]
%\vspace{-0.2in}
\caption[acceptance-percentages]{Percentage of total selected MC
single top quark events (i.e., all channels shown in the table) 
for each jet multiplicity and number of
$b$-tagged jets, and the associated signal-to-background ratios, for
the electron and muon channels combined. The values shown in bold type
are for the channels used in this analysis.}
\label{acceptance-percentages}
\begin{ruledtabular}
\begin{tabular}{l||ccccc}
\multicolumn{6}{c}
{\hspace{0.5in}\underline{Distribution of Signal Events}}
\vspace{0.05in}\\
 & 1 jet & 2 jets & 3 jets & 4 jets & $\ge5$ jets \\
\hline
 & \multicolumn{5}{c}{s-channel $tb$} \\
0 $b$ tags~~&    8\%   &     19\%   &     9\%  &      3\%   &   1\%   \\
            & 1:11,000 &   1:1,600 &   1:1,200 &    1:1,100 & 1:1,000 \\
1 $b$ tag   &    6\%   & {\bf 24\%} & {\bf 12\%} & {\bf 3\%}  &   1\%  \\
            & 1:270    & {\bf 1:55} & {\bf 1:73}  & {\bf 1:130} & 1:200   \\
2 $b$ tags  &   ---    & {\bf  9\%} & {\bf  4\%} & {\bf 1\%}  &   0\%   \\
            &   ---    & {\bf 1:12} & {\bf 1:27}  & {\bf 1:92}  & 1:110   \\
\hline
 & \multicolumn{5}{c}{t-channel $tqb$}                        \\
0 $b$ tags  &   10\%   &     27\%   &    13\%  &      4\%   &   1\%   \\
            & 1:4,400  &   1:520    &   1:400  &   1:360    & 1:300   \\
1 $b$ tag   &    6\%   & {\bf 20\%} & {\bf 11\%} & {\bf 4\%}  &   1\%   \\
            & 1:150    & {\bf 1:32} & {\bf 1:37}  & {\bf 1:58}  & 1:72    \\
2 $b$ tags  &   ---    & {\bf  1\%} & {\bf  2\%} & {\bf 1\%}  &   0\%   \\
            &   ---    & {\bf 1:100} & {\bf 1:36} & {\bf 1:65}  & 1:70
\end{tabular}
\end{ruledtabular}
\end{table}

\vspace{-0.2in}
\subsection{Differences from Previous Searches}
\label{search-differences}

We summarize here the changes and improvements made to the analysis
since the previously published D0 result that used 230~pb$^{-1}$ of
data~\cite{run2-d0-plb,run2-d0-prd}. The most important difference is
that we have analyzed a dataset four times as large. Other changes
include the following: (i) use of an improved model for the t-channel
$tqb$ signal from the package {\singletop}~\cite{singletop-mcgen},
based on {\comphep}~\cite{comphep}, which better reproduces NLO-like
parton kinematics; (ii) use of an improved model for the {\ttbar} and
$W$+jets backgrounds from the {\sc alpgen} package~\cite{alpgen} that
has parton-jet matching~\cite{jet-matching} implemented with {\sc
pythia}~\cite{pythia} to avoid duplicate generation of some
initial-state and final-state jet kinematics; (iii) determination from
data of the ratio of $W$ boson plus {\bbbar} or {\ccbar} jets to the
total rate of $W$+jets production; (iv) omission of a separate
calculation of the diboson backgrounds $WW$ and $WZ$ since they are
insignificant; (v) differences in electron, muon, and jet
identification requirements and minimum \pt's; (vi) use of a
significantly higher efficiency $b$-tagging algorithm based on a
neural network; (vii) splitting of the analysis by jet and $b$-tag
multiplicity so as not to dilute the strength of high-acceptance, good
signal-to-background channels by mixing them with poorer ones; (viii)
simplification of the treatment of the smallest sources of systematic
uncertainty (since the analysis precision is statistics dominated);
(ix) use of improved multivariate techniques to separate signal from
background; and (x) optimization of the search to find the combined
single top quark production from both the s- and t-channels,
$tb$+$tqb$.

%---------------------------------------------------------------------
%---------------------------------------------------------------------
\section{The D0 Detector}
\label{detector}

The D0 detector~\cite{d0-nim} consists of three major parts: a
tracking system to determine the trajectories and momenta of charged
particles, a calorimeter to measure the energies of electromagnetic
and hadronic showers, and a system to detect muons, which are the only
charged particles that are typically not contained within the
calorimeter. The first element at the core of the detector is a
tracking system that consists of a silicon microstrip tracker (SMT)
and a central fiber tracker (CFT), both located within a 2~T
superconducting solenoidal magnet. The SMT has six barrel modules in
the central region, each comprising four layers arranged axially
around the beam pipe, and 16 radial disks interspersed with and beyond
the central barrels. Ionization charge is collected by $\approx
800,000$ $p$- or $n$-type silicon strips of pitch between $50$ and
$150~\mu\rm m$ that are used to measure the positions of the
hits. Tracks can be reconstructed up to pseudorapidities~\cite{eta} of
$|\deta|\approx 3.0$.

The CFT surrounds the SMT with eight thin coaxial barrels, each
supporting two doublets of overlapping scintillating fibers of
0.835~mm diameter, one doublet being parallel to the beam axis, and
the other alternating by $\pm 3^{\circ}$ relative to the
axis. Visible-light photon counters (VLPCs) collect the light signals
from the fibers, achieving a cluster resolution of about $100~\mu\rm
m$ per doublet layer.

Central and forward preshower detectors contribute to the
identification of electrons and photons. The central preshower
detector is located just outside of the superconducting coil and the
forward ones are mounted in front of the endcap calorimeters. The
preshower detectors comprise several layers of scintillator strips
that are read out using wavelength-shifting fibers and VLPCs.

Three finely grained uranium/liquid-argon sampling calorimeters
constitute the primary system used to identify electrons, photons, and
jets. The central calorimeter (CC) covers $|\deta|$ up to $\approx
1.1$. The two end calorimeters (EC) extend the coverage to
$|\deta|\approx 4.2$. Each calorimeter contains an electromagnetic
(EM) section closest to the interaction region with approximately 20
radiation lengths of material, followed by fine and coarse hadronic
sections with modules that increase in size with distance from the
interaction region and ensure particle containment with approximately
six nuclear interaction lengths. In addition to the preshower
detectors, scintillators between the CC and EC provide sampling of
developing showers in the cryostat walls for $1.1<|\deta|<1.4$.

The three-layer muon system is located beyond the calorimetry, with
1.8~T iron toroids after the first layer to provide a stand-alone
muon-system momentum measurement. Each layer comprises tracking
detectors and scintillation trigger counters. Proportional drift tubes
10~cm in diameter allow tracking in the region $|\deta| < 1$, and 1~cm
mini drift tubes extend the tracking to $|\deta| < 2$.

Additionally, plastic scintillator arrays covering $2.7 < |\deta| <
4.4$ are used to measure the rate of inelastic collisions in the D0
interaction region and calculate the Tevatron instantaneous and
integrated luminosities.

We select the events to be studied offline with a three-tiered trigger
system. The first level of the trigger makes a decision based on
partial information from the tracking, calorimeter, and muon
systems. The second level of the trigger uses more refined information
to further reduce the rate. The third trigger level is based on
software filters running in a farm of computers that have access to
all information in the events.

%---------------------------------------------------------------------
%---------------------------------------------------------------------
\section{Triggers and Data}
\label{triggers-data}

The data were collected between August 2002 and December 2005, with
$913 \pm 56$~pb$^{-1}$ and $871 \pm 53$~pb$^{-1}$ of good quality
events in the electron and muon channels respectively.

As the average instantaneous luminosity of the Tevatron has increased
over time, the triggers used to collect the data have been
successively changed to maintain background rejection. The
requirements at the highest trigger level are the following, with the
associated integrated luminosity included in parentheses:

\vspace{0.15in}
{\bf Electron+jets triggers}
\vspace{-0.08in}
\begin{myenumerate}
\item One electron with $\pt>15$~GeV and two jets with
$\pt>15$~GeV (103~pb$^{-1}$)
\item One electron with $\pt>15$~GeV and two jets with $\pt>20$~GeV
(227~pb$^{-1}$)
\item One electron with $\pt>15$~GeV, one jet with $\pt>25$~GeV, and
a second jet with $\pt>20$~GeV (289~pb$^{-1}$)
\item One electron with $\pt>15$~GeV and two jets with $\pt>30$~GeV
(294~pb$^{-1}$)
\end{myenumerate}

{\bf Muon+jets triggers}
\vspace{-0.08in}
\begin{myenumerate}
\item One lower-trigger-level muon with no $\pt$ threshold and one
jet with $\pt>20$~GeV (107~pb$^{-1}$)
\item One lower-trigger-level muon with no $\pt$ threshold and one
jet with $\pt>25$~GeV (278~pb$^{-1}$)
\item One muon with $\pt>3$~GeV and one jet with $\pt>30$~GeV
(252~pb$^{-1}$)
\item One isolated muon with $\pt>3$~GeV and one jet with
$\pt>25$~GeV (21~pb$^{-1}$)
\item One muon with $\pt>3$~GeV and one jet with $\pt>35$~GeV
(214~pb$^{-1}$)
\end{myenumerate}

The average efficiency of the electron+jets triggers is 87\% for $tb$
events and 86\% for $tqb$ events that pass the final selection
cuts. The average efficiency of the muon+jets triggers is 87\% for
$tb$ and 82\% for $tqb$ events.

Note that for the electron+jets triggers, the electron usually
satisfies one of the jet requirements, and thus there are usually only
two independent objects required in each event (one electron and one
jet).

%---------------------------------------------------------------------
%---------------------------------------------------------------------
\section{Event Reconstruction}
\label{event-reco}
\vspace{-0.05in}

Physics objects are reconstructed from the digital signals recorded in
each part of the detector. Particles can be identified by certain
patterns and, when correlated with other objects in the same event,
they provide the basis for understanding the physics that produced
such signatures in the detector.

\vspace{-0.1in}
\subsection{Primary Vertices}
\label{primary-vertex-id}
\vspace{-0.05in}

The location of the hard-scatter interaction point is reconstructed by
means of an adaptive primary vertex algorithm~\cite{adaptive-pv}. This
algorithm first selects tracks coming from different interactions by
clustering them according to their $z$~position along the nominal beam
line. In the second step, the location and width of the beam in the
transverse plane (perpendicular to the beam line) are determined and
then used to re-fit tracks, and each cluster of tracks is associated
with a vertex using the ``adaptive'' technique that gives all tracks a
weight and iterates the fit. The third and last step consists of
choosing the vertex that has the lowest probability of coming from a
minimum bias interaction (a {\ppbar} scatter event), based on the
$p_T$ values of the tracks assigned to each vertex. The hard-scatter
vertex is distinguished from soft-interaction vertices by the higher
average $p_T$ of its tracks. In multijet data, the position resolution
of the primary vertex in the transverse plane is around 40~$\mu$m,
convoluted with a typical beam spot size of around 30~$\mu$m.

\vspace{-0.1in}
\subsection{Electrons}
\label{electron-id}
\vspace{-0.05in}

Electron candidates are defined as clusters of energy depositions in
the electromagnetic section of the central calorimeter ($|\deta| <
1.1$) consistent in shape with an electromagnetic shower. At least
90\% of the energy of the cluster must be contained in the
electromagnetic section of the calorimeter, $f_{\mathrm{EM}} > 0.9$,
and the cluster must satisfy the following isolation criterion:
\begin{equation}
\label{eq:emiso}
\frac{E_{\mathrm{total}}(\mathcal{R}<0.4) -
      E_{\mathrm{EM}}(\mathcal{R}<0.2)}
     {E_{\mathrm{EM}}(\mathcal{R}<0.2)} < 0.15,
\end{equation}
where $E$ is the electron candidate's energy measured in the
calorimeter, and $\mathcal{R}=\sqrt{(\Delta\phi)^2+(\Delta\eta)^2}$ is
the radius of a cone defined by the azimuthal angle $\phi$ and the
pseudorapidity $\eta$, centered on the electron candidate's track if
there is an associated track, or the calorimeter cluster if there is
not. Two classes of electrons are subsequently defined and used in
this analysis:
\begin{myitemize}
\item {\bf Loose electron}\\
A loose electron must pass the identification requirements listed
above. In addition, the energy deposition in the calorimeter must be
matched with a charged particle track from the tracking detectors with
$p_T>5$~GeV. Finally, a shower-shape chi-squared, based on seven
variables that compare the values of the energy deposited in each
layer of the electromagnetic calorimeter with average distributions
from simulated electrons, has to satisfy $\chi^2_{\mathrm{cal}} < 50$.
\item {\bf Tight electron}\\
A tight electron must pass the loose requirements, and have a value of
a seven-variable EM-likelihood $\mathcal{L} > 0.85$. The following
variables are used in the likelihood: (i)~$f_{\mathrm{EM}}$;
(ii)~$\chi^2_{\mathrm{cal}}$;
(iii)~$E_T^{\mathrm{cal}}/p_T^{\mathrm{track}}$, the transverse energy
of the cluster divided by the transverse momentum of the matched
track; (iv)~the $\chi^2$ probability of the match between the track
and the calorimeter cluster; (v)~the distance of closest approach
between the track and the primary vertex in the transverse plane; (vi)
the number of tracks inside a cone of $\mathcal{R} = 0.05$ around the
matched track; and (vii)~the $\sum{p_T}$ of tracks within an
$\mathcal{R} = 0.4$ cone around the matched track. The average tight
electron identification efficiency in data is around $75\%$.
\end{myitemize}

\vspace{-0.1in}
\subsection{Muons}
\label{muon-id}
\vspace{-0.05in}

Muons are identified by combining tracks in the muon spectrometer with
central detector tracks. Muons are reconstructed up to $|\deta| = 2$
by first finding hits in all three layers of the muon spectrometer and
requiring that the timing of these hits be consistent with the muon
originating in the center of the detector from the correct
proton-antiproton bunch crossing, thereby rejecting cosmic
rays. Secondly, all muon candidates must be matched to a track in the
central tracker, where the central track must pass the following
criteria: (i)~$\chi^2$ per degree of freedom must be less than 4; and
(ii)~the distance of closest approach between the track and the
primary vertex must be less than 0.2~mm if the track has SMT hits and
less than 2~mm if it does not. Two classes of muons are then defined
for this analysis:
\begin{myitemize}
\item {\bf Loose-isolated muon}\\
A loose muon must pass the identification requirements given above.
Loose muons must in addition be isolated from jets. The distance
between the muon and any jet axis in the event has to satisfy
$\mathcal{R}({\mathrm{muon,~jet}}) > 0.5$.
\item {\bf Tight-isolated muon}\\
A tight muon must pass the loose-isolation requirement and additional
isolation criteria as follows: (i)~the transverse momenta of all
tracks within a cone of radius $\mathcal{R} = 0.5$ around the muon
direction, except the track matched to the muon, must add up to less
than $20\%$ of the muon $p_T$; and (ii)~the energy deposited in a cone
of radius $0.1 < \mathcal{R} < 0.4$ around the muon direction must be
less that $20\%$ of the muon $p_T$.
\end{myitemize}

\subsection{Jets}
\label{jet-id}

We reconstruct jets based on calorimeter cell energies, using the
midpoint cone algorithm~\cite{jet-definition} with radius
$\mathcal{R}=0.5$. Noisy calorimeter cells are ignored in the
reconstruction algorithm by only selecting cells whose energy is at
least four standard deviations above the average electronic noise and
any adjacent cell with at least two standard deviations above the
average electronic noise.

To reject poor quality or noisy jets, we require all jets to have the
following: (i)~$0.05 < f_{\mathrm{EM}} < 0.95$ in the central region,
with the lower cut looser in the intercryostat and forward regions;
(ii)~fraction of jet $p_T$ in the coarse hadronic calorimeter layers
$< 0.4$ in the central region, with looser requirements in the forward
regions; and (iii)~at least $50\%$ of the $p_T$ of the jet, not
including the coarse hadronic layers, matched to energy depositions in
towers in Level~1 of the trigger in a cone of radius $\mathcal{R} =
0.5$ around the jet axis in the central region, with looser
requirements in the forward regions.

Jet energy scale corrections are applied to convert reconstructed jet
energies into particle-level energies. The energy of each jet
containing a muon within $\mathcal{R}({\mathrm{muon,~jet}}) < 0.5$
(considered to originate from a semileptonic $c$- or $b$-quark decay)
is corrected to account for the energy of the muon and the
accompanying neutrino (because that energy is not deposited in the
calorimeter and so would otherwise be undermeasured). For this
correction, it is assumed that the neutrino has the same energy as the
muon.

Jets that have the same $\eta$ and $\phi$ as a reconstructed electron
are removed from the list of jets to avoid double-counting objects.

\subsection{Missing Transverse Energy}
\label{met-id}

Neutrinos carry away momentum that can be inferred using momentum
conservation in the transverse plane. The sum of the transverse
momenta of undetected neutrinos is equal to the negative of the sum of
the transverse momenta of all particles observed in the detector. In
practice, we compute the missing transverse energy by adding up
vectorially the transverse energies in all cells of the
electromagnetic and fine hadronic calorimeters. Cells in the coarse
hadronic calorimeter are only added if they are part of a jet. This
raw quantity is then corrected for the energy corrections applied to
the reconstructed objects and for the momentum of all muons in the
event, corrected for their energy loss in the calorimeter.

\subsection{$\mathbi{b}$ Jets}
\label{b-tagging}

Given that single top quark events have at least one $b$~jet in the
final state, we use a $b$-jet tagger to identify jets originating from
$b$~quarks. In addition to the jet quality criteria described in
previous sections, a ``taggability'' requirement is applied. This
requires the jets to have at least two good quality tracks with $p_T >
1$~GeV and $p_T > 0.5$~GeV respectively, that include SMT hits and
which point to a common origin. A neural network (NN) tagging
algorithm is used to identify jets originating from a $b$~quark. The
tagger and its performance in the data is described in detail in
Ref.~\cite{thesis-scanlon}. We summarize briefly here its main
characteristics. The NN tagger uses the following variables, ranked in
order of separation power, to discriminate $b$~jets from other jets:
(i)~decay length significance of the secondary vertex reconstructed by
the secondary vertex tagger (SVT); (ii)~weighted combination of the
tracks' impact parameter significances; (iii)~jet lifetime probability
(JLIP), the probability that the jet originates from the primary
vertex~\cite{thesis-greder}; (iv)~$\chi^2$ per degree of freedom of
the SVT secondary vertex; (v)~number of tracks used to reconstruct the
secondary vertex; (vi)~mass of the secondary vertex; and (vii)~number
of secondary vertices found inside the jet.

For this analysis, we require the NN output to be greater than 0.775
for the jet to be considered $b$~tagged. The average probability for a
light jet in data to be falsely tagged at this operating point is
$0.5\%$, and the average $b$-tagging efficiency in data is $47\%$ for
jets with $|\deta| < 2.4$.

%---------------------------------------------------------------------
%---------------------------------------------------------------------
\section{Simulated Event Samples}
\label{mc-samples}

\subsection{Event Generation}
\label{evt-gen}

For this analysis, we generate single top quark events with the
{\comphep}-{\singletop}~\cite{comphep,singletop-mcgen} Monte Carlo
event generator. {\singletop} produces events whose kinematic
distributions match those from NLO
calculations~\cite{singletop-xsec-sullivan}. The top quark mass is set
to 175~GeV, the set of parton distribution functions (PDF) is
CTEQ6L1~\cite{cteq}, and the renormalization and factorization scales
are $m_{\mathrm{top}}^2$ for the s-channel and
$(m_{\mathrm{top}}/2)^2$ for the t-channel. These scales are chosen
such that the LO cross sections are closest to the NLO cross
sections~\cite{wbb-boos}. The top quarks and the $W$~bosons from the
top quark decays are decayed in {\singletop} to ensure the spins are
properly transferred. {\pythia}~\cite{pythia} is used to add the
underlying event, initial-state and final-state radiation, and for
hadronization. {\tauola}~\cite{tauola} is used to decay tau leptons,
and {\evtgen}~\cite{evtgen} to decay $b$~hadrons. To calculate the
expected number of signal events, these samples are normalized to the
NLO cross sections~\cite{singletop-xsec-sullivan} for a top quark mass
of 175~GeV: $0.88 \pm 0.14$~pb for the s-channel and $1.98 \pm
0.30$~pb for the t-channel.

The $W$+jets and {\ttbar} samples are generated using
{\alpgen}~\cite{alpgen}. The version we use includes a parton-jet
matching algorithm that follows the MLM
prescription~\cite{jet-matching,jet-matching-top}. For the {\ttbar}
samples, the top quark mass is set to 175~GeV, the scale is
$m_{\mathrm{top}}^2+\sum p_T^2({\mathrm{jets}})$, and the PDF set is
CTEQ6L1. For the $W$+jets events, the PDF is also CTEQ6L1 and the
scale is $m_W^2+p_T^2(W)$. The $W$+jets events include separate
generation of each jet multiplicity from $W$+ 0 light partons to $W$+
at least 5 light partons for events with no heavy-flavor partons (we
refer to these samples as $Wjj$). Those with ${\bbbar}$ and
${\ccbar}$ partons have separately generated samples with between 0
and 3 additional light partons. The {\ttbar} events include separate
samples with additional jets from 0 to 2 light partons.

For the $W$+jets sets, we remove events with heavy flavor jets added
by {\pythia} so as not to duplicate the phase space of those generated
already by {\alpgen}. The $Wcj$ subprocesses are included in the $Wjj$
sample with massless charm quarks.

Since the $W$+jets background is normalized to data (see
Sec.~\ref{matrix-method}), it implicitly includes all sources of
$W$+jets, $Z$+jets, and diboson events with similar jet-flavor
composition, in particular $Z$+jets events where one of the leptons
from the $Z$~boson decay is not identified.

\begin{figure}[!h!btp]
\includegraphics[width=0.38\textwidth]{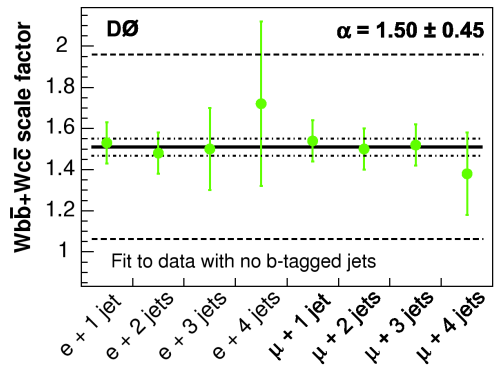}
\vspace{-0.1in}
\caption[alpha]{Measurements of the scale factor $\alpha$ used to
convert the fraction of $W{\bbbar}$ and $W{\ccbar}$ events in the
$W$+jets background model from leading order to higher order. The
points are the measured correction factor in each dataset. The solid
line is the average of these values. The dot-dash inner band shows the
uncertainty from the fit to the eight data points. The dashed outer
line shows the uncertainty on $\alpha$ used in the analysis to allow
for the assumption that the scale factor should be the same for
$W{\bbbar}$ and $W{\ccbar}$, and for small differences in the shapes
of distributions between the $W$ + heavy flavor and $W$ + light flavor
jets.}
\label{alpha-scalefactor}
\end{figure}

The proportions of $W{\bbbar}$ and $W{\ccbar}$ in the $W$+jets model
are set by {\alpgen} at leading order precision. However, higher order
calculations~\cite{wbb-ellis,wbb-cordero,wb-campbell} indicate that
there should be a higher fraction of events with heavy-flavor jets. We
measure a scale factor for the $W{\bbbar}$ and $W{\ccbar}$ subsamples
using several untagged data samples (with zero $b$-tagged jets) that
have negligible signal content. We obtain:
\begin{equation}
\begin{array}{c}
 \alpha(N_{Wb\bar{b}} + N_{Wc\bar{c}}) + N_{Wjj} + N_{t\bar{t}} + 
        N_{\mathrm{multijets}} =
 N_{\mathrm{data}}^{\mathrm{zero{\mbox{\small -}}tag}} \\
 \alpha = 1.50 \pm 0.45
\end{array}
\end{equation}
where the numbers of events $N_i$ for each background component
correspond to the expected number of events after event selection
(described in Sec.~\ref{event-selection}) and background normalization
(described in Sec.~\ref{background}) and removing events with one or
more $b$-tagged jets. Additionally, we check that the same value of
$\alpha = 1.5$ is obtained from the complementary $W$ + 1 jet sample,
where we require the only jet to be $b$~tagged.
Figure~\ref{alpha-scalefactor} illustrates the measurement of the
scale factor $\alpha$.

We examine the distributions expected to suffer the largest shape
dependence from higher order corrections, such as the invariant mass
of the two leading jets and the $p_T$ of the $b$-tagged jet, and find
good agreement between the shapes of the data and the background
model, not only in the signal region, but also in samples enriched
with $W$+jets events.

Table~\ref{mcstats} shows the cross sections, branching fractions,
initial numbers of events, and integrated luminosities of the
simulated samples used in this analysis.

\begin{table*}[!h!tbp]
\begin{center}
\vspace{-0.2in}
\begin{minipage}{5in}
\caption[mcstatistics]{Cross sections, branching fractions,
initial numbers of events, and integrated luminosities of the
simulated event samples. Here, $\ell$ means $e$, $\mu$, and $\tau$.}
\label{mcstats}
\begin{ruledtabular}
\begin{tabular}{l||cccc}
\multicolumn{5}{c}{\hspace{1in}\underline{Statistics of the Simulated Samples}} \vspace{0.05in}\\
            &  Cross Section  &  Branching       &  Number   &  Integrated\vspace{-0.04in} \\
Event Type  &       [pb]      &  Fraction        & of Events &  Luminosity [fb$^{-1}$]     \\
\hline
{\bf Signals}         &                 &                     &         &        \\
~~$tb\to e$+jets    & $0.88 \pm 0.14$ & $0.1111 \pm 0.0022$ &  92,620 &    947 \\
~~$tb\to\mu$+jets   & $0.88 \pm 0.14$ & $0.1111 \pm 0.0022$ & 122,346 &  1,251 \\
~~$tb\to\tau$+jets  & $0.88 \pm 0.14$ & $0.1111 \pm 0.0022$ &  76,433 &    782 \\
~~$tqb\to e$+jets   & $1.98 \pm 0.30$ & $0.1111 \pm 0.0022$ & 130,068 &    591 \\
~~$tqb\to\mu$+jets  & $1.98 \pm 0.30$ & $0.1111 \pm 0.0022$ & 137,824 &    626 \\
~~$tqb\to\tau$+jets & $1.98 \pm 0.30$ & $0.1111 \pm 0.0022$ & 117,079 &    532 \\
~~{\bf{Signal total}}  & $\mathbf{2.86 \pm 0.45}$ & $\mathbf{0.3333 \pm 0.0067}$ & {\bf{676,370}} &  \\
{\bf Backgrounds}              &               &                     &           &     \\
~~${\ttbar}\to \ell$+jets   & $6.8 \pm 1.2$ & $0.4444 \pm 0.0089$ &   474,405 & 157 \\
~~${\ttbar}\to \ell\ell$    & $6.8 \pm 1.2$ & $0.1111 \pm 0.0089$ &   468,126 & 620 \\
~~{\bf{Top pairs total~~~~~~}}  & $\mathbf{6.8 \pm 1.2}$ & $\mathbf{0.5555 \pm 0.0111}$ & {\bf{942,531}} &  \\
~~$W{\bbbar}\to\ell\nu bb$  & $142$         & $0.3333 \pm 0.0066$ & 1,335,146 &  28 \\
~~$W{\ccbar}\to\ell\nu cc$  & $583$         & $0.3333 \pm 0.0066$ & 1,522,767 &   8 \\
~~$Wjj\to\ell\nu jj$        & $18,734$      & $0.3333 \pm 0.0066$ & 8,201,446 &   1 \\
~~{$\mathbf{W}${\bf+jets total}}  & $\mathbf{19,459}$   & $\mathbf{0.3333 \pm 0.0067}$ & {\bf{11,059,359}} & 
\end{tabular}
\end{ruledtabular}
\end{minipage}
\end{center}
\end{table*}

\vspace{-0.05in}
\subsection{Correction Factors}

\begin{figure*}[!h!btp]
\includegraphics[width=0.4\textwidth]
{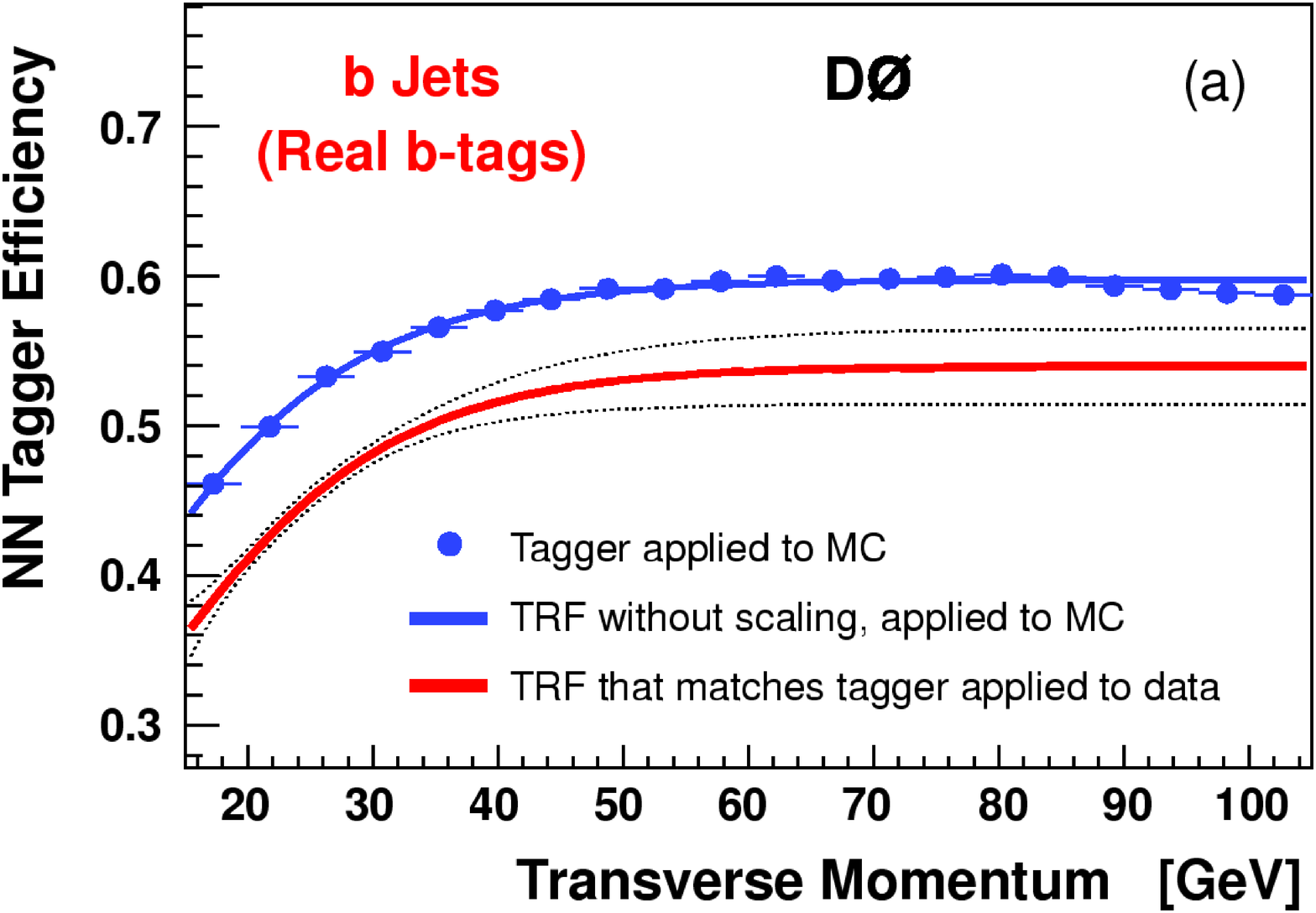}
\includegraphics[width=0.4\textwidth]
{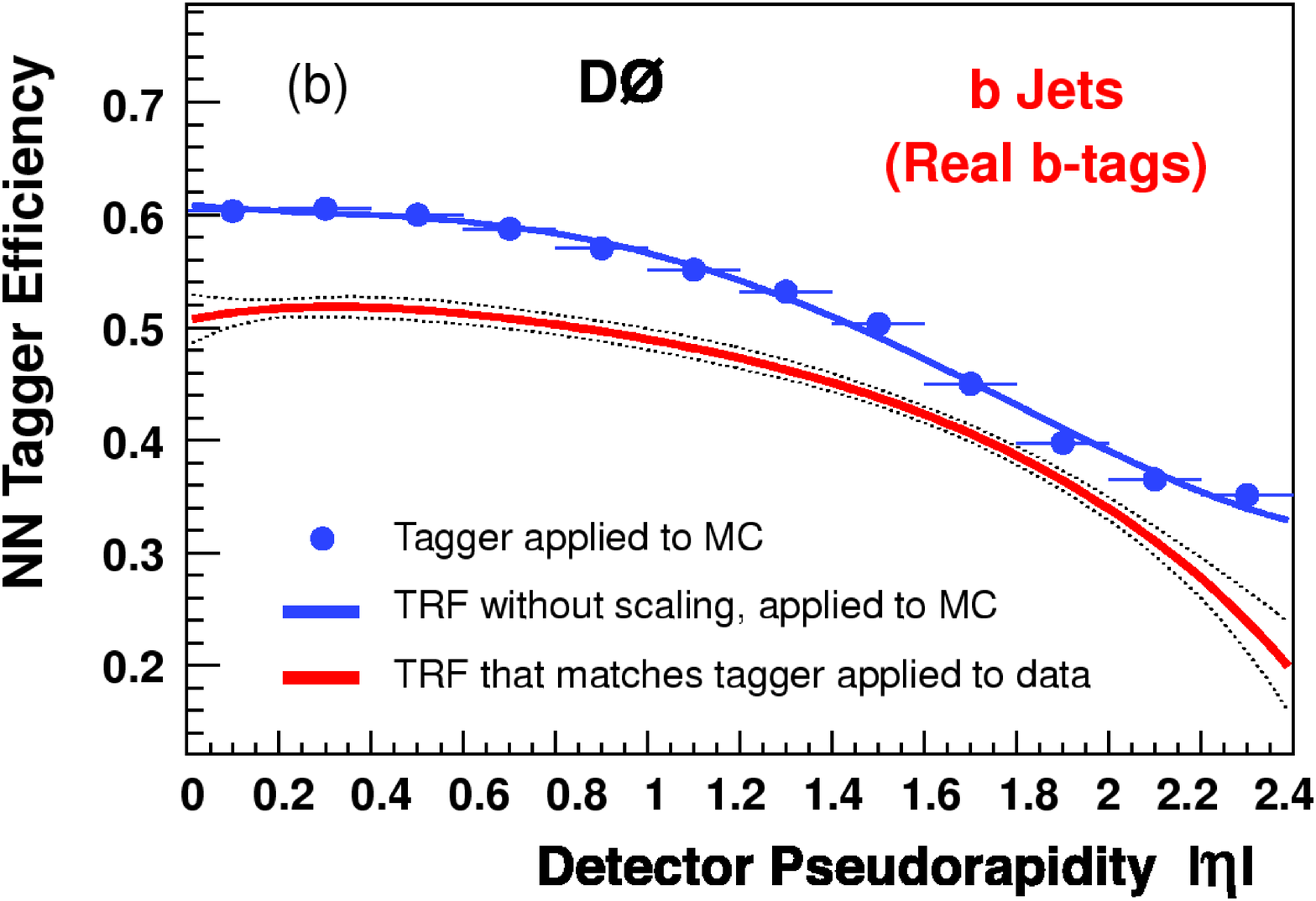}

\includegraphics[width=0.4\textwidth]
{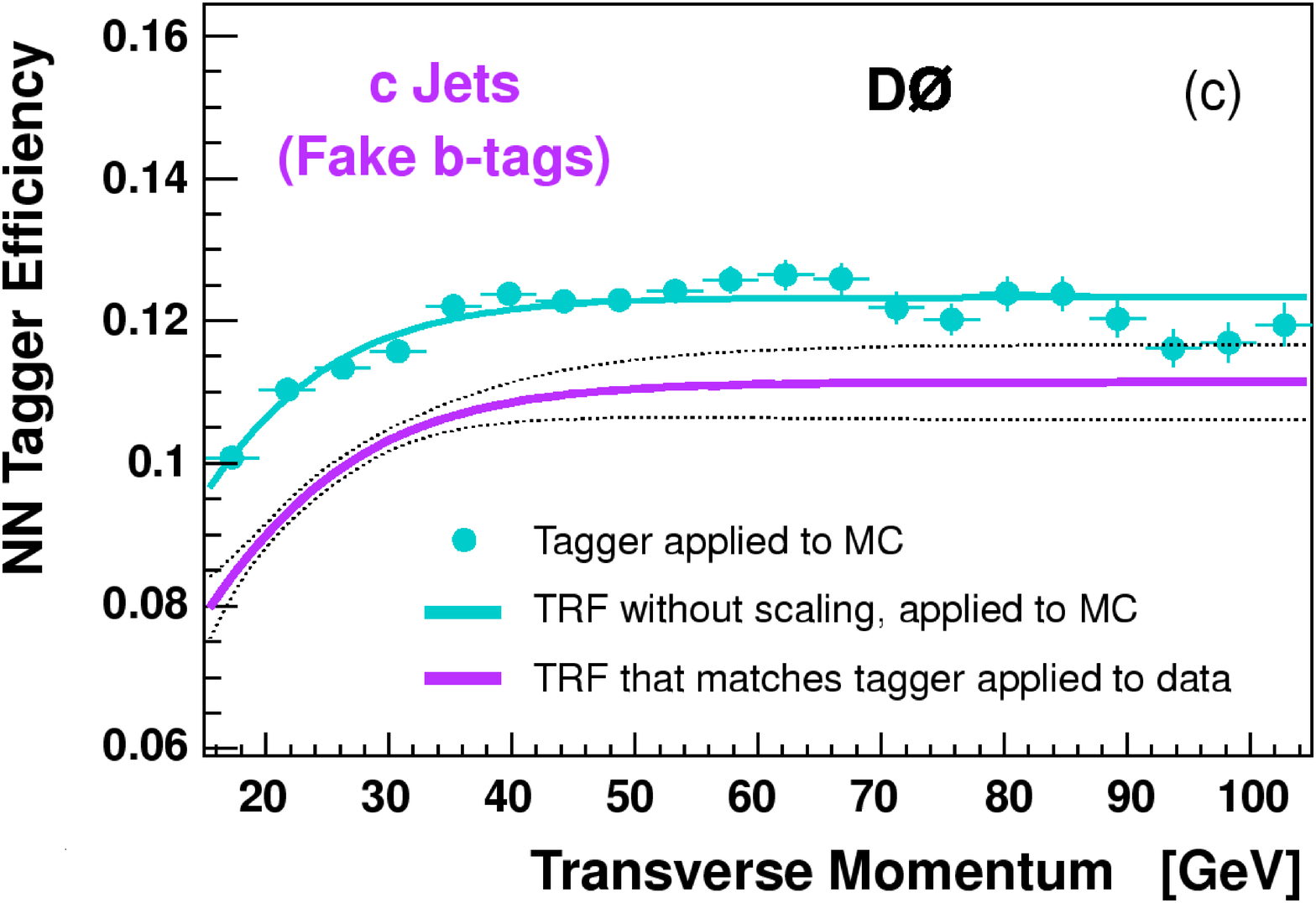}
\includegraphics[width=0.4\textwidth]
{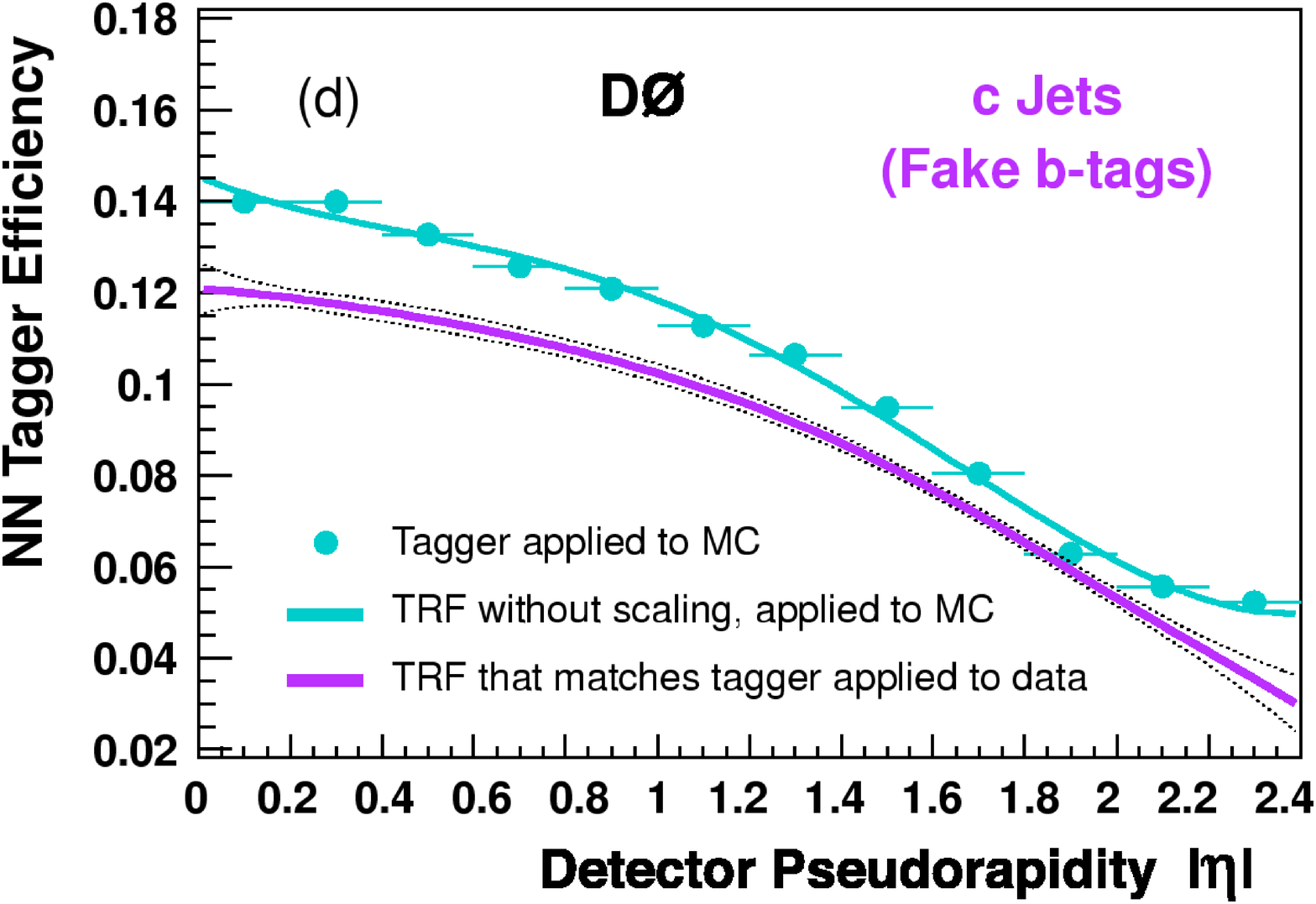}

\includegraphics[width=0.4\textwidth]
{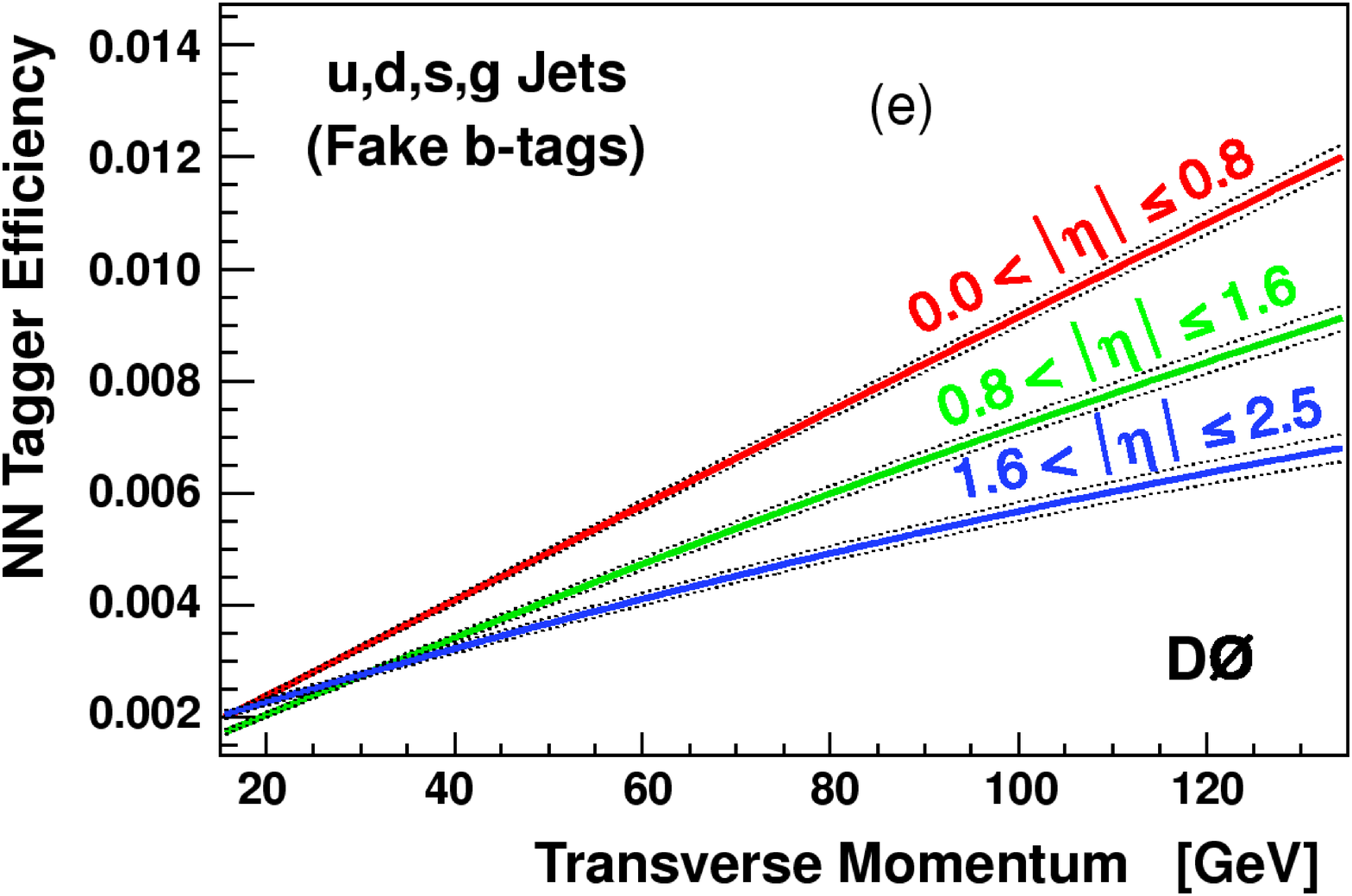}

\vspace{-0.1in}
\caption[tagratefuncs]{The tag-rate functions (TRFs) used to weight
the MC events according to the probability that they should be
$b$~tagged. In plots (a)--(d), the points show the neural network
$b$~tagging algorithm (the ``tagger'') applied directly to the MC
events. The upper line that passes through the points is the result of
the tag-rate functions, before scaling-to-data, being applied to the
MC events to reproduce the result from the tagger. The lower line,
with dotted error band, shows the tag-rate functions after they have
been scaled to match the efficiency of the NN $b$~tagging algorithm
applied to data. In plot (e), the lines show the (scaled) tag-rate
functions that are applied to MC events.}
\label{tag-rate-functions}
\end{figure*}

We pass the simulated events through a {\geant}-based
model~\cite{geant} of the D0 detector. The simulated samples
then have correction factors applied to ensure that the reconstruction
and selection efficiencies match those found in data. Generally the
efficiency to reconstruct, identify, and select objects in the
simulated samples is higher than in data, so the following scale
factors are used to correct for that difference:

\begin{myitemize}
\item {\bf Trigger efficiency correction factors}\\
The probability for each simulated event to fire the triggers detailed
in Sec.~\ref{triggers-data} is calculated as a weight applied to each
object measured in the event. Electron and jet efficiencies, for all
levels of the trigger architecture, are parametrized as functions of
$p_T$ and $\deta$. Muon efficiencies are parametrized as functions of
$\deta$ and $\phi$. These corrections are measured using data obtained
with triggers different from those used in this search to avoid
biases.

\item {\bf Electron identification efficiency correction factors}\\
We correct each simulated event in the electron channel with a factor
that accounts for the differences in electron cluster finding
identification, $f_{\rm EM}$, and isolation efficiencies in the
simulation and data. This correction factor is measured in $Z{\rar}ee$
data and simulated events, and parametrized as a function of
$\deta$. A second scale factor is applied to account for the
differences between the data and the simulation in the $\chi^2_{\rm
cal}$, track matching, and EM-likelihood efficiencies. This second
scale factor is also derived from $Z{\rar}ee$ data and simulated
events and parametrized as a function of $\deta$ and $\phi^{\rm det}$.

\item {\bf Muon identification and isolation efficiency correction factors}\\
We correct each simulated event in the muon channel for the muon
identification, track match, and isolation efficiencies. The
identification correction factor is parametrized as a function of
$\deta$ and $\phi$, track match as a function of track-$z$ and
$\deta$, and isolation as a function of the number of jets in the
event. These corrections are measured in $Z{\rar}\mu\mu$ data and
simulated events.

\item {\bf Jet reconstruction efficiency and energy resolution correction factors}\\
Simulated jets need to be corrected for differences in the
reconstruction and identification efficiency and for the worse energy
resolution found in data than in the simulation. The jet energy scale
correction is applied to the simulation as in the data, but then
simulated jets are corrected for the jet reconstruction efficiency and
smeared to match the jet energy resolution found in back-to-back
photon+jet events.

\item {\bf Taggability and $\mathbi{b}$-tagging efficiency correction
factors}\\ In data, the taggability and $b$-tagging requirements are
applied directly, as described in Sec.~\ref{b-tagging}. For simulated
samples, \emph{taggability-rate functions} and \emph{tag-rate
functions} are applied instead of the direct selection because the
modeling of the detector is not sufficiently accurate. The
taggability-rate function is parametrized in jet $p_T$, $\eta$, and
primary vertex $z$, and is measured in the selected data sample
(Sec.~\ref{event-selection}) with one loose-isolated lepton. We check
that the efficiency is the same as in the data sample with one
tight-isolated lepton within the uncertainties. The average
taggability for central high-$p_T$ jets is around $90\%$.

The $b$-jet efficiency correction is measured in data using a
muon-in-jet sample and a $b$-jet enriched subset where one jet is
required to have a small JLIP value, and in an admixture of $Z{\rar}
{\bbbar}$ and ${\ttbar}$ simulated events where the $b$-jets are
required to contain a muon. The $b$-tag efficiency correction for
$c$-quark jets is derived in a combined MC sample with $Z$~boson,
multijets, and ${\ttbar}$ decays to $c$ quarks, and assuming that the
MC-to-data scale factor is the same as for the $b$-jet efficiency. The
$b$-tag efficiency correction for light jets is derived from multijet
data. All these $b$-tagging corrections are parametrized as functions
of the jet $p_T$ and ${\eta}$. Figure~\ref{tag-rate-functions}
illustrates the tag-rate functions used in this analysis.
\end{myitemize}

\clearpage

%---------------------------------------------------------------------
%---------------------------------------------------------------------
\section{Event Selection}
\label{event-selection}

\subsection{Selection Requirements}
\label{selection-cuts}

We apply a loose event selection to find $W$-like events containing an
isolated lepton, missing transverse energy, and two to four jets with
high transverse momentum. The samples after this selection, which we
call ``pretagged,'' (i.e., \emph{before} tagging has been applied),
are dominated by $W$+jets events, with some {\ttbar} contribution that
becomes more significant for higher jet multiplicities. The final
selection improves the signal-to-background ratio significantly by
requiring the presence of one or two $b$-tagged jets.

\vspace{0.15in}
{\bf Common selections} for both $e$ and $\mu$ channels
\vspace{-0.08in}
\begin{myitemize}
\item Good quality (for data with all subdetectors working properly)
\item Pass trigger: offline electrons and muons in the data are
matched to the object that fired the appropriate trigger for that
run period
\item Good primary vertex: $|z_{\mathrm{PV}}| < 60$~cm with at least
      three tracks attached
\item Missing transverse energy: $15 < {\met} < 200$~GeV
\item Two, three, or four jets with $p_T > 15$~GeV and
      ${\meta} < 3.4$
\item Leading jet $p_T > 25$~GeV and
      ${\meta} < 2.5$
\item Second leading jet $p_T > 20$~GeV
\item Jet triangle cut  $|\Delta\phi(\mathrm{leading~jet,~{\met}})|
      \mathrm{~vs.~}{\met}$ (see Fig.~8 in Ref.~\cite{run2-d0-prd}
      for a pictorial view of these cuts):

     $|\Delta\phi|<1.5+(\pi-1.5)${\met}(GeV)$/35$~rad
\item One or two $b$-tagged jets
\end{myitemize}

{\bf Electron channel selection}
\vspace{-0.07in}
\begin{myitemize}
\item Only one tight electron with $p_T > 15$~GeV and ${\mdeta} < 1.1$
\item No tight muon with $p_T > 18$~GeV and ${\mdeta} < 2.0$ 
\item No second loose electron with $p_T > 15$~GeV and any ${\deta}$
\item Electron coming from the primary vertex:
      $|\Delta z(e,\mathrm{PV})| < 1$~cm
\item Electron triangle cuts $|\Delta\phi(e,{\met})|\mathrm{~vs.~}
      {\met}$ (see Fig.~8 in Ref.~\cite{run2-d0-prd}):
   \vspace{-0.07in} 
   \begin{enumerate}
   \item $|\Delta\phi(e,{\met})|>2-2${\met}(GeV)$/40$~rad	
   %\item from 2 to 0~rad when ${\met} = 0$~GeV, and {\met} from 0
   %to 40~GeV when $|\Delta\phi| = 0$~rad
   \item $|\Delta\phi(e,{\met})|>1.5-1.5${\met}(GeV)$/50$~rad
   %\item from 1.5 to 0~rad when ${\met} = 0$~GeV, and {\met} from 0
   %to 50~GeV when $|\Delta\phi| = 0$~rad
   \item $|\Delta\phi(e,{\met})|<2+(\pi-2)${\met}(GeV)$/24$~rad
   %\item  from 2 to $\pi$~rad when ${\met} = 0$~GeV, and {\met}
   %from 0 to 24~GeV when $|\Delta\phi| = \pi$~rad
   \end{enumerate}
\end{myitemize}

{\bf Muon channel selection}
\vspace{-0.07in}
\begin{myitemize}
\item Only one tight muon with $p_T > 18$~GeV and ${\mdeta} < 2.0$ 
\item No tight electron with $p_T > 15$~GeV and ${\mdeta} < 2.5$
\item Muon coming from the primary vertex:
      $|\Delta z(\mu,\mathrm{PV})| < 1$~cm
\item Muon triangle cuts $|\Delta\phi(\mu, {\met})|\mathrm{~vs.~}
      {\met}$ (see Fig.~8 in Ref.~\cite{run2-d0-prd}): 
   \vspace{-0.07in}
   \begin{enumerate}
   \item  $|\Delta\phi(\mu, {\met})|>1.1-1.1${\met}(GeV)$/80$~rad
   %\item  from 1.1 to 0~rad when ${\met} = 0$~GeV, and {\met}
   %from 0 to 80~GeV when $|\Delta\phi| = 0$~rad
   \item  $|\Delta\phi(\mu, {\met})|>1.5-1.5${\met}(GeV)$/50$~rad	
   %\item  from 1.5 to 0~rad when ${\met} = 0$~GeV, and {\met}
   %from 0 to 50~GeV when $|\Delta\phi| = 0$~rad
   \item  $|\Delta\phi(\mu, {\met})|<2.5+(\pi-2.5)${\met}(GeV)$/30$~rad	
   %\item from 2.5 to $\pi$~rad when ${\met} = 0$~GeV, and {\met}
   %from 0 to 30~GeV when $|\Delta\phi| = \pi$~rad
   \end{enumerate}
\end{myitemize}

Some of the selection criteria listed above are designed to remove
areas of the data that are difficult to model. In particular, the
upper {\met} selection gets rid of a few events where the muon $p_T$
fluctuated to a large value. The ``triangle cuts'' are very efficient
in removing multijet events where a misreconstructed jet creates fake
missing energy aligned or anti-aligned in azimuth with the lepton or
jet.

\vspace{0.10in}
{\bf Background-data selection} for measuring the multijet background
\vspace{-0.07in}
\begin{myitemize}
\item All the same selection criteria as listed above except for the
      tight lepton requirements
\item Electron channel --- only one loose-but-not-tight electron
\item Muon channel --- only one loose-but-not-tight muon
\end{myitemize}
\vspace{-0.07in}
The definitions of loose and tight electrons and muons are in
Secs.~\ref{electron-id} and \ref{muon-id}.

\subsection{Numbers of Events After Selection}
\label{event-numbers}

Table~\ref{numbers-of-events} shows the numbers of events in the
signal and background samples and in the data after applying the
selection criteria. Note that these numbers are just counts of events
used later in the analysis, and not signal or background yields after
normalizations and corrections have been applied.

\begin{table*}[!h!tbp]
\begin{center}
\begin{minipage}{6.5in}
\caption[numbersofevents]{Numbers of events for the electron and muon
channels after selection. The MC samples include events coming from
$\tau$ decays, $\tau\to\ell\nu$ where $\ell=e$ in the electron channel
and $\ell=\mu$ in the muon channel.}
\label{numbers-of-events}
\begin{ruledtabular}
\begin{tabular}{l||ccccc|ccccc}
\multicolumn{11}{c}{\hspace{1in}\underline{Numbers of Events After Selection}} \vspace{0.05in} \\
& \multicolumn{5}{c|}{Electron Channel} & \multicolumn{5}{c}{Muon Channel} \\
                            & 1 jet & 2 jets & 3 jets & 4 jets & $\ge5$ jets
                            & 1 jet & 2 jets & 3 jets & 4 jets & $5$ jets \\
\hline			                   
{\bf Signal MC}               &         &        &        &        &        &         &        &        &        &       \\
~~$tb$                      &   6,908 & 19,465 &  9,127 &  2,483 &   595  &   3,878 & 12,852 &  6,458 &  1,809 &   401 \\
~~$tqb$                     &   8,971 & 22,758 & 12,080 &  3,797 &  1,092 &   8,195 & 21,066 & 11,193 &  3,489 &   835 \\
{\bf Background MC}~~          &         &        &        &        &        &         &        &        &        &       \\
~~${\ttbar}{\rar}\ell\ell$  &   7,671 & 29,537 & 26,042 & 12,068 &  5,396 &   5,509 & 24,595 & 21,803 &  9,788 & 3,442 \\
~~${\ttbar}{\rar}\ell$+jets &     522 &  5,659 & 22,477 & 27,319 & 14,298 &     232 &  3,376 & 16,293 & 22,680 & 8,658 \\ 
~~$Wb\bar{b}$               &  26,611 & 13,914 &  9,011 &  3,848 &  1,434 &  27,764 & 14,488 &  9,427 &  3,874 & 1,204 \\
~~$Wc\bar{c}$               &  21,765 & 13,453 &  7,562 &  2,252 &    591 &  32,712 & 19,047 & 10,141 &  3,051 &   663 \\
~~$Wjj$                     & 134,660 & 61,497 & 34,162 &  8,290 &  1,750 & 147,842 & 66,201 & 36,673 &  9,169 & 1,502 \\
{\bf Pretag data}              &         &        &        &        &        &         &        &        &        &       \\
~~Multijets                 &  11,565 &  6,993 &  4,043 &  1,317 &    431 &     897 &    658 &    462 &    151 &    48 \\
~~Signal data               &  27,370 &  8,220 &  3,075 &    874 &    223 &  17,816 &  6,432 &  2,590 &    727 &   173 \\
{\bf One-tag data}             &         &        &        &        &        &         &        &        &        &       \\
~~Multijets                 &     246 &    322 &    226 &     93 &     34 &      31 &     51 &     49 &     21 &     8 \\
~~Signal data               &     445 &    357 &    207 &     97 &     35 &     289 &    287 &    179 &    100 &    38 \\
{\bf Two-tags data}            &         &        &        &        &        &         &        &        &        &       \\
~~Multijets                 &     --- &     12 &     15 &     14 &      7 &     --- &      3 &      4 &      1 &     4 \\
~~Signal data               &     --- &     30 &     37 &     22 &     10 &     --- &     23 &     32 &     27 &    10 
\end{tabular}
\end{ruledtabular} 
\end{minipage}
\end{center}
\end{table*}

%---------------------------------------------------------------------
%---------------------------------------------------------------------
\section{Background Model}
\label{background}

\subsection{$\mathbf{W}$+Jets and Multijets Backgrounds}
\label{matrix-method}

The $W$+jets background is modeled using the parton-jet matched
{\alpgen} simulated samples described in Sec.~\ref{mc-samples}. This
background is normalized to data before $b$~tagging, using a procedure
explained below. Because we normalize to data and do not use theory
cross sections, small components of the total background from $Z$+jets
and diboson processes ($WW$, $WZ$, and $ZZ$, which amount to less than
4\% of the total background expectation after tagging) are implicitly
included in the $W$+jets part of the background model. This
simplification does not affect the final results because of the low
rate from these processes in the final selected dataset, and because
the kinematics of the events are similar to those in $W$+jets
events. They are thus identified together with $W$+jets events by the
multivariate discriminants.

The multijet background is modeled using datasets that contain
misidentified leptons, as described at the end of
Sec.~\ref{selection-cuts}. These datasets provide the shape for the
multijet background component in each analysis channel. They are
normalized to data as part of the $W$+jets normalization process.

We normalize the $W$+jets and multijet backgrounds to data before
tagging using the matrix method~\cite{mm-explained}, which lets us
estimate how many events in the pretagged samples contain a
misidentified lepton (originating from multijet production) and how
many events have a real isolated lepton (originating from $W$+jets or
$\ttbar$). Two data samples are defined, the \emph{tight} sample,
which is the signal sample after all selection cuts have been applied,
and the \emph{loose} sample, where the same selection has been applied
but requiring only loose lepton quality. The tight data sample, with
$N_{{\rm tight}}$ events, is a subset of the loose data sample with
$N_{\rm loose}$ events. The loose sample contains $N_{\rm loose}^{\rm
real{\mbox{\small -}}\ell}$ events with a real lepton (signal-like
events, mostly $W$+jets and $\ttbar$) and $N_{\rm loose}^{\rm
fake{\mbox{\small -}}\ell}$ fake lepton events, which is the number of
multijet events in the loose sample.

We measure the probability $\varepsilon_{\rm real{\mbox{\small
-}}\ell}$ for a real isolated lepton to pass the tight lepton
selection in $Z\to\ell\ell$ data events. The probability for a
fake-isolated lepton to pass the tight-isolated lepton criteria,
$\varepsilon_{\rm fake{\mbox{\small -}}\ell}$, is measured in a sample
enriched in multijet events with the same selection as the signal data
but requiring $\met<10$~GeV. In the electron channel, these
probabilities are parametrized as $\varepsilon_{\rm real{\mbox{\small
-}}e}(p_T,\eta)$ and $\varepsilon_{\rm fake{\mbox{\small -}}e}(N_{\rm
jets},{\rm trigger~period})$. In the muon channel, they are
parametrized as $\varepsilon_{\rm real{\mbox{\small -}}\mu}(N_{\rm
jets},p_T)$ and $\varepsilon_{\rm fake{\mbox{\small -}}\mu}(\eta)$.
With these definitions, the matrix method is applied using the
following two equations:
\begin{eqnarray}
N_{\rm loose}  & = & N_{\rm loose}^{\rm{fake{\mbox{\small -}}\ell}}
                   + N_{\rm loose}^{{\rm real{\mbox{\small -}}\ell}}\\
N_{{\rm tight}}& = & N_{{\rm tight}}^{{\rm fake{\mbox{\small -}}\ell}}
                   + N_{{\rm tight}}^{{\rm real{\mbox{\small -}}\ell}}
                                                          \nonumber \\
               & = & \varepsilon_{{\rm fake{\mbox{\small -}}\ell}} \;
                     N_{\rm loose}^{{\rm fake{\mbox{\small -}}\ell}}
                   + \varepsilon_{{\rm real{\mbox{\small -}}\ell}} \;
                     N_{\rm loose}^{{\rm real{\mbox{\small -}}\ell}},
\end{eqnarray}
and solving for $N_{\rm loose}^{\rm fake{\mbox{\small -}}\ell}$ and
$N_{\rm loose}^{\rm real{\mbox{\small -}}\ell}$ so that the multijet
and the $W$-like contributions in the tight sample $N_{\rm tight}^{\rm
fake{\mbox{\small -}}\ell}$ and $N_{\rm tight}^{\rm real{\mbox{\small
-}}\ell}$ can be determined.

\begin{figure}[!h!btp]
\includegraphics[width=0.38\textwidth]
{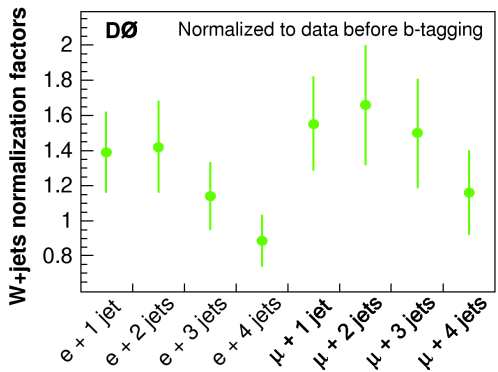}
\vspace{-0.1in}
\caption[mmfactors]{The factors used to normalize the $W$+jets
background model to pretagged data in each analysis channel.}
\label{wjets-mm-factors}
\end{figure}

The results of the matrix method normalization, which we apply
separately in each jet multiplicity bin, are shown in
Table~\ref{mm-numbers}. The values shown for $\varepsilon_{\rm
real{\mbox{\small -}}\ell}$ and $\varepsilon_{\rm fake{\mbox{\small
-}}\ell}$ are averages for illustration only. The pretagged
background-data sample is scaled to $N_{\rm tight}^{\rm
fake{\mbox{\small -}}\ell}$, and the $W$+jets simulated samples
($Wb\bar{b}$+$Wc\bar{c}$+$Wjj$) are scaled to $N_{\rm tight}^{\rm
real{\mbox{\small -}}\ell}$, after subtracting the expected number of
$\ttbar$ events in each jet multiplicity bin of the tight
sample. These normalization factors are illustrated in
Fig.~\ref{wjets-mm-factors}.

\begin{table*}[!h!tbp]
\begin{center}
\begin{minipage}{6.5in}
\caption[mmnumbers]{Matrix method normalization values in the electron
and muon channels for the loose and tight selected samples, and the
expected contribution from multijet and $W$-like events.}
\label{mm-numbers}
\begin{ruledtabular}
\begin{tabular}{l||ccccc|ccccc}
\multicolumn{11}{c}{\hspace{1in}\underline{Normalization of $W$+Jets and Multijets Backgrounds to Data}}\vspace{0.05in} \\
& \multicolumn{5}{c|}{Electron Channel} & \multicolumn{5}{c}{Muon Channel}    \\
                               & 1 jet  & 2 jets & 3 jets & 4 jets & $\ge5$ jets
                               & 1 jet  & 2 jets & 3 jets & 4 jets & $5$ jets \\
\hline
$N_{{\rm loose}}$              & 38,935 & 15,213 &  7,118 &  2,191 &   654  & 18,714 &  7,092 &  3,054 &   878  &   221 \\
$N_{{\rm tight}}$              & 27,370 &  8,220 &  3,075 &    874 &   223  & 17,816 &  6,432 &  2,590 &   727  &   173 \\
$\varepsilon_{\rm real{\mbox{\small -}}\ell}$    &  0.873 &  0.874 &  0.874 &  0.875 & 0.875  &  0.991 &  0.989 &  0.987 & 0.961  & 0.878 \\
$\varepsilon_{\rm fake{\mbox{\small -}}\ell}$    &  0.177 &  0.193 &  0.188 &  0.173 & 0.173  &  0.408 &  0.358 &  0.342 & 0.309  & 0.253 \\
$N_{\rm tight}^{\rm fake{\mbox{\small -}}\ell}$~~&  1,691 &  1,433 &    860 &    256 &    86  &    498 &    329 &    223 &    56  &    10 \\
$N_{\rm tight}^{\rm real{\mbox{\small -}}\ell}$  & 25,679 &  6,787 &  2,215 &    618 &   137  & 17,319 &  6,105 &  2,369 &   669  &   162 
\end{tabular}
\end{ruledtabular}
\end{minipage}
\end{center}
\end{table*}

\subsection{Top-Quark Pairs Background}
\label{ttbar-background}

Background from the {\ttbar} process is modeled using the parton-jet
matched {\alpgen} simulated samples described in
Sec.~\ref{mc-samples}. These events are normalized to the theoretical
cross section~\cite{ttbar-xsec-kidonakis} at $m_{\rm top} = 175$~GeV
(chosen to match the value used to generate the samples), which is
6.8~pb.

%---------------------------------------------------------------------
%---------------------------------------------------------------------
\section{Signal Acceptances}
\label{signal-acceptances}

Table~\ref{acceptances} shows the percentage of each signal that
remains after selection. We achieve roughly 30\% higher acceptances in
this analysis compared to our previously published
analysis~\cite{run2-d0-plb,run2-d0-prd} from the use of the more
efficient neural network $b$-tagging algorithm. The total acceptance
for the s-channel $tb$ process is $(3.2 \pm 0.4)\%$ and for the
t-channel $tqb$ process it is $(2.1 \pm 0.3)\%$.

\begin{table*}[!h!btp]
\begin{center}
\begin{minipage}{6.5in}
\caption[acceptances]{Signal acceptances after selection.}
\label{acceptances}
\begin{ruledtabular}
\begin{tabular}{l||ccccc|ccccc}
\multicolumn{11}{c}{\hspace{1in}\underline{Signal Acceptances}} \vspace{0.05in} \\
& \multicolumn{5}{c|}{Electron Channel} & \multicolumn{5}{c}{Muon Channel} \\
                     & 1 jet & 2 jets & 3 jets & 4 jets & $\ge5$ jets
                     & 1 jet & 2 jets & 3 jets & 4 jets & $5$ jets \\
\hline
{\bf Before tagging}~~  &        &        &        &        &        &        &        &        &        &        \\
~~$tb$               & 0.55\% & 1.77\% & 0.83\% & 0.23\% & 0.06\% & 0.33\% & 1.36\% & 0.69\% & 0.19\% & 0.05\% \\
~~$tqb$              & 0.52\% & 1.49\% & 0.79\% & 0.25\% & 0.07\% & 0.36\% & 1.17\% & 0.64\% & 0.20\% & 0.05\% \\
{\bf One-tag}          &        &        &        &        &        &        &        &        &        &        \\
~~$tb$               & 0.24\% & 0.82\% & 0.39\% & 0.11\% & 0.03\% & 0.15\% & 0.64\% & 0.32\% & 0.09\% & 0.02\% \\
~~$tqb$              & 0.18\% & 0.61\% & 0.34\% & 0.11\% & 0.03\% & 0.13\% & 0.50\% & 0.28\% & 0.09\% & 0.02\% \\
{\bf Two-tags}         &        &        &        &        &        &        &        &        &        &        \\
~~$tb$               &  ---   & 0.29\% & 0.14\% & 0.04\% & 0.02\% &  ---   & 0.24\% & 0.12\% & 0.03\% & 0.01\% \\
~~$tqb$              &  ---   & 0.02\% & 0.05\% & 0.02\% & 0.01\% &  ---   & 0.01\% & 0.04\% & 0.02\% & 0.01\%
\end{tabular}
\end{ruledtabular}
\end{minipage}
\end{center}
\end{table*}

%---------------------------------------------------------------------
%---------------------------------------------------------------------
\section{Event Yields}
\label{event-yields}

We use the term ``yield'' to mean the number of events of the signal
or background in question predicted to be in the 0.9~fb$^{-1}$ of
data analyzed here. Tables~\ref{pretag-yields}, \ref{onetag-yields},
and \ref{twotag-yields} show these yields for all signals and
backgrounds separated by lepton flavor and jet multiplicity within
each table, and by the numbers of $b$-tagged jets between the
tables. Because the $W$+jets and multijet backgrounds are normalized
to data before tagging, the sum of the backgrounds is constrained to
equal the number of events observed in the data, as seen in the first
table. The yield values shown in these tables have been rounded to
integers for clarity, so that the sums of the components will not
always equal exactly the values given for these sums. All calculations
however have been done with full-precision values.

Only events with two, three and four jets are used in this analysis,
but we show the acceptances and the yields for events with one and for
five or more jets in these tables to demonstrate the consistency of
the analysis in those channels.  Tables~\ref{onetag-yields} and
\ref{twotag-yields} show that most of the signal is contained in the
two and three jet bins. However, as discussed in
Sec.~\ref{sec:expected-sensitivity}, our maximum predicted sensitivity
is obtained by including events with 2--4 jets.
 
\begin{table*}[hbtp]
\begin{center}
\begin{minipage}{5.8in}
\caption[pretagyields]{Predicted yields after selection and before
$b$~tagging.}
\label{pretag-yields}
\begin{ruledtabular}
\begin{tabular}{l||ccccc|ccccc}
\multicolumn{11}{c}{\hspace{1in}\underline{Yields Before $b$-Tagging}} \vspace{0.05in} \\
& \multicolumn{5}{c|}{Electron Channel} & \multicolumn{5}{c}{Muon Channel} \\
                            & 1 jet & 2 jets & 3 jets & 4 jets & $\ge5$ jets
                            & 1 jet & 2 jets & 3 jets & 4 jets & $5$ jets  \\
\hline
{\bf Signals}                 &        &       &       &       &      &        &       &       &      &      \\
~~$tb$                      &      4 &    14 &     7 &     2 &    0 &      3 &    10 &     5 &    1 &    0 \\
~~$tqb$                     &      9 &    27 &    14 &     5 &    1 &      6 &    20 &    11 &    3 &    1 \\
{\bf Backgrounds}              &        &       &       &       &      &        &       &       &      &      \\
~~${\ttbar}{\rar}\ell\ell$  &      9 &    35 &    28 &    10 &    4 &      5 &    27 &    22 &    8 &    3 \\
~~${\ttbar}{\rar}\ell$+jets &      2 &    26 &   103 &   128 &   67 &      1 &    14 &    71 &   99 &   43 \\
~~$Wb\bar{b}$               &    659 &   358 &   149 &    42 &    5 &    431 &   312 &   161 &   47 &   10 \\
~~$Wc\bar{c}$               &  1,592 &   931 &   389 &    93 &   10 &  1,405 & 1,028 &   523 &  131 &   21 \\
~~$Wjj$                     & 23,417 & 5,437 & 1,546 &   343 &   51 & 15,476 & 4,723 & 1,591 &  385 &   85 \\
~~Multijets                 &  1,691 & 1,433 &   860 &   256 &   86 &    498 &   329 &   223 &   58 &   10 \\
\hline
{\bf Background Sum}~~         & 27,370 & 8,220 & 3,075 &   874 &  223 & 17,816 & 6,434 & 2,592 &  727 &  172 \\
\hline
{\bf Data}                   & 27,370 & 8,220 & 3,075 &   874 &  223 & 17,816 & 6,432 & 2,590 &  727 &  173
\end{tabular}
\end{ruledtabular}
\end{minipage}
\end{center}
\end{table*}

\begin{table*}[htbp]
\begin{center}
\begin{minipage}{5.8in}
\caption[onetagyields]{Predicted yields after selection for events
with exactly one $b$-tagged jet.}
\label{onetag-yields}
\begin{ruledtabular}
\begin{tabular}{l||ccccc|ccccc}
\multicolumn{11}{c}{\hspace{1in}\underline{Yields With One $b$-Tagged Jet}} \vspace{0.05in} \\
& \multicolumn{5}{c|}{Electron Channel} & \multicolumn{5}{c}{Muon Channel} \\
                            & 1 jet & 2 jets & 3 jets & 4 jets & $\ge5$ jets
                            & 1 jet & 2 jets & 3 jets & 4 jets & $5$ jets  \\
\hline
{\bf Signals}                 &      &       &      &      &      &      &      &      &      &      \\
~~$tb$                      &    2 &     7 &    3 &    1 &    0 &    1 &    5 &    2 &    1 &    0 \\
~~$tqb$                     &    3 &    11 &    6 &    2 &    1 &    2 &    9 &    5 &    2 &    0 \\
{\bf Backgrounds}              &      &       &      &      &      &      &      &      &      &      \\
~~${\ttbar}{\rar}\ell\ell$  &    4 &    16 &   13 &    5 &    2 &    2 &   13 &   10 &    4 &    1 \\
~~${\ttbar}{\rar}\ell$+jets &    1 &    11 &   47 &   58 &   30 &    0 &    6 &   32 &   45 &   20 \\
~~$Wb\bar{b}$               &  188 &   120 &   50 &   14 &    2 &  131 &  110 &   56 &   16 &    4 \\
~~$Wc\bar{c}$               &   81 &    74 &   36 &    9 &    1 &   64 &   74 &   46 &   13 &    2 \\
~~$Wjj$                     &  175 &    61 &   20 &    5 &    1 &  125 &   58 &   23 &    6 &    2 \\
~~Multijets                 &   36 &    66 &   48 &   18 &    7 &   17 &   26 &   24 &    8 &    2 \\
\hline
{\bf Background Sum}~~         &  484 &   348 &  213 &  110 &   43 &  340 &  286 &  191 &   93 &   30 \\
\hline
{\bf Data}                   &  445 &   357 &  207 &   97 &   35 &  289 &  287 &  179 &  100 &   38
\end{tabular}
\end{ruledtabular}
\end{minipage}
\end{center}
\end{table*}

\begin{table*}[hbtp]
\begin{center}
\begin{minipage}{5.8in}
\caption[twotagyields]{Predicted yields after selection for events
with exactly two $b$-tagged jets.}
\label{twotag-yields}
\begin{ruledtabular}
\begin{tabular}{l||ccccc|ccccc}
\multicolumn{11}{c}{\hspace{1in}\underline{Yields With Two $b$-Tagged Jets}} \vspace{0.05in} \\
& \multicolumn{5}{c|}{Electron Channel} & \multicolumn{5}{c}{Muon Channel} \\
                            & 1 jet & 2 jets & 3 jets & 4 jets & $\ge5$ jets
                            & 1 jet & 2 jets & 3 jets & 4 jets & $5$ jets  \\
\hline
{\bf Signals}                 &      &       &       &       &      &      &       &       &       &       \\
~~$tb$                      &  --- &   2.3 &   1.1 &   0.3 &  0.1 &  --- &   1.9 &   0.9 &   0.3 &  0.1  \\
~~$tqb$                     &  --- &   0.3 &   0.8 &   0.4 &  0.2 &  --- &   0.2 &   0.7 &   0.4 &  0.1  \\
{\bf Backgrounds}              &      &       &       &       &      &      &       &       &       &       \\
~~${\ttbar}{\rar}\ell\ell$  &  --- &   5.5 &   4.6 &   1.7 &  0.7 &  --- &   4.6 &   3.8 &   1.4 &  0.5  \\
~~${\ttbar}{\rar}\ell$+jets &  --- &   1.7 &  13.6 &  21.8 & 11.7 &  --- &   1.0 &  10.2 &  18.0 &  8.1  \\
~~$Wb\bar{b}$               &  --- &  16.2 &   6.8 &   1.8 &  0.3 &  --- &  15.3 &   8.2 &   2.3 &  0.6  \\
~~$Wc\bar{c}$               &  --- &   1.6 &   1.1 &   0.4 &  0.1 &  --- &   1.6 &   1.5 &   0.5 &  0.1  \\
~~$Wjj$                     &  --- &   0.1 &   0.1 &   0.0 &  0.0 &  --- &   0.1 &   0.1 &   0.0 &  0.0  \\
~~Multijets                 &  --- &   2.5 &   3.2 &   2.7 &  1.4 &  --- &   1.5 &   1.9 &   0.4 &  0.8  \\
\hline
{\bf Background Sum}~~         &  --- &  27.5 &  29.4 &  28.4 & 14.2 &  --- &  24.1 &  25.7 &  22.7 & 10.1  \\
\hline
{\bf Data}                   &  --- &   30  &   37  &   22  &  10  &  --- &   23  &   32  &   27  &  10
\end{tabular}
\end{ruledtabular}
\end{minipage}
\end{center}
\end{table*}

Table~\ref{yields-errors} summarizes the signals, summed backgrounds,
and data for each channel, showing the uncertainties on the signals
and backgrounds, and the signal-to-background ratios.
Table~\ref{allchans-yields} shows the signal and background yields
summed over electron and muon channels and 1- and 2-tagged jets in the
2-jet, 3-jet, and 4-jet bins, and for the 2-, 3-, and 4-jet bins
combined.

\begin{table*}[htbp]
\begin{center}
\begin{minipage}{6.5in}
\caption[yieldserrors]{Summed signal and background yields after
selection with total uncertainties, the numbers of data events, and
the signal-to-background ratio in each analysis channel. Note that the
signal includes both s-channel and t-channel single top quark
processes.}
\label{yields-errors}
\begin{ruledtabular}
\begin{tabular}{l||cccc|cccc}
\multicolumn{9}{c}{\hspace{1in}\underline{Summary of Yields with Uncertainties}} \vspace{0.05in}  \\
                & \multicolumn{4}{c}{Electron Channel} &  \multicolumn{4}{c}{Muon Channel}   \\
                &       1 jet        &      2 jets     &     3 jets      &    4 jets      
                &       1 jet        &      2 jets     &     3 jets      &    4 jets      \\
\hline		
{\bf Zero-tag}    &                    &                 &                 &                
                &                    &                 &                 &                \\
~~Signal Sum    &      $9 \pm 2$     &    $21 \pm 4$   &    $10 \pm 2$   &    $3 \pm 1$   
                &      $5 \pm 1$     &    $15 \pm 3$   &     $7 \pm 2$   &    $3 \pm 1$   \\
~~Bkgd Sum      & $26,886 \pm 626$   & $7,845 \pm 336$ & $2,832 \pm 144$ &  $735 \pm 60$  
                & $17,476 \pm 515$   & $6,124 \pm 351$ & $2,375 \pm 178$ &  $610 \pm 50$  \\
~~Data          &  29,925            &  7,833          &  2,831          &   752           
                &  17,527            &  6,122          &  2,378          &   599          \\ 
~~Signal:Bkgd~~ &       1:3,104      &      1:378      &      1:286      &     1:259      
                &       1:3,253      &      1:407      &      1:320      &     1:292      \\
{\bf One-tag}     &                    &                 &                 &                                 
                &                    &                 &                 &                \\
~~Signal Sum    &      $5 \pm 1$     &    $18 \pm 3$   &     $9 \pm 2$   &    $3 \pm 1$   
                &      $3 \pm 1$     &    $14 \pm 3$   &     $7 \pm 2$   &    $2 \pm 1$   \\
~~Bkgd Sum      &    $484 \pm 86$    &   $348 \pm 61$  &   $213 \pm 30$  &  $110 \pm 16$  
                &    $340 \pm 63$    &   $286 \pm 58$  &   $191 \pm 34$  &   $93 \pm 15$  \\
~~Data          &     445            &    357          &    207          &    97           
                &     289            &    287          &    179          &   100          \\ 
~~Signal:Bkgd~~ &       1:95         &      1:20       &      1:23       &     1:38       
                &       1:101        &      1:21       &      1:26       &     1:42       \\
{\bf Two-tags}     &                    &                 &                 &                                
                &                    &                 &                 &                \\
~~Signal Sum    &         ---        &   $2.6 \pm 0.6$ &   $1.9 \pm 0.4$ &  $0.7 \pm 0.2$ 
                &         ---        &   $2.1 \pm 0.5$ &   $1.6 \pm 0.4$ &  $0.6 \pm 0.2$ \\
~~Bkgd Sum      &         ---        &  $27.5 \pm 6.5$ &  $29.4 \pm 5.7$ & $28.4 \pm 6.0$ 
                &         ---        &  $24.1 \pm 6.1$ &  $25.7 \pm 5.5$ & $22.7 \pm 5.4$ \\
~~Data          &         ---        &   30            &   37            &  22            
                &         ---        &   23            &   32            &  27            \\
~~Signal:Bkgd~~ &         ---        &      1:10       &      1:15       &     1:39       
                &         ---        &      1:12       &      1:16       &     1:37       
\end{tabular}
\end{ruledtabular}
\end{minipage}
\end{center}
\end{table*}

\begin{table*}[hbtp]
\begin{center}
\begin{minipage}{4.5in}
\caption[allchansyields]{Yields after selection for the analysis
channels combined.}
\label{allchans-yields}
\begin{ruledtabular}
\begin{tabular}{l|@{\extracolsep{\fill}}r@{\extracolsep{0pt}$\pm$}l@{}%
                  @{\extracolsep{\fill}}r@{\extracolsep{0pt}$\pm$}l@{}%
                  @{\extracolsep{\fill}}r@{\extracolsep{0pt}$\pm$}l@{}%
		  @{\extracolsep{\fill}}r@{\extracolsep{0pt}$\pm$}l@{}}
\multicolumn{9}{c}{\hspace{1in}\underline{Summed Yields}} \vspace{0.05in} \\
& \multicolumn{8}{c}{$e$+$\mu$ + 1+2 tags} \\
                            & \multicolumn{2}{c}{2 jets} & \multicolumn{2}{c}{3 jets} 
	                    & \multicolumn{2}{c}{4 jets} & \multicolumn{2}{c}{2,3,4 jets} \\
\hline			                                              		  
{\bf Signals}                 &  \multicolumn{8}{c}{}                         \\
~~$tb$                      &  16 &   3 &   8 &  2 &   2 &  1 &     25 &  6  \\
~~$tqb$                     &  20 &   4 &  12 &  3 &   4 &  1 &     37 &  8  \\
{\bf Backgrounds}              &  \multicolumn{8}{c}{}                          \\
~~${\ttbar}{\rar}\ell\ell$  &  39 &   9 &  32 &  7 &  11 &  3 &     82 & 19  \\
~~${\ttbar}{\rar}\ell$+jets &  20 &   5 & 103 & 25 & 143 & 33 &    266 & 63  \\
~~$Wb\bar{b}$               & 261 &  55 & 120 & 24 &  35 &  7 &    416 & 87  \\
~~$Wc\bar{c}$               & 151 &  31 &  85 & 17 &  23 &  5 &    259 & 53  \\
~~$Wjj$                     & 119 &  25 &  43 &  9 &  12 &  2 &    174 & 36  \\
~~Multijets                 &  95 &  19 &  77 & 15 &  29 &  6 &    202 & 39  \\
\hline
{\bf Background Sum}           & 686 & 131 & 460 & 75 & 253 & 42 &  1,398 & 248 \\
\hline
{\bf Backgrounds+Signals}~~     & 721 & 132 & 480 & 76 & 260 & 43 &  1,461 & 251 \\
\hline
{\bf Data}                   &  \multicolumn{2}{c}{697} & \multicolumn{2}{c}{455} &
			       \multicolumn{2}{c}{246} & \multicolumn{2}{c}{1,398}      
\end{tabular}
\end{ruledtabular}
\end{minipage}
\end{center}
\end{table*}

\begin{figure*}[!h!tbp]
\includegraphics[width=0.32\textwidth]
{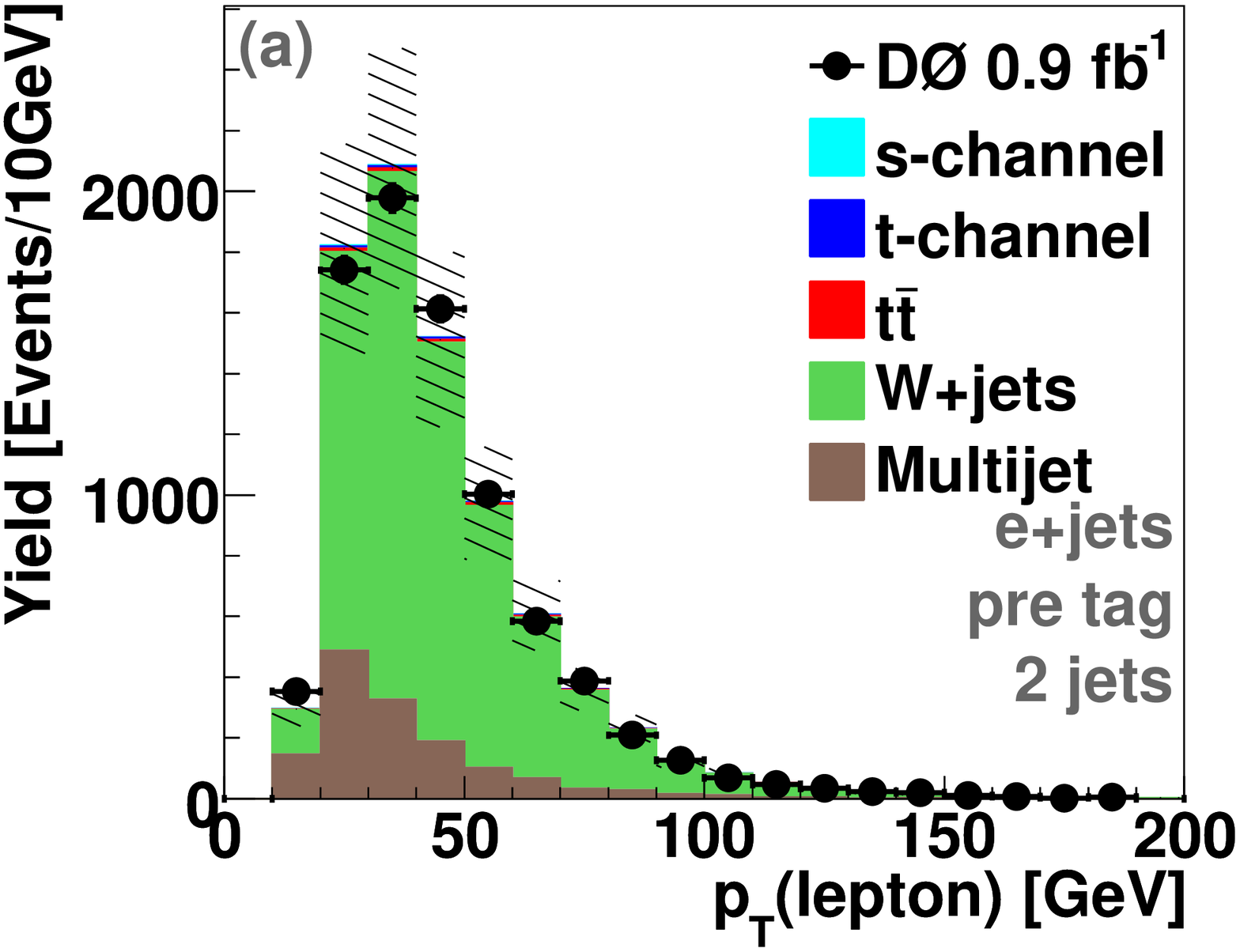}
\includegraphics[width=0.32\textwidth]
{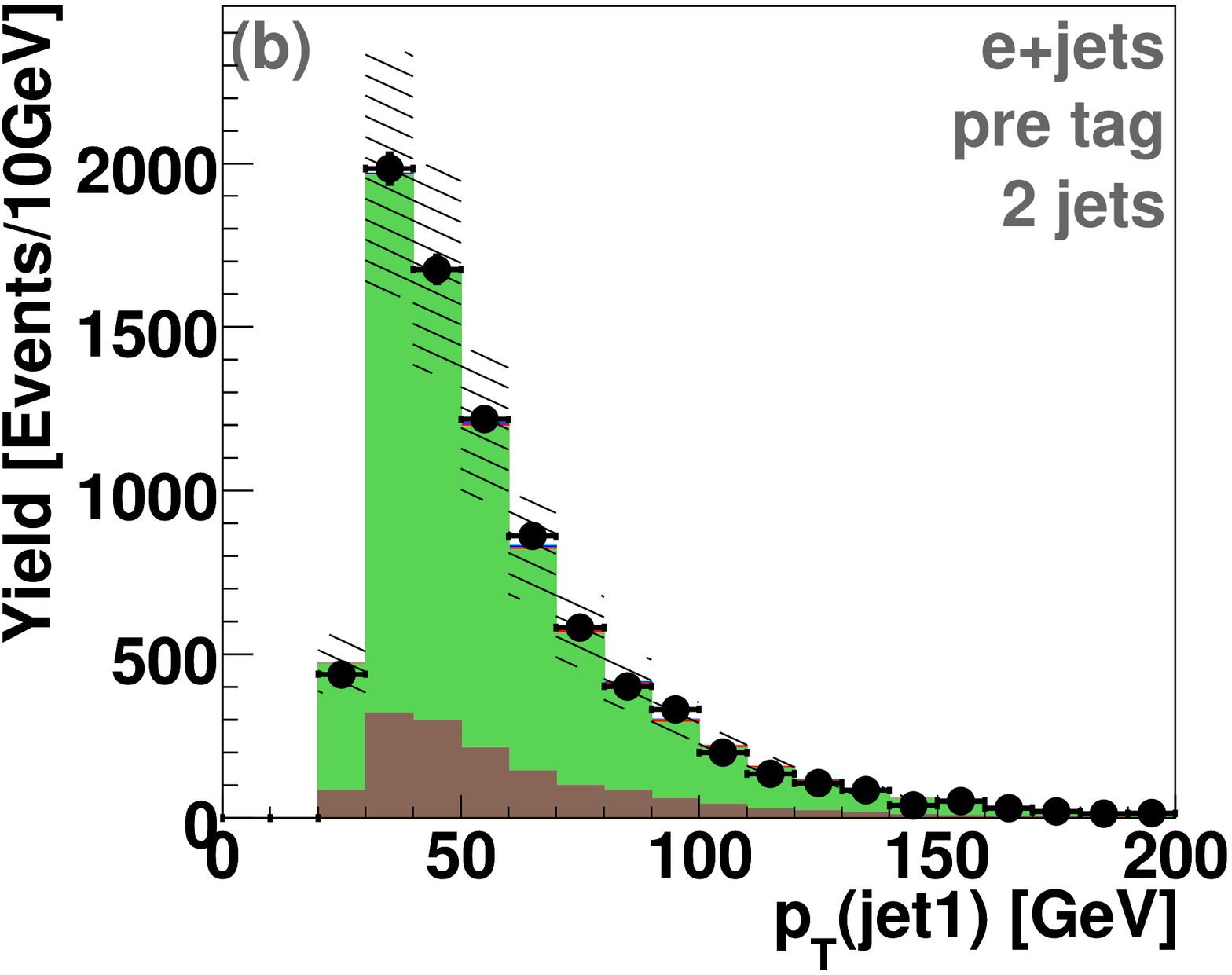}
\includegraphics[width=0.32\textwidth]
{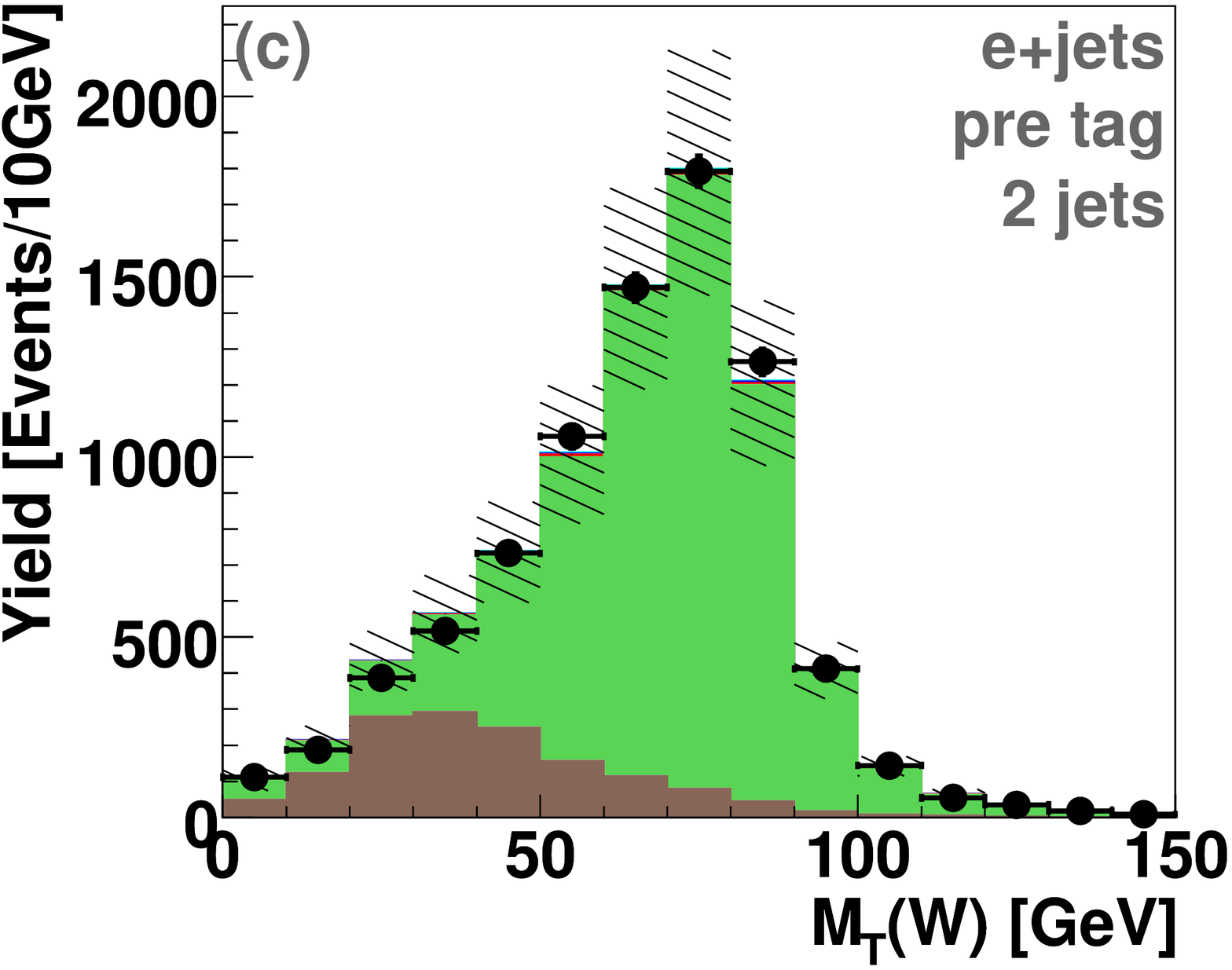}
\includegraphics[width=0.32\textwidth]
{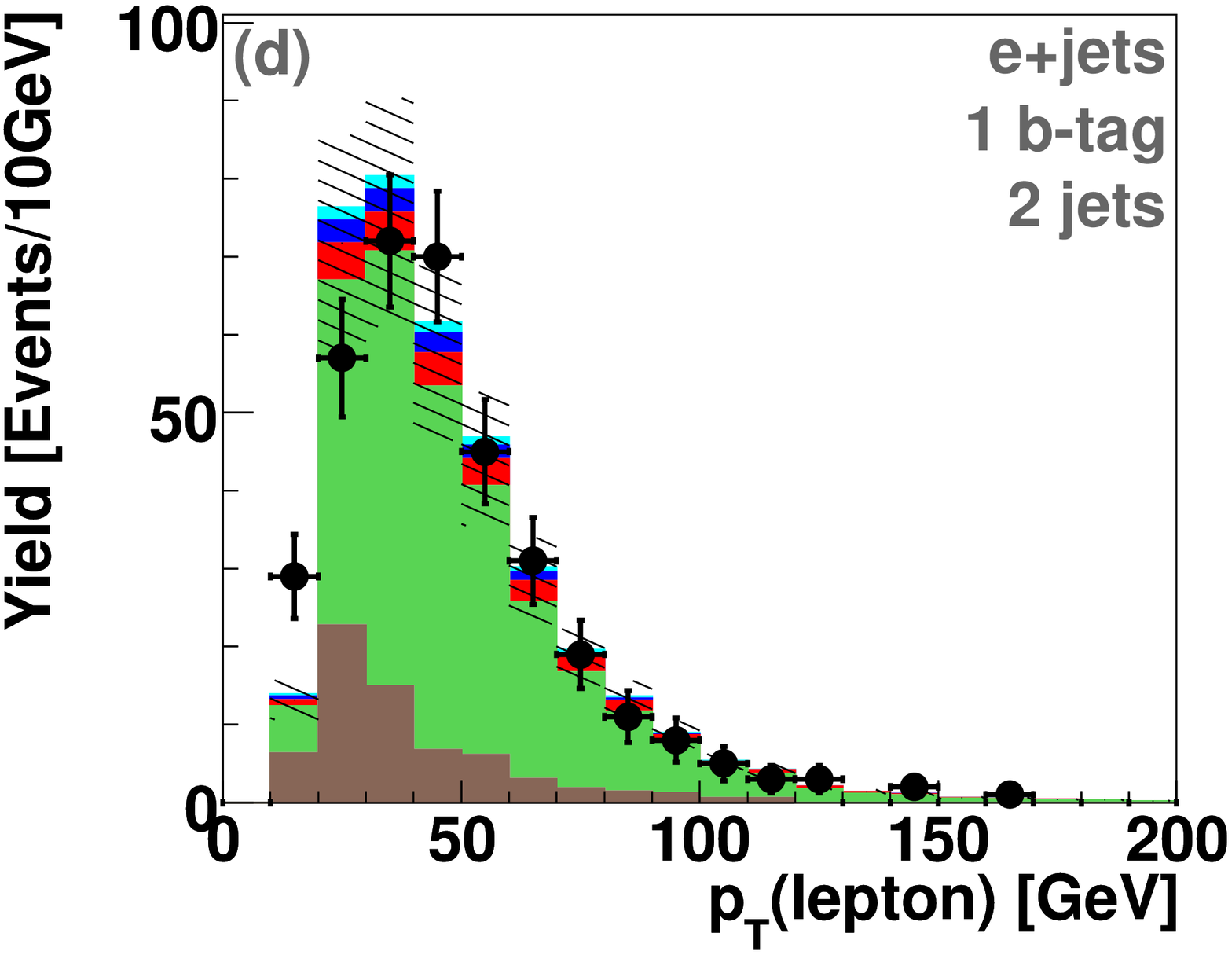}
\includegraphics[width=0.32\textwidth]
{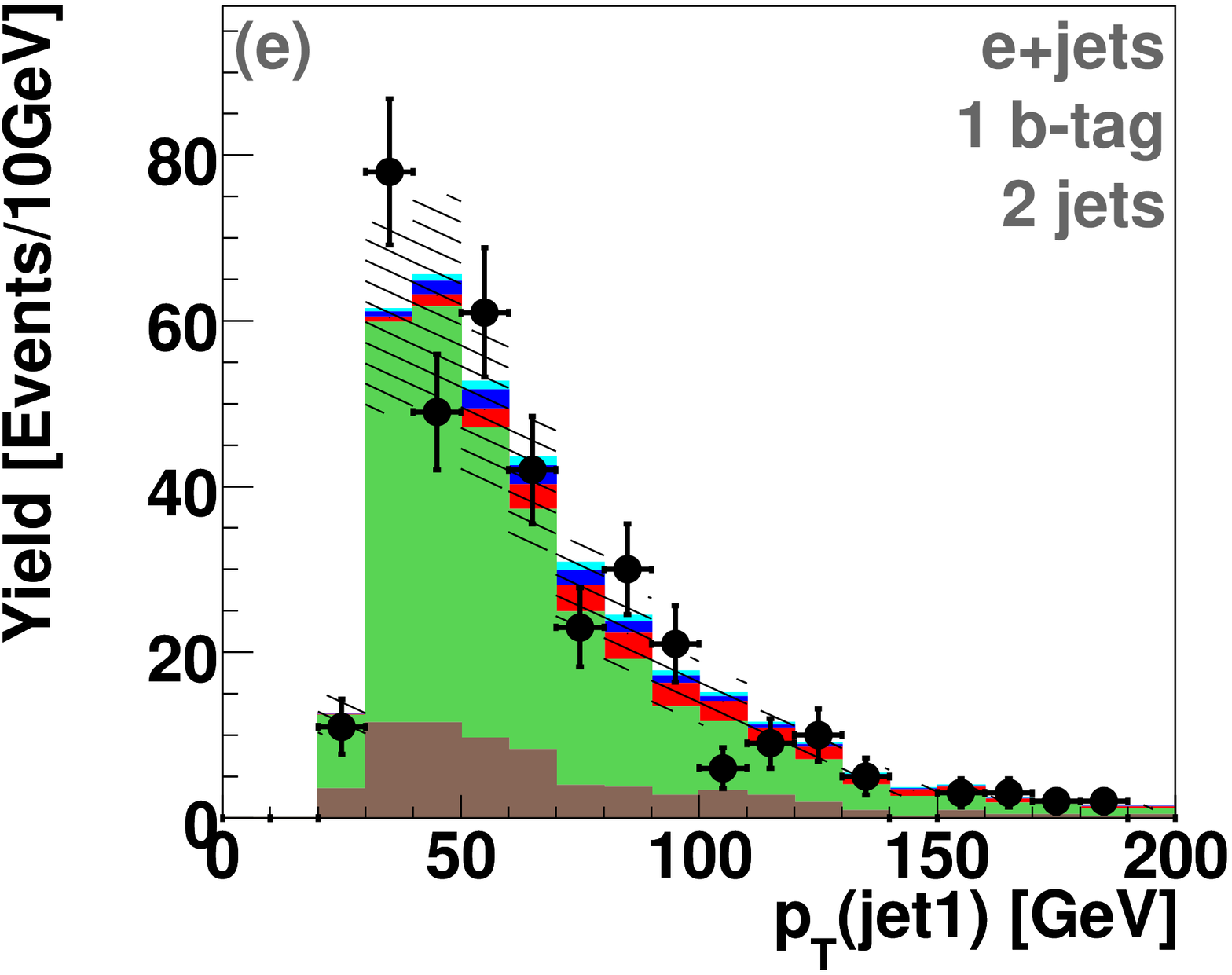}
\includegraphics[width=0.32\textwidth]
{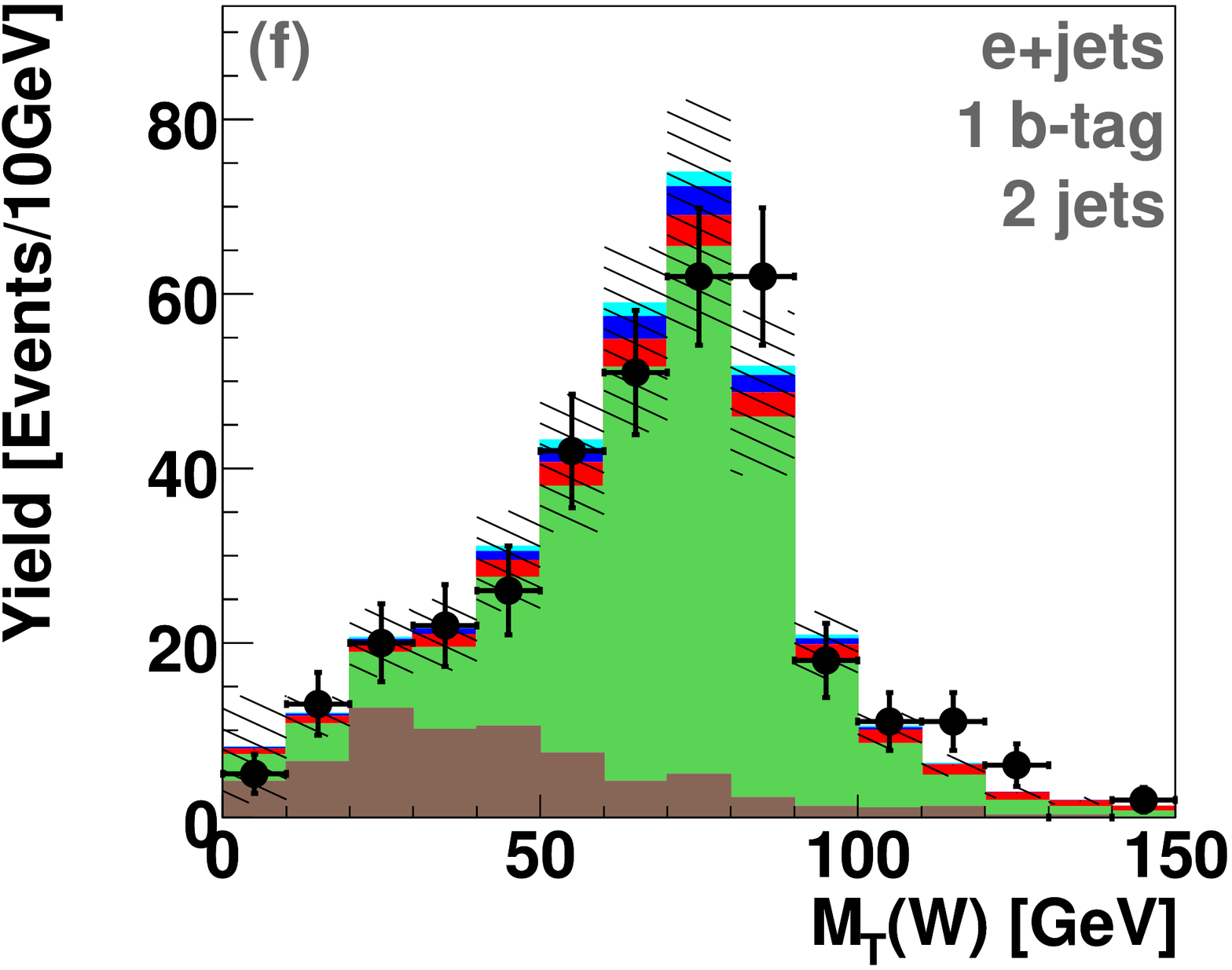}
\vspace{-0.1in}
\caption[agreement-electron]{The first row shows pretagged
distributions for the $p_T$ of the electron, the $p_T$ of the leading
jet, and the reconstructed $W$~boson transverse mass. The second row
shows the same distributions after tagging for events with exactly one
$b$-tagged jet. The hatched area is the $\pm 1\sigma$ uncertainty on
the total background prediction.}
\label{agreement-electron}
\end{figure*}

\begin{figure*}[!h!tbp]
\includegraphics[width=0.32\textwidth]
{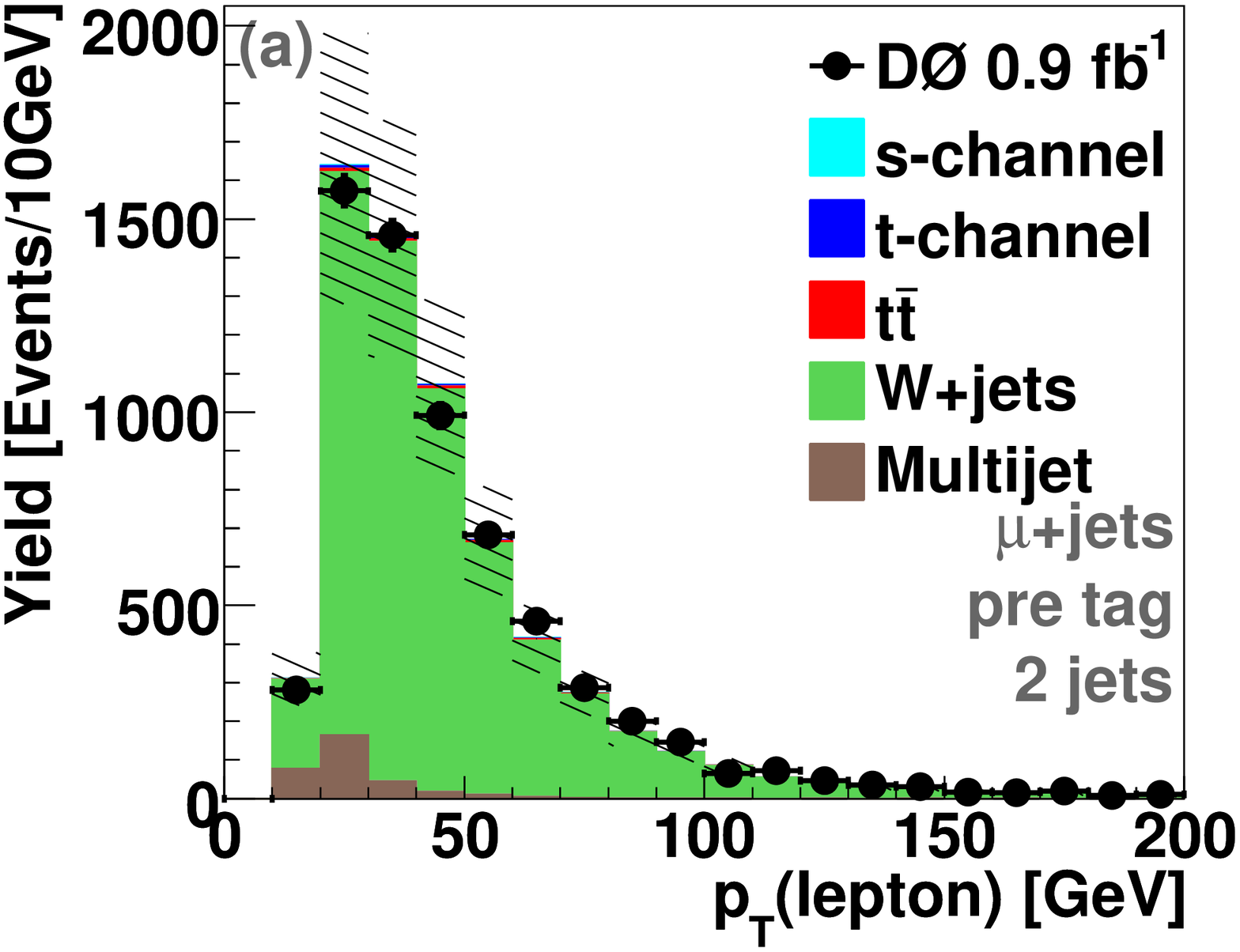}         
\includegraphics[width=0.32\textwidth]
{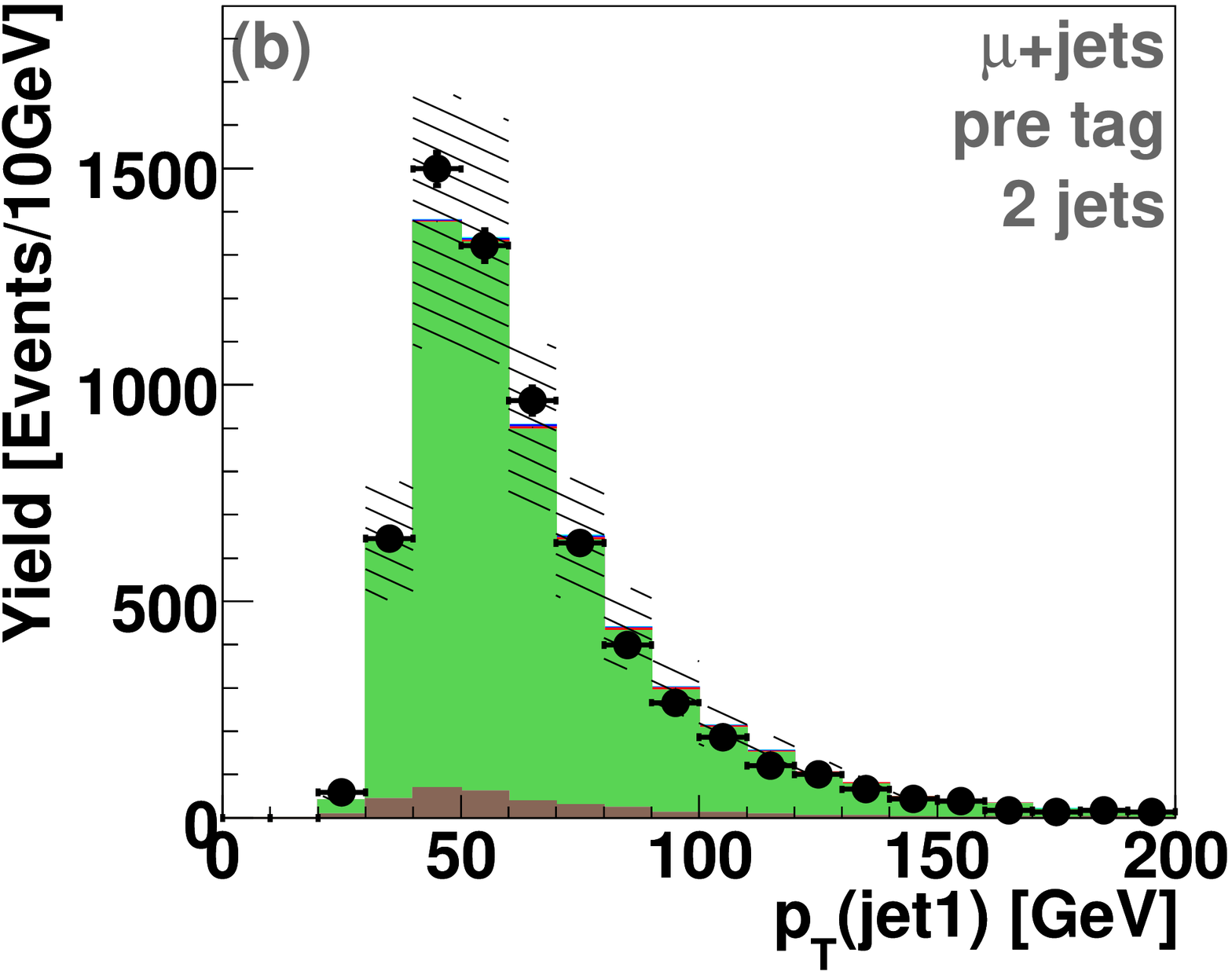}           
\includegraphics[width=0.32\textwidth]
{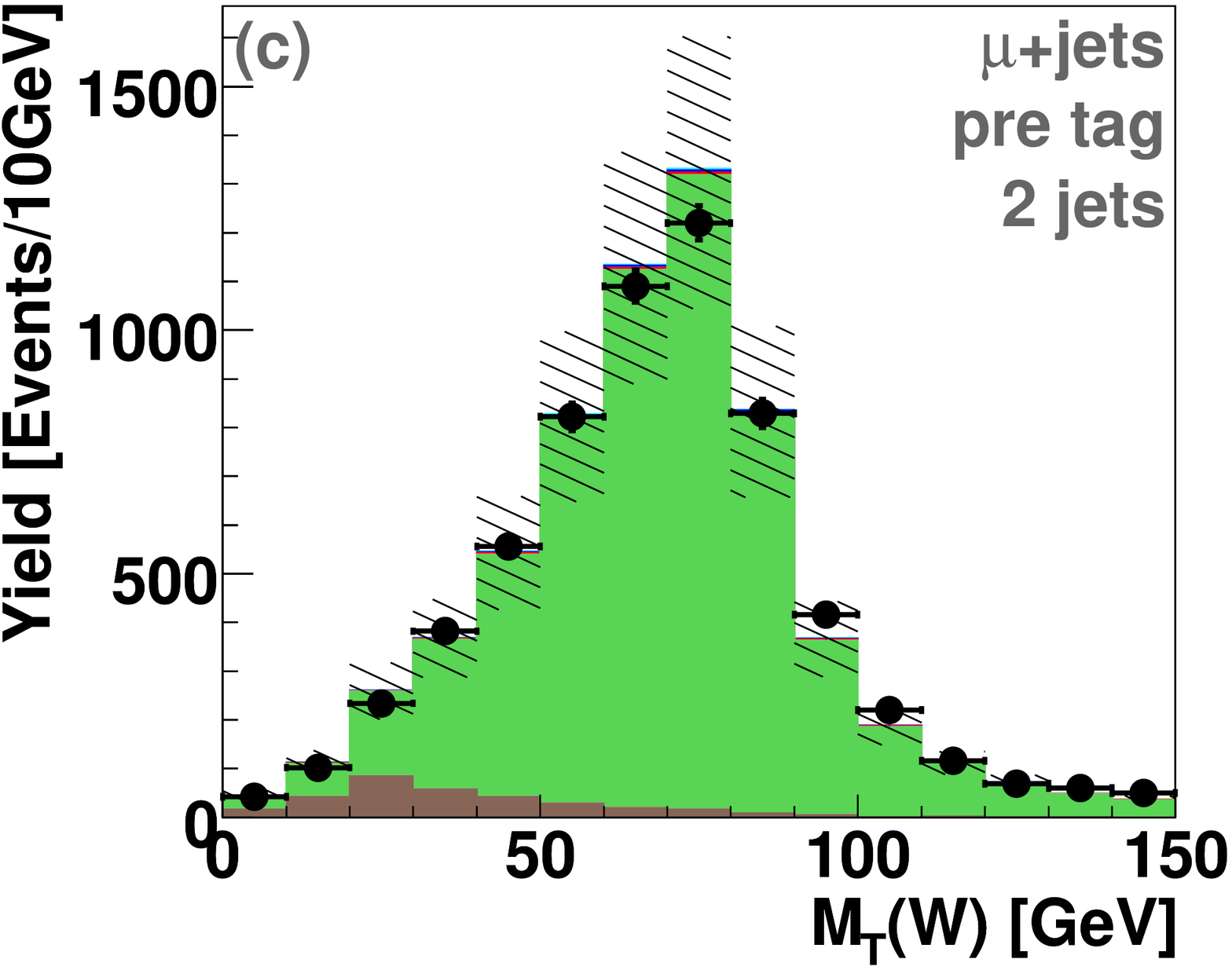}  
\includegraphics[width=0.32\textwidth]
{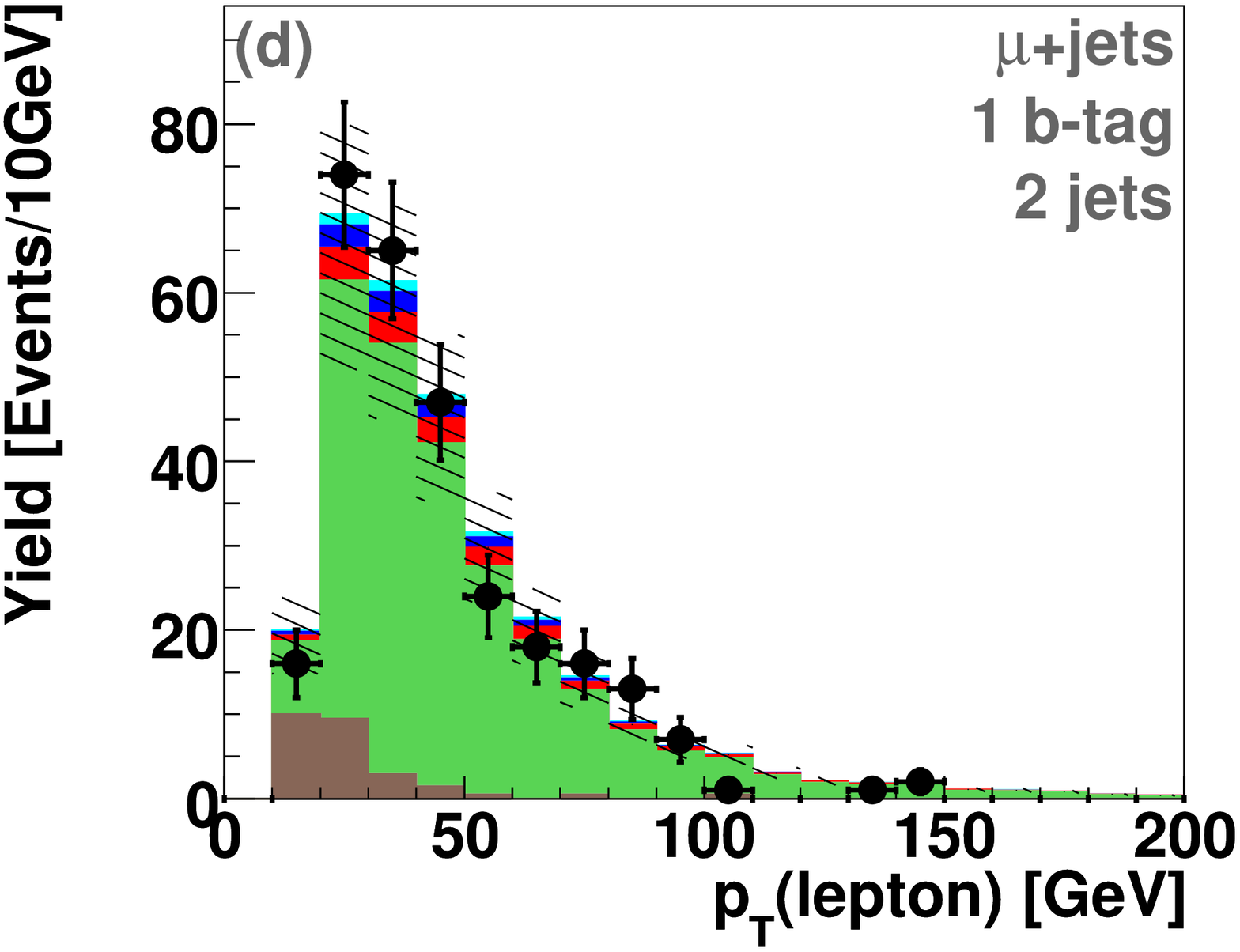}       
\includegraphics[width=0.32\textwidth]
{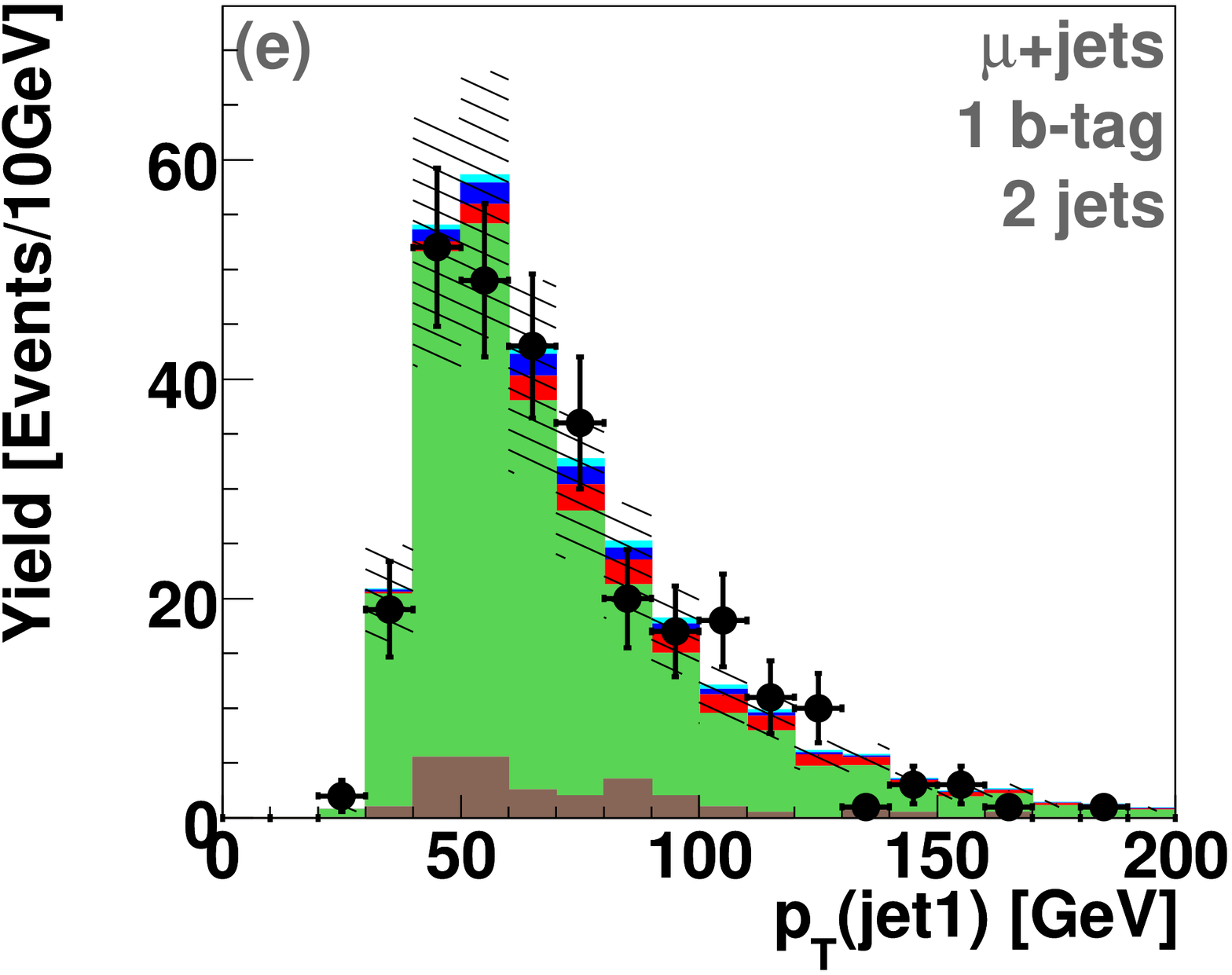}         
\includegraphics[width=0.32\textwidth]
{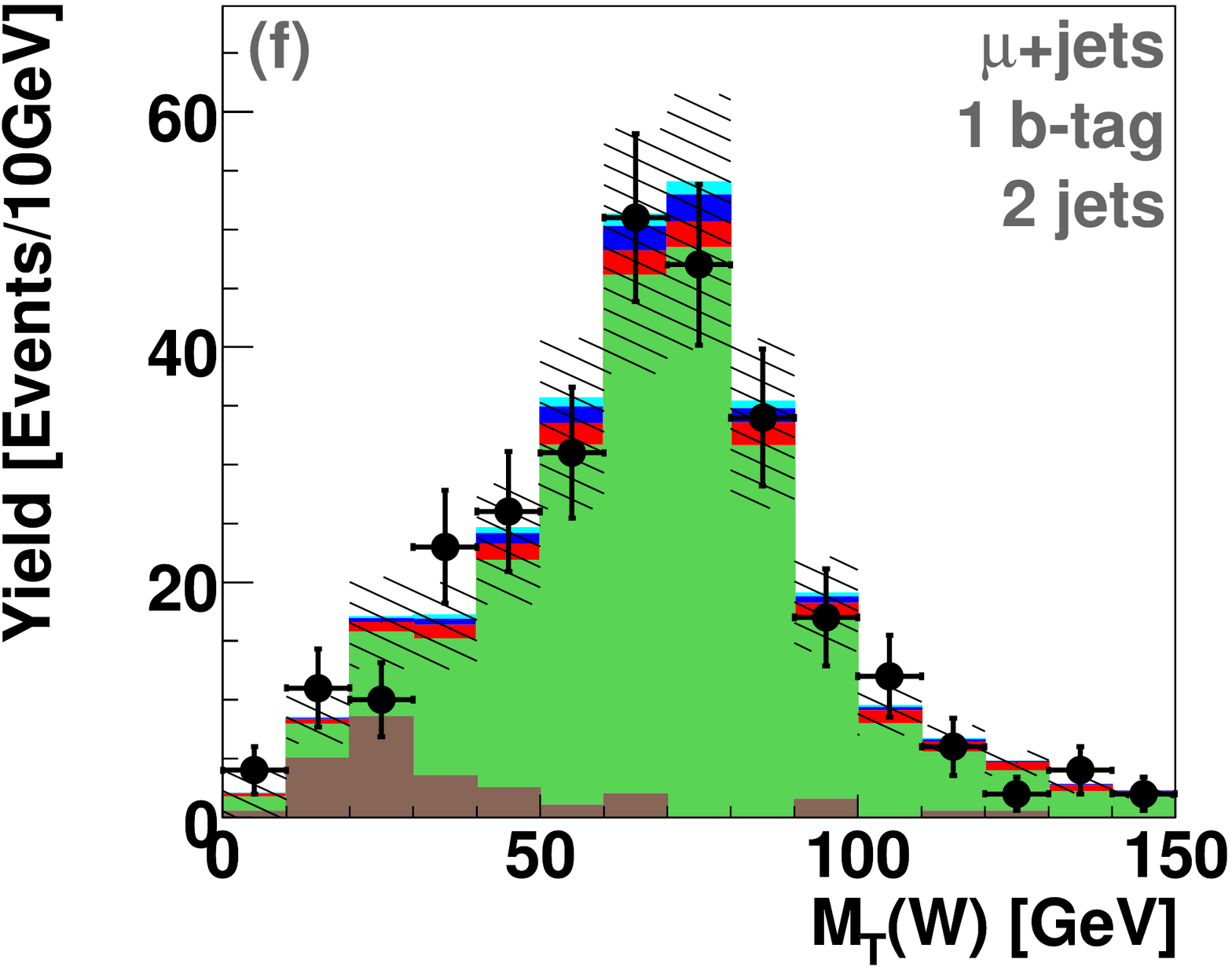}
\vspace{-0.1in}
\caption[agreement-muon]{The first row shows pretagged
distributions for the $p_T$ of the muon, the $p_T$ of the leading jet,
and the reconstructed $W$~boson transverse mass. The second row shows
the same distributions after tagging for events with exactly one
$b$-tagged jet. The hatched area is the $\pm 1\sigma$ uncertainty on
the total background prediction.}
\label{agreement-muon}
\end{figure*}

Some basic kinematic distributions are shown for electron channel
events in Fig.~\ref{agreement-electron} and for muon channel events in
Fig.~\ref{agreement-muon}. Since the yields are normalized before
$b$~tagging, in each case the pretagged distributions are shown in the
first row of distributions and the one-tag distributions are shown in
the second row.

\clearpage
%---------------------------------------------------------------------
%---------------------------------------------------------------------
\section{Systematic Uncertainties}
\label{systematics}

We consider several sources of systematic uncertainty in this analysis
and propagate them separately for each signal and background source
throughout the calculation. Systematic uncertainties enter the
analysis in two ways: as uncertainty on the normalization of the
background samples and as effects that change the shapes of
distributions for the backgrounds and the expected signals. The effect
of these uncertainties on the discriminant outputs and how they affect
the cross section measurement is described in
Sec.~\ref{sec:priordensity}. Table~\ref{systematics-values} summarizes
the relative uncertainties on each of the sources described below.

The first uncertainties listed here affect only the {\ttbar}
background normalization.

\begin{myitemize}
\item {\bf Integrated luminosity} \\ 
At 6.1\%~\cite{luminosity-uncertainty}, this is a small contribution
to the {\ttbar} yield uncertainty.

\item {\bf Theoretical cross section} \\ 
The uncertainty on the {\ttbar} cross section includes components for
the choice of scale and PDF, and also, more significantly, a large
component from the top quark mass uncertainty (i.e., using 175~GeV in
this analysis when the latest world average value is
$170.9\pm1.8$~GeV). The combined uncertainty on the cross section is
taken as 18\%.
\end{myitemize}

The following uncertainties arise from the correction factors and
functions applied to the simulated samples to make them match data,
and thus affect both the signal acceptances and the {\ttbar}
background yield.

\begin{myitemize}
\item {\bf Trigger efficiency} \\
Functions that represent the trigger efficiency for each object type
and trigger level as a function of $p_T$, $\deta$, and $\phi$ are used
to weight simulated events. The functions are shifted up and down by
one standard deviation of the statistical error arising from the data
samples used to calculate the functions and the weight of each event
is recalculated. Fixed uncertainties of 3\% in the electron channel
and 6\% in the muon channel are chosen since they encompass all the
small variations seen in each analysis channel.

\item {\bf Primary vertex selection efficiency} \\
The primary vertex selection efficiency in data and the simulation are
not the same. We assign a systematic uncertainty of 3\% for the
difference between the beam profile along the longitudinal
direction in data and the simulated distribution.

\item {\bf Electron reconstruction and basic identification efficiency} \\
The electron reconstruction and basic identification correction
factors are parametrized as a function of \deta. The 2\% uncertainty
in the efficiency accounts for its dependence on variables other than
\deta, and as a result of limited data statistics used to determine
the correction factors.

\item {\bf Electron shower shape, track match, and likelihood efficiency} \\
The electron shower shape, track match, and likelihood correction
factors are parametrized as a function of $\deta$ and $\phi$. The 5\%
uncertainty in the efficiency accounts for the dependence on other
variables, such as the number of jets and the instantaneous
luminosity, and as a result of limited data statistics in determining
these correction factors.

\item {\bf Muon reconstruction and identification efficiency} \\
The correction factor uncertainty of 7\% includes contributions from
the method used to determine the correction functions, from the
background subtraction, and from the limited statistics in the
parametrization as a function of the $\deta$ and $\phi$ of the muon.

\item {\bf Muon track matching and isolation} \\
The muon tracking correction functions have an uncertainty that
includes contributions from the method used to measure the functions,
from the background subtraction, luminosity and timing bias, and from
averaging over $\phi$ and the limited statistics in each bin used to
calculate the functions. The muon isolation correction uncertainty is
estimated based on its dependence on the number of jets, and covers
the dependences not taken into account such as $p_T$ and $\deta$. The
overall value of these uncertainties combined is 2\%.

\item {\bf Jet fragmentation} \\
This systematic uncertainty covers the lack of certainty in the jet
fragmentation model (and is measured as the difference between
{\pythia} and {\herwig}~\cite{herwig} fragmentation) as well as the
uncertainty in the modeling of initial-state and final-state
radiation. It is 5\% for $tb$ and $tqb$ and 7\% for {\ttbar}.

\item {\bf Jet reconstruction and identification} \\
The efficiency to reconstruct jets is similar in data and simulated
events, but the efficiency of the simulated jets is nevertheless
corrected by a parametrization of this discrepancy as a function of
jet $p_T$. We assign a 2\% error to the parametrization based on the
statistics of the data sample.

\item {\bf Jet energy scale and jet energy resolution} \\
The jet energy scale (JES) is raised and lowered by one standard
deviation of the uncertainty on it and the whole analysis repeated. In
the data, the JES uncertainty contains the jet energy resolution
uncertainty. But in the simulation, the jet energy resolution
uncertainty is not taken into account in the JES uncertainty. To
account for this, the energy smearing in the simulated samples is
varied by the size of the jet energy resolution. This uncertainty
affects the acceptance and the shapes of the distributions. The value
of this uncertainty varies from $1\%$ to $20\%$, depending on the
analysis channel, with typical values between $6\%$ and $10\%$.
\end{myitemize}

The uncertainty on the $W$+jets and multijets background yields
comes from the normalization to data. The $W$+jets yield is 100\% anticorrelated
with the multijets yield.

\begin{myitemize}
\item {\bf Matrix-method normalization} \\
The determination of the number of real-lepton events in data is
affected by the uncertainties associated with the determination of the
probabilities for a loose lepton to be (mis)identified as a (fake)
real lepton, $\varepsilon_{\rm fake{\mbox{\small -}}\ell}$ and
$\varepsilon_{\rm real{\mbox{\small -}}\ell}$. The
normalization is also affected by the limited statistics of the data
sample as described in Sec.~\ref{matrix-method}. The combined
uncertainties on the $W$+jets and multijets yields vary between 17\%
and 28\%, depending on the analysis channel.

\item {\bf Heavy flavor ratio} \\
The uncertainty on the scale factor applied to set the $Wb\bar{b}$ and
$Wc\bar{c}$ fractions of the $W$+jets sample, as described in
Sec.~\ref{mc-samples}, is estimated to cover several effects:
dependence on the $b$-quark $\pt$, the difference between the zero-tag
samples where it is estimated and the signal samples where it is used,
and the intrinsic uncertainty on the value of the LO cross section it
is being applied to. This uncertainty is 30\%. It is included in the
matrix method uncertainty described above.
\end{myitemize}

There is one source of uncertainty that affects the signal
acceptances, and both the {\ttbar} and $W$+jets background yields.

\begin{myitemize}
\item {\bf $\mathbf{b}$-tag modeling}\\
The uncertainty associated with the taggability-rate and tag-rate
functions is evaluated by raising and lowering the tag rate by one
standard deviation separately for both the taggability and the tag
rate components and determining the new event tagging weight. These
uncertainties originate from several sources as follows: statistics of
the simulated event sets; the assumed fraction of heavy flavor in the
simulated multijet sample used for the mistag rate determination; and
the choice of parametrizations. The $b$-tag modeling uncertainty
varies from 2\% to 16\%, depending on the analysis channel, and we
include the variation on distribution shapes, as well as on sample
normalization.
\end{myitemize}

\begin{table}[!h!tbp]
\begin{center}
\begin{minipage}{3in}
\caption{Summary of the relative systematic uncertainties. The ranges
shown represent the different samples and channels.}
\label{systematics-values}
\begin{ruledtabular}
\begin{tabular}{l|c}
\multicolumn{2}{c}
{\underline{Relative Systematic Uncertainties} \vspace{0.05in}} \\
\hline
Integrated luminosity              &  $6\%$  \\
{\ttbar} cross section             & $18\%$  \\
Electron trigger                   &  $3\%$  \\
Muon trigger                       &  $6\%$  \\
Primary vertex                     &  $3\%$  \\
Electron reconstruction \& identification~~ &  $2\%$  \\
Electron track match \& likelihood  &  $5\%$  \\
Muon reconstruction \& identification  &  $7\%$  \\
Muon track match \& isolation      &  $2\%$  \\
Jet fragmentation                  & (5--7)$\%$\\
Jet reconstruction and identification &  $2\%$  \\
Jet energy scale                   & (1--20)$\%$\\
Tag-rate functions                 & (2--16)$\%$\\
Matrix-method normalization        &(17--28)$\%$\\
\hline
Heavy flavor ratio                 & $30\%$  \\ 
$\varepsilon_{\rm real{\mbox{\small -}}e}$       &  $2\%$  \\
$\varepsilon_{\rm real{\mbox{\small -}}\mu}$     &  $2\%$  \\
$\varepsilon_{\rm fake{\mbox{\small -}}e}$       &(3--40)$\%$\\ 
$\varepsilon_{\rm fake{\mbox{\small -}}\mu}$     &(2--15)$\%$
\end{tabular}
\end{ruledtabular}
\end{minipage}
\end{center}
\end{table}
%---------------------------------------------------------------------
%---------------------------------------------------------------------
\section{Multivariate Analyses}
\label{sec:multivariate-intro}

The search for single top quark production is significantly more
challenging than the search for {\ttbar} production. The principal
reasons are the smaller signal-to-background ratio for single top
quarks and the large overlap between the signal distributions and
those of the backgrounds. We therefore concluded from the outset
that optimal signal-background discrimination would be necessary to
have any chance of extracting a single top quark signal from the
available dataset.

Optimal event discrimination is a well-defined problem with a
well-defined and unique solution. Given the probability
\begin{equation}
p(S|\mathbf{x}) = 
\frac{p(\mathbf{x}|S) p(S)}{p(\mathbf{x}|S) p(S)
+ p(\mathbf{x}|B) p(B)}
\end{equation}
that an event described by the variables $\mathbf{x}$ is of the signal
class, $S$, the signal can be extracted optimally, that is, with the
smallest possible uncertainty~\cite{classification-barlow}, by
weighting events with $p(S|\mathbf{x})$, or, as we have done, by
fitting the sum of distributions of $p(S|\mathbf{x})$ for signal and
background to data, as described in
Sec.~\ref{sec:cross_section_measurements}. In practice, since {\em
any} one-to-one function of $p(S|\mathbf{x})$ is equivalent to
$p(S|\mathbf{x})$, it is sufficient to construct an approximation to
the discriminant
\begin{equation}
\label{eq:D}
D(\mathbf{x}) =
\frac{p(\mathbf{x}|S)}{p(\mathbf{x}|S) + p(\mathbf{x}|B)}
\end{equation}
built using equal numbers of signal and background events, that is,
with $p(S) = p(B)$. Each of the three analyses we have undertaken is
based on a different numerical method to approximate the discriminant
$D(\mathbf{x})$. From this perspective, they are conceptually
identical.

In this paper, we present results from three different multivariate
techniques applied to the selected dataset: boosted decision trees
(DT) in Sec.~\ref{sec:DecisionTree}, Bayesian neural networks (BNN) in
Sec.~\ref{sec:BayesianNN}, and matrix elements (ME) in
Sec.~\ref{sec:MatrixElement}. The DT analysis approximates the
discriminant $D(\mathbf{x})$ using an average of many piece-wise
approximations to $D(\mathbf{x})$. The BNN analysis uses nonlinear
functions that approximate $D(\mathbf{x})$ directly, that is, without
first approximating the densities $p(\mathbf{x}|S)$ and
$p(\mathbf{x}|B)$. The ME method approximates the densities
$p(\mathbf{x}|S)$ and $p(\mathbf{x}|B)$ semi-analytically, starting
with leading-order matrix elements, and computes $D(\mathbf{x})$ from
them.

The three analyses also differ by the choice of variables used. The
basic observables are:
\begin{myenumerate}
\item missing transverse energy 2-vector $(\met, \phi)$,
\item lepton 4-vector $(E_T, \eta, \phi)$, assuming massless leptons,
\item jet 4-vector $(E_T, \eta, \phi)$, assuming massless
jets, and jet-type, that is, whether it is a $b$~jet or not, for each
jet.
\end{myenumerate}
These, essentially, are the observables used in the matrix element
analysis. The other analyses, however, make use of physically
motivated variables~\cite{variables-dudko,variables-boos} derived from
the fundamental observables. Of course, the derived variables contain
no more information than is contained in the original degrees of
freedom. However, for some numerical approximation methods, it may
prove easier to construct an accurate approximation to $D(\mathbf{x})$
if it is built using carefully chosen derived variables than one
constructed directly in terms of the underlying degrees of freedom. It
may also happen that a set of judiciously chosen derived variables,
perhaps one larger than the set of fundamental observables, yields
better performing discriminants simply because the numerical
approximation algorithm is better behaved or converges faster.

The complete set of variables used in the DT and BNN analyses is shown
in Table~\ref{variables}. Jets are sorted in $p_T$ and index~1 refers
to the leading jet in a jet category: ``jet$n$'' ($n$=1,2,3,4)
corresponds to each jet in the event, ``tag$n$'' to $b$-tagged jets,
``untag$n$'' to non-$b$-tagged jets, and ``notbest$n$'' to all but the
best jet. The ``best'' jet is defined as the one for which the
invariant mass $M(W,{\rm jet})$ is closest to $m_{\mathrm{top}} =
175$~GeV.

Aplanarity, sphericity, and centrality are variables that describe the
direction and shape of the momentum flow in the
events~\cite{eventshape-barger-phillips}. The variable $H$ is the
scalar sum of the energy in an event for the jets as shown. $H_T$ is
the scalar sum of the transverse energy of the objects in the
event. $M$ is the invariant mass of various combinations of
objects. $M_T$ is the transverse mass of those objects. $Q$ is the
charge of the electron or muon.

\begin{figure*}[!h!tbp]
\includegraphics[width=0.30\textwidth]
{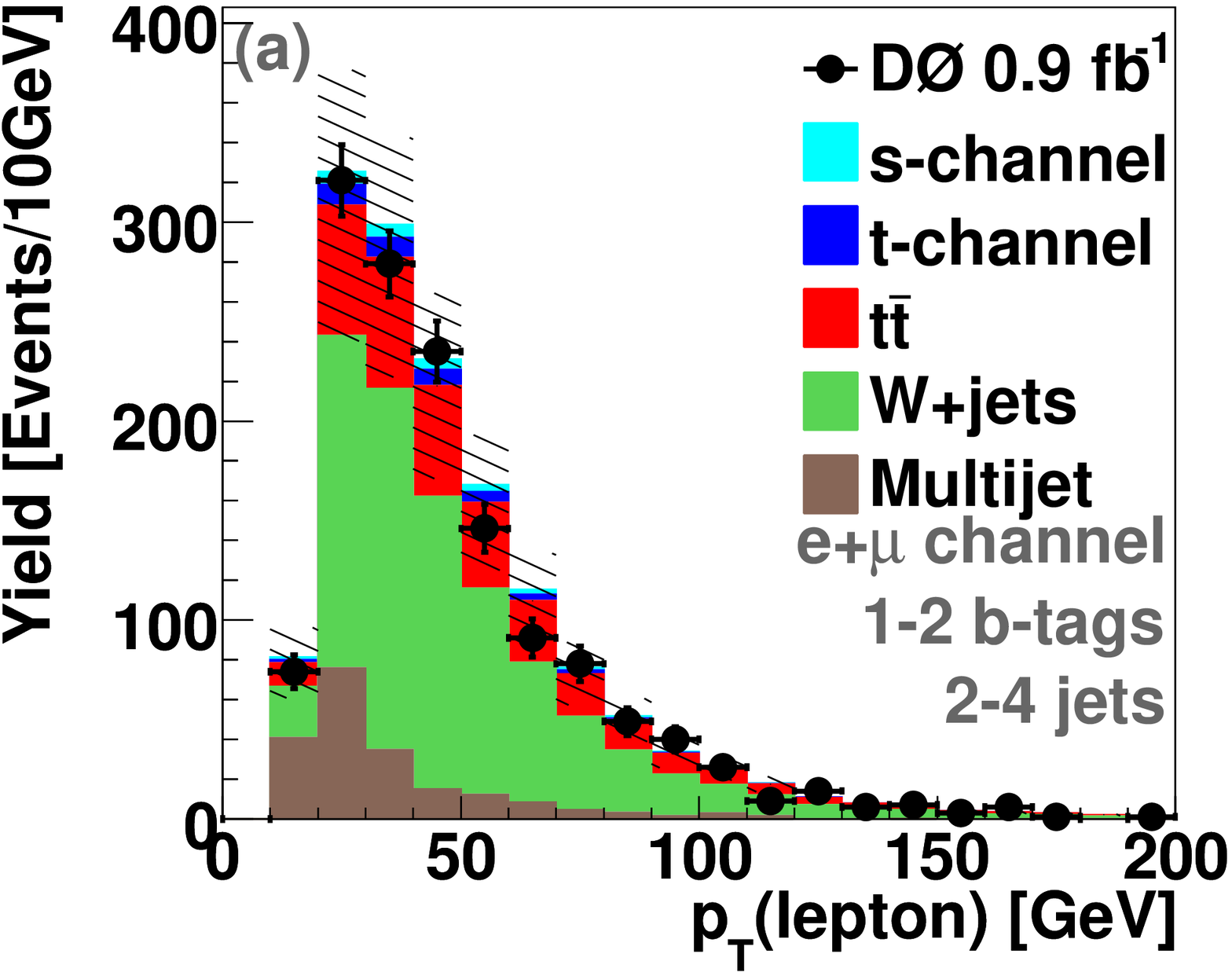}
\includegraphics[width=0.30\textwidth]
{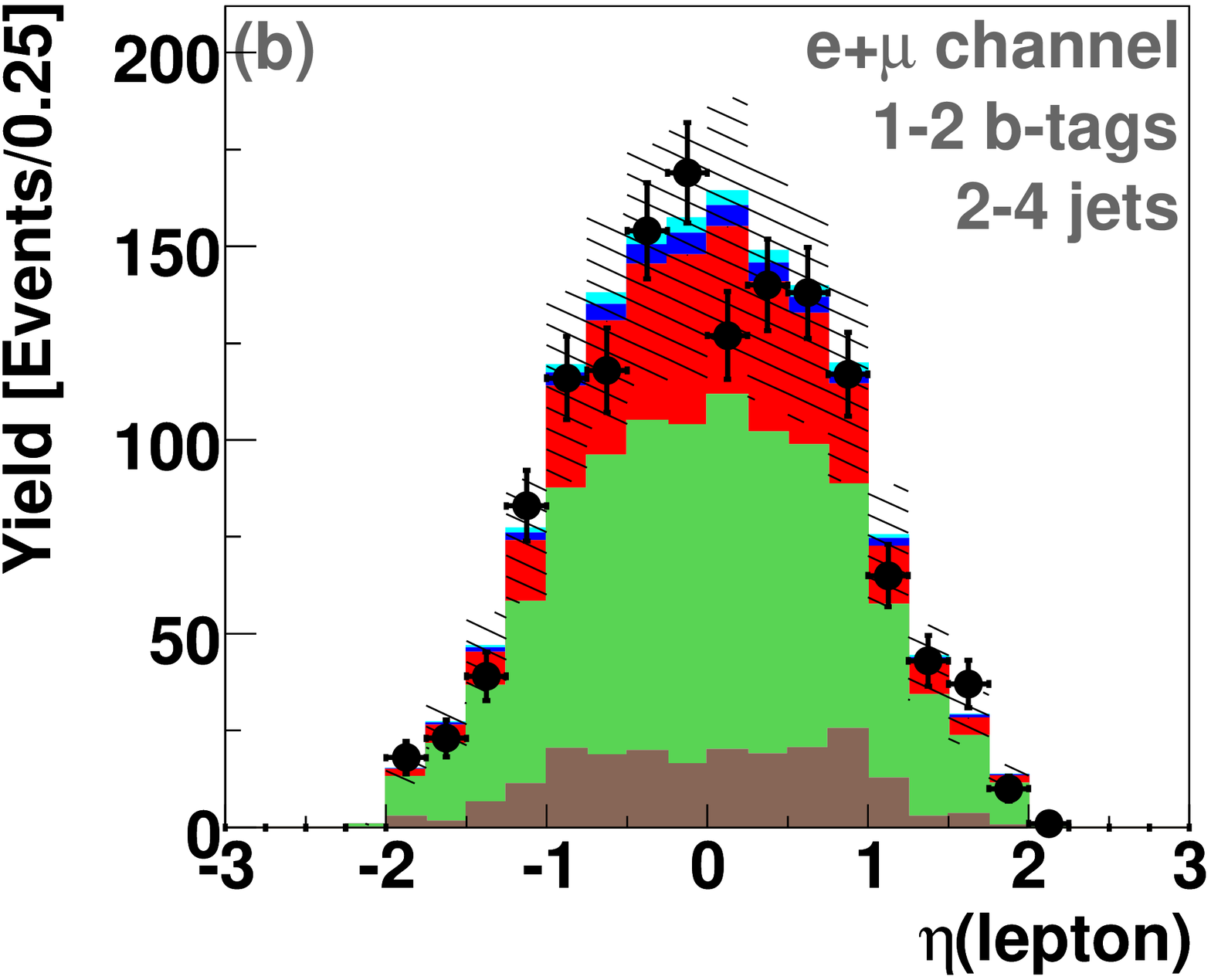}
\includegraphics[width=0.30\textwidth]
{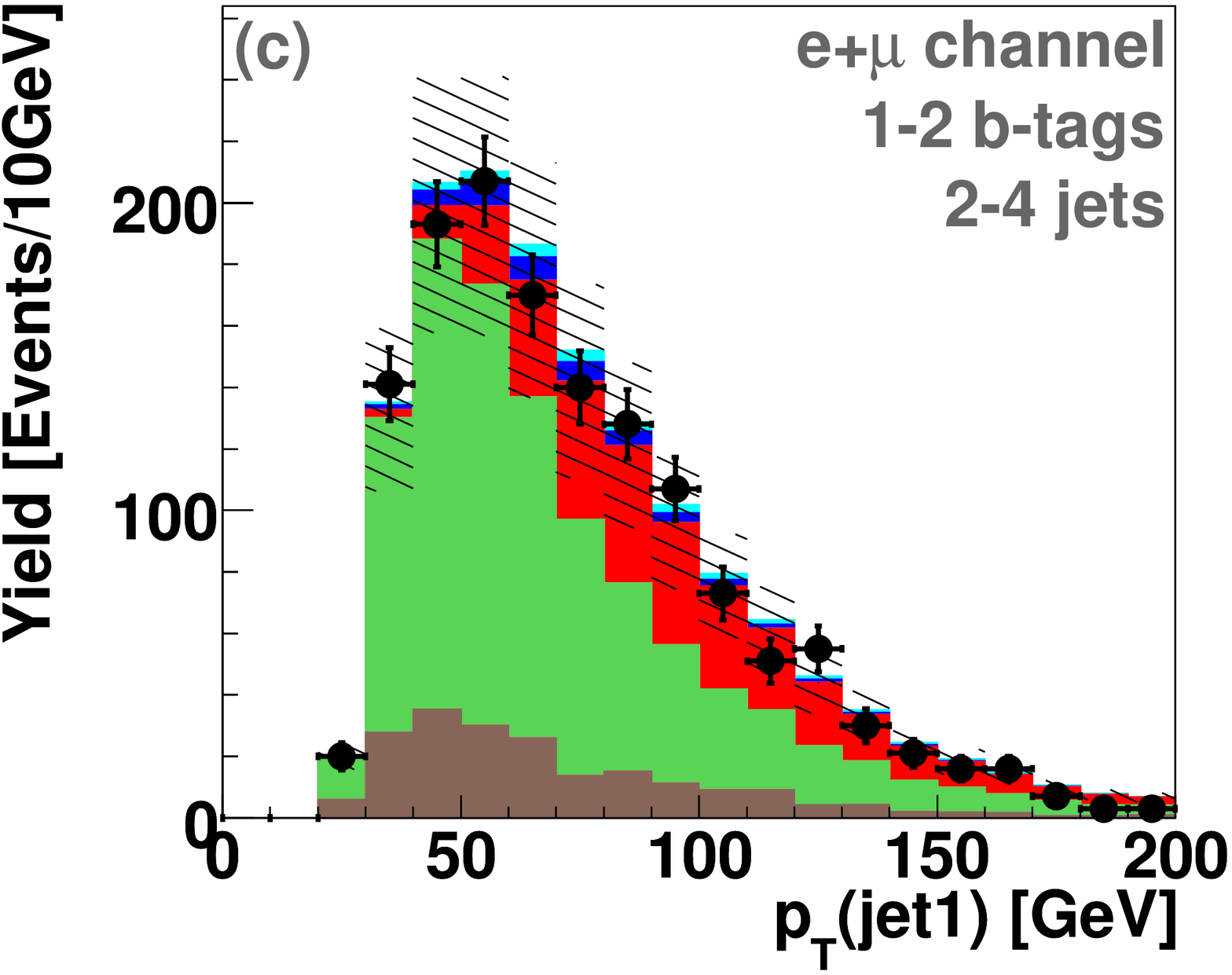}

\includegraphics[width=0.30\textwidth]
{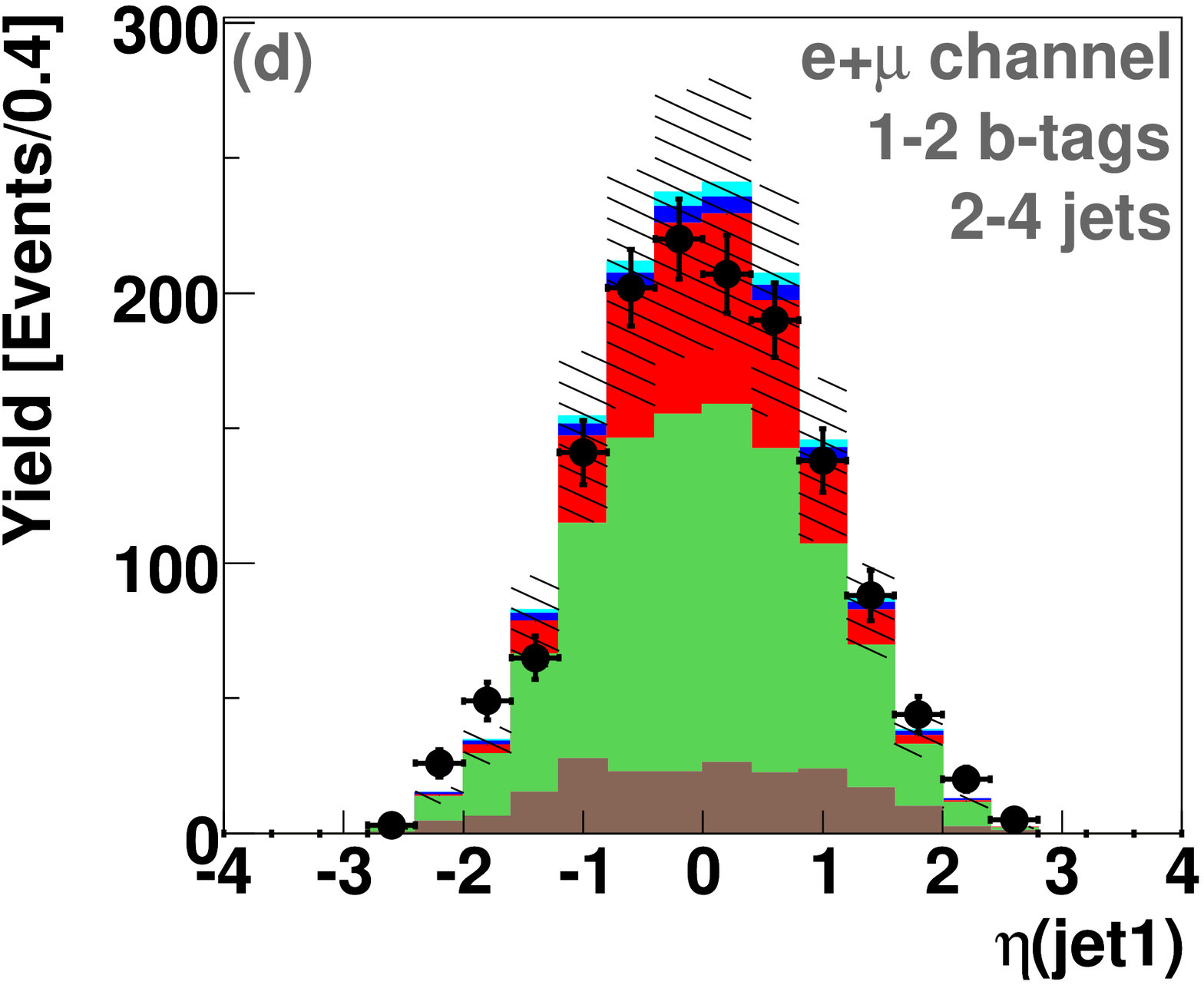}
\includegraphics[width=0.30\textwidth]
{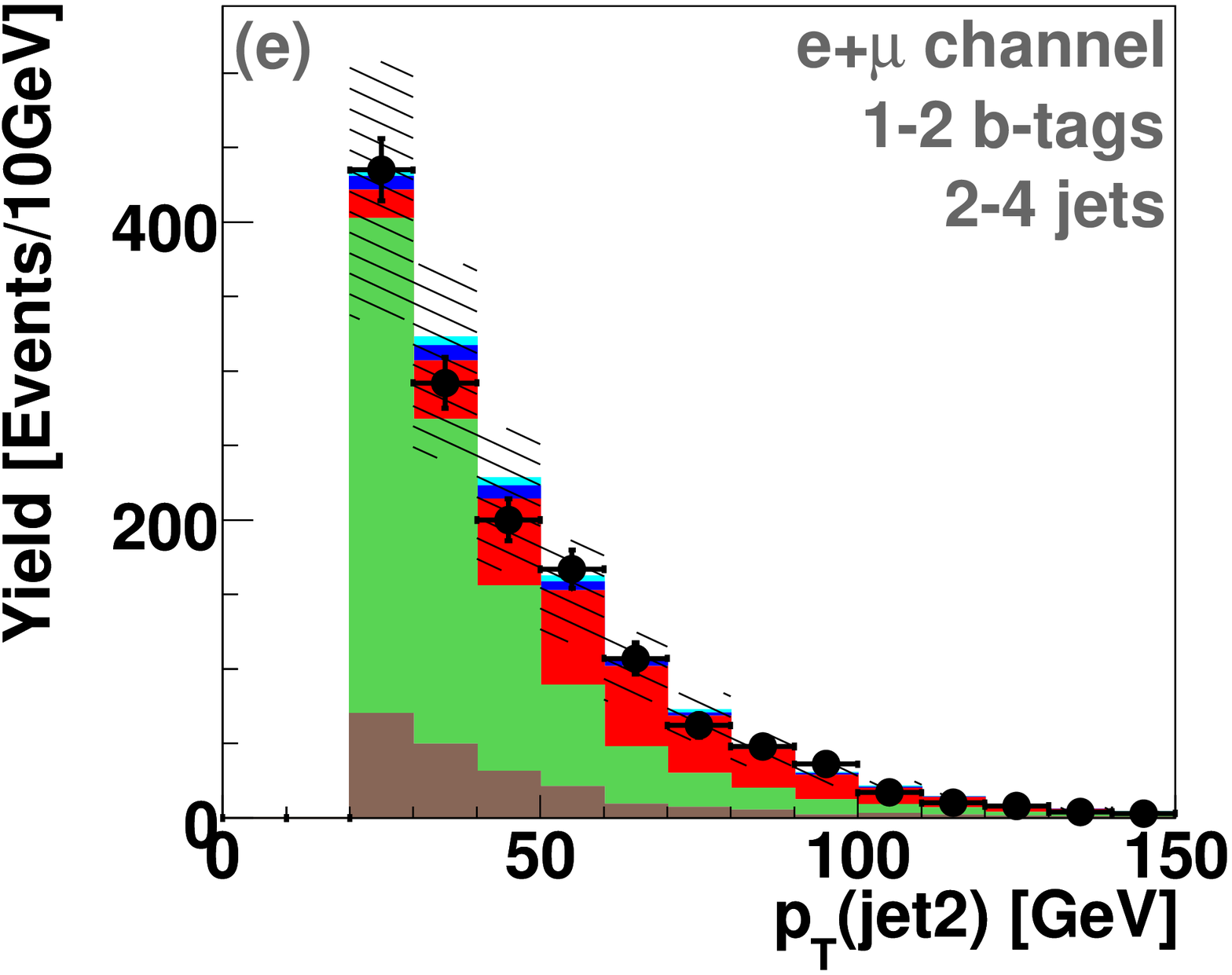}
\includegraphics[width=0.30\textwidth]
{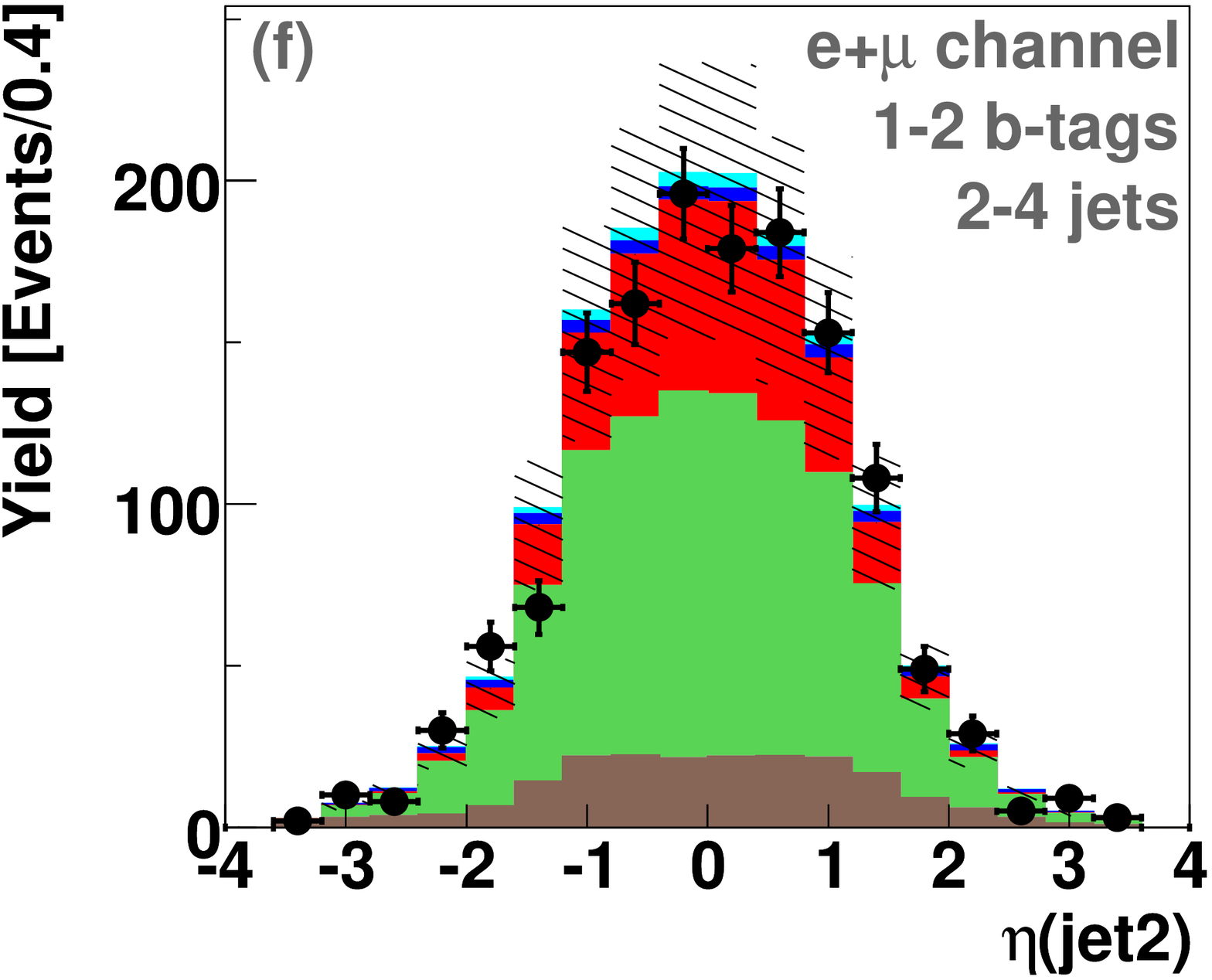}

\vspace{-0.1in}
\caption[vars_obj]{Comparison of SM signal, backgrounds, and data
after selection and requiring at least one $b$-tagged jet for six
discriminating individual object variables. Electron and muon channels
are combined. The plots show (a)~the lepton transverse momentum and
(b)~pseudorapidity, (c)~the leading jet transverse momentum and
(d)~pseudorapidity, (e)~the second leading jet transverse momentum and
(f)~pseudorapidity. The hatched area is the $\pm 1\sigma$ uncertainty
on the total background prediction.}
\label{fig:vars_obj}
\end{figure*}

\begin{figure*}[!h!btp]
\vspace{0.7in}
\includegraphics[width=0.30\textwidth]
{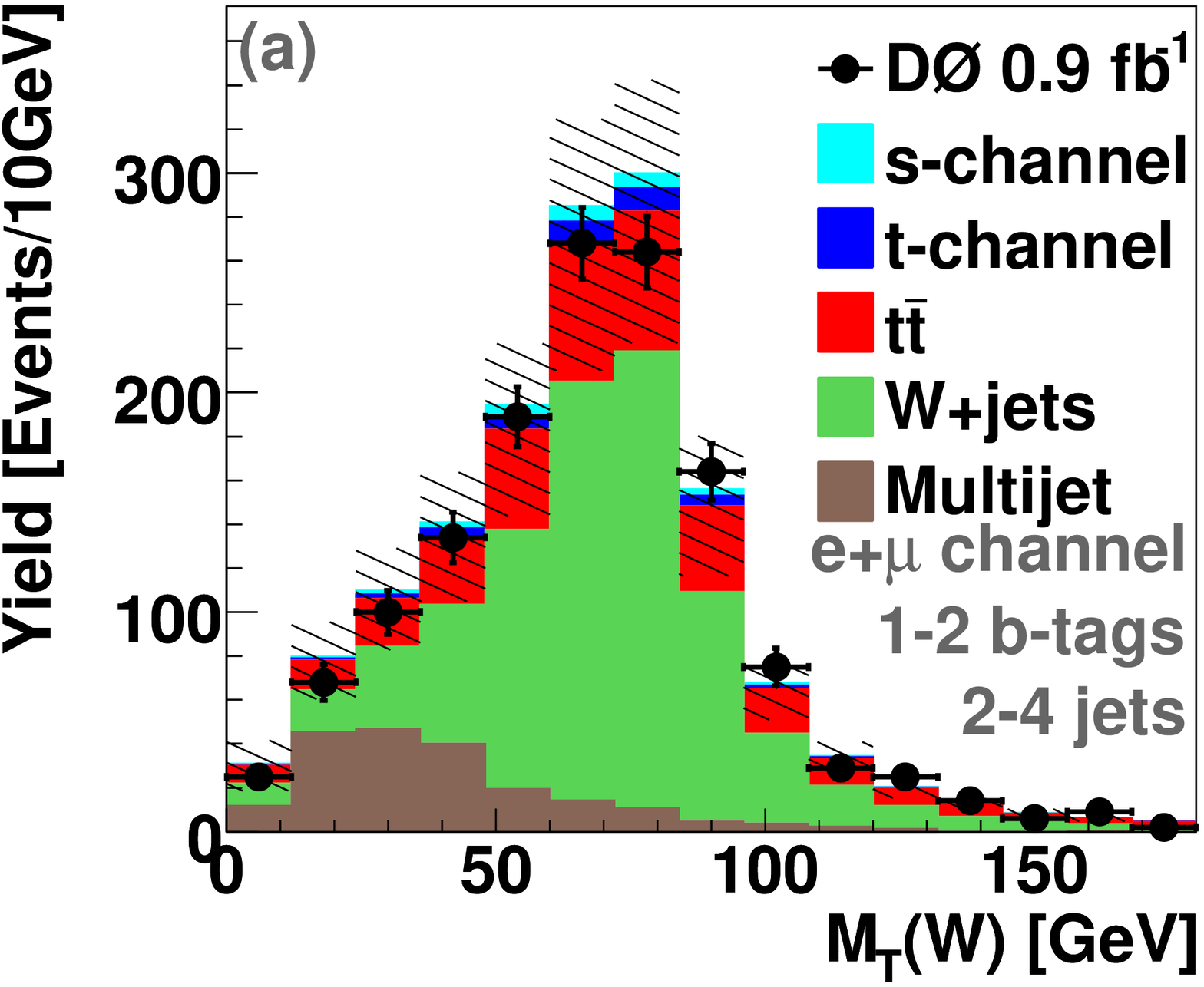}
\includegraphics[width=0.30\textwidth]
{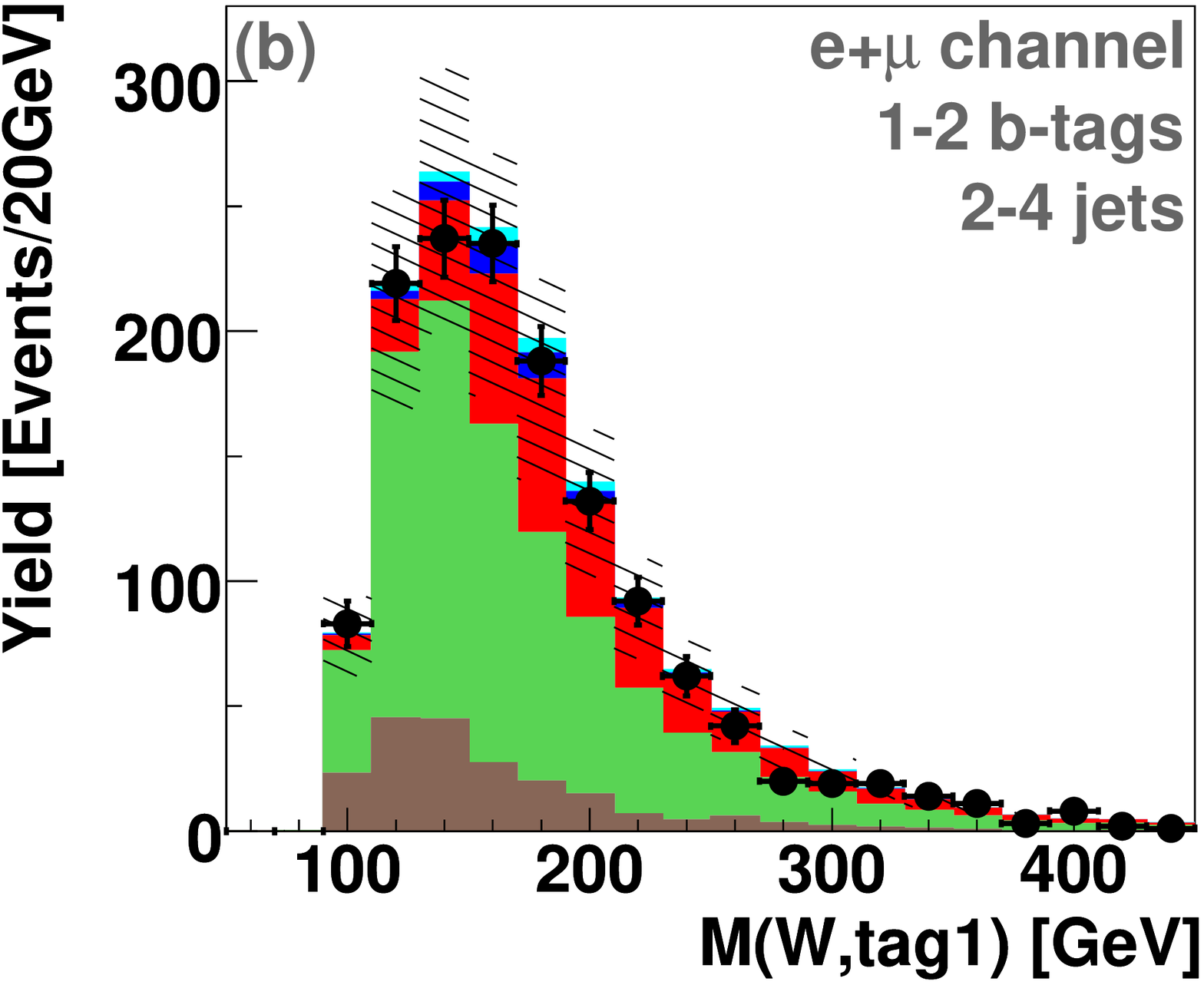}
\includegraphics[width=0.30\textwidth]
{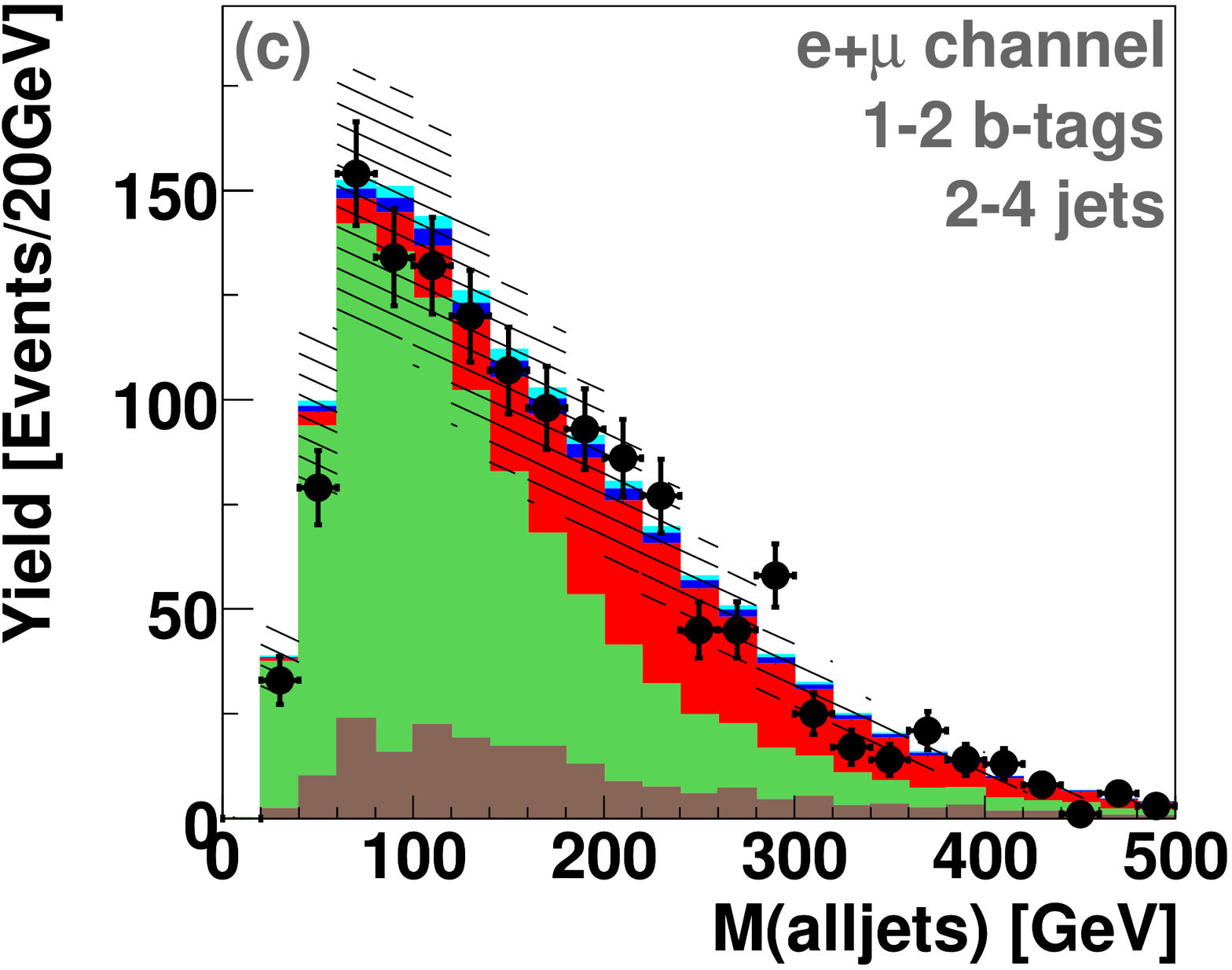}

\includegraphics[width=0.30\textwidth]
{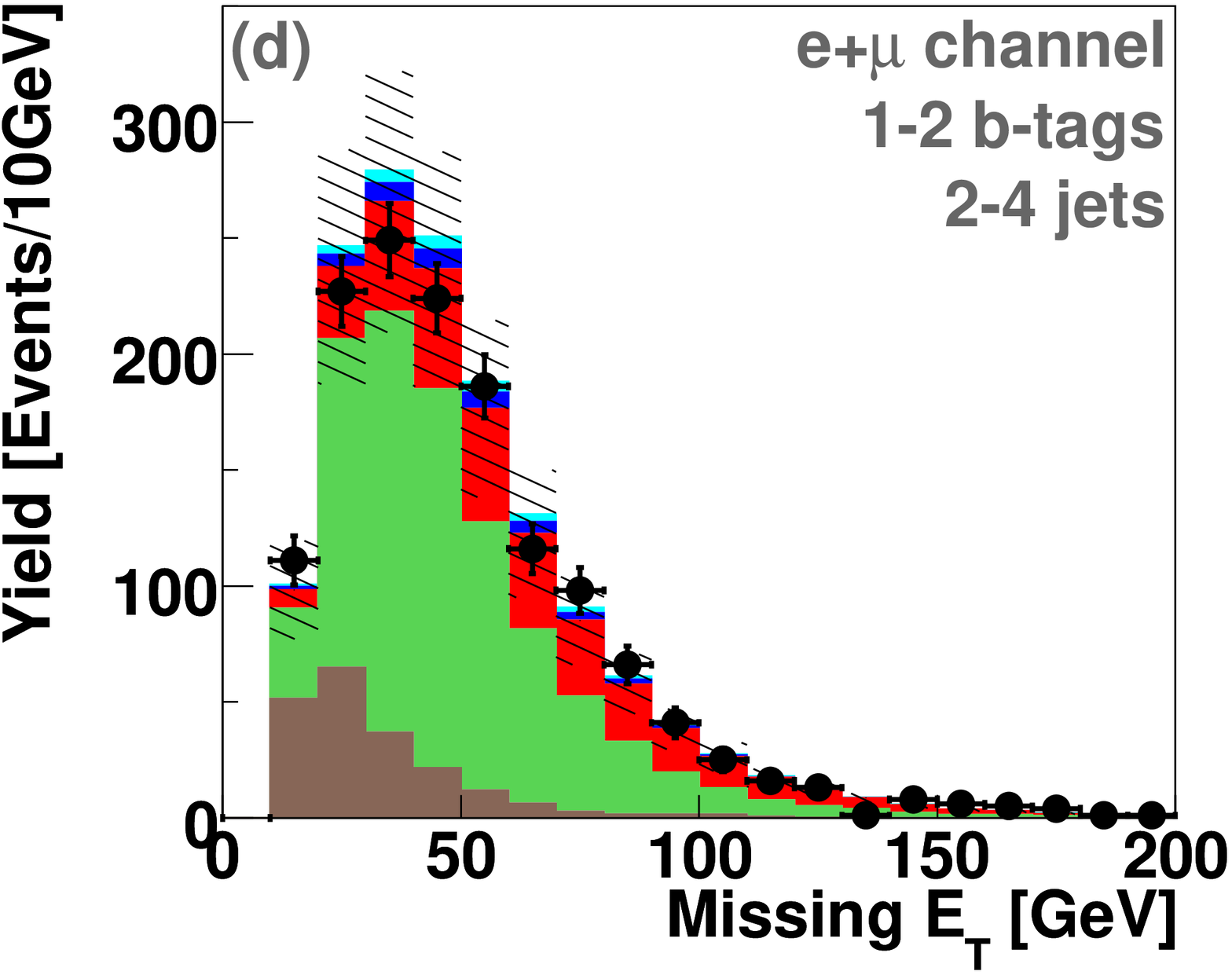}
\includegraphics[width=0.30\textwidth]
{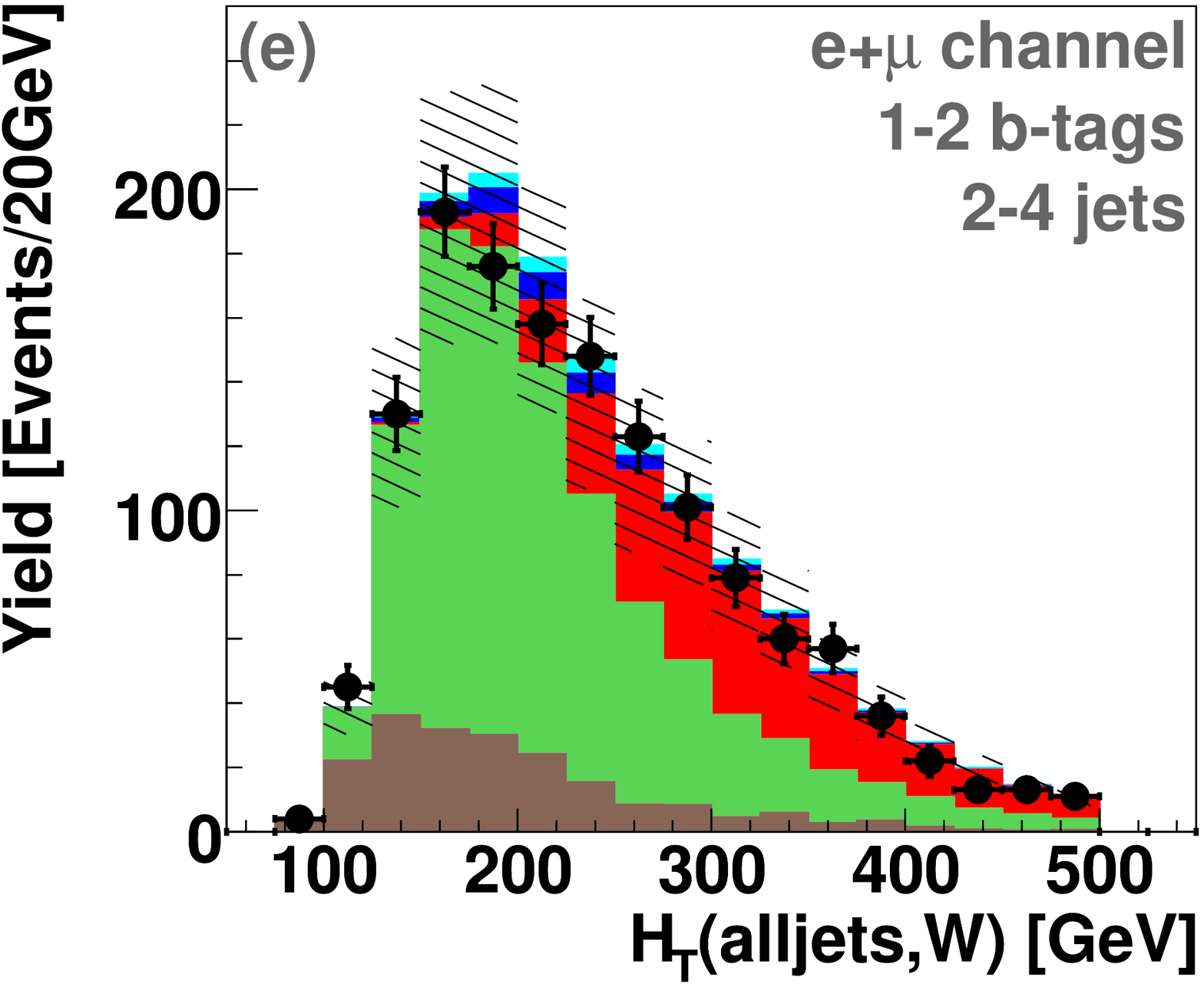}
\includegraphics[width=0.30\textwidth]
{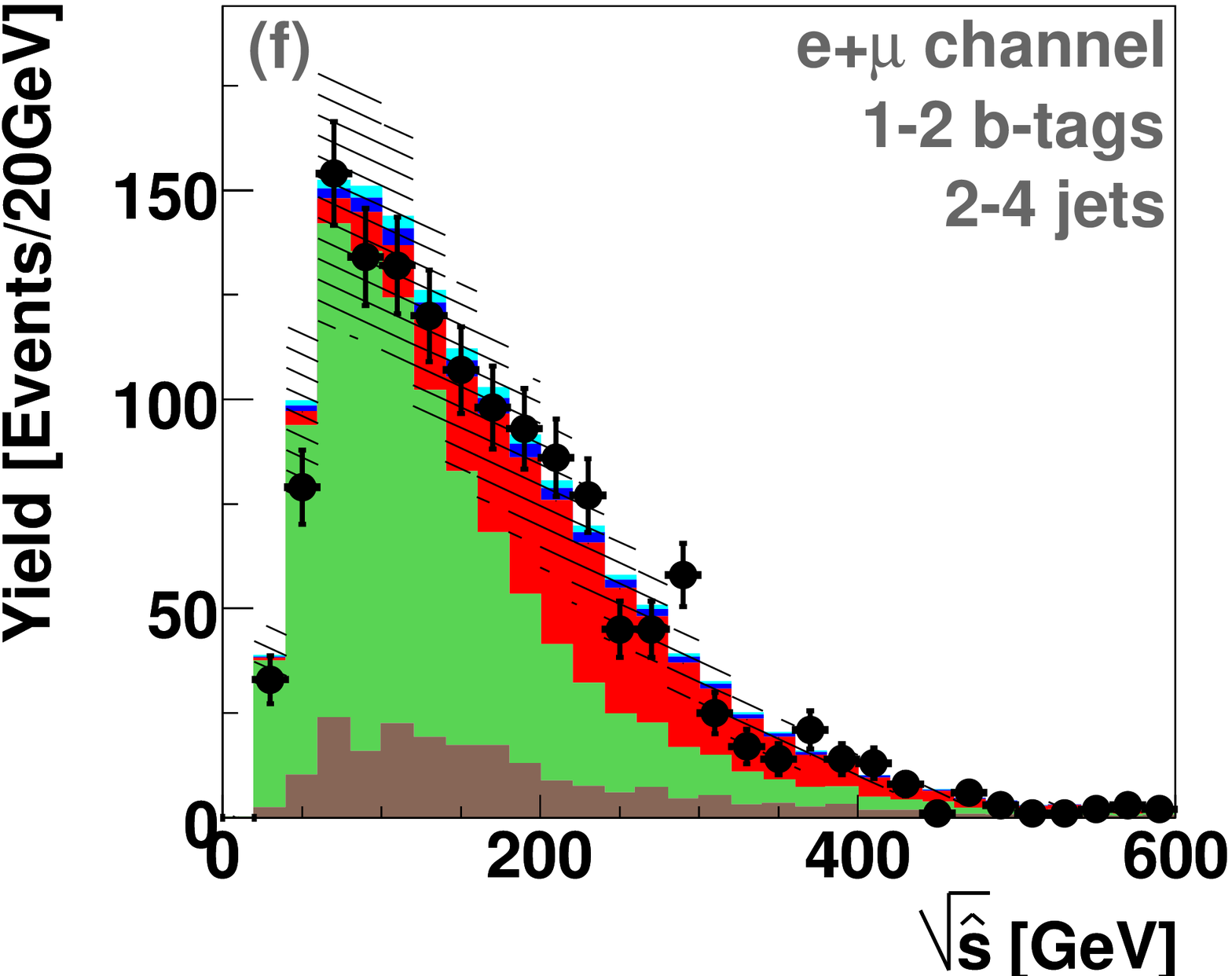}

\vspace{-0.1in}
\caption[vars_evt]{Comparison of SM signal, backgrounds, and data
after selection and requiring at least one $b$-tagged jet for six
discriminating event kinematic variables. Electron and muon channels
are combined. Shown are (a)~the invariant transverse mass of the
reconstructed $W$~boson, (b)~the invariant mass of the $b$-tagged jet
and the $W$~boson, (c)~the invariant mass of all jets, (d)~the missing
transverse energy, (e)~the scalar sum of the transverse momenta of
jets, lepton and neutrino, (f)~the invariant mass of all final state
objects. The hatched area is the $\pm 1\sigma$ uncertainty on the
total background prediction.}
\label{fig:vars_evt}
\end{figure*}

\begin{figure*}[!h!t!bp]
\includegraphics[width=0.30\textwidth]
{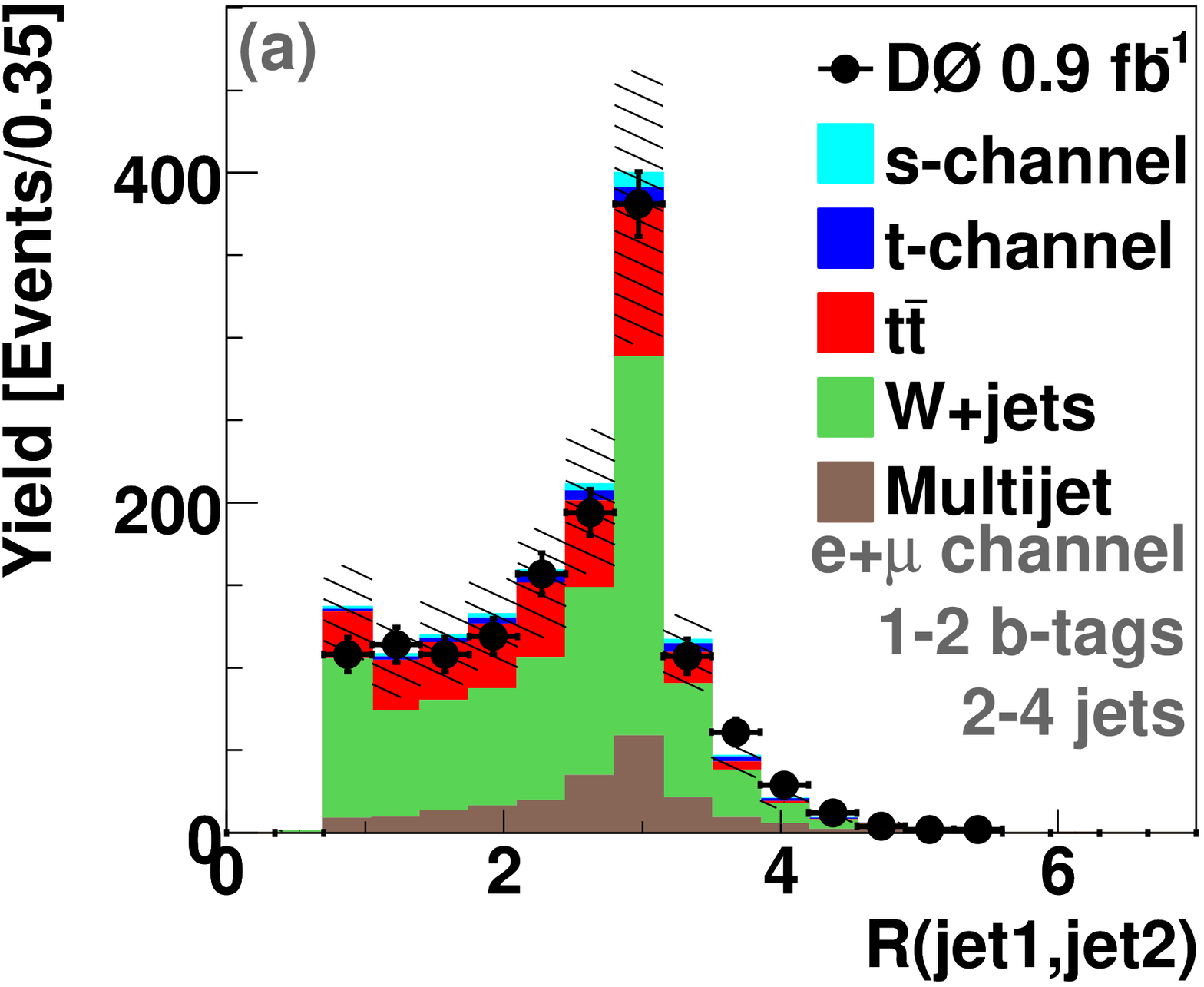}
\includegraphics[width=0.30\textwidth]
{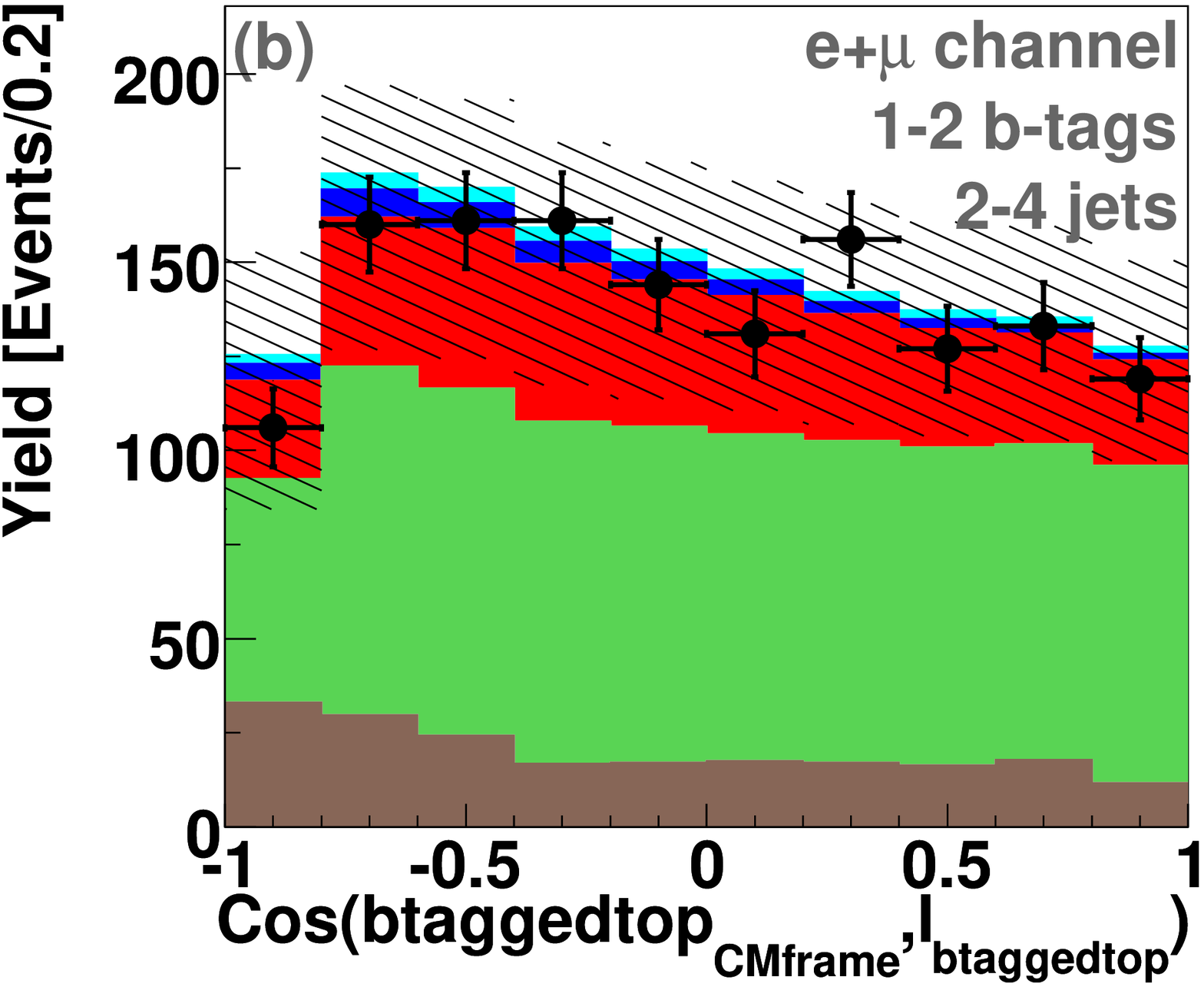}
\includegraphics[width=0.30\textwidth]
{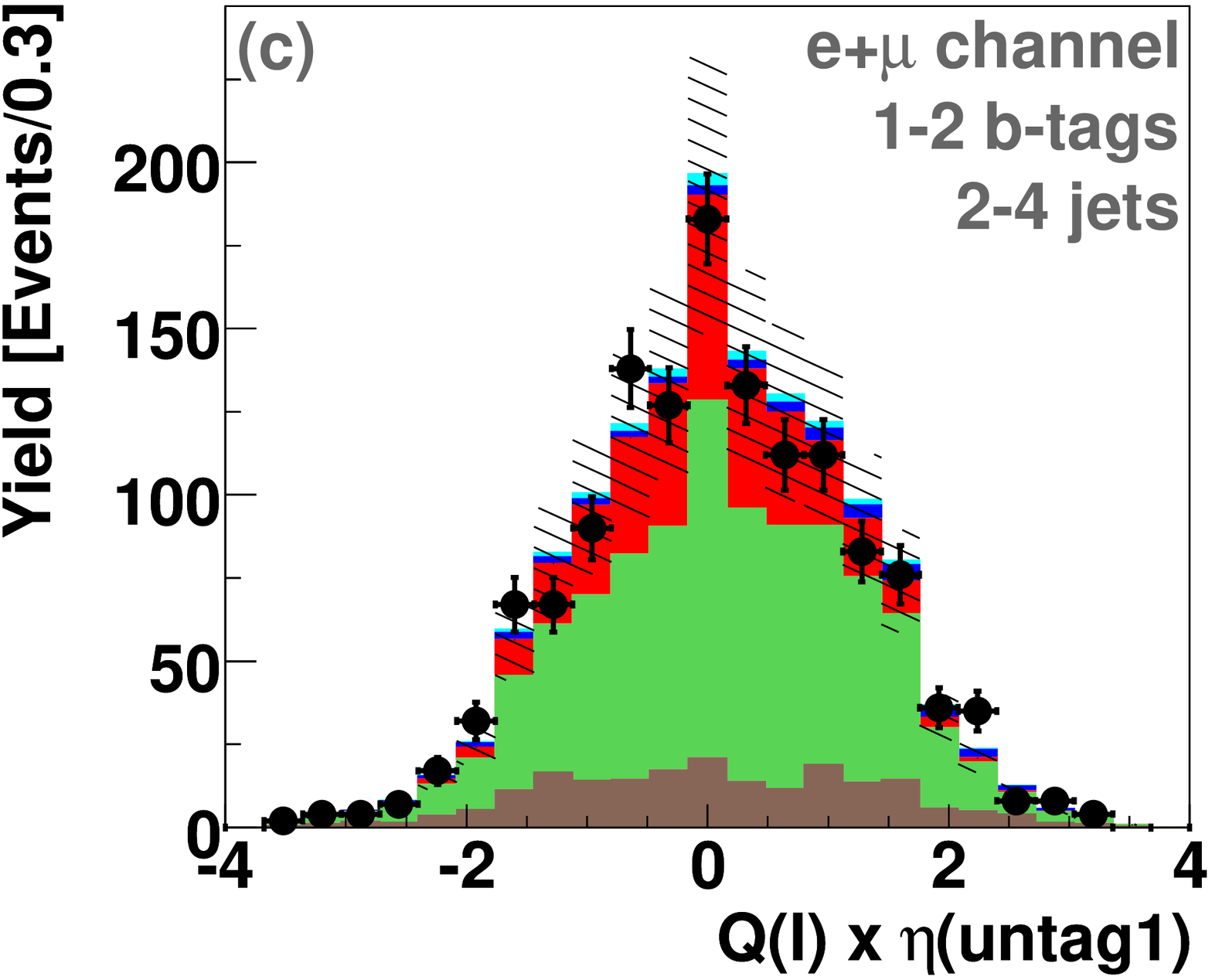}
\vspace{-0.1in}
\caption[vars_ang]{Comparison of SM signal, backgrounds, and data
after selection and requiring at least one $b$-tagged jet for three
discriminating angular correlation variables. Electron and muon
channels are combined. Shown are (a) the angular separation between
the two leading jets, (b) the cosine of the angle between the
reconstructed $b$-tagged top quark in the center-of-mass rest frame
and the lepton in the $b$-tagged top quark rest frame, and (c) the
charge of the lepton multiplied by the pseudorapidity of the leading
untagged jet. The hatched area is the $\pm 1\sigma$ uncertainty on the
total background prediction.}
\label{fig:vars_ang}
\end{figure*}

\begin{figure*}[!h!btp]
\includegraphics[width=0.30\textwidth]
{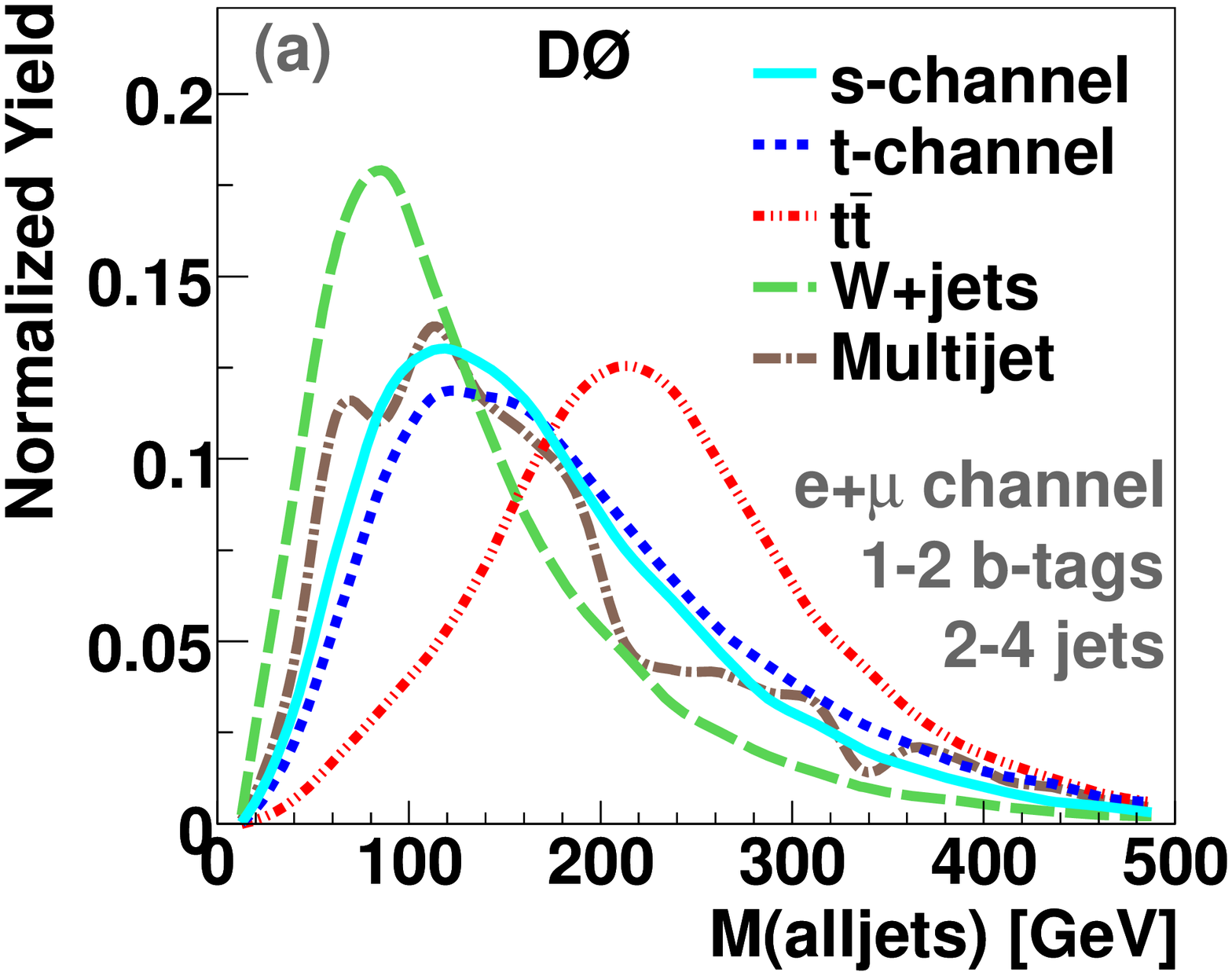}
\includegraphics[width=0.30\textwidth]
{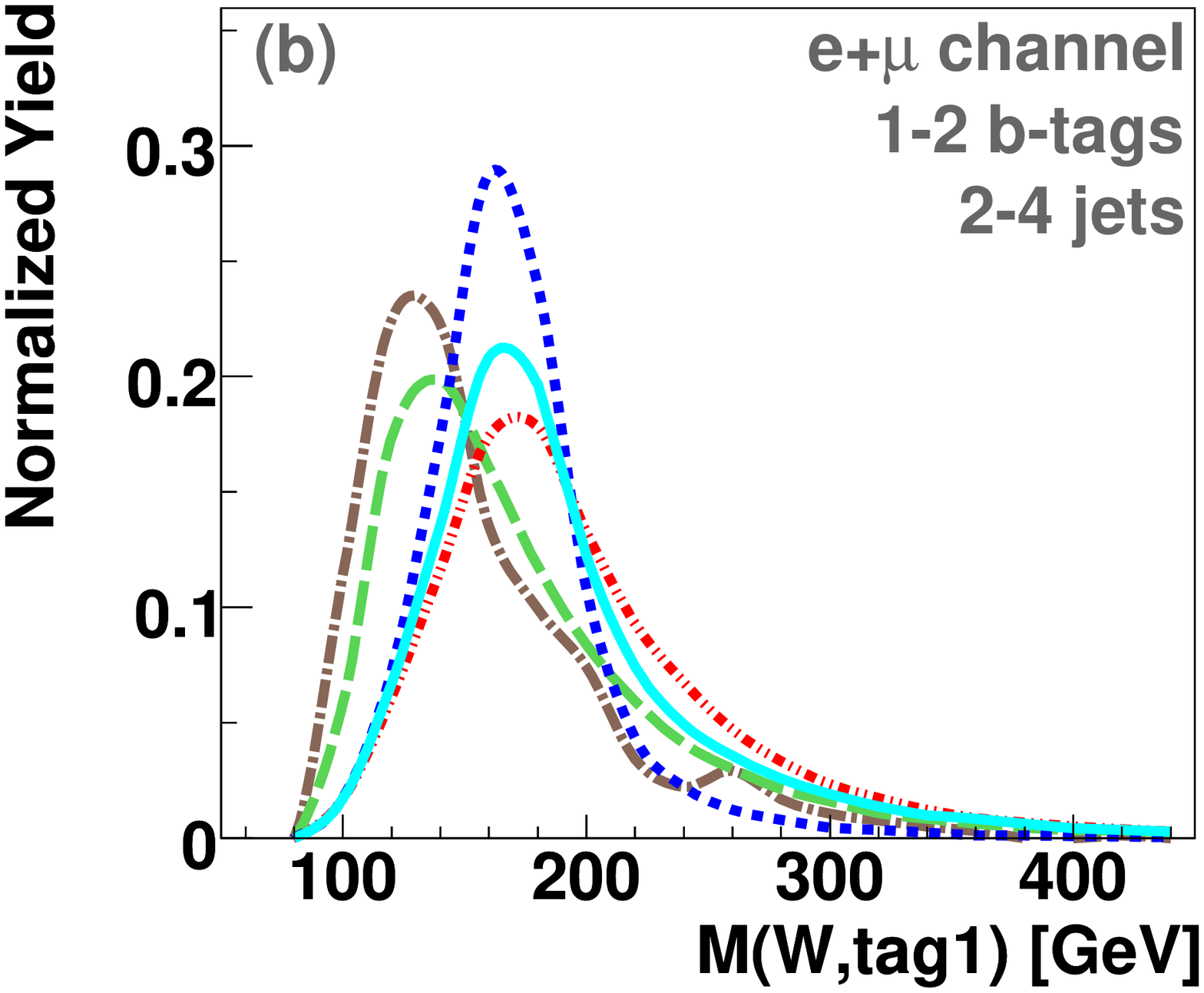}
\includegraphics[width=0.30\textwidth]
{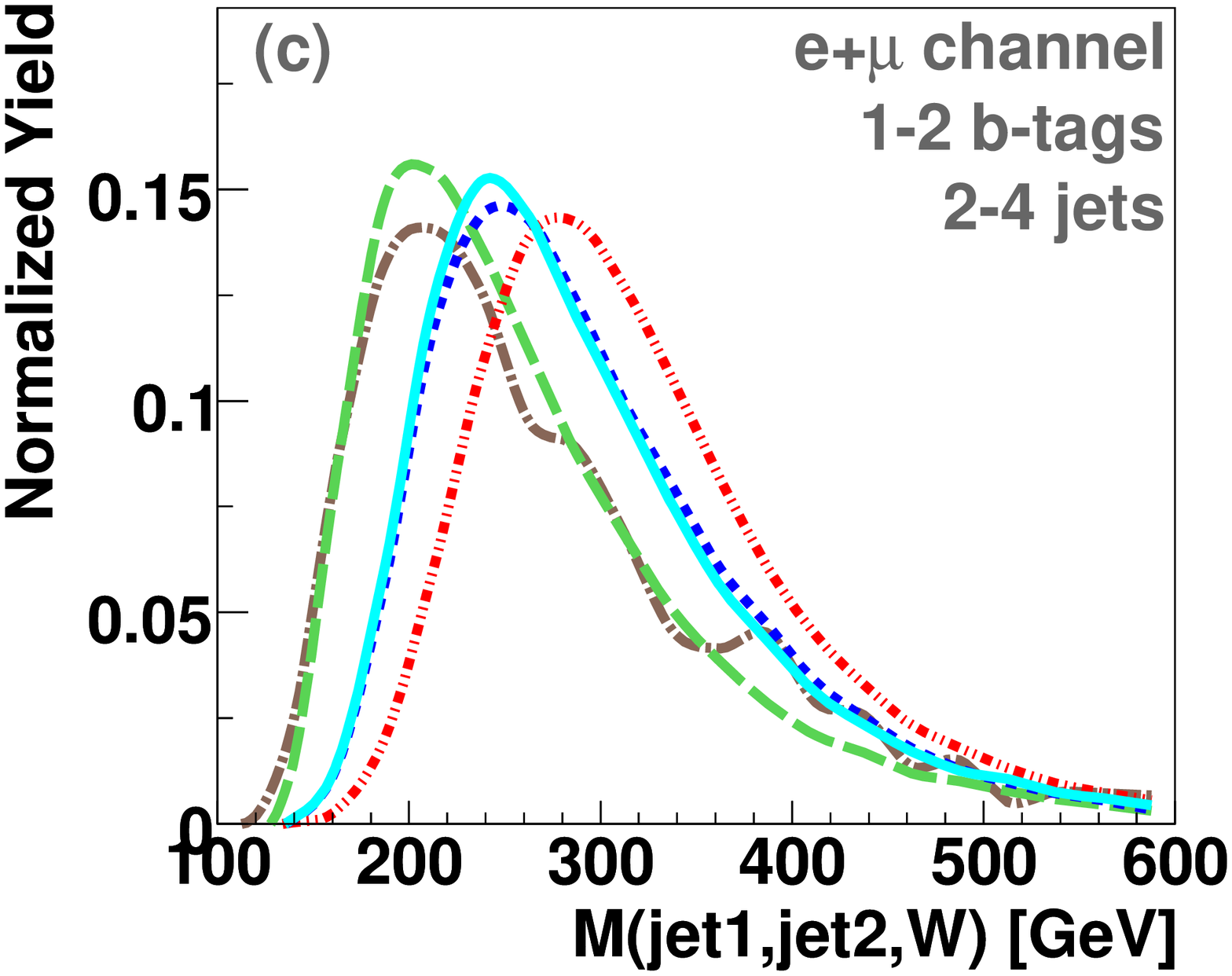}

\includegraphics[width=0.30\textwidth]
{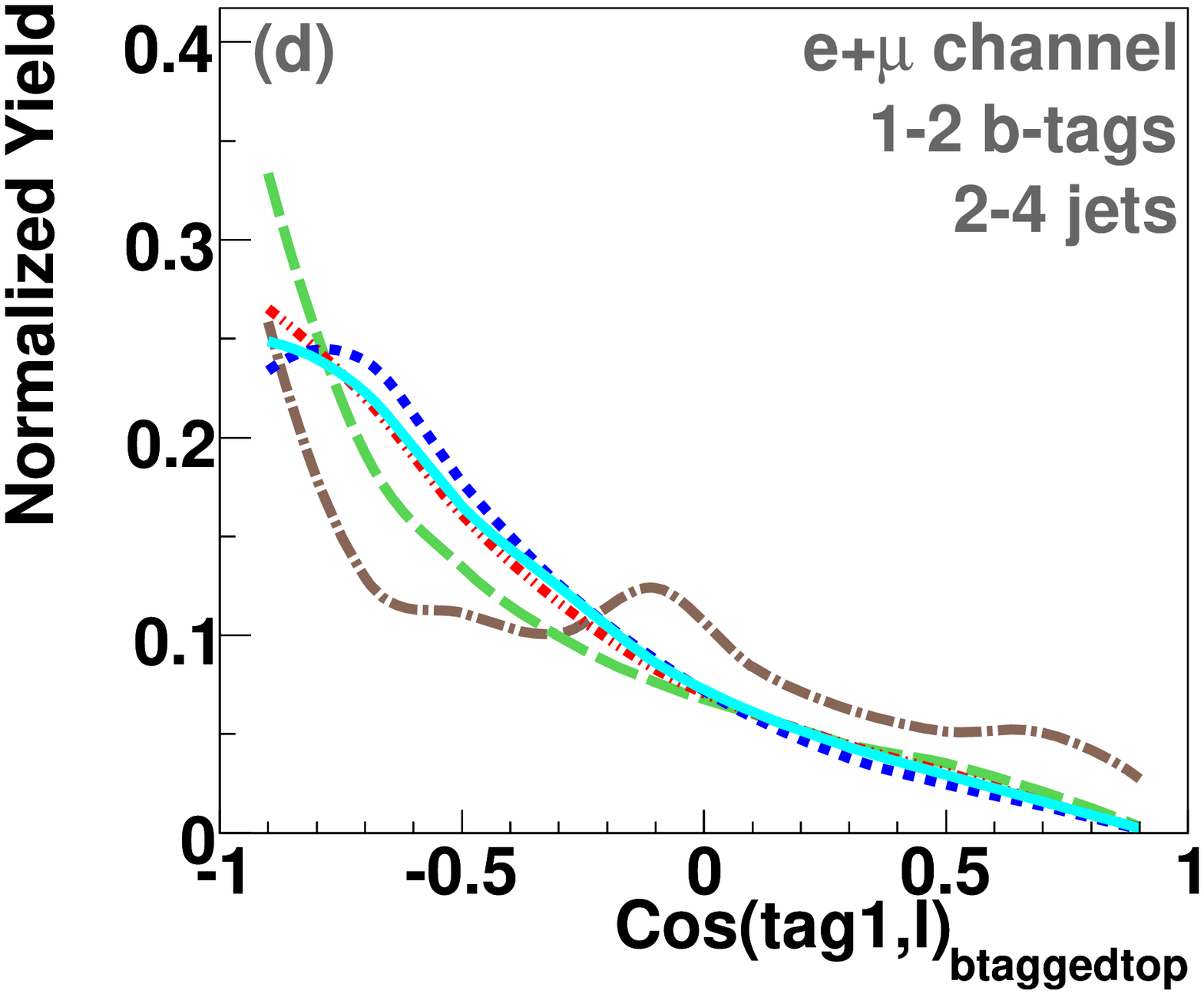}
\includegraphics[width=0.30\textwidth]
{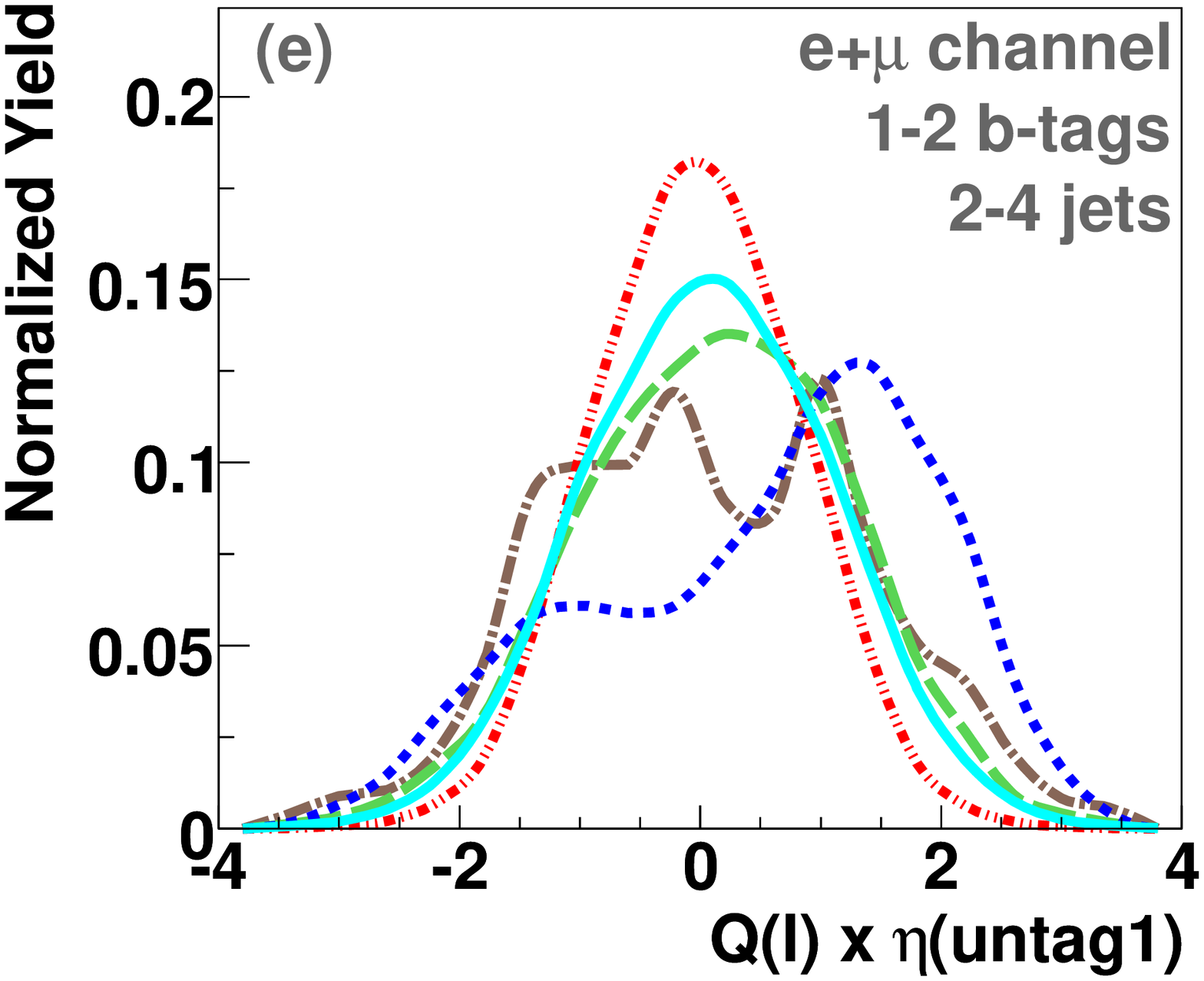}
\includegraphics[width=0.30\textwidth]
{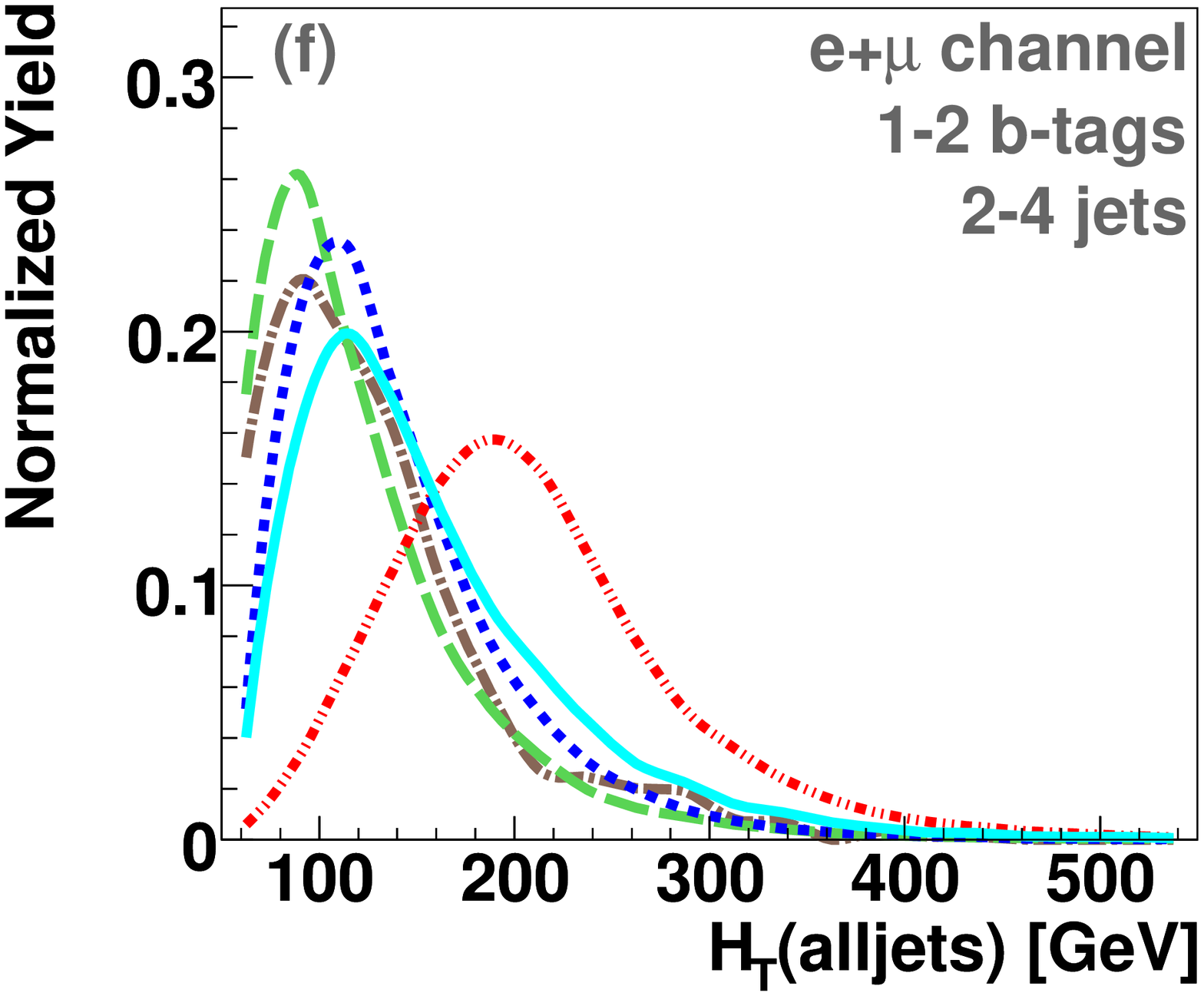}

\includegraphics[width=0.30\textwidth]
{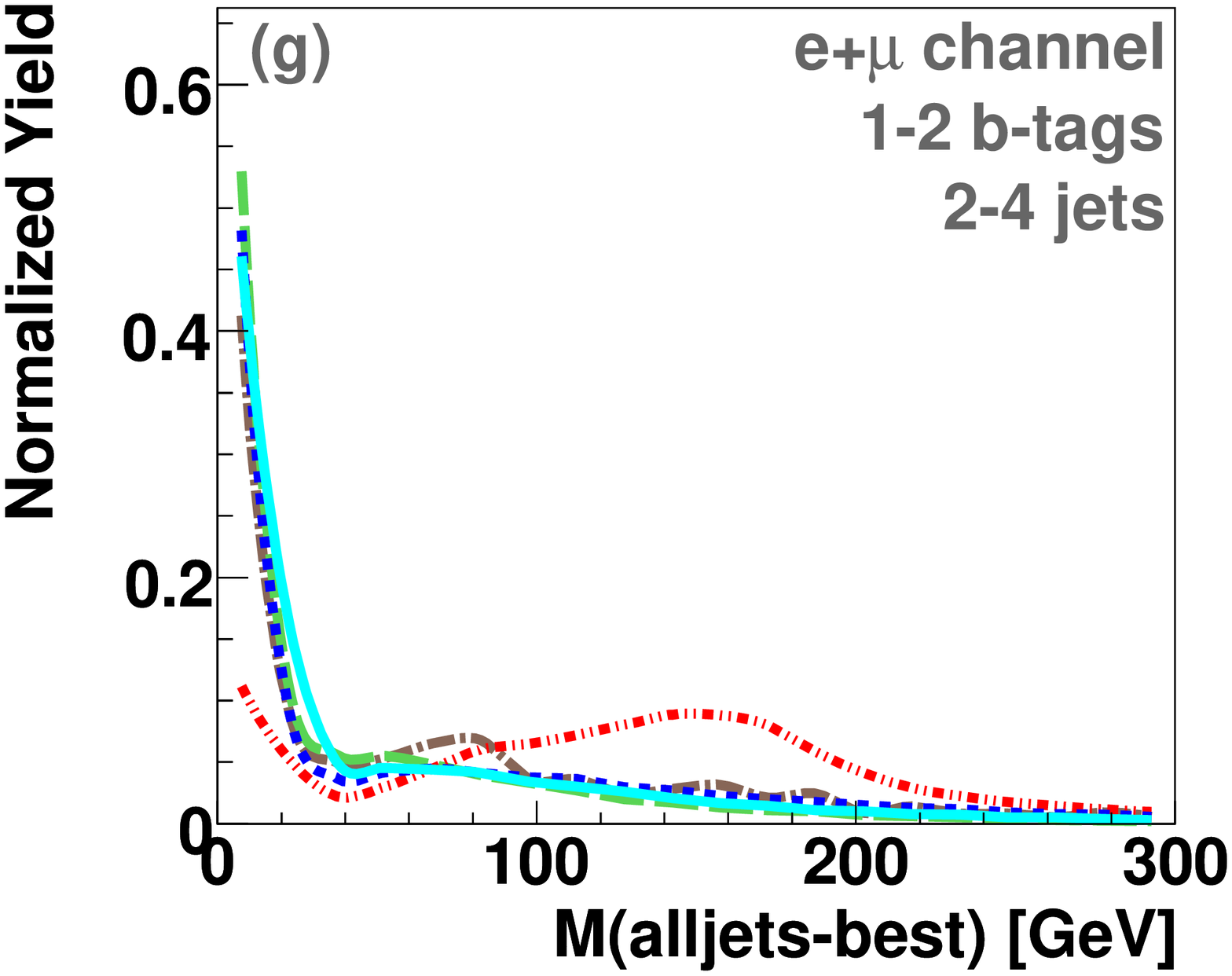}
\includegraphics[width=0.30\textwidth]
{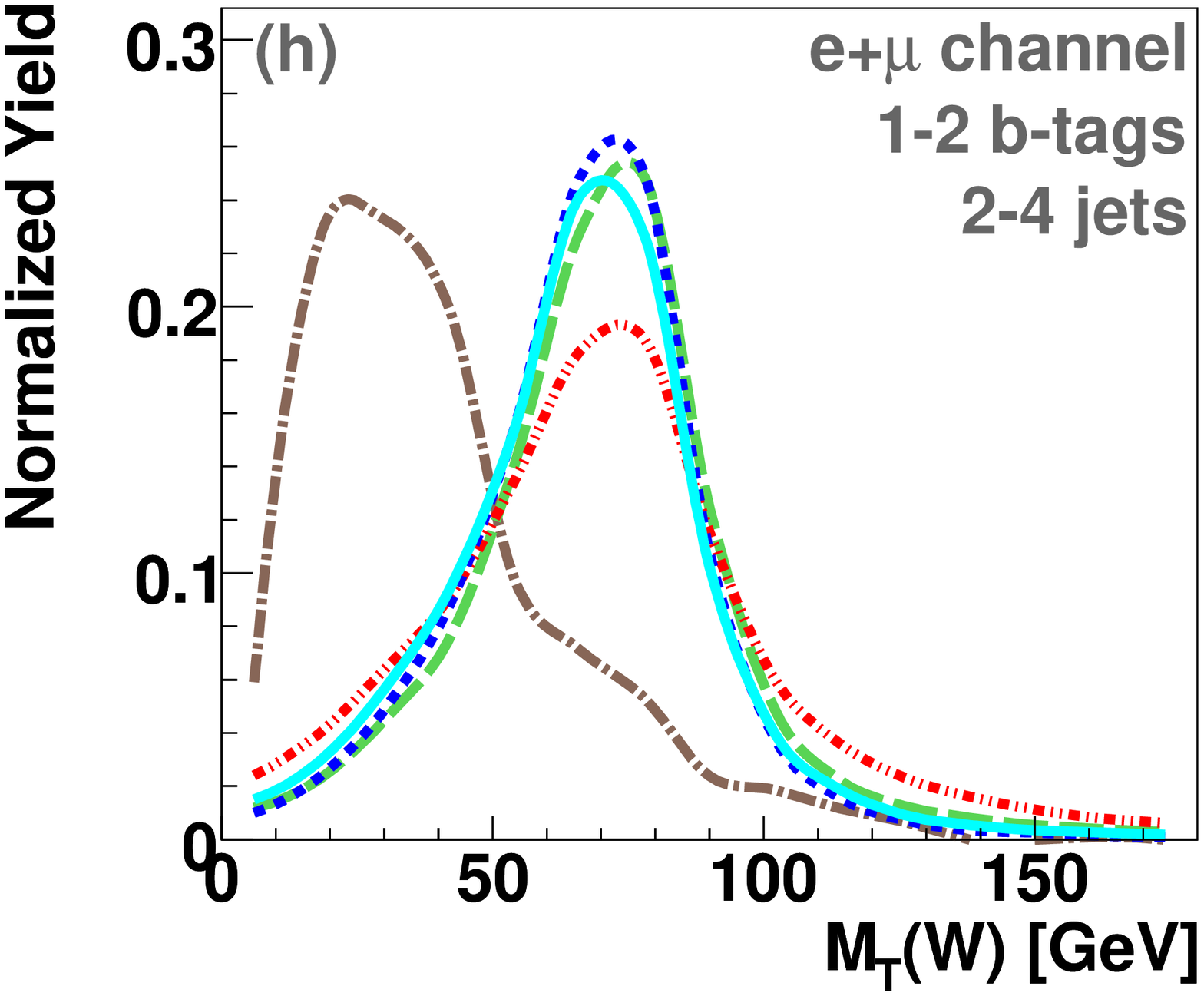}
\includegraphics[width=0.30\textwidth]
{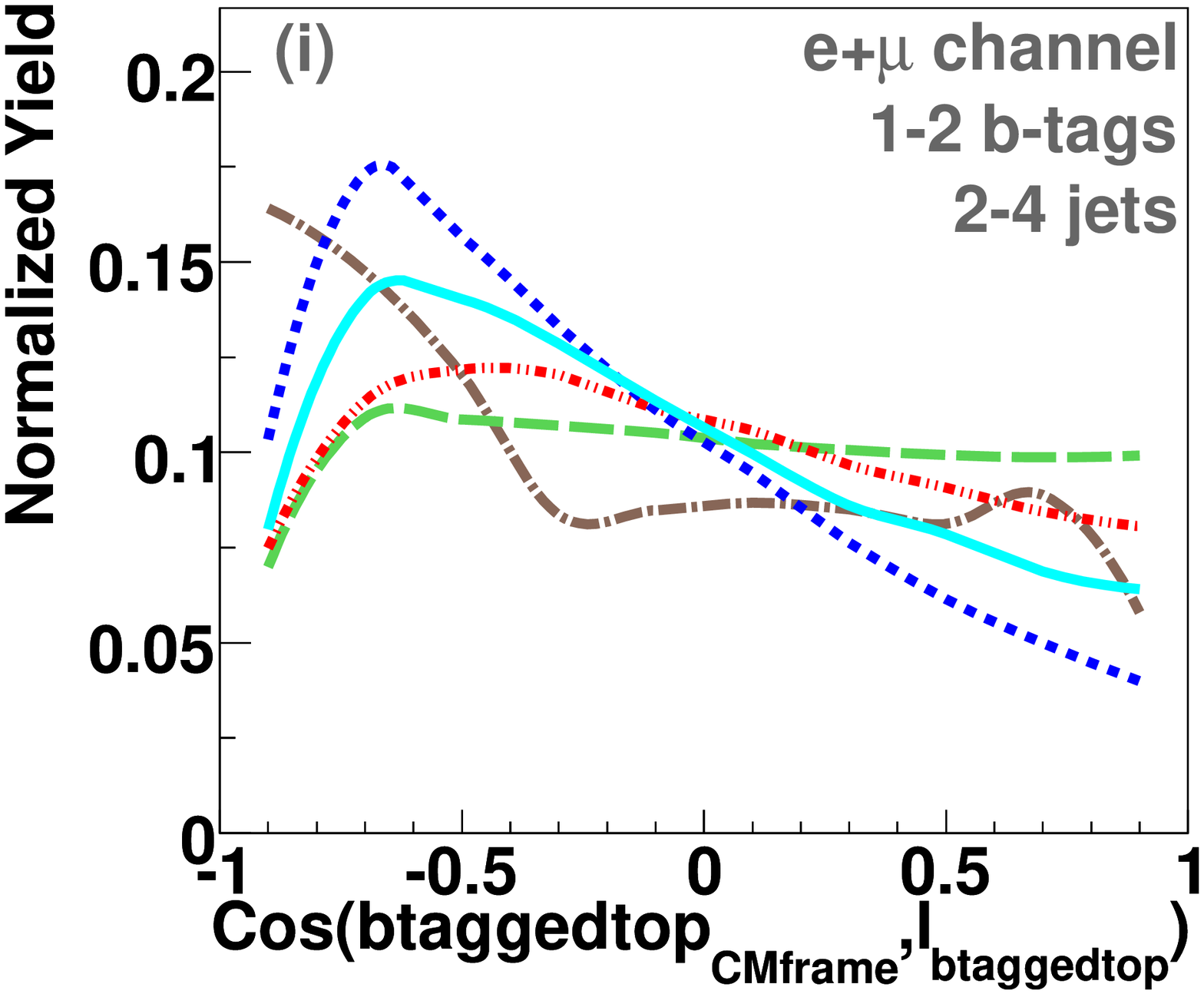}

\vspace{-0.1in}
\caption[vars_separation]{Shape comparison between the s- and
t-channel signals and the backgrounds in the most discriminating
variables for the decision tree analysis chosen from the
$e$+2jets/1tag channel. Shown are (a)~the invariant mass of all jets,
(b)~the invariant mass of the $b$-tagged jet and the $W$~boson,
(c)~the invariant mass of the two leading jets and the $W$~boson,
(d)~the cosine of the $b$-tagged jet and the lepton in the
reconstructed $b$-tagged top quark rest frame, (e)~the charge of the
lepton multiplied by the pseudorapidity of the leading untagged jet,
(f)~the scalar sum of the transverse momenta of all the jets, (g)~the
invariant mass of all jets minus the best jet, (h)~the invariant
transverse mass of the reconstructed $W$~boson, and (i)~the cosine of
the angle between the reconstructed $b$-tagged top quark in the
center-of-mass rest frame and the lepton in the $b$-tagged top quark
rest frame. All histograms are normalized to unit area.}
\label{fig:vars_separation}
\end{figure*}

A selection of these variables is shown in
Figs.~\ref{fig:vars_obj},~\ref{fig:vars_evt}, and~\ref{fig:vars_ang}
for the sum of all channels: electron plus muon channels, two to four
jets, and one or two $b$-tagged jets. Figure~\ref{fig:vars_separation}
shows distributions for some of the variables from
Table~\ref{variables} for SM signals and the background components,
normalized to unit area, so that the differences in shapes may be
seen.

\begin{table}[!h!tbp]
\caption[variables]{Variables used with the decision trees and
Bayesian neural networks analyses, in three categories: object
kinematics; event kinematics; and angular variables. For the angular
variables, the subscript indicates the reference frame. $\star$
indicates variables that were only used for the DT analysis.
$\dagger$ indicates variables only used by the BNN analysis.}
\label{variables}
\begin{ruledtabular}
\begin{tabular}{l|l}
\multicolumn{2}{c}{\underline{DT and BNN Input Variables}}\vspace{0.05in}\\
\hline
\bf{Object Kinematics}                                      & \bf{Event Kinematics}                  \\
~~$p_T$(jet1)                                           & ~~Aplanarity(alljets,$W$)          \\
~~$p_T$(jet2)                                           & ~~Sphericity(alljets,$W$)          \\
~~$p_T$(jet3)                                           & ~~Centrality(alljets)$^\dagger$    \\
~~$p_T$(jet4)                                           & ~~${\met}$                         \\
~~$p_T$(best)                                           & ~~$H$(alljets)$^\dagger$           \\
~~$p_T$(notbest1)$^\star$                               & ~~$H$(jet1,jet2)$^\dagger$         \\
~~$p_T$(notbest2)$^\star$                               & ~~$H_T$(alljets)                   \\
~~$p_T$(tag1)                                           & ~~$H_T$(alljets$-$best)$^\star$    \\
~~$p_T$(untag1)                                         & ~~$H_T$(alljets$-$tag1)$^\star$    \\
~~$p_T$(untag2)                                         & ~~$H_T$(alljets,$W$)               \\
~~$p_T$($\ell$)$^\dagger$                               & ~~$H_T$(jet1,jet2)                 \\
                                          		& ~~$H_T$(jet1,jet2,$W$)             \\
\bf{Angular Variables}                                      & ~~$M$(alljets)                     \\
~~$\cos$(jet1,alljets)$_{\rm alljets}$                  & ~~$M$(alljets$-$best)$^\star$      \\
~~$\cos$(jet2,alljets)$_{\rm alljets}$                  & ~~$M$(alljets$-$tag1)$^\star$      \\
~~$\cos$(notbest1,alljets)$_{\rm alljets}$              & ~~$M$(jet1,jet2)                   \\
~~$\cos$(tag1,alljets)$_{\rm alljets}^\star$         	& ~~$M$(jet1,jet2,$W$)               \\
~~$\cos$(untag1,alljets)$_{\rm alljets}$                & ~~$M$($W$,best)                    \\
~~$\cos$(best,notbest1)$_{\rm besttop}$                 & ~~~~(i.e., ``best'' $m_{\rm top}$) \\
~~$\cos$(best,$\ell$)$_{\rm besttop}$                  	& ~~$M$($W$,tag1)                    \\
~~$\cos$(notbest1,$\ell$)$_{\rm besttop}$               & ~~~~(i.e., ``$b$-tagged'' $m_{\rm top}$)\\
~~$\cos$(\qtimesz,$\ell$)$_{\rm besttop}$               & ~~$M_T$(jet1,jet2)                 \\
~~$\cos$(besttop$_{\rm CMframe},\ell_{\rm besttop}$)    & ~~$p_T$($W$)$^\dagger$             \\
~~$\cos$(jet1,$\ell$)$_{\rm btaggedtop}$                & ~~$p_T$(alljets$-$best)            \\
~~$\cos$(jet2,$\ell$)$_{\rm btaggedtop}$                & ~~$p_T$(alljets$-$tag1)            \\
~~$\cos$(tag1,$\ell$)$_{\rm btaggedtop}$                & ~~$p_T$(jet1,jet2)                 \\
~~$\cos$(untag1,$\ell$)$_{\rm btaggedtop}$              & ~~$M_T$($W$)                       \\
~~$\cos$(btaggedtop$_{\rm CMframe},\ell_{\rm btaggedtop}$) & ~~$Q$($\ell$)$\times \eta$(untag1) \\
~~$\cos$(jet1,$\ell$)$_{\rm lab}^\dagger$               & ~~$\sqrt{\hat{s}}$                 \\
~~$\cos$(jet2,$\ell$)$_{\rm lab}^\dagger$              	& \\
~~$\cos$(best,$\ell$)$_{\rm lab}^\dagger$              	& \\
~~$\cos$(tag1,$\ell$)$_{\rm lab}^\dagger$               & \\
~~$\mathcal{R}$(jet1,jet2)$^\star$                      &
\end{tabular}
\end{ruledtabular}
\end{table}

\clearpage
%---------------------------------------------------------------------
%---------------------------------------------------------------------
\section{Boosted Decision Trees Analysis}
\label{sec:DecisionTree} 

A decision tree~\cite{decisiontrees-breiman,decisiontrees-bowserchao}
employs a machine-learning technique that effectively extends a simple
cut-based analysis into a multivariate algorithm with a continuous
discriminant output. Boosting is a process that can be used on any
weak classifier (defined as any classifier that does a little better
than random guessing). In this analysis, we apply the boosting
procedure to decision trees in order to enhance separation of signal
and background.

\subsection{Decision Tree Algorithm}
\label{sec:dtalgo}

\begin{figure}[!h!tbp]
\includegraphics[width=0.38\textwidth]{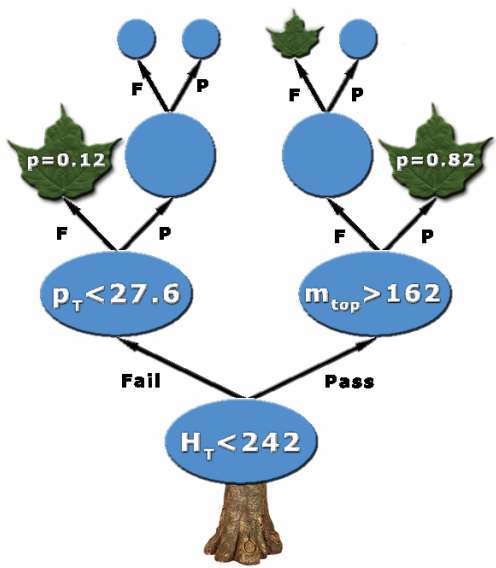}
\vspace{-0.1in}
\caption[dt]{Graphical representation of a decision tree. 
Nodes with their associated splitting test are shown as (blue) circles
and terminal nodes with their purity values are shown as (green)
leaves. An event defined by variables ${\bf{x_i}}$ of which $H_T <
242$~GeV and $m_{\mathrm{top}} > 162$~GeV will return $D({\bf{x_i}}) =
0.82$, and an event with variables ${\bf{x_j}}$ of which $H_T \ge
242$~GeV and $p_T \ge 27.6$~GeV will have $D({\bf{x_j}}) = 0.12$. All
nodes continue to be split until they become leaves. (color online)}
\label{fig:dt}
\end{figure}

A decision tree classifies events based on a set of cumulative
selection criteria (cuts) that define several disjoint subsets of
events, each with a different signal purity. The decision tree is
built by creating two {\em branches} at every nonterminal node, i.e.,
splitting the sample of events under consideration into two subsets
based on the most discriminating selection criterion for that sample.
Terminal nodes are called {\em leaves} and each leaf has an assigned
purity value $p$. A simple decision tree is illustrated in
Fig.~\ref{fig:dt}. An event defined by variables $\mathbf{x}$ will
follow a unique path through the decision tree and end up in a
leaf. The associated purity $p$ of this leaf is the decision tree
discriminant output for the event: $D(\mathbf{x})=p$, with
$D(\mathbf{x})$ given in Eq.~\ref{eq:D}.
 
One of the primary advantages of decision trees over a cut-based
analysis is that events which fail an individual cut continue to be
considered by the algorithm. Limitations of decision trees include the
instability of the tree structure with respect to the training sample
composition, and the piecewise nature of the output. Training on
different samples may produce very different trees with similar
separation power. The discrete output comes from the fact that the
only possible values are the purities of each leaf and the number of
leaves is finite.
 
Decision tree techniques have interesting features, as follows: the
tree has a human-readable structure, making it possible to know why a
particular event was labeled signal or background; training is fast
compared to neural networks; decision trees can use discrete variables
directly; and no preprocessing of input variables is necessary. In
addition, decision trees are relatively insensitive to extra
variables: unlike neural networks, adding well-modeled variables that
are not powerful discriminators does not degrade the performance of
the decision tree (no additional noise is added to the system).

\subsubsection{Training}
\label{sec:dttraining}

The process in which a decision tree is created is usually referred to
as {\em decision tree training}. Consider a sample of known signal and
background events where each event is defined by a weight~$w$ and a
list of variables~$\mathbf{x}$. The following algorithm can be applied
to such a sample in order to create a decision tree:
\begin{myenumerate}

\item Initially normalize the signal training sample to the
background training sample such that $\sum w_{\rm signal} = \sum
w_{\rm background}$.

\item Create the first node, containing the full sample.

\item Sort events according to each variable in turn. For each
variable, the splitting value that gives the best signal-background
separation is found (more on this in the next section). If no split
that improves the separation is found, the node becomes a leaf.

\item The variable and split value giving the best separation are 
selected and the events in the node are divided into two subsamples
depending on whether they pass or fail the split criterion. These
subsamples define two new child nodes.

\item If the statistics are too low in any node, it becomes a leaf.

\item Apply the algorithm recursively from Step 3 until all
nodes have been turned into leaves.

\end{myenumerate}

Each of the leaves is assigned the purity value
\begin{equation}
p=\frac{s}{s+b},
\end{equation}
where $s$ ($b$) is the weighted sum of signal (background) events in
the leaf. This value is an approximation of the discriminant
$D(\mathbf{x})$ defined in Eq.~\ref{eq:D}.

\subsubsection{Splitting a Node}
\label{sec:dtsplitting}

Consider an impurity measure $i(t)$ for node $t$. Desirable features
of such a function are that it should be maximal for an equal mix of
signal and background (no separation), minimal for nodes with either
only signal or only background events (perfect separation), symmetric
in signal and background purity, and strictly concave in order to
reward purer nodes. Several such functions exist in the literature.
We have not found a significant advantage to any specific choice and
hence use the common ``Gini index''~\cite{gini-index}.

The impurity measure, or Gini index, is defined as
\begin{equation}
i_\mathrm{Gini} = 2p(1-p) = \frac{2sb}{(s+b)^2},
\end{equation}
where $s$ ($b$) is the sum of signal (background) weights in a
node. One can now define the decrease of impurity (goodness of split)
associated with a split $S$ of node $t$ into children $t_P$ and $t_F$:
\begin{equation}
\Delta i_{\rm Gini}(S,t)
     = i_{\rm Gini}(t)- p_P \cdot
       i_{\rm Gini}(t_P) - p_F\cdot i_{\rm Gini}(t_F),
\end{equation}
where $p_P$ ($p_F$) is the fraction of events that passed (failed)
split $S$. The goal is to find the split $S^*$ that maximizes the
decrease of impurity, which corresponds to finding the split that
minimizes the overall tree impurity.

\subsubsection{Boosting}
\label{sec:dtboosting}
 
A powerful technique to improve the performance of any weak classifier
was introduced a decade ago: boosting~\cite{boosting-freund}. Boosting
was recently used in high energy physics with decision trees by the
MiniBooNE experiment~\cite{boosting-roe,boosting-yang}.

The basic principle of boosted decision trees is to train a tree,
minimize some error function, and create a tree $T_{n+1}$ as a
modification of tree $T_n$. The boosting algorithm used in D0's single
top quark search is adaptive boosting, known in the literature as
AdaBoost~\cite{boosting-freund}.

Once a tree indexed by $n$ is built with associated discriminant
$D_n(\mathbf{x})$, its associated error $\epsilon_n$ is calculated as
the sum of the weights of the {\em misclassified} events. An event is
considered misclassified if $|D_n(\mathbf{x})-y|>0.5$ where $y$ is 1
for a signal event and 0 for background. The tree weight is calculated
according to
\begin{equation}
\label{eq:TreeWeight}
\alpha_n=\beta \times \ln \frac{1-\epsilon_n}{\epsilon_n},
\end{equation}
where $\beta$ is the boosting parameter. For each misclassified event,
its weight $w_i$ is scaled by the factor $e^{\alpha_n}$ (which will be
greater than 1). Hence misclassified events will get higher weights. A
new tree indexed by $n+1$ is created from the reweighted training
sample now working harder on the previously misclassified events. This
is repeated $N$ times, where $N$, the number of boosting cycles, is a
parameter specified by the user. The final boosted decision tree
result for event $i$ is
\begin{equation}
\label{eq:BDToutput}
D(x_i) = \frac{1}{\sum^{N}_{n=0}\alpha_n}
         \sum^{N}_{n=0}\alpha_nD_n(x_i).
\end{equation}

In all of our tests, boosting improves performance. Another advantage
of boosting decision trees is that averaging produces smoother
approximations to $D(\mathbf{x})$. In this analysis 20 boosted trees
are used for each analysis channel, which improves the performance by
5 to 10\%. The increase in performance saturates in the region of 20
boosting cycles, varying slightly from channel to channel.

\subsubsection{Decision Tree Parameters}

Several internal parameters can influence the development of a
decision tree.

\begin{myitemize}

\item Initial normalization. Step 1 in Sec.~\ref{sec:dttraining}. In
this analysis, we normalize both signal and background such that their
sums of weights are both 0.5.

\item Criteria to decide when to stop the splitting procedure 
due to too low statistics (Step 3 in Sec.~\ref{sec:dttraining}).
In this analysis the minimum node size is 100 events per node. 

\item Impurity function to use to find the best split. We use 
the Gini index as mentioned in Sec.~\ref{sec:dtsplitting}.

\item Number of boosting cycles. For this analysis we use 20
boosting cycles.

\item Value of the boosting parameter $\beta$. We find
$\beta = 0.2$ gives the best expected separation.

\end{myitemize}

\subsection{Variable Selection}
\label{sec:dtvariables}

A list of sensitive variables has been derived based on an analysis of
the signal and background Feynman
diagrams~\cite{wbb-boos,variables-dudko,variables-boos}, from studies
of single top quark production at next-to-leading
order~\cite{singletop-xsec-NLO-1,singletop-xsec-NLO-2}, and from other
analyses~\cite{t-channel-yuan,spin-variables-mahlon}. The variables
fall into three categories: individual object kinematics, global event
kinematics, and variables based on angular correlations. The complete
list of 49 variables is shown in Table~\ref{variables}.

Previous iterations of the single top quark analysis at
D0~\cite{run1-d0-plb,run2-d0-plb,run2-d0-prd} have always used far
fewer input variables. One of the main reasons was that the
discriminant was computed with neural networks. Introducing too many
variables can degrade the performance of a network, and testing each
combination of variables is time-consuming. However, we observe that
adding more variables does not degrade the DT performance. If newly
introduced variables have some discriminative power, they improve the
performance of the tree. If they are not discriminative enough, they
are ignored. We tested this empirical observation by training
different trees using several subsets of variables from the list of 49
variables. Adding more variables to the training sets never degraded
the performance of the trees. Therefore, rather than producing
separately optimized lists of variables for each analysis channel, the
full list of 49 variables is used in all cases.

\subsection{Decision Tree Training}

We train the decision trees on one third of the available simulated
events and keep the rest of the events to measure the acceptances. As
a cross check, we have also trained on one half and on two thirds of
the sample and have found consistent results with those obtained from
using only one third. We therefore only present results with one third
of the sample used for training.

Three signals are considered:
\vspace{-0.05in}
\begin{myitemize}
\item s-channel single top quark process only ($tb$)
\item t-channel single top quark process only ($tqb$)
\item s- and t-channel single top quark processes combined
      ($tb$+$tqb$)
\end{myitemize}
For simplicity, and because the decision trees are expected to deal
well with all components at once, trees are trained against all
backgrounds together rather than making separate trees for each
background. The background includes simulated events for
${\ttbar}{\rar}\ell$+jets, ${\ttbar}{\rar}\ell\ell$+jets, and $W$+jets
(with three separate components for $W{\bbbar}$, $W{\ccbar}$ and
$Wjj$). Each background component is represented in proportion to its
expected fraction in the background model. This leads to three
different decision trees: ($tb$, $tqb$, $tb$+$tqb$ against ${\ttbar}$,
$W$+jets) for each training. In the $tb$+$tqb$ training, the s- and
t-channel components of the signal are taken in their SM proportions.

Samples are split by lepton flavor, jet multiplicity, and number of
$b$-tagged jets. The current analysis uses the following samples: one
isolated electron or muon; 2, 3 or 4 jets; and 1 or 2 $b$~tags. Each
sample is treated independently with its own training for each signal,
leading to 36 different trees (3 signals $\times$ 2 lepton flavors
$\times$ 3 jet multiplicities $\times$ 2 $b$-tagging possibilities).

%---------------------------------------------------------------------
%---------------------------------------------------------------------
\section{Bayesian Neural Networks Analysis}
\label{sec:BayesianNN}

\subsection{Introduction}
\label{sec:BNNIntroduction}

A neural network (NN)
$n(\mathbf{x},\mathbf{w})$~\cite{neuralnetworks-bishop} is a nonlinear
function, with adjustable parameters $\mathbf{w}$, which is capable of
modeling any real function of one or more
variables~\cite{neuralnetworks-blum}. In particular, it can model the
discriminant $D(\mathbf{x})$ in Eq.~\ref{eq:D}. Typically, one finds a
single point $\mathbf{w_0}$ in the network parameter space for which
$D(\mathbf{x}) \approx n(\mathbf{x}, \mathbf{w_0})$. This can be
achieved by minimizing an error function that measures the discrepancy
between the value of the function $n(\mathbf{x}, \mathbf{w})$ and the
desired outcome for variables $\mathbf{x}$: 1 for a signal event and 0
(or $-1$) if $\mathbf{x}$ pertain to a background event. If the error
function is built using equal numbers of signal and background events,
the minimization yields the result $D(\mathbf{x}) = n(\mathbf{x},
\mathbf{w_0})$~\cite{bayesNNs-ruck,bayesNNs-wan} provided that the
function $n(\mathbf{x}, \mathbf{w})$ is sufficiently flexible and that
a sufficient number of training events are used.

One shortcoming of the minimization is its tendency, unless due care
is exercised, to pick a point $\mathbf{w_0}$ that fits the function
$n(\mathbf{x}, \mathbf{w})$ too tightly to the training data. This
over-training can yield a function, $n(\mathbf{x}, \mathbf{w_0})$,
that is a poor approximation to the discriminant $D(\mathbf{x})$
(Eq.~\ref{eq:D}). In principle, the over-training problem can be
mitigated, and more accurate and robust estimates of $D(\mathbf{x})$
constructed, by recasting the task of finding the best approximation
to $D(\mathbf{x})$ as one of inference from a Bayesian
viewpoint~\cite{bayesNNs-neal,bayesNNs-bhat}. The task is to infer the
set of parameters $\mathbf{w}$ that yield the best approximation of
$n(\mathbf{x}, \mathbf{w})$ to $D(\mathbf{x})$.

Given training data $T$, which comprise an equal admixture of signal
and background events, one assigns a probability $p(\mathbf{w}|T) dw$
to each point in the parameter space of the network. Since each point
$w$ corresponds to a network with a specific set of parameter values,
the probability $p(\mathbf{w}|T) dw$ quantifies the degree to which
the network is a good fit to the training data $T$. However, instead
of finding the best single point $\mathbf{w_0}$, one averages
$n(\mathbf{x}, \mathbf{w})$ over every possible point $\mathbf{w}$,
weighted by the probability of each point. A Bayesian neural network
(BNN)~\cite{bayesNNs-neal,bayesNNs-bhat} is defined by the function
\begin{equation}
\label{eq:BNN}
 n(\mathbf{x}) = \int n(\mathbf{x},\mathbf{w})\, p(\mathbf{w}|T)\, dw,
\end{equation}
that is, it is a weighted average over all possible network functions
of a given architecture. The calculation is Bayesian because one is
performing an integration over a parameter space. If the function
$p(\mathbf{w}|T)$ is sufficiently smooth, one would expect the
averaging in Eq.~\ref{eq:BNN} to yield a more robust and more accurate
estimate of the discriminant $D(\mathbf{x})$ than from a single best
point $\mathbf{w_0}$.

There is however a practical difficulty with Eq.~\ref{eq:BNN}: it
requires the evaluation of a complicated high-dimension integral.
Fortunately, this is feasible using sophisticated numerical methods,
such as Markov Chain Monte
Carlo~\cite{bayesNNs-neal,markovchain-duane,markovchain-berg}. We use
this method to sample from the posterior density $p(\mathbf{w}|T)$ and
to approximate Eq.~\ref{eq:BNN} by the sum
\begin{equation}
\label{eq:BNNAverage}
 n(\mathbf{x}) \approx \frac{1}{K} \sum_{k=1}^K
                       n (\mathbf{x}, \mathbf{w}_k),
\end{equation}
where $K$ is the sample size.

We perform the Bayesian neural network calculations for this
analysis using the ``Software for Flexible Bayesian Modeling''
package~\cite{bayesNNs-software}.

\subsubsection{BNN Posterior Density}
\label{sec:BNNposterior}

Given training event $T = \mathbf{t}, \mathbf{x}$, where $\mathbf{t}$
denotes the {\em targets} --- 1 for signal and 0 for background ---
and $\mathbf{x}$ denotes the set of associated variables, we construct
the posterior probability density $p(\mathbf{w}|T)$ via Bayes' theorem
\begin{eqnarray}
\label{eq:post}
 p(\mathbf{w}|T)
 & = & \frac{p(T|\mathbf{w}) p(\mathbf{w})}{p(T)}\nonumber \\
 & = & \frac{p(\mathbf{t}|\mathbf{x},\mathbf{w}) 
             p(\mathbf{x}|\mathbf{w})p(\mathbf{w})}
            {p(\mathbf{t}|\mathbf{x})p(\mathbf{x})}\, \nonumber \\
 & = & \frac{p(\mathbf{t}|\mathbf{x},\mathbf{w})p(\mathbf{w})}
            {p(\mathbf{t}|\mathbf{x})}, \,
\end{eqnarray}
with $p(\mathbf{x}|\mathbf{w}) = p(\mathbf{x})$. We see that there
are two functions to be defined: the likelihood $p(\mathbf{t}|
\mathbf{x}, \mathbf{w})$ and the prior probability density
$p(\mathbf{w})$. For this analysis, the neural network functions have
the form
\begin{eqnarray}
\label{eq:BNNfunc}
 n(\mathbf{x},\mathbf{w})
 & = & \frac{1}{1+\exp[-f(\mathbf{x},\mathbf{w})]} , \\
 \mathrm{where} \nonumber \\
\label{eq:F}
 f(\mathbf{x},\mathbf{w})
 & = & b + \sum_{h=1}^H v_h \, 
       \tanh(a_h + \sum_{i=1}^I u_{hi} \, x_i) \, .
\end{eqnarray}
$H$ is the number of hidden nodes and $I$ is the number of input
variables, $\mathbf{x}$. The adjustable parameters $\mathbf{w}$ of the
networks are $u_{hi}$ and $v_h$ (the weights) and $a_h$ and $b$ (the
biases).

\subsubsection{BNN Likelihood}

If $\mathbf{x}$ are the variables for an event, then the event's
probability to be signal is $n(\mathbf{x},\mathbf{w})$; if it is a
background event, then its probability is
$1-n(\mathbf{x},\mathbf{w})$. Therefore, the probability of the
training event set is
\begin{equation}
\label{eq:like1}
 p(\mathbf{t}|\mathbf{x},\mathbf{w})
 = \prod_{j=1}^N n^{t_j} \, (1-n)^{1-t_j} \, ,
\end{equation}
where $t_j = 1$ for signal and $t_j = 0$ for background, and $n$ is
the total number of events. The BNN likelihood is proportional to this
probability.

\subsubsection{BNN Prior Density}
\label{sec:BNNprior}

The last ingredient needed to complete the Bayesian calculation is a
prior probability density. This is the most difficult function to
specify. However, experience suggests that for each network
parameter, a Gaussian centered at the origin of the parameter space
produces satisfactory results. Moreover, the widths of the Gaussian
should be chosen to favor parameter values close to the origin, since
smaller parameter values yield smoother approximations to
$D(\mathbf{x})$. Conversely, large parameter values yield jagged
approximations. However, since one does not know \emph{a priori} what widths
are appropriate, initially we allowed their values to adapt according
to the noise level in the training data. Subsequently, we found that
excessive noise in the training data can cause the parameter values to
grow too large. Therefore, we now keep the widths fixed to a small set
of values determined using single neural networks. This change is an
improvement over the method used in Ref.~\cite{run2-d0-prl-evidence}.

\subsubsection{Sampling the BNN Posterior Density}
\label{sec:BNNsampling}

To compute the average in Eq.~\ref{eq:BNNAverage} requires a sample of
points $\mathbf{w}$ from the posterior density, $p(\mathbf{w}|T)$.
These points are generated using a Markov Chain Monte Carlo method. We
first write the posterior density as
\begin{equation}
 p(\mathbf{w}|T) = \exp{[-V(\mathbf{w})]},
\end{equation}
where $V(\mathbf{w}) = -\ln p(\mathbf{w}|T)$ may be thought of as a
``potential'' through which a ``particle'' moves. We then add a
``kinetic energy'' term $T(\mathbf{p}) = \frac{1}{2} \mathbf{p}^2$,
where $\mathbf{p}$ is a vector of the same dimensionality as
$\mathbf{w}$, which together with the potential yields the particle's
``Hamiltonian'' ${\cal{H}} = T + V$. For a system governed by a
Hamiltonian, every phase space point $(\mathbf{w}, \mathbf{p})$ will
be visited eventually in such a way that the phase space density of
points is proportional to $\exp(-{\cal{H}})$. The phase space is
traversed by alternating between long deterministic trajectories and
stochastic changes in momentum. After every random change, one decides
whether or not to accept the new phase space point: the new state is
accepted if the energy has decreased, and accepted with a probability
less than one if the contrary is true. This algorithm yields a Markov
chain $\mathbf{w_1}, \mathbf{w_2}, \ldots \mathbf{w_K}$ of points,
which converges to a sequence of points that constitute a faithful
sample from the density $p(\mathbf{w}|T)$. In our calculations, each
deterministic trajectory comprises 100 steps, followed by a
randomization of the momentum. This creates a point that could be used
in Eq.~\ref{eq:BNNAverage}. However, since the correlation between
adjacent points is high, this pair of actions is repeated 20 times,
which constitutes one iteration of the algorithm, and a point is saved
after each iteration.

\subsection{BNN Training}
\label{sec:BNNtraining}

\begin{figure}[!h!tbp] 
\includegraphics[width=0.45\textwidth] 
{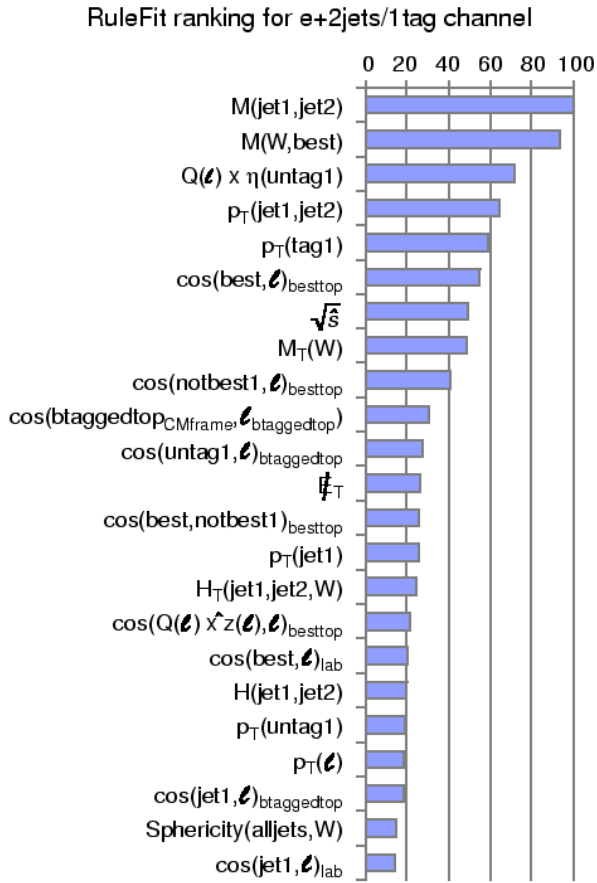}
\vspace{-0.1in}
\caption[e1tag]{BNN input variables according to their RuleFit
ranking for the electron+2jets/1tag channel.} 
\label{fig:e1tag} 
\end{figure}

The Bayesian neural networks are trained separately for each of the 12
analysis channels, with different sets of variables in each
channel. The variables are selected using an algorithm called
RuleFit~\cite{rulefit-friedman} that orders them according to their
discrimination importance (on a scale of 1 to 100). Variables with
discrimination importance greater than 10 are used, which results in
the selection of between 18 and 25 variables in the different
channels. For example, the variables for the electron+2jets/1tag
channel are shown in Fig.~\ref{fig:e1tag}. Each network contains a
single hidden layer with 20 nodes, with the sample size $K$ set to
100. The number of signal and background events used in the training
is 10,000 each.

\vspace{1in}
%---------------------------------------------------------------------
%---------------------------------------------------------------------
\section{Matrix Elements Analysis}
\label{sec:MatrixElement}

The main idea behind the matrix element (ME) technique is that the
physics of a collision, including all correlations, is contained in
the matrix element $\mathcal{M}$, where
\begin{equation}
\label{golden}
 d\sigma = \frac{(2\pi)^4|\mathcal{M}|^2}{F}d\Phi,
\end{equation}
and $d\sigma$ is the differential cross section, $F$ is the flux
factor, and $d\Phi$ is the
Lorentz invariant phase space factor. The ME analysis builds a
discriminant directly using Eq.~\ref{golden}, thereby potentially
making use of all the available kinematic information in the event. In
particular, the method uses
\begin{equation}
\label{px}
 p(\mathbf{x} | \mathrm{process}_i) = \frac{1}{\sigma_i}\,
 \totalderiv{\sigma_i}{\mathbf{x}}
\end{equation}
where $\mathbf{x}$ is the configuration of the event, and
$p(\mathbf{x} | \mathrm{process}_i)$ is the probability density to
observe $\mathbf{x}$ given that the physics process is
$\mathrm{process}_i$ to build the discriminant given in
Eq.~\ref{eq:D}.

For each data and simulated event, two discriminant values are
calculated: a t-channel discriminant and an s-channel discriminant.
The t-channel discriminant uses the t-channel matrix elements when
calculating $p(\mathbf{x} | S)$ as in Eq.~\ref{eq:D}, while the
s-channel discriminant uses s-channel matrix elements. For each
analysis channel, these discriminant values are plotted in a
two-dimensional histogram, out of which a cross section measurement is
extracted, as will be discussed in
Sec.~\ref{sec:cross_section_measurements}. The ME analysis only uses
events with two or three jets and one or two $b$-tags, and given the
two types of leptons, that results in eight independent analysis
channels.

The matrix element method was developed by D0 to measure the top quark
mass~\cite{topmass-d0-me-1} and has been used by
D0~\cite{topmass-d0-me-2} and
CDF~\cite{topmass-cdf-me-1a,topmass-cdf-me-1b,topmass-cdf-me-2} for
subsequent measurements. The ME method has also been used to measure
the longitudinal $W$~boson helicity fraction in top quark
decays~\cite{Whelicity-d0-me}. The result detailed here marks the
first use of the method to separate signal from background in a
particle search~\cite{run2-d0-prl-evidence}.

\vspace{-0.1in}
\subsection{Event Probability Density Functions}
\label{sec:diffcross}
\vspace{-0.1in}

The event configuration, $\mathbf{x}$, represents the set of
reconstructed four-momenta for all selected final state objects, plus
any extra reconstruction-level information, such as whether a jet is
$b$~tagged, if there is a muon in a jet, the quality of the muon
track, and so on. However, the matrix element, $\mathcal{M}$, depends
on the parton-level configuration of the event, which we label
$\mathbf{y}$. The differential cross section, $d\sigma/d\mathbf{x}$,
can be related to the parton-level variant, $d\sigma/d\mathbf{y}$, by
integrating over all the possible parton values, using the parton
distribution functions to relate the initial state partons to the
proton and antiproton, and using a \emph{transfer function} to relate
the outgoing partons to the reconstructed objects:
\begin{widetext}
\begin{equation}
\label{dsigma}
 \totalderiv{\sigma}{\mathbf{x}} = \sum_j \int d\mathbf{y}
 \left[ f_{1, j}(q_{1}, Q^{2})\, f_{2, j}(q_{2}, Q^{2})\,
 \totalderiv{\sigma_{\mathrm{HS},j}}{\mathbf{y}}\,
 W(\mathbf{x}|\mathbf{y},j)\,
 \Theta_{\mathrm{parton}}(\mathbf{y}) \right] 
\end{equation}
\end{widetext}
where
\begin{myitemize}

\item $\sum_j$ is the sum over different configurations that
contribute to the differential cross section: it is the discrete
analogue to $\int d\mathbf{y}$. Specifically, this summation includes
summing over the initial parton flavors in the hard scatter collision
and the different permutations of assigning jets to partons.

\item $\int d\mathbf{y}$ is an integration over the phase space:
\begin{equation}
\label{dy}
 \int d\mathbf{y}
 = \int dq_1dq_2d^3p_{\ell}d^3p_{\nu}d^3p_{q_1}d^3p_{q_2}\ldots.
\end{equation}
Many of these integrations are reduced by delta functions.

\item $f_{n,j}(q, Q^{2})$ is the parton distribution function in the
proton or antiproton ($n$ = 1 or 2, respectively) for the initial
state parton associated with configuration $j$, carrying momentum $q$,
evaluated at the factorization scale $Q^2$. We use the same
factorization scales as used when the simulated samples were
generated. This analysis uses CTEQ6L1~\cite{cteq} leading-order parton
distribution functions via {\sc{lhapdf}}~\cite{lhapdf}.

\item $d\sigma_{\mathrm{HS}}/d\mathbf{y}$ is the differential cross
section for the hard scatter (HS) collision. It is proportional to the
square of the leading-order matrix element as given by
(c.f.,~Eq.~\ref{golden}):
\begin{equation}
\label{dsigma_hs}
 d\sigma_{\mathrm{HS},j}
 = \frac{(2\pi)^4|{\cal M}|^{2}}
        {4\sqrt{(q_{1} \cdot q_{2})^2 - m_{1}^2m_{2}^2}} d\Phi
\end{equation}
where $q$ and $m$ are the four-momenta and masses of the initial-state
partons.

\item $W(\mathbf{x}\,|\,\mathbf{y}, j)$ is called the transfer
function; it represents the conditional probability to observe
configuration $\mathbf{x}$ in the detector given the original parton
configuration ($\mathbf{y}$, $j$). The transfer function is divided
into two parts:
\begin{equation}
\label{wbreakdown}
 W(\mathbf{x}\,|\,\mathbf{y}, j)
 = W_\mathrm{perm}(\mathbf{x}\,|\,\mathbf{y}, j)~
   W_\mathrm{reco}(\mathbf{x}\,|\,\mathbf{y}, j)
\end{equation}
where $W_\mathrm{perm}(\mathbf{x}\,|\,\mathbf{y}, j)$, discussed in
Sec.~\ref{sec:perm}, is the weight assigned to the given jet-to-parton
permutation and $W_\mathrm{reco}(\mathbf{x}\,|\,
\mathbf{y}, j)$, discussed in Sec.~\ref{sec:tf}, relates the
reconstructed value to parton values for a given permutation.
 
\item $\Theta_{\mathrm{parton}}(\mathbf{y})$ represents the
parton-level cuts applied in order to avoid singularities in the
matrix element evaluation.

\end{myitemize}

{\sc Vegas} Monte Carlo integration is used, as implemented in the GNU
Scientific Library~\cite{vegas-lepage,gnu-scientific-library}.

The probability to observe a particular event given a process
hypothesis, Eq.~\ref{px}, also requires the total cross section
($\times$ branching fraction) as a normalization. The total cross
section ($\sigma$) is just an integration of Eq.~\ref{dsigma}:
\begin{equation}
\label{cross} 
 \sigma = \int d\mathbf{x} \,\totalderiv{\sigma}{\mathbf{x}}\,
 \Theta_\mathrm{reco}(\mathbf{x}).
\end{equation}
The term $\Theta_\mathrm{reco}(\mathbf{x})$ approximates the selection
cuts. While conceptually simple, Eq.~\ref{cross} represents a large
integral: 13 dimensions for two-jet events, 17 dimensions for
three-jet events other than {\ttbar}, and 20 dimensions for {\ttbar}
events. However, this integral needs to be calculated only once, not
once per event, so the actual integration time is insignificant.

\subsubsection{Matrix Elements}
\label{sec:mes}

The matrix elements used in this analysis are listed in
Table~\ref{elements}. The code to calculate the matrix elements is
taken from the {\madgraph}~\cite{madgraph-madevent} leading-order
matrix-element generator and uses the {\helas}~\cite{helas} routines
to evaluate the diagrams. In Table~\ref{elements}, for the single top
quark processes, the top quark is assumed to decay leptonically:
$t{\rar}Wb{\rar}\ell^+ \nu b$. For the $W+\mathrm{jets}$ processes,
the $W$ boson is also assumed to decay leptonically:
$W^{+}{\rar}\ell^{+}\nu$. Charge-conjugate processes are included. The
same matrix elements are used for both the electron and muon
channels. Furthermore, we use the same matrix elements for heavier
generations of incoming quarks, assuming a diagonal CKM matrix. In
other words, for the $t\bar{b}$ process, we use the same matrix
element for $u\bar{d}$ and $c\bar{s}$ initial-state partons.

\begin{table*}[!h!tbp]
\begin{ruledtabular}
\begin{minipage}{4.5in}
\caption[matrix-elements]{The matrix elements used in this analysis.
The numbers in parentheses specify the number of Feynman diagrams
included in each process. For simplicity, only processes that contain
a positively-charged lepton in the final state are shown. The
charge-conjugated processes are also used.}
\label{elements}
\begin{tabular}{ll|ll}
\multicolumn{4}{c}{\underline{Matrix Elements}} \vspace{0.05in} \\
\multicolumn{2}{l}{~~~~~~~~~~Two Jets}    & \multicolumn{2}{l}{~~~~~~~~~~Three Jets}    \\
Name        & Process                       & Name        & Process                         \\
\hline
{\bf Signals} &                               &{\bf Signals}  &                                 \\
~~$tb$      & $u\bar{d}{\rar}t\bar{b}$ (1)  & ~~$tbg$     & $u\bar{d}{\rar}t\bar{b}g$ (5)   \\
~~$tq$      & $ub{\rar}td$ (1)              & ~~$tqg$     & $ub{\rar}tdg$ (5)               \\
            & $\bar{d}b{\rar}t\bar{u}$ (1)  &             & $\bar{d}b{\rar}t\bar{u}g$ (5)   \\
            &                               & ~~$tqb$     & $ug{\rar}td\bar{b}$ (4)         \\
            &                               &             & $\bar{d}g{\rar}t\bar{u}\bar{b}$ (4) \\
{\bf Backgrounds} &                            & {\bf Backgrounds} &                             \\
~~$Wbb$     & $u\bar{d}{\rar}Wb\bar{b}$ (2) & ~~$Wbbg$    & $u\bar{d}{\rar}Wb\bar{b}g$ (12) \\
~~$Wcg$     & $\bar{s}g{\rar}W\bar{c}g$ (8) & ~~$Wcgg$    & $\bar{s}g{\rar}W\bar{c}gg$ (54) \\
~~$Wgg$     & $u\bar{d}{\rar}Wgg$ (8)~~     & ~~$Wggg$    & $u\bar{d}{\rar}Wggg$ (54)       \\
            &                               & ~~{\lepjets}
                           & $q\bar{q}{\rar}t\bar{t}{\rar}\ell^{+}\nu b\bar{u}d\bar{b}$ (3) \\
            &                               &             
                           & $gg{\rar}t\bar{t}{\rar}\ell^{+}\nu b\bar{u}d\bar{b}$ (3)
\end{tabular}
\end{minipage}
\end{ruledtabular}
\end{table*}

New to the analysis after the result published in
Ref.~\cite{run2-d0-prl-evidence} is an optimization of the three-jet
analysis channel. For these events, a significant fraction of the
background is $t\bar{t}{\rar}\ell$+jets, as can be seen from the yield
tables (see, e.g., Table~\ref{allchans-yields}). While no new
processes are added to the two-jet analysis, $tqg$, $Wcgg$, $Wggg$,
and {\lepjets} are now included in the three-jet analysis.

\subsubsection{Top Pairs Integration}
\label{sec:integ}

For the {\lepjets} integration, we cannot assume one-to-one matching
of parton to reconstructed object. The final state has four quarks, so
one-to-one matching would lead to a four-jet event. We are interested,
however, in using the \lepjets\ matrix element in the three-jet
bin. The \ttbar\ events therefore have to ``lose'' one jet to enter
this bin. One way that a jet could be lost is by having its
reconstructed \pt\ be below the selection threshold, which is
15\,GeV. Another way to lose a jet is if it is merged with another
nearby jet. The jet could also be outside the $\eta$ acceptance of the
analysis with $|\eta| > 3.4$. There is in addition a general
reconstruction inefficiency that can cause a jet to be lost, but it is
a small effect.

A study of {\elepjets} simulated events before tagging shows that
$80\%$ of the time when a jet is lost, there is no other jet that
passes the selection cuts within $\mathcal{R} < 0.5$, that is, it has
not been merged with another jet. The transverse momentum of quarks
not matched to a jet passing the selection cuts is peaked at around
15~GeV, indicating that the jet is often lost because it falls below
the jet $p_T$ threshold. This study shows that the light-quark jets,
which have a softer $p_T$ spectrum, are 1.7 times as likely to be lost
owing to the $p_T$ cut as the heavy-quark jets. This observation
motivated the following simplification: assume that the lost jet is
from a light quark coming from the hadronically decaying $W$ boson. In
the most common case, the probability assigned to losing a jet given
parton transverse energy $E_T^{\mathrm{parton}}$ is the probability
that the jet is reconstructed to have $E_T^{\mathrm{reco}} < 15$~GeV,
which can be calculated from the jet transfer function
$W_{\mathrm{jet}}$ (discussed in Sec.~\ref{sec:tf}):
\begin{equation}
 \mathrm{max}\left\{\int_0^{15} dE_T^{\mathrm{reco}} W_{\mathrm{jet}}
 (E_T^{\mathrm{reco}} | E_T^{\mathrm{parton}}),\,
 0.05\right\}.
\end{equation}
A minimum probability of 5\% is used to account for other
inefficiencies in reconstructing a jet. A random number determines
which of the two quarks coming from the $W$ boson is lost for a
particular sample point in the MC integration. Other special cases
considered are when the two light quarks have $\mathcal{R}(q_1,q_2) <
0.6$, in which case they are assumed to merge, or if the
pseudorapidity of the quark is outside our acceptance, in which case
it is assumed lost.

\subsubsection{Assignment Permutations}
\label{sec:perm}

The (discrete) summation over different configurations incorporated in
Eq.~\ref{dsigma} includes the summation over the different ways to
assign the partons to the jets. A weight for each permutation is
included as the $W_\mathrm{perm}$ part of the transfer function. This
analysis uses two pieces of information to determine the weight,
namely $b$~tagging and muon charge (the muon from $b$~decay):
\begin{equation}
 W_\mathrm{perm} = W_{b\mathrm{tag}} W_{\mu\mathrm{charge}}.
\end{equation}

The $b$-tagging weight is assumed to factor by jet:
\begin{equation}
\label{btagtf}
 W_{b\mathrm{tag}} = \prod_{i} w_{b\mathrm{tag}}
 (\mathrm{tag}_i\,|\,\alpha_i, p_{Ti}, \eta_i),
\end{equation}
where $\alpha_i$ is the flavor of quark $i$ and $\mathrm{tag}_i$ is
true or false depending on whether the jet is $b$~tagged or not. The
weights assigned to cases with and without a $b$~tag are:
\begin{eqnarray} 
 w_{b\mathrm{tag}} (\mathrm{tag}\,|\,\alpha, p_T, \eta)
 & = & P^\mathrm{taggable}(p_T, \eta)
       \varepsilon_\alpha(p_T, \eta) \nonumber \\
 w_{b\mathrm{tag}} (\mathrm{notag}\,|\,\alpha, p_T, \eta)
 & = & 1 - P^\mathrm{taggable}(p_T, \eta)
 \varepsilon_\alpha(p_T, \eta) \nonumber
\end{eqnarray}
where $\varepsilon_\alpha$ is the tag-rate function for the particular
quark flavor and $P^\mathrm{taggable}$ is the taggability-rate
function, which is the probability that a jet is taggable.

For the s-channel matrix element and for the {\lepjets} matrix
element, there are both a $b$~quark and a $\bar{b}$~quark in the final
state. Furthermore, the matrix element is not symmetric with respect
to the interchange of the $b$ and $\bar{b}$ quarks, so it is helpful
to be able to distinguish between $b$~jets and $\bar{b}$~jets to make
the correct assignment. In the case of muonic decays of the $b$ or
$\bar{b}$ quark, it is possible to distinguish between the jets by the
charge of the decay muon. One complication is that a charm quark may
also decay muonically, and the charge of the muon differs between
$b{\rar}c\mu^{-}\nu$ and
$b{\rar}cX\bar{X'}{\rar}s\mu^{+}\bar{\nu}X\bar{X'}$. However, because
$p_T^{\mathrm{rel}}$, the muon transverse momentum relative to the jet
axis, differs in the two cases, the charge of the muon still provides
information. Similarly to $W_{b\mathrm{tag}}$, we assign the muon
charge weight $W_{\mu\mathrm{charge}}$ based on whether the jet, if it
is assumed to be a $b$ or $\bar{b}$ in the given permutation, contains
a muon of the appropriate charge. The weight is calculated by the
probability that a $b$ or a $\bar{b}$ quark decays directly into a
muon given that there is a muon in the jet, parametrized as a function
of $p_T^{\mathrm{rel}}$ of the muon.

\subsubsection{Object Transfer Functions}
\label{sec:tf} 

We assume that the parton-level to reconstruction-level transfer
function, $W_\mathrm{reco}$ in Eq.~\ref{wbreakdown}, can be factorized
into individual per-object transfer functions:
\begin{equation}
 W_\mathrm{reco}(\mathbf{x}\,|\, \mathbf{y}, j)
 = \prod_i W_{ij}(x_i\,|\, y_i),
\end{equation}
where $W_{ij}(x_i\,|\, y_i)$ is a transfer function for one object ---
a jet, a muon, an electron --- and $x_i$ and $y_i$ are reconstructed
and parton-level information, respectively, for that object. We assume
that angles are well measured, so the only transfer functions that are
not delta functions are those for energy (for jets and electrons) and
$1/p_T$ (for muons). The jet transfer functions, which give the
probability to measure a jet energy given a certain parton energy, are
parametrized as double Gaussians in four pseudorapidity ranges, for
light jets, for $b$~jets with a muon within the jet, and for $b$~jets
with no muon in the jet. The electron and muon transfer functions are
parametrized as single Gaussians.

The jet and muon transfer functions are measured in {\pythia}
{\lepjets} simulated events. The electron transfer functions are based
on the electron resolution measured in single electron and $Z$~boson
peak simulated events.

\subsection{Single Top Quark Discriminants}
\label{disc-def}

We build separate s-channel and t-channel discriminants, $D_s$ and
$D_t$. The signal probability densities for the various channels are:
\begin{eqnarray}
 p(\mathbf{x} \,|\, \mathrm{2jet,s})
 & = &\frac{1}{\sigma_{tb}}\,\totalderiv{\sigma_{tb}}{\mathbf{x}} \\ 
 p(\mathbf{x} \,|\, \mathrm{2jet,t})
 & = &\frac{1}{\sigma_{tq}}\, \totalderiv{\sigma_{tq}}{\mathbf{x}} \\ 
 p(\mathbf{x} \,|\, \mathrm{3jet,s})
 & = &\frac{1}{\sigma_{tbg}}\,\totalderiv{\sigma_{tbg}}{\mathbf{x}}\\ 
 p(\mathbf{x} \,|\, \mathrm{3jet,t})
 & = &\frac{1}{(\sigma_{tqb} + \sigma_{tqg})}\,
   \totalderiv{(\sigma_{tqb} + \sigma_{tqg})}{\mathbf{x}} 
\label{eq:3t}.
\end{eqnarray}
Equation~\ref{eq:3t} can also be written as:
\begin{equation}
 p(\mathbf{x} \,|\, \mathrm{3jet,t})
 = w_{tqb} p(\mathbf{x} | tqb) + w_{tqg}
 p(\mathbf{x} | tqg),
\end{equation}
where $w_{tqb}$ and $w_{tqg}$ are the relative yields of the two
signal processes. Calculating the yield fractions using
Eq.~\ref{cross}, for single-tagged events we use $w_{tqb} = 0.6$ and
$w_{tqg} = 0.4$, while for double-tagged events we use $w_{tqb} = 1$
and $w_{tqg} = 0$.

We apply the same methodology of using weights based on yield fraction
for the $p(\mathbf{x}|\mathrm{background})$ calculations. We do not
use a matrix element for every background that exists, however, so the
yield fractions cannot be determined as for the signal
probabilities. Some, such as $u\bar{d}{\rar}Wc\bar{c}$, are not
included because they have similar characteristics to ones that are
included, such as $u\bar{d}{\rar}Wb\bar{b}$. Therefore, we use the
yields as determined from the simulated samples and consider what
background the matrix elements are meant to discriminate against. We
find the performance of the discriminant to be not very sensitive to
the chosen weights if the weights are reasonable, and have used the
weights given in Table~\ref{frac}.

\begin{table}[!h!tbp]
\caption[Background weights]{Background weights chosen for each
analysis channel in two-jet and three-jet events.}
\label{frac}
\begin{ruledtabular}
\begin{tabular}{l||cc|cc}
\multicolumn{5}{c}
{\hspace{0.5in}\underline{Background Fractions}}\vspace{0.05in} \\
 & \multicolumn{2}{c}{1 tag} & \multicolumn{2}{c}{2 tags}\\
Weight        & Electron & Muon~~ & Electron & Muon \\
\hline
{\bf Two-Jet Events}&       &        &        &        \\
~~$w_{Wbb}$     &  0.55  &  0.60  &  0.83  &  0.87  \\
~~$w_{Wcg}$     &  0.15  &  0.15  &  0.04  &  0.04  \\
~~$w_{Wgg}$     &  0.35  &  0.30  &  0.13  &  0.09  \\
{\bf Three-Jet Events}~~&    &        &        &        \\
~~$w_{Wbbg}$    &  0.35  &  0.45  &  0.30  &  0.40  \\
~~$w_{Wcgg}$    &  0.10  &  0.10  &  0.02  &  0.03  \\
~~$w_{Wggg}$    &  0.30  &  0.25  &  0.13  &  0.10  \\
~~$w_{t\bar{t}{\rar}\ell+\mathrm{jets}}$
                &  0.25  &  0.20  &  0.55  &  0.47
\end{tabular}
\end{ruledtabular}
\end{table}

\clearpage
%---------------------------------------------------------------------
%---------------------------------------------------------------------
\section{Multivariate Output Distributions}
\label{sec:outputs}
\vspace{-0.05in}

\begin{figure*}[!h!tbp] 
\includegraphics[width=0.32\textwidth] 
{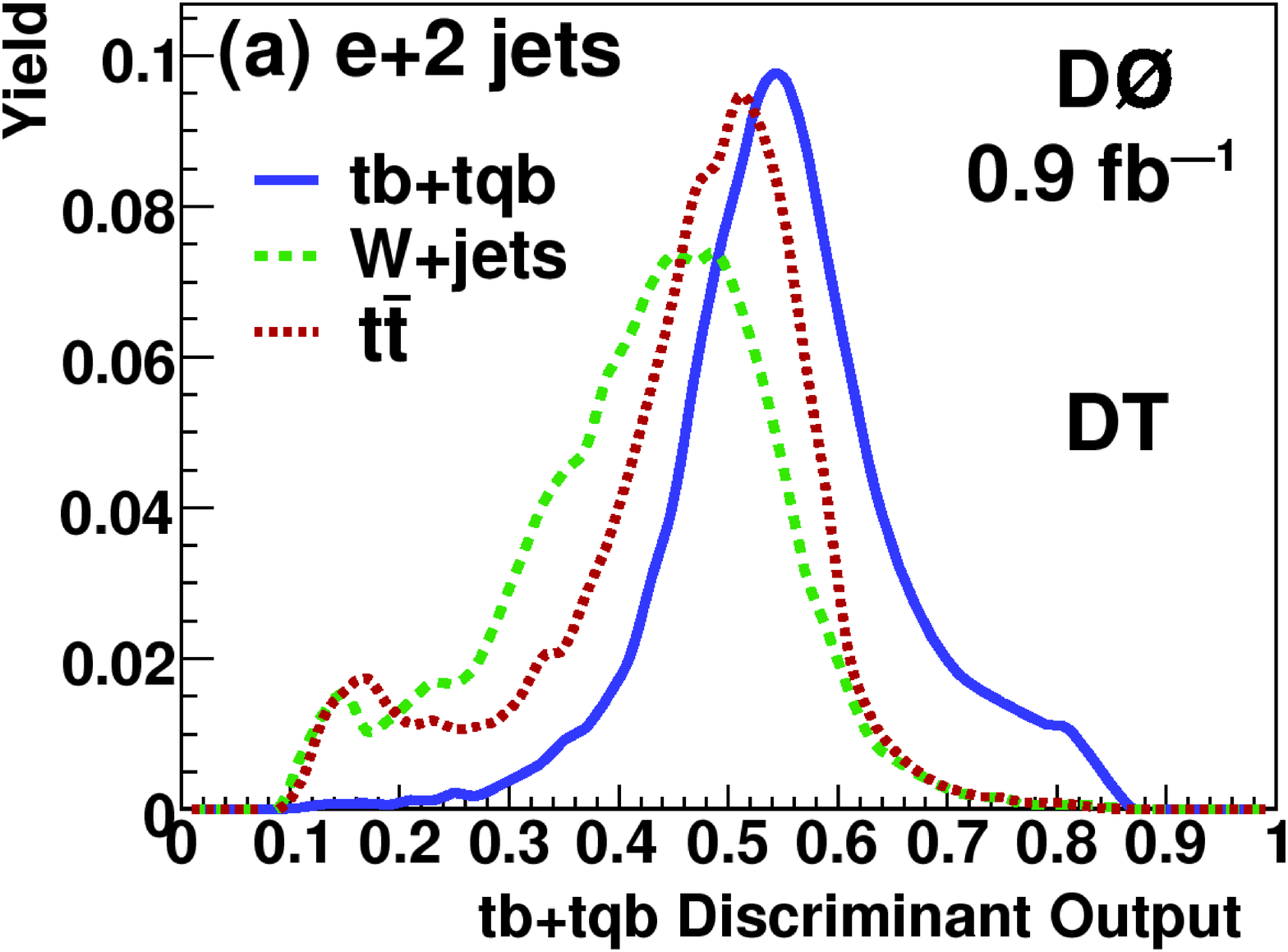}
\includegraphics[width=0.32\textwidth] 
{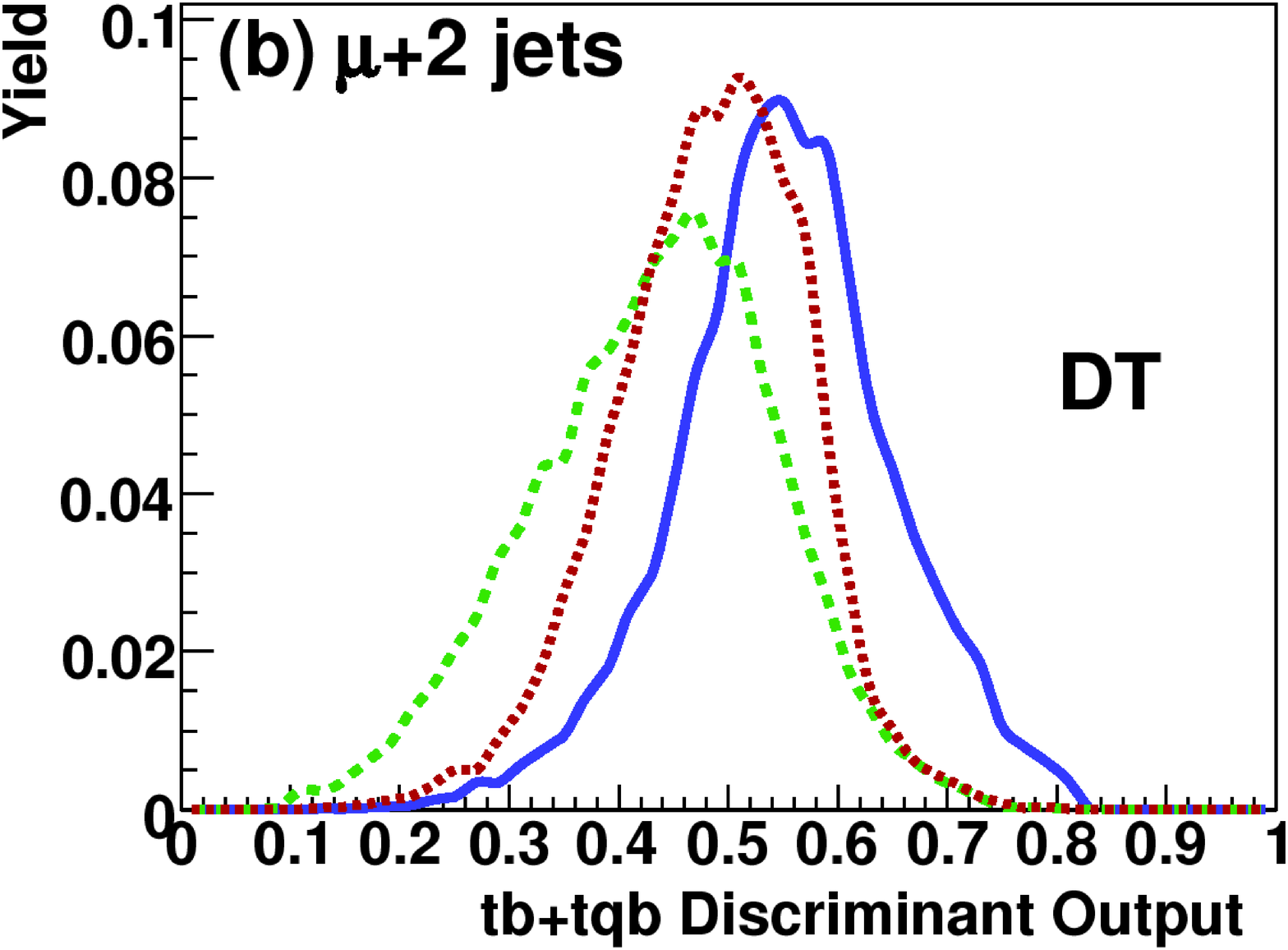}

\includegraphics[width=0.32\textwidth]
{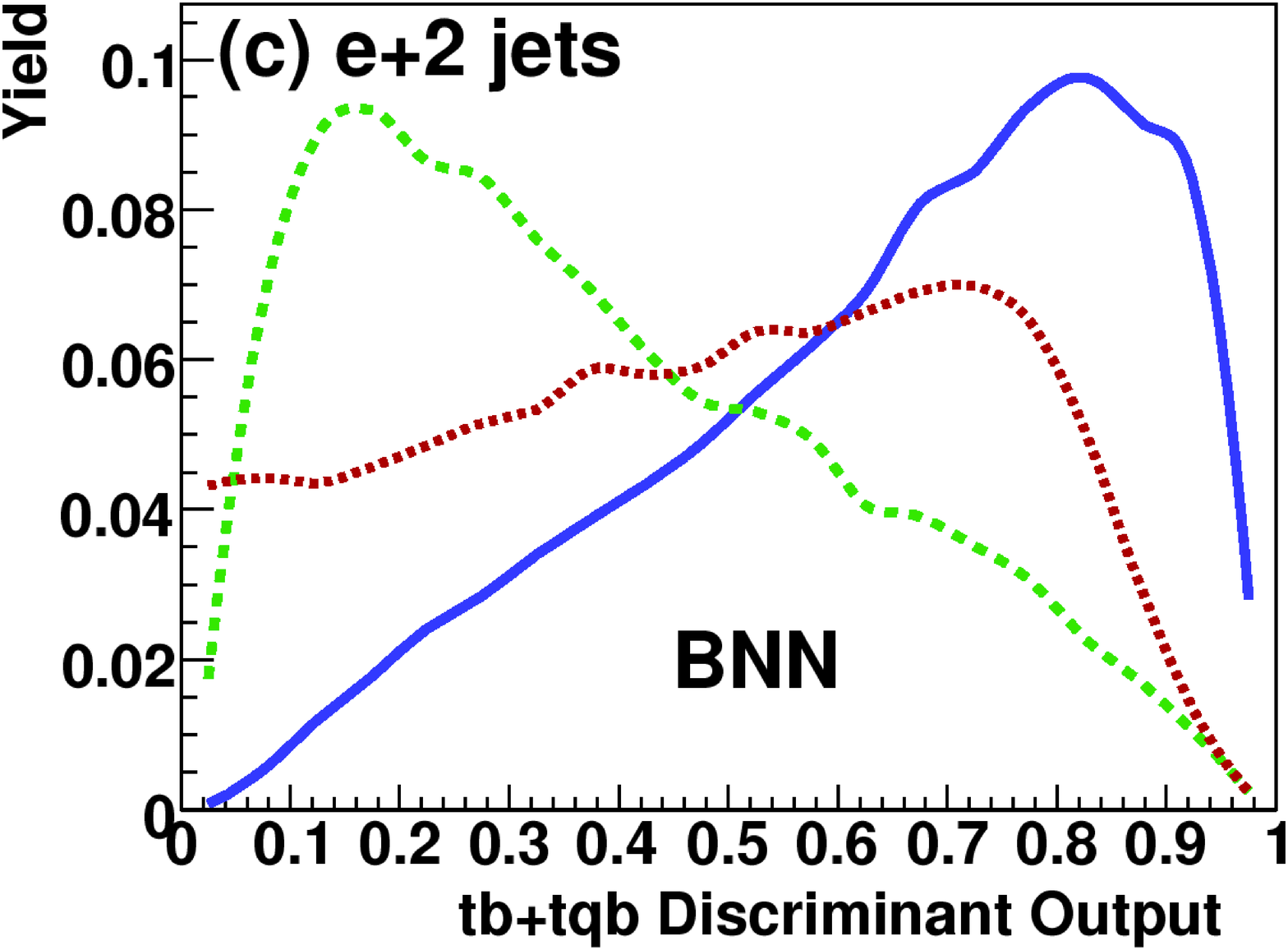}
\includegraphics[width=0.32\textwidth]
{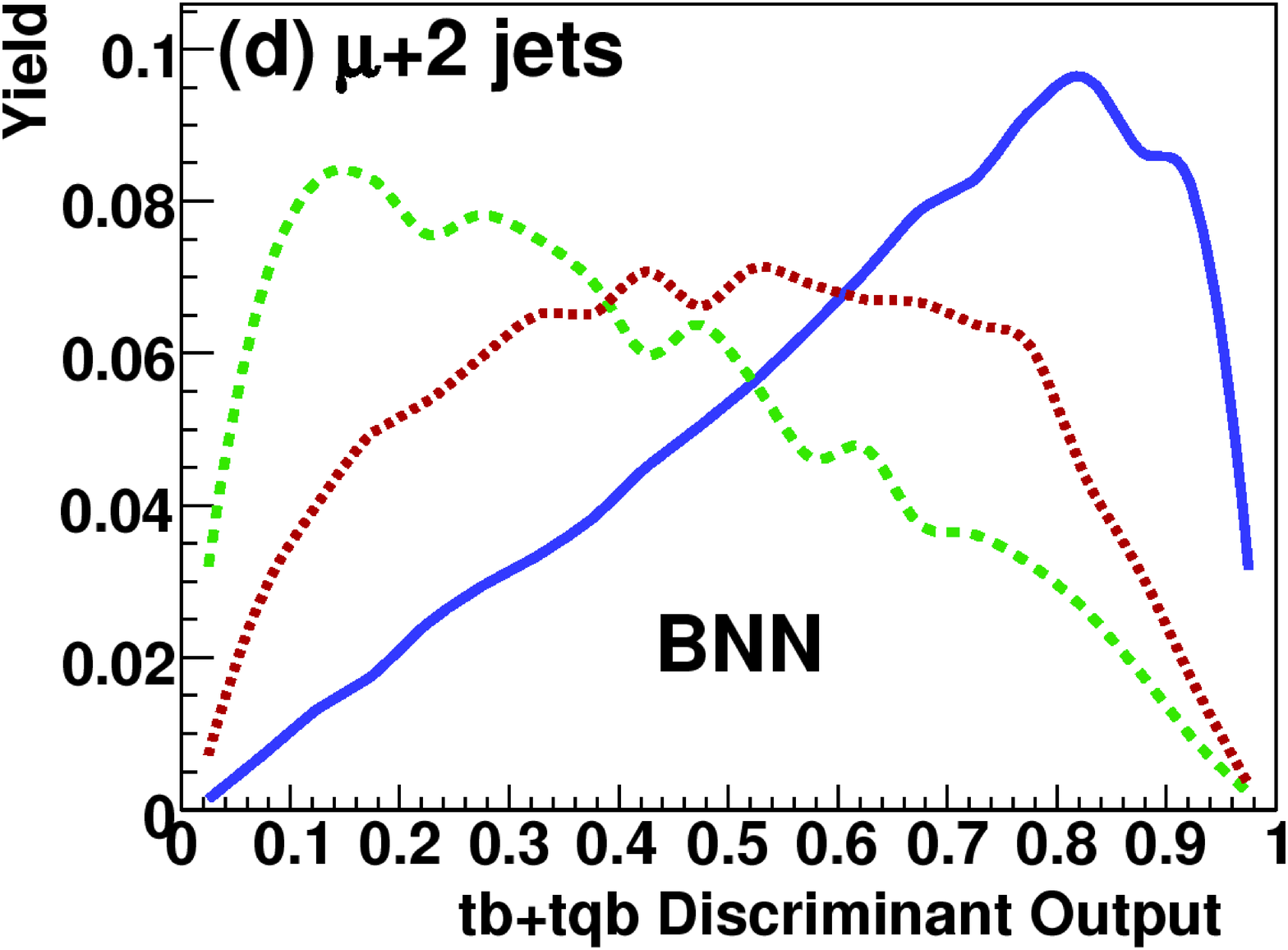}

\includegraphics[width=0.32\textwidth]
{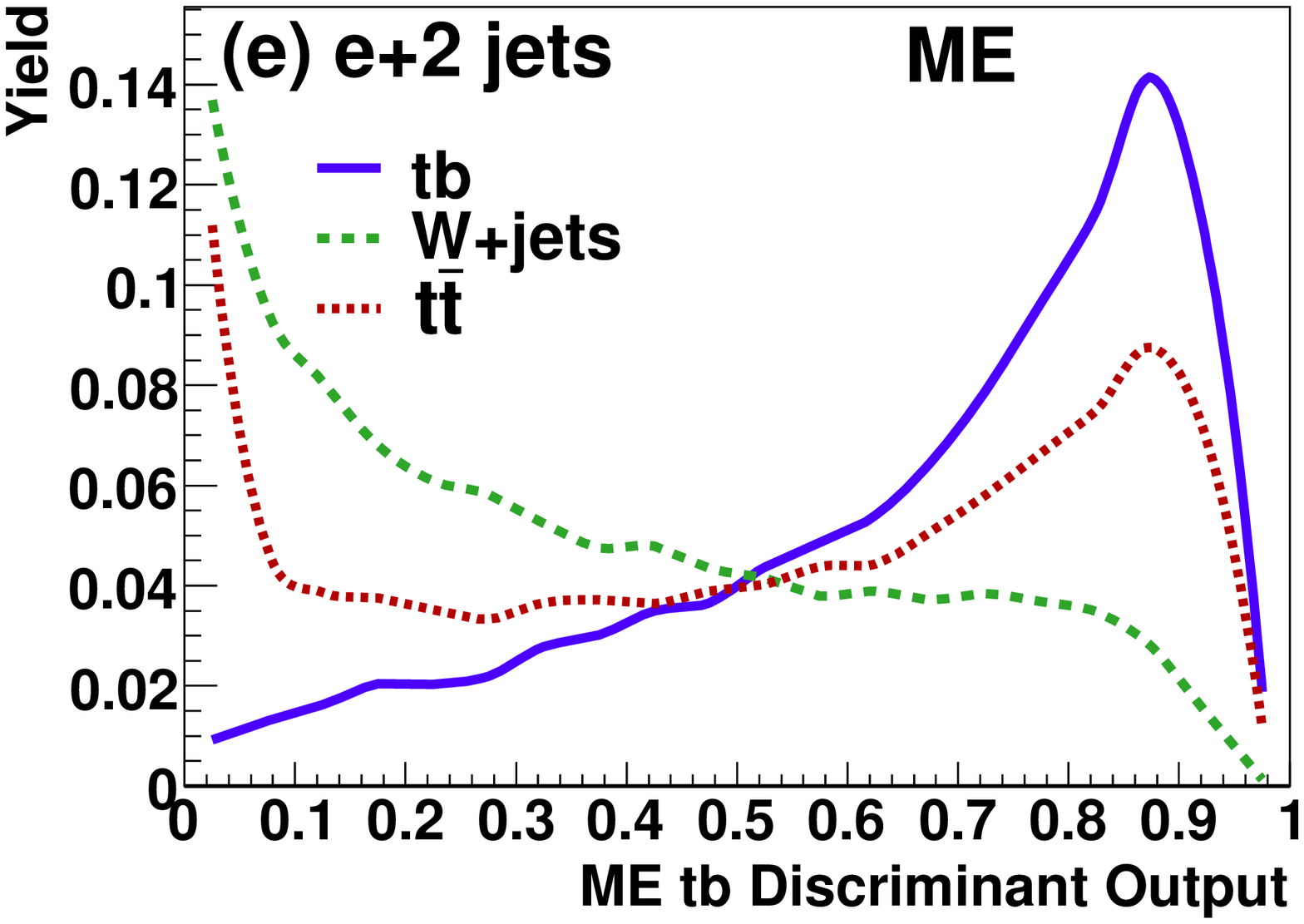}
\includegraphics[width=0.32\textwidth]
{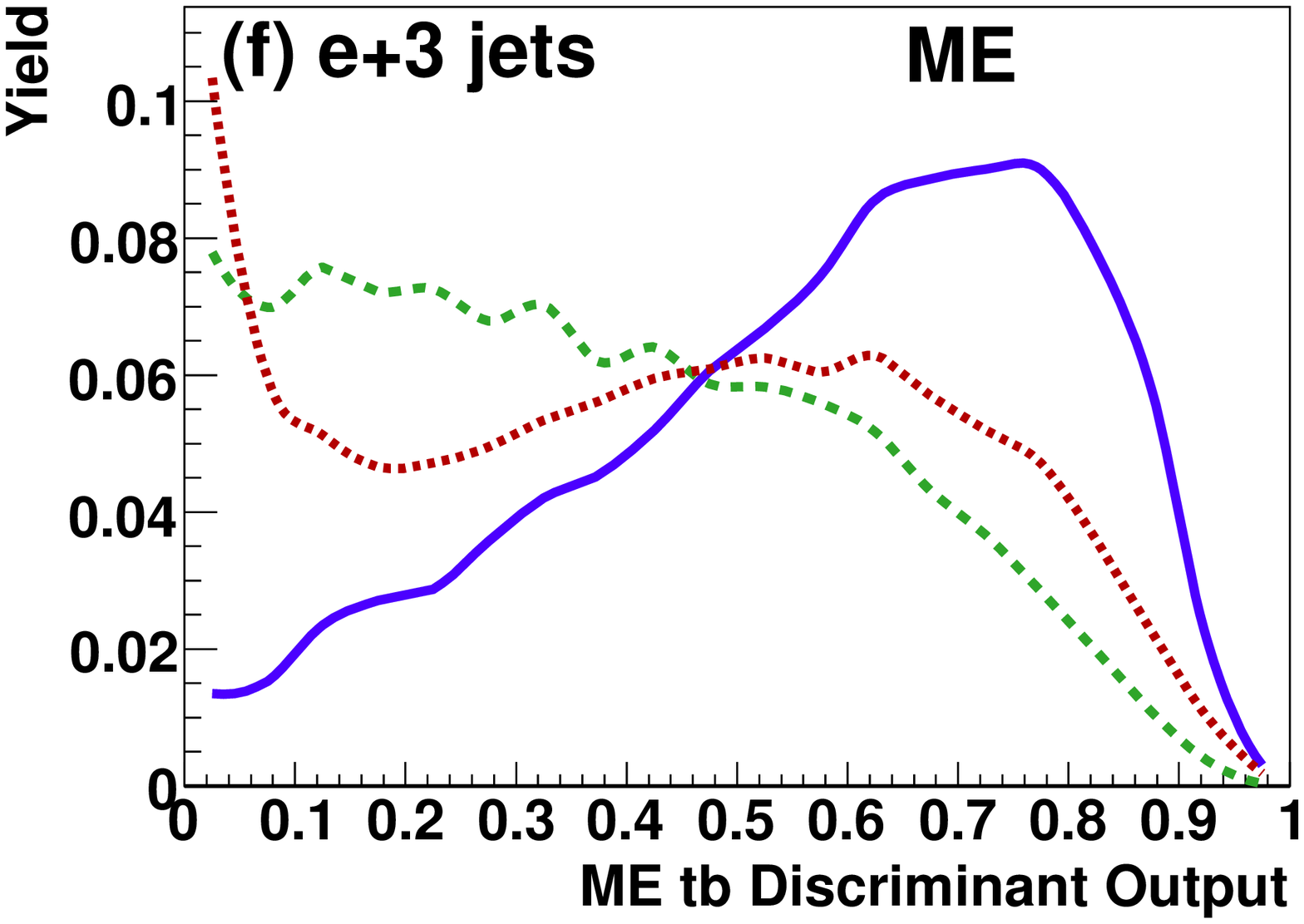}

\includegraphics[width=0.32\textwidth]
{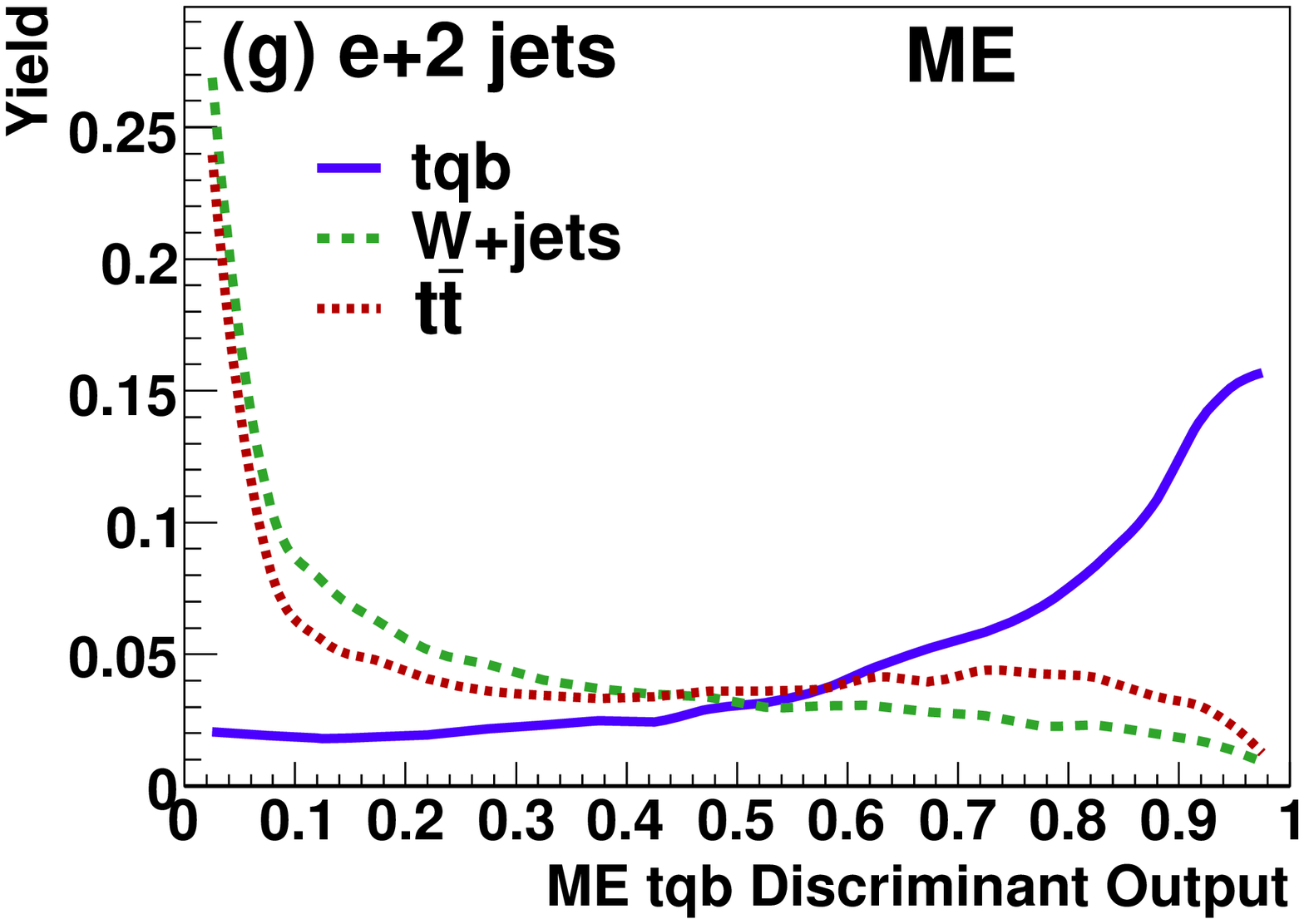}
\includegraphics[width=0.32\textwidth]
{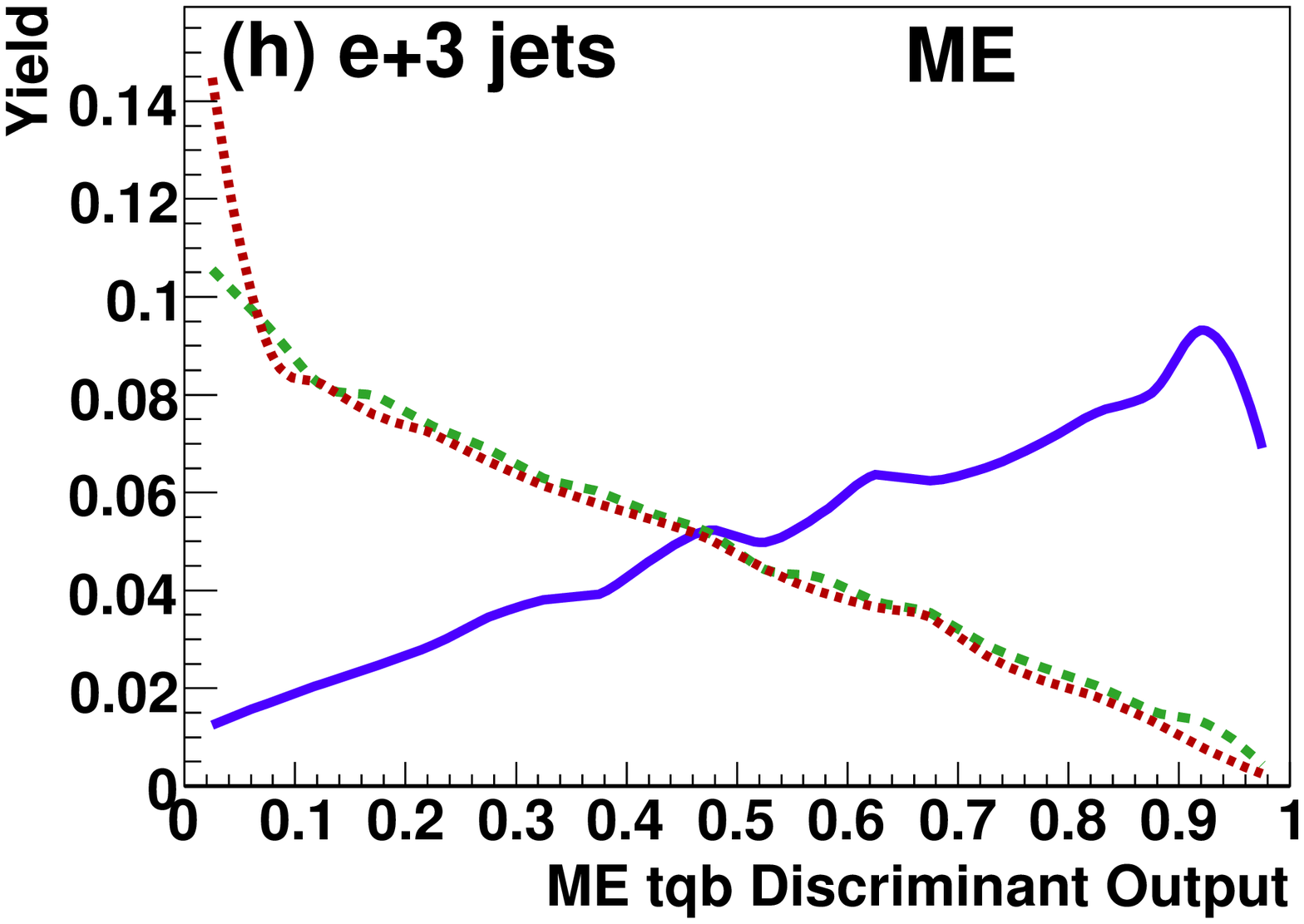}
 
\vspace{-0.1in}
\caption[discrim-outputs]{For plots (a)--(d), DT and BNN discriminant
outputs for $tb$+$tqb$ in the $e$+jets channel (left column) and
$\mu$+jets channel (right column) for events with two jets of which
one is $b$~tagged. Plots (e) and (f) show the ME discriminant outputs
for $tb$ in the $e$+jets channel, for two-jet and three-jet events
respectively. Plots (g) and (h) show the ME discriminants for $tqb$ in
the $b$~tagged $e$+jets channel, for two-jet and three-jet events
respectively.  All histograms are normalized to unity. }
\label{fig:DiscriminantOutputs}
\end{figure*}

\begin{figure*}[!h!tbp] 
\includegraphics[width=0.30\textwidth] 
{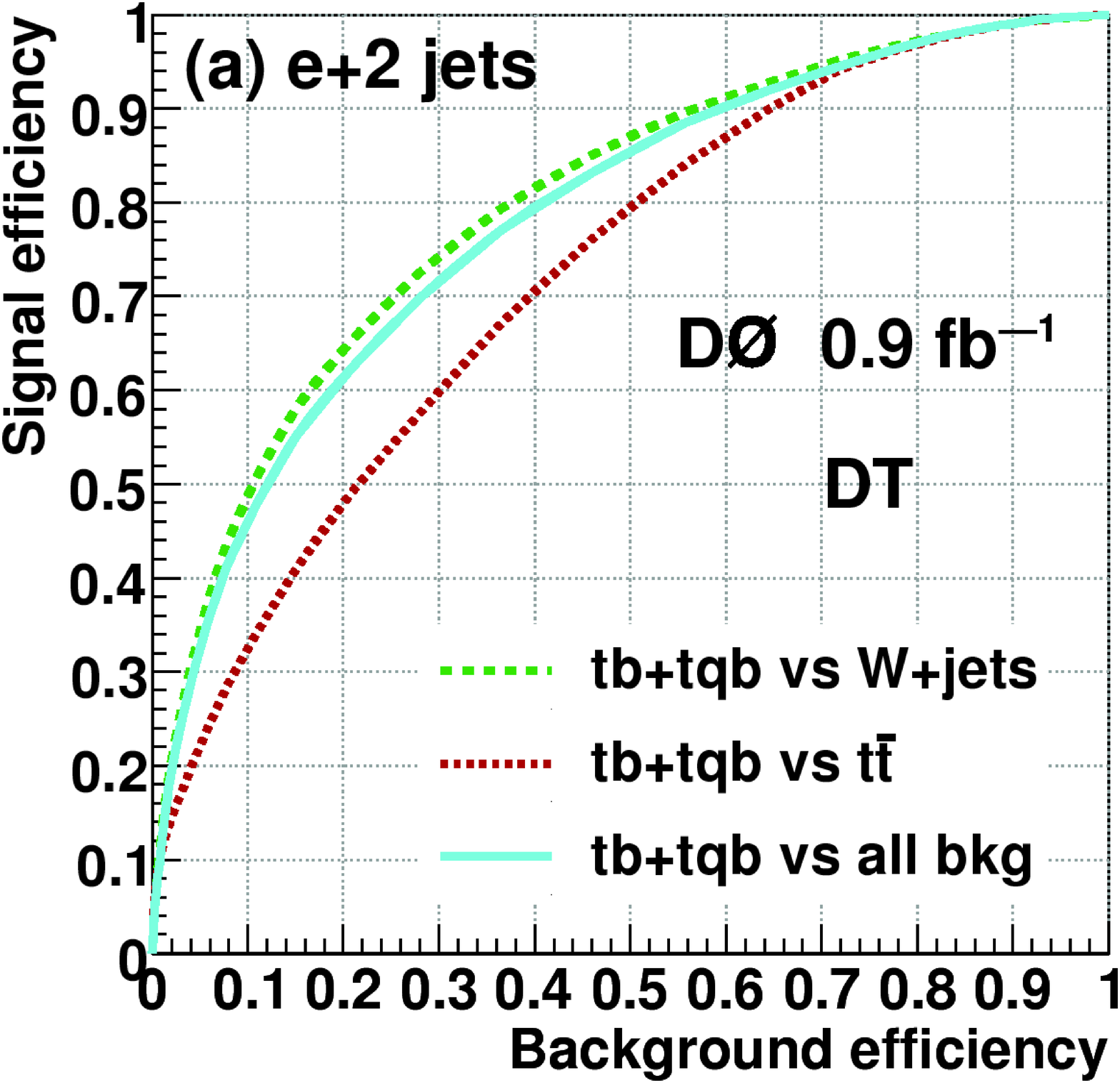}
\includegraphics[width=0.30\textwidth] 
{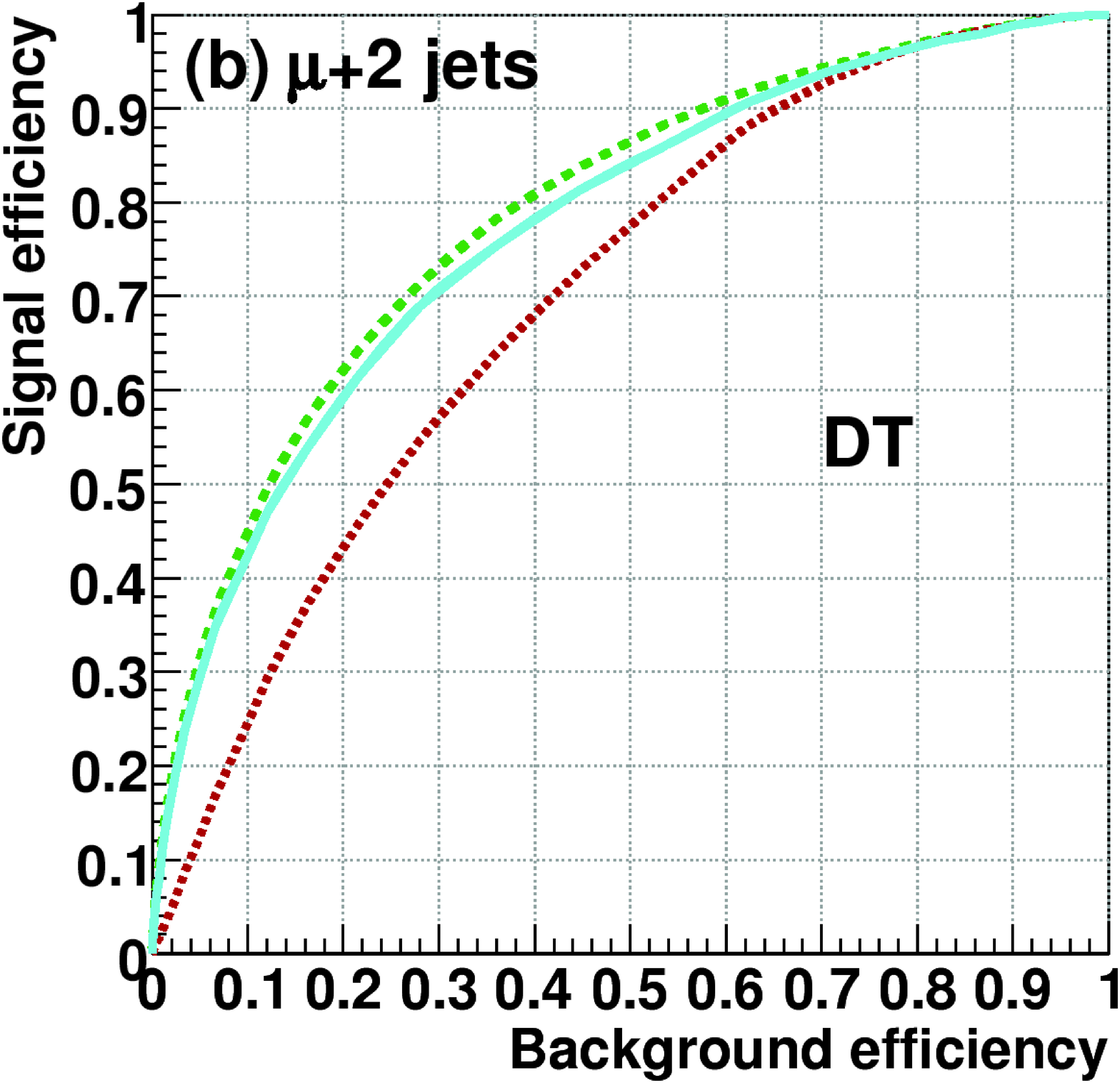}

\includegraphics[width=0.30\textwidth] 
{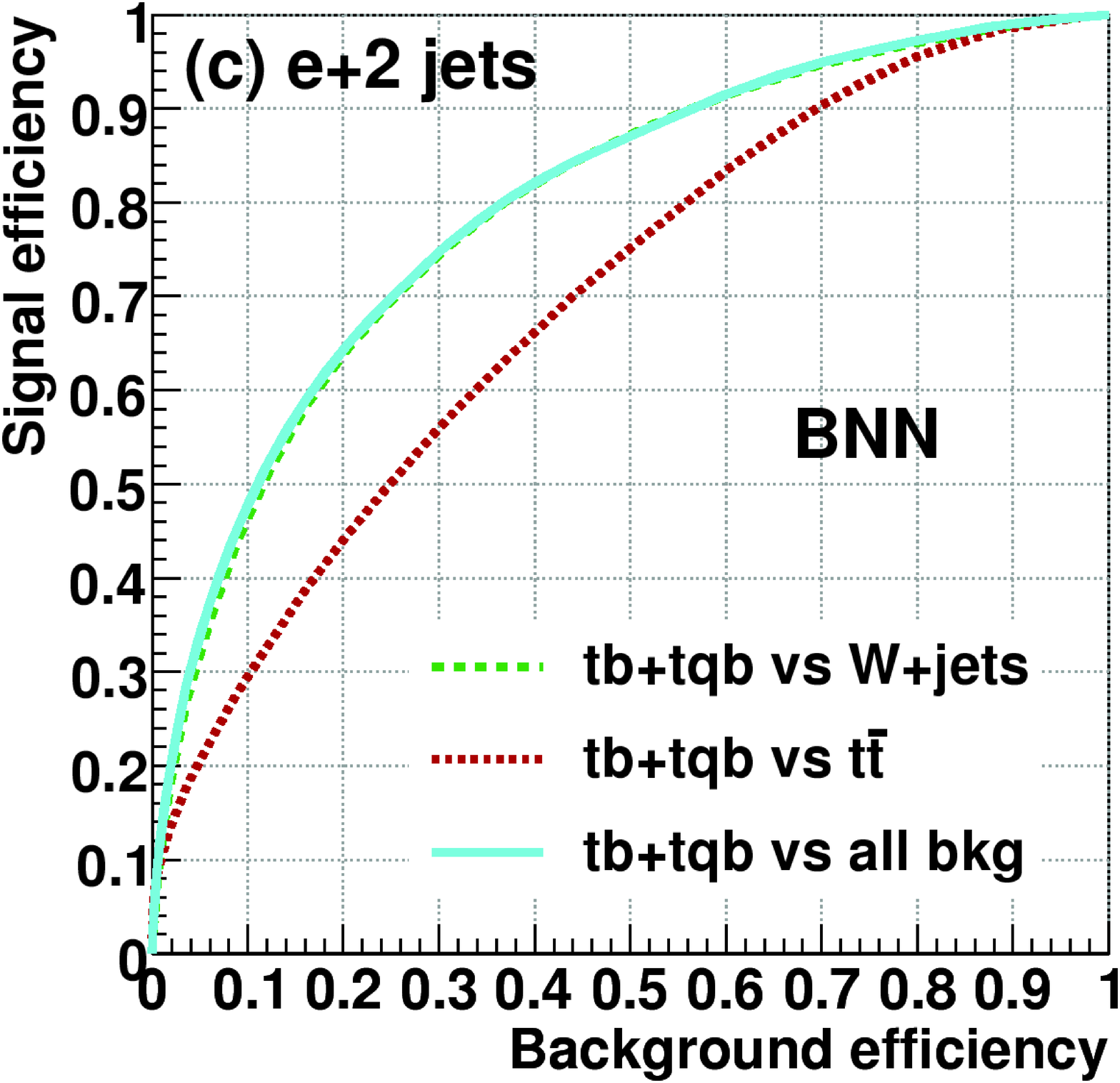}
\includegraphics[width=0.30\textwidth] 
{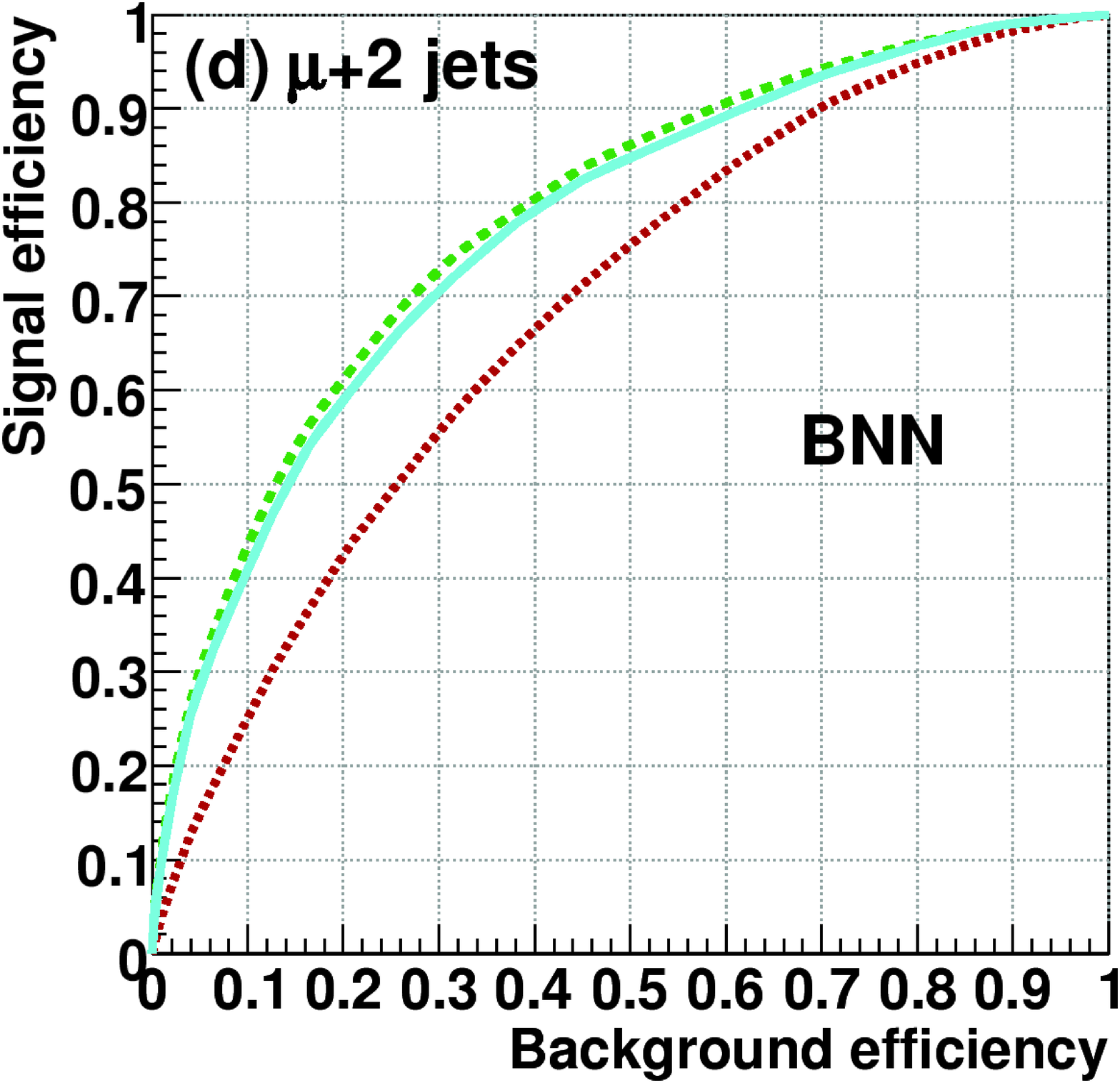}

\includegraphics[width=0.30\textwidth] 
{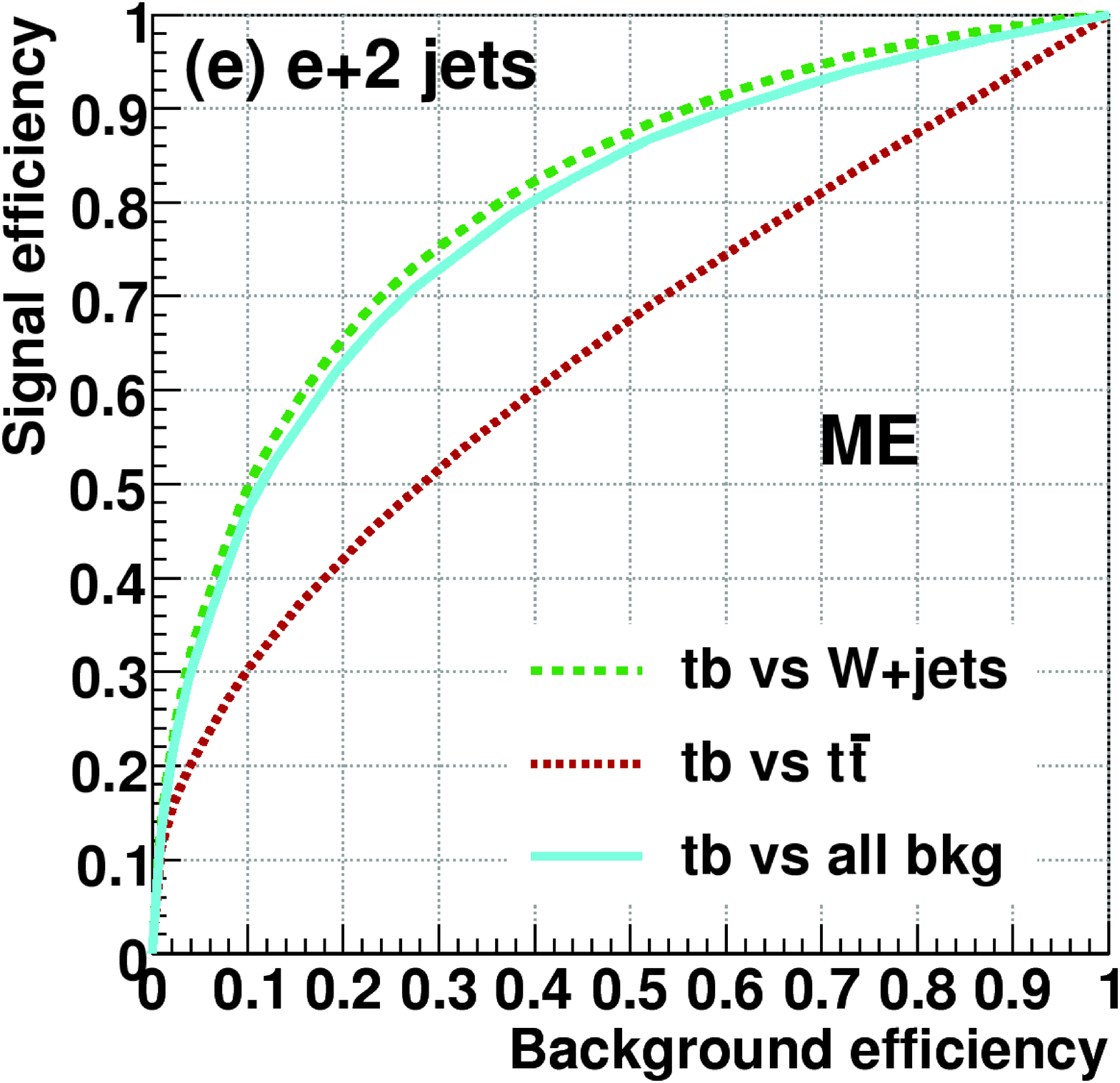}
\includegraphics[width=0.30\textwidth] 
{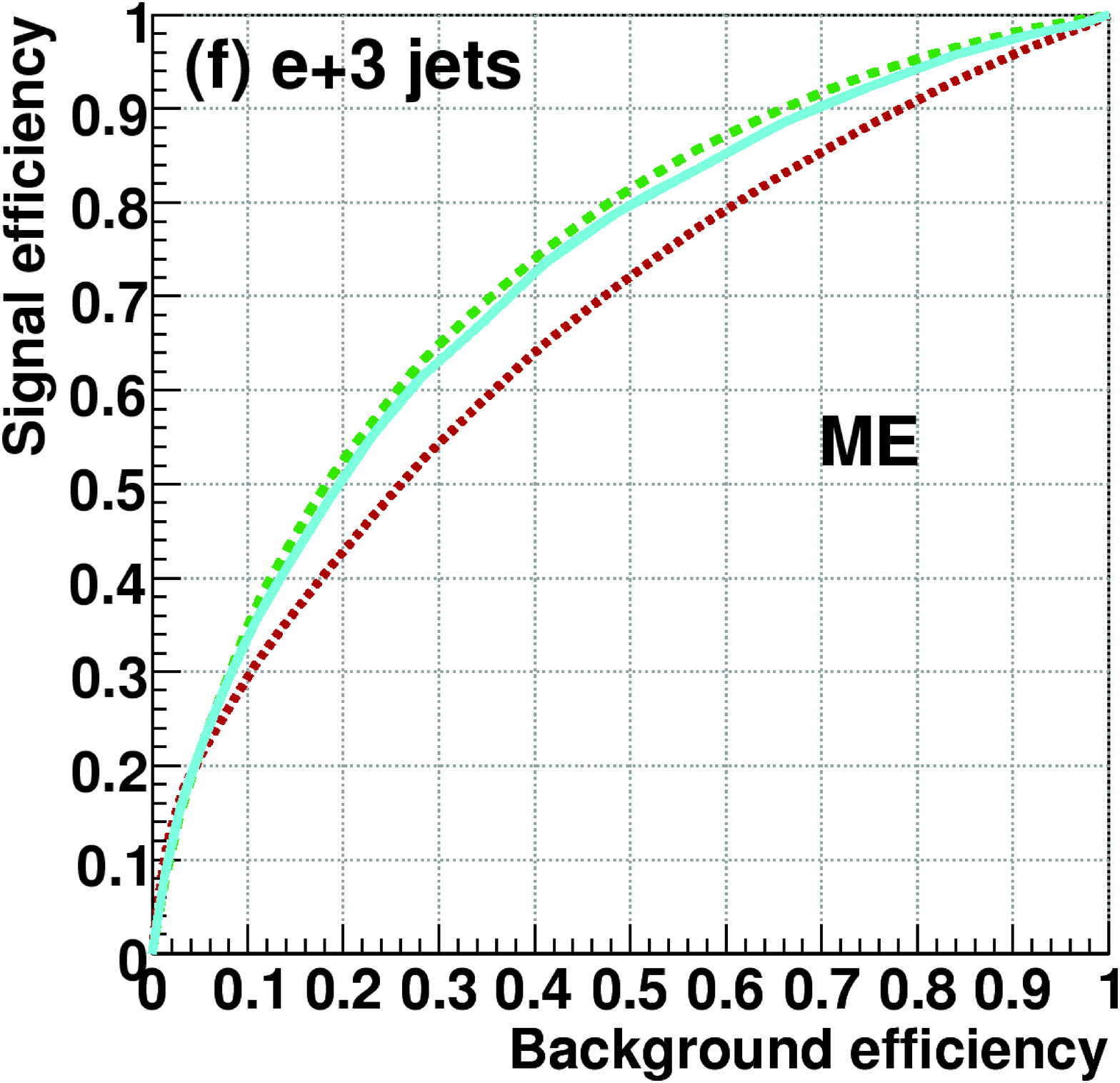}

\includegraphics[width=0.30\textwidth] 
{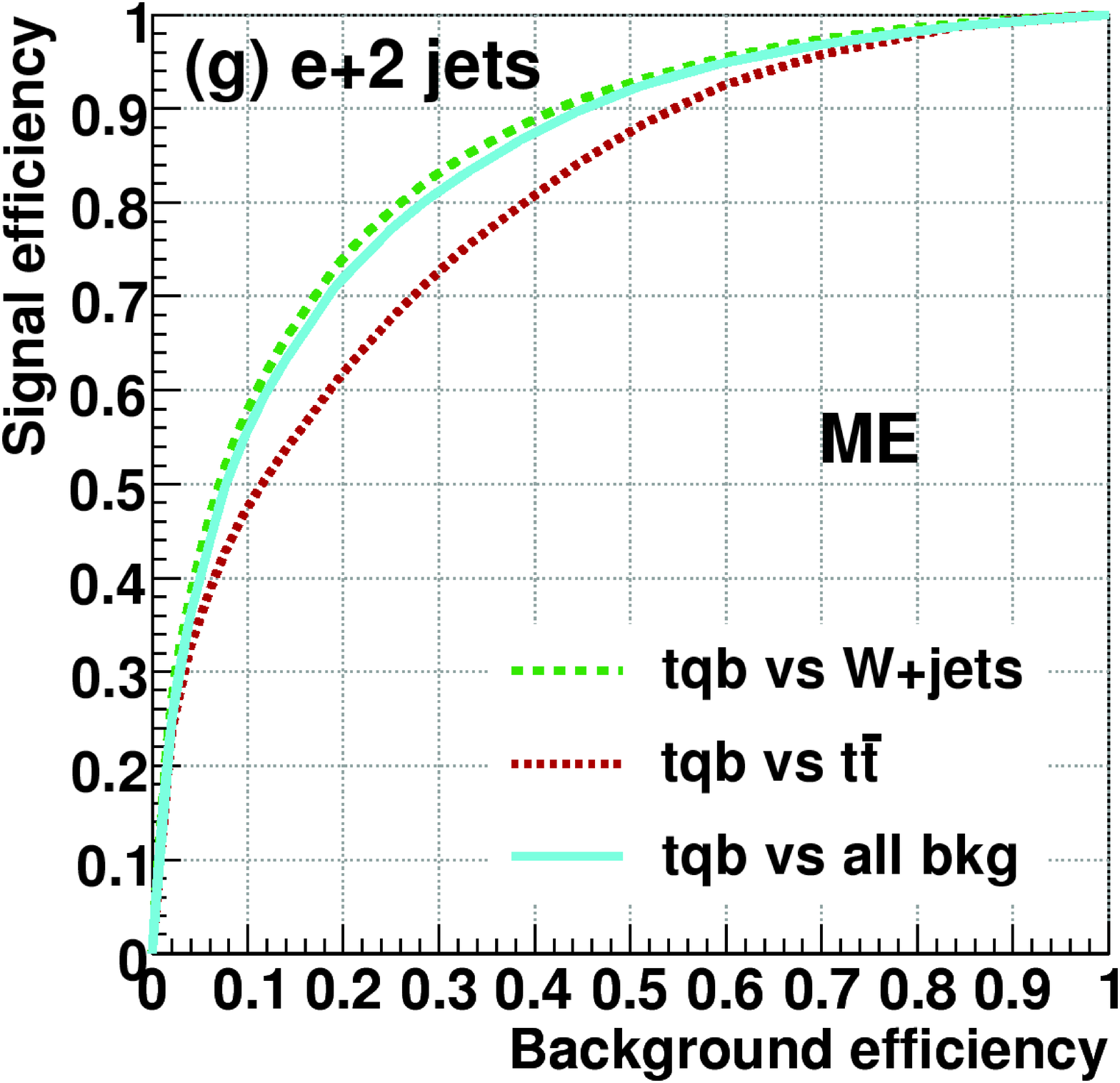}
\includegraphics[width=0.30\textwidth] 
{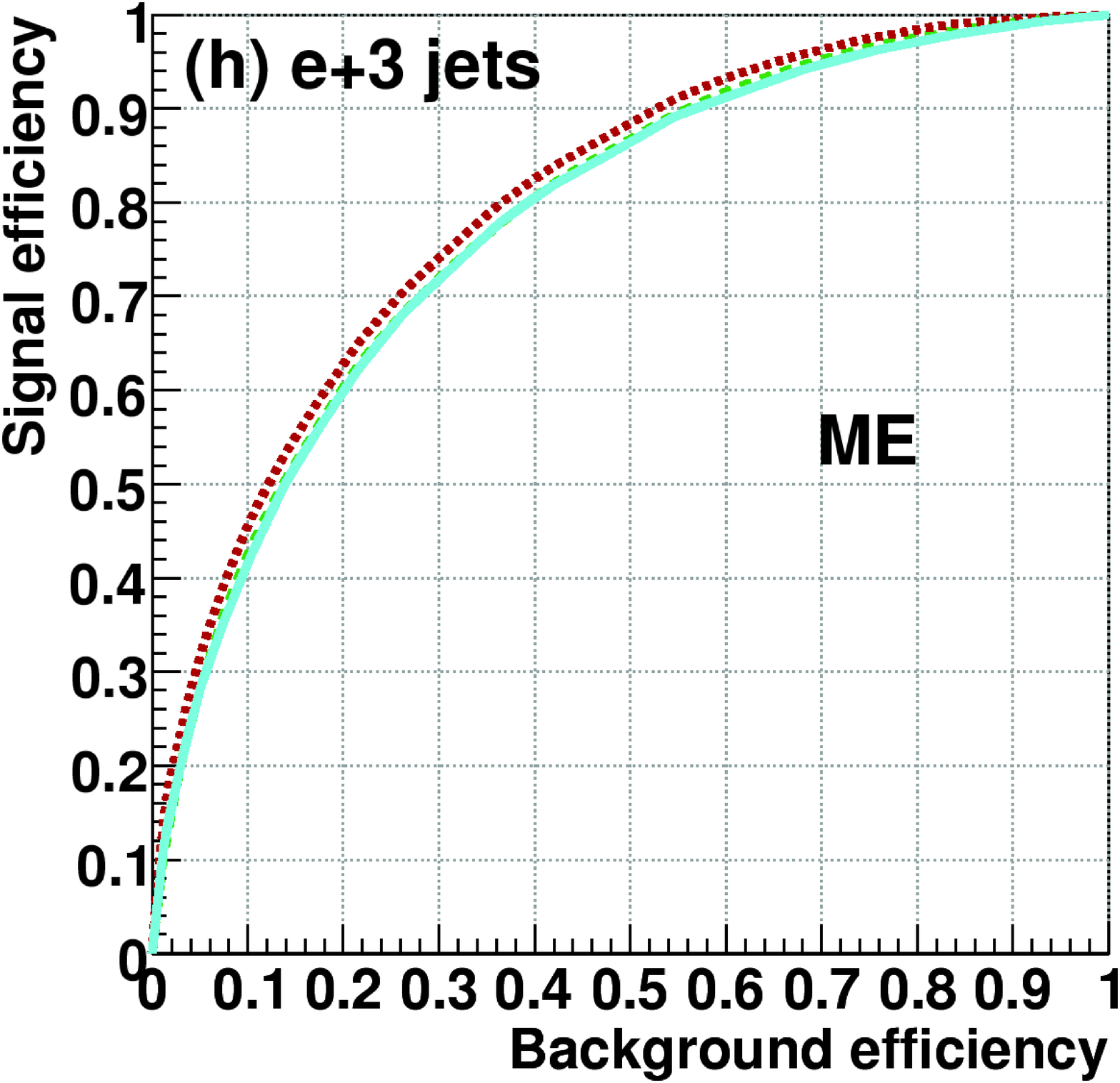}
 
\vspace{-0.1in}
\caption[discrim-efficiency]{For plots (a)--(d), DT and BNN $tb$+$tqb$
signal efficiency versus background efficiency in the $e$+jets channel
(left column) and $\mu$+jets channel (right column) for events with
two jets of which one is $b$~tagged. Plots (e)--(h) show the ME signal
versus background efficiency for $tb$ signal (third row) and $tqb$
signal (fourth row), for $b$~tagged $e$+jets events with two jets
(left column) and three jets (right column). These curves are derived
from the discriminants shown in Fig.~\ref{fig:DiscriminantOutputs}.}
\label{fig:SigBkgEff}
\end{figure*}

Discriminant output shapes for signal and different background
components are shown in Fig.~\ref{fig:DiscriminantOutputs},
demonstrating the ability of the three analyses to separate signal
from background. The DT discriminant is narrower and more central
owing to the averaging effect of boosting (according to
Eq.~\ref{eq:BDToutput}). The separation powers of the discriminants
shown in Fig.~\ref{fig:DiscriminantOutputs} are more directly
visualized in Fig.~\ref{fig:SigBkgEff}.

\begin{figure*}[!h!tbp]
\includegraphics[width=0.32\textwidth]
{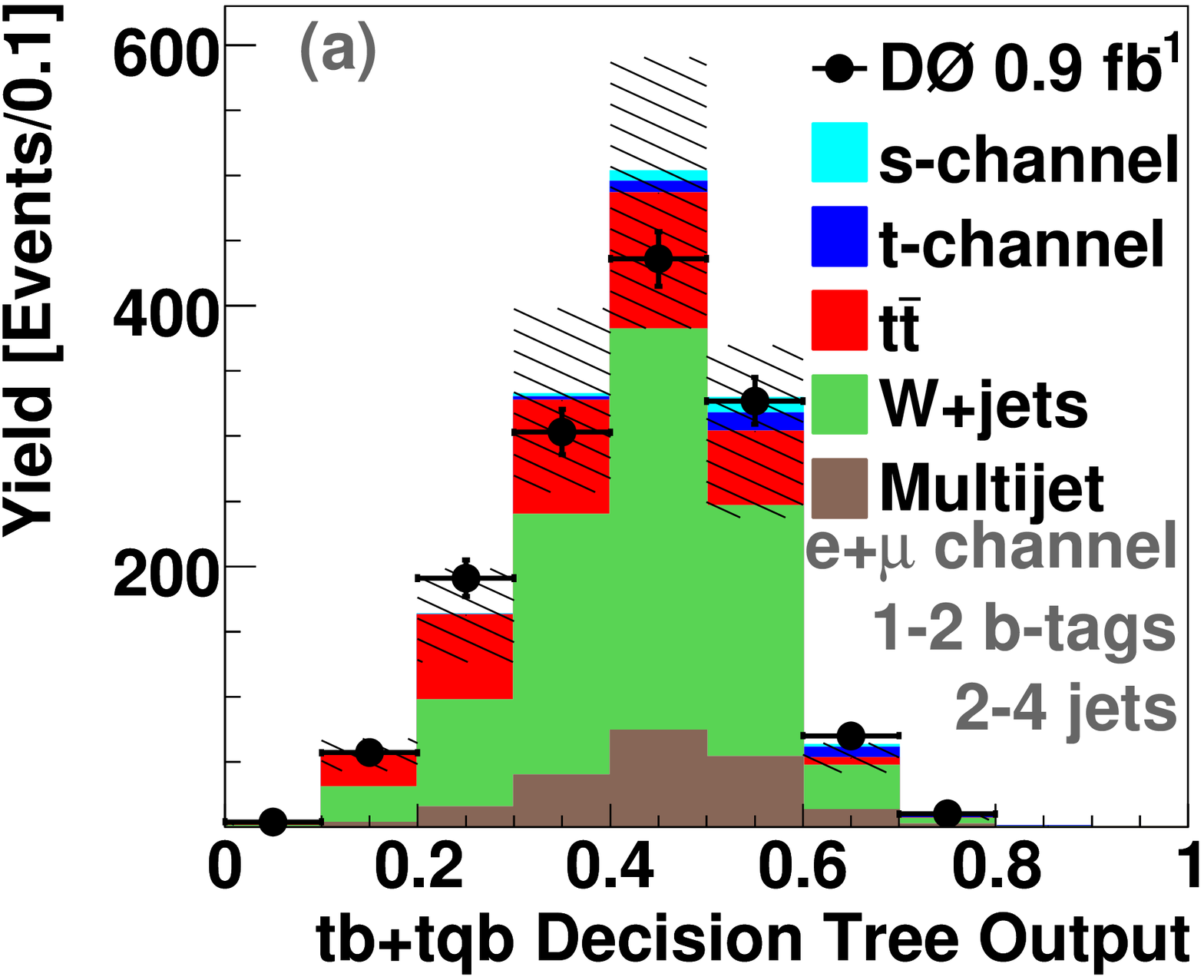}
\includegraphics[width=0.32\textwidth]
{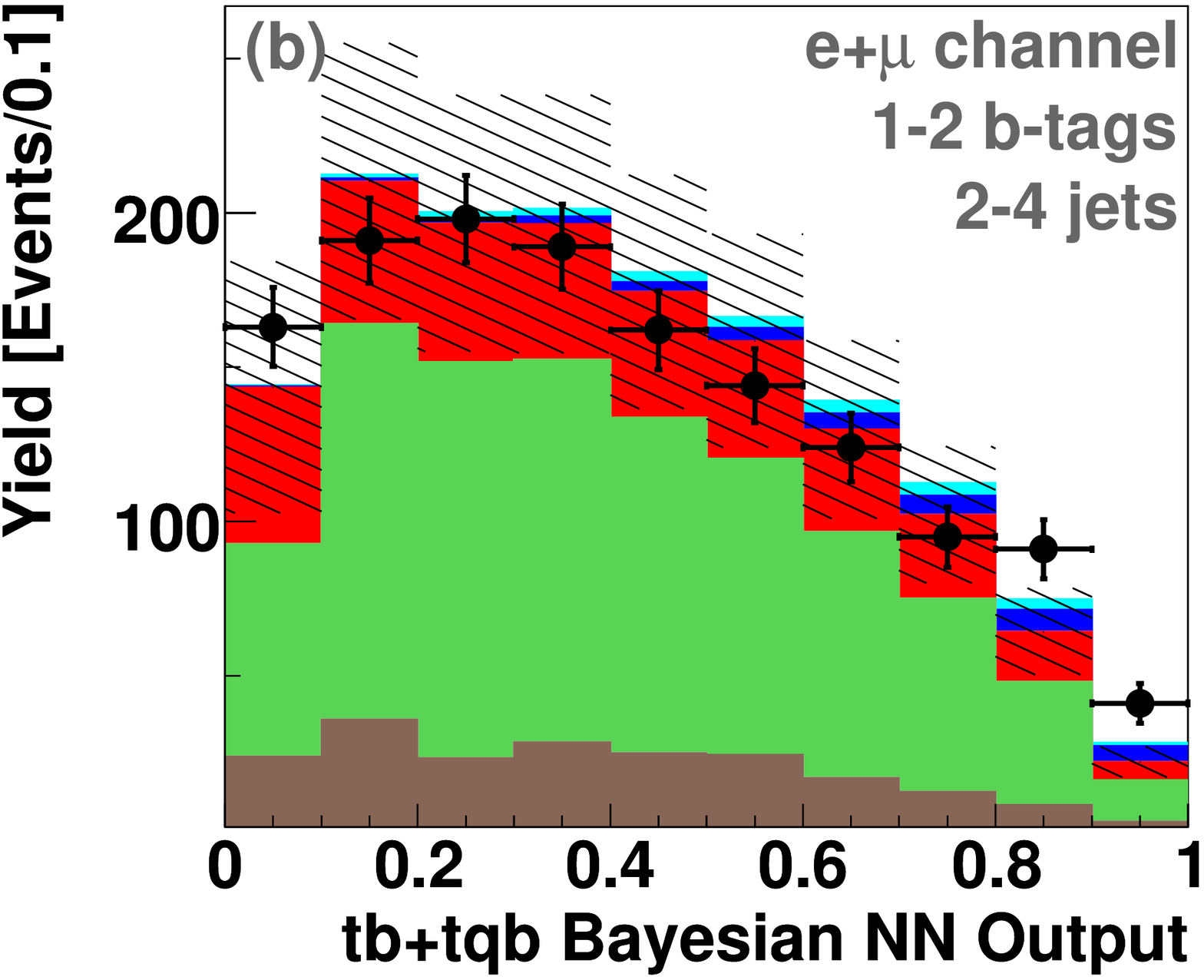}

\includegraphics[width=0.32\textwidth]
{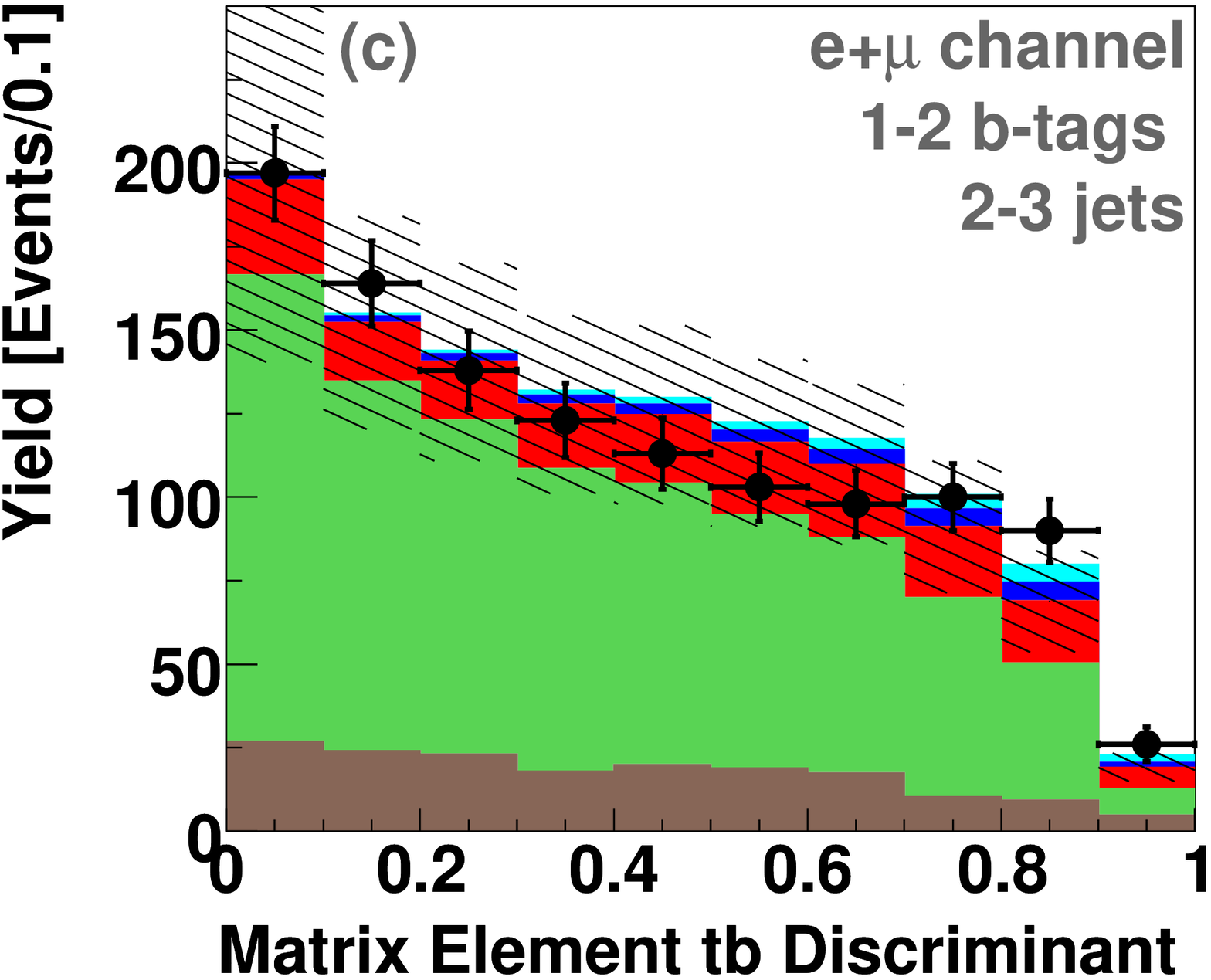}
\includegraphics[width=0.32\textwidth]
{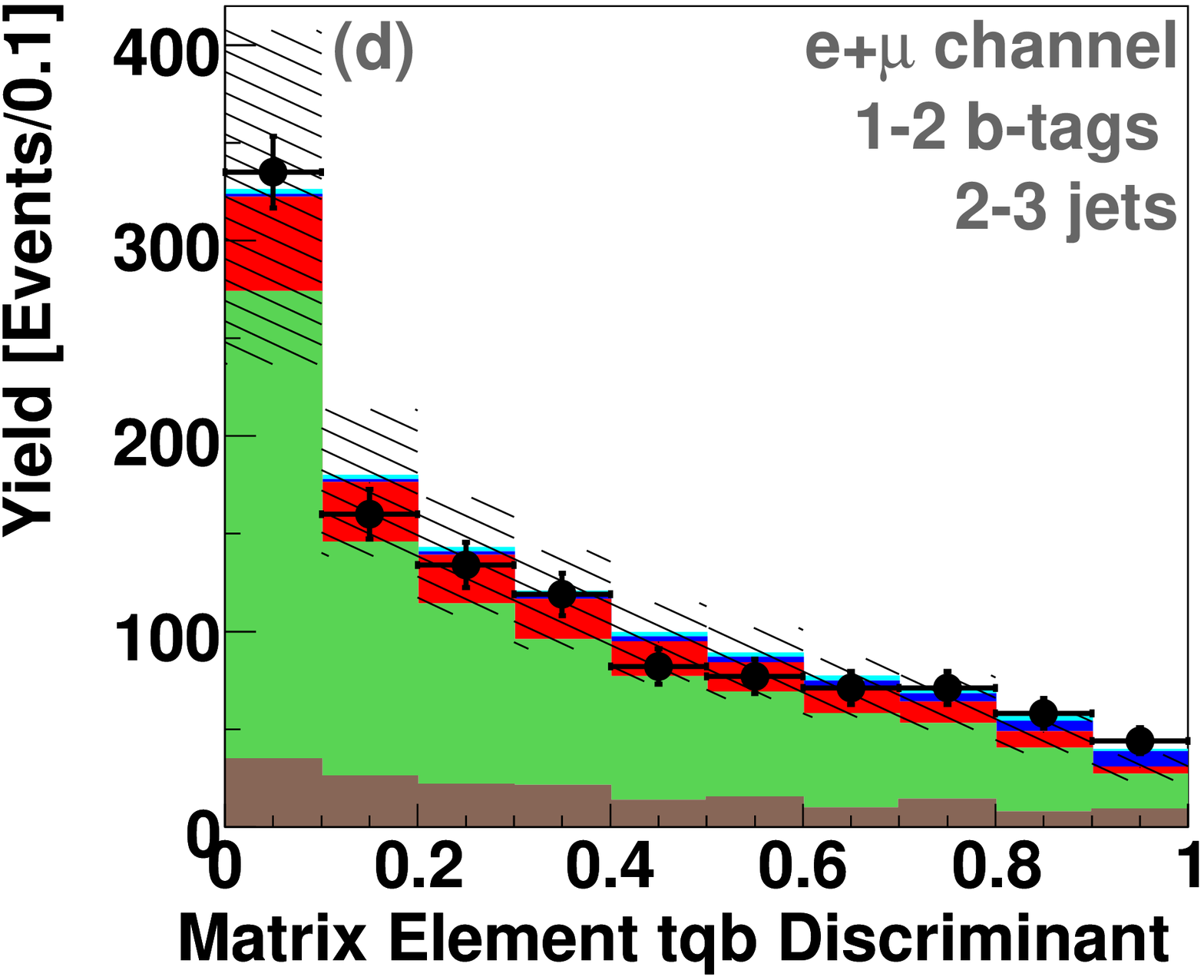}
\vspace{-0.1in}
\caption[disc]{The discriminant outputs of the three
multivariate discriminants: (a)~DT, (b)~BNN, (c)~ME s-channel, and
(d)~ME t-channel discriminants. The signal components are
normalized to the expected standard model cross sections of 0.88~pb and 1.98~pb
for the s- and t-channels, respectively.
The hatched bands show the 1~$\sigma$ uncertainty on the background.}
\label{fig:disc}
\end{figure*}

The discriminant outputs for the data and the expected standard model
contributions are shown in Fig.~\ref{fig:disc} for the three
multivariate techniques. The outputs show good agreement between data
and backgrounds, except in the high discriminant regions, where an
excess of data over the background prediction is observed.

%---------------------------------------------------------------------
%---------------------------------------------------------------------
\vspace{-0.15in}
\section{Ensembles and Bias Studies}
\label{sec:ensembles}
\vspace{-0.05in}

We have described three sophisticated analyses (DT, BNN, ME), each of
which produces a posterior density for the single top quark production
cross section. When applied to real data, we obtain well-behaved
posterior densities. However, this does not guarantee that these
methods are trustworthy and perform as advertised. In order to
validate the methods, it is necessary to study their behavior on
ensembles of pseudodatasets with characteristics as close as possible
to those of the real data.  We can use such ensembles to determine,
for example, whether an analysis is able to extract a cross section
from a signal masked by large backgrounds. We can also determine
whether the claimed accuracy is warranted. Moreover, by running the
three analyses on exactly the same ensembles, we can study in detail
the correlations across analyses and the frequency properties of
combined results and their significance.

We generate pseudodatasets from a pool of weighted signal and
background events, separately for the electron and muon channels. For
example, out of 1.3 million electron events, we calculate a total
background yield of 756 events in the selected data. We randomly
sample a count $N$ from a Poisson distribution of mean $n = 756$ and
select $N$ events, with replacement, from the pool of 1.3 million
weighted events so that events are selected with a frequency
proportional to their weight.  The sample contains the appropriate
admixture of signal and background events, as well as the correct
Poisson statistics. Moreover, we take into account the fact that the
multijets and $W$+jets sample sizes are $100\%$ anticorrelated. The
sample is then partitioned according to the $b$~tag and jet
multiplicities, mirroring what is done to the real data.  The Poisson
sampling, followed by sampling with replacement, is repeated to
generate as many pseudodatasets as needed. Each pseudodataset is then
analyzed in exactly the same way as real data.

We have performed studies using many different ensembles, of which the
most important ones are:
\begin{myitemize}

\item {\bf Background only (i.e., zero signal) ensemble with systematics} ---
the background is set to the estimated background yield value; the
signal cross section is set to 0~pb; these Poisson-smeared means are
further randomized to represent the effects of all systematic
uncertainties.

\item {\bf Standard model signal ensemble with systematics} ---
the background is set to the estimated background yield value; the
signal cross section is set to the standard model value of 2.86~pb;
these Poisson-smeared means are further randomized to represent the
effects of all systematic uncertainties.

\item {\bf Ensembles with different signal cross sections} ---
the background is set to the estimated background yield value; the
signal cross section is set to a fixed value between 0~pb and a few
times the standard model value in each ensemble; only Poisson-smearing
for statistical effects is applied.

\end{myitemize}
We use the zero-signal ensemble (with systematics) to calculate the
$p$-value, a measure of the significance of the observed excess. The
$p$-value is the probability that we obtain a measured cross section
greater than or equal to the observed cross section, if there were no
signal present in the data.

We use the SM signal ensemble (with systematics) to determine the
correlations between the three analysis methods so we can combine
their results. We also use this ensemble to calculate the
compatibility of our measured result with the SM prediction, by
determining how many pseudodatasets have a measured cross section at
least as high as the result measured with data.

\begin{figure*}[!h!btp]
\begin{center}
\includegraphics[width=0.38\textwidth]{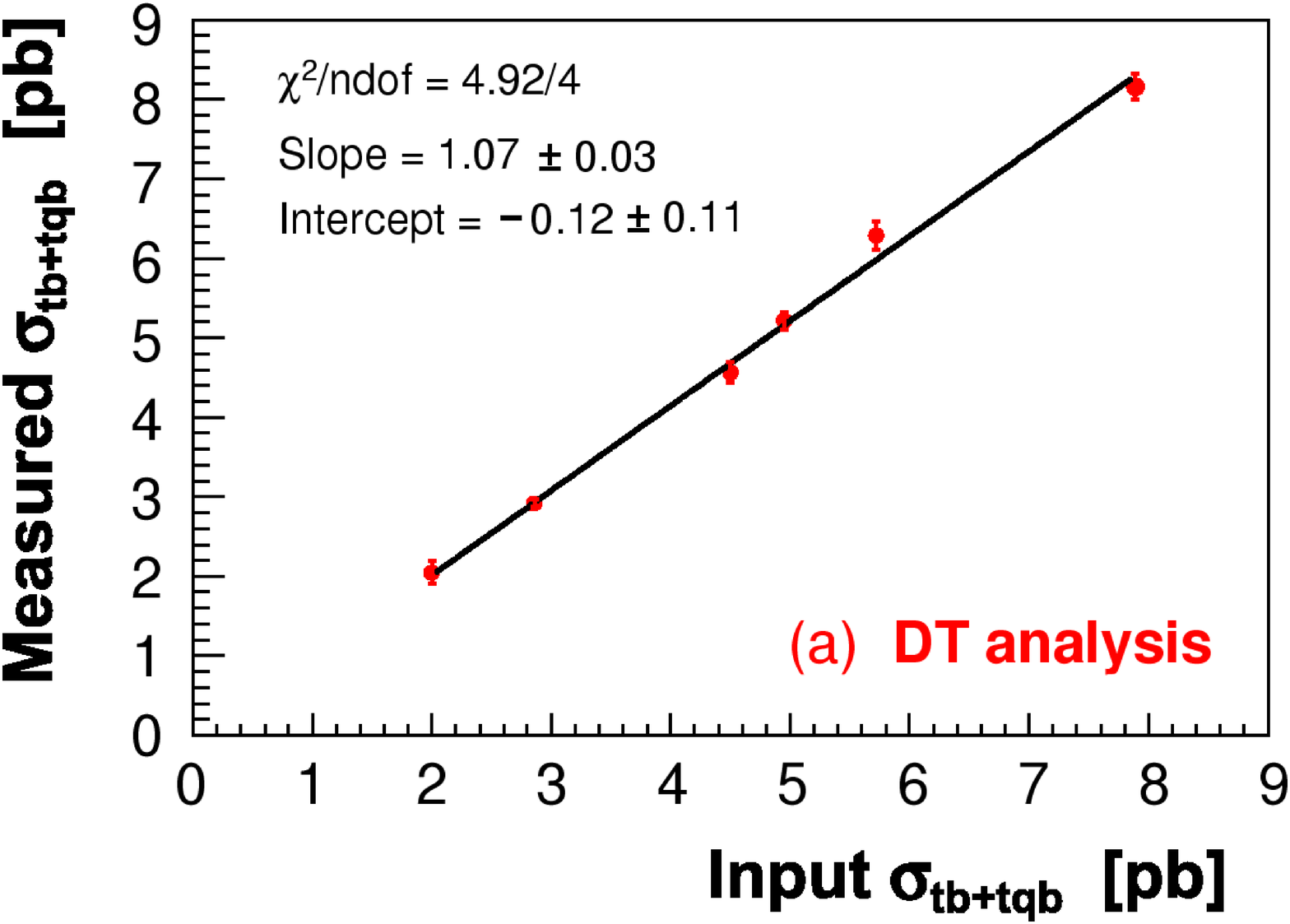} 
\includegraphics[width=0.38\textwidth]{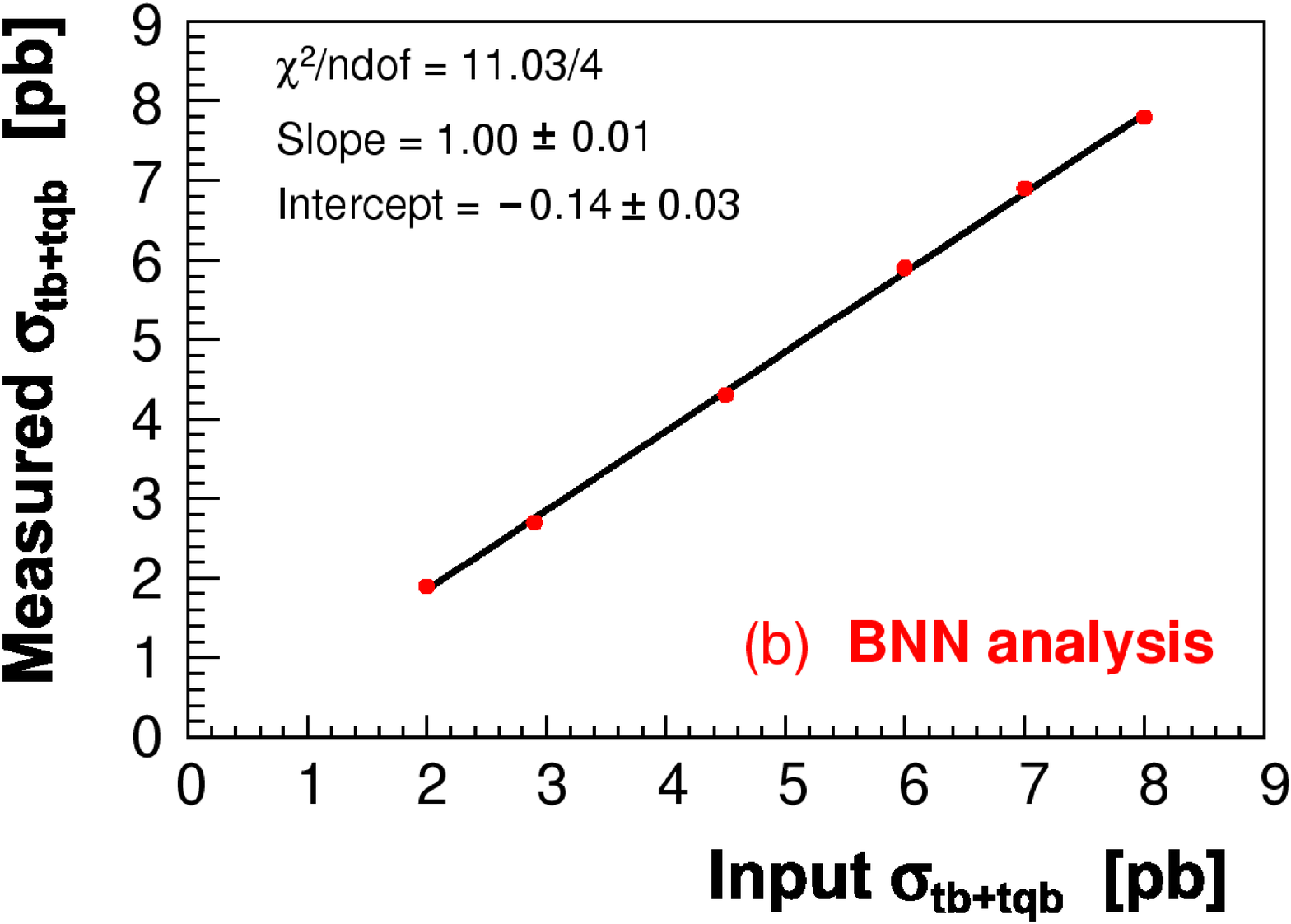} 

\includegraphics[width=0.38\textwidth]{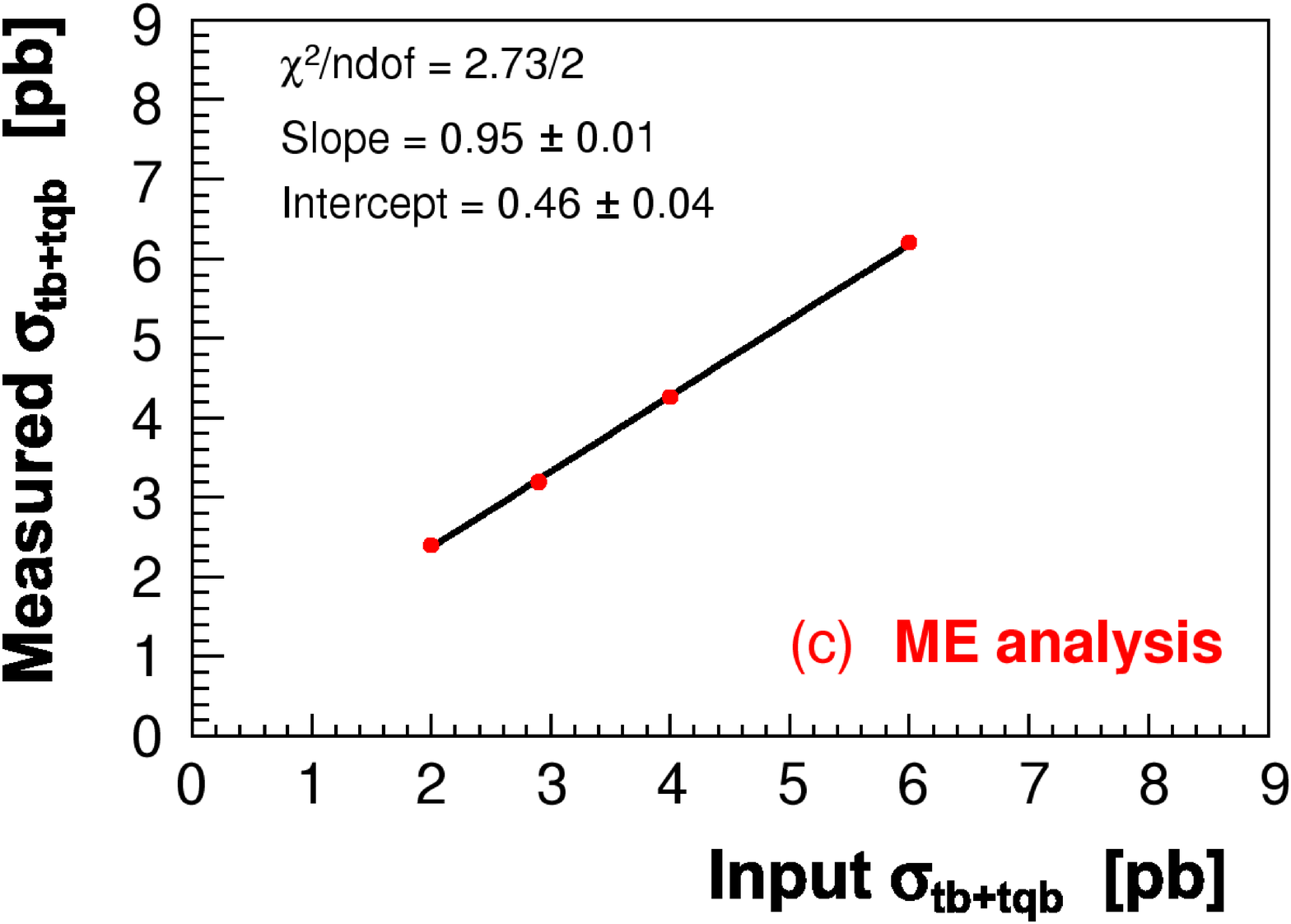} 
\vspace{-0.1in} 
\caption[ensembleSMst]{Ensemble average of measured cross section
as a function of the input single top quark cross section for the
(a)~DT, (b)~BNN, and (c)~ME analyses.}
\label{fig:calibration} 
\end{center}
\end{figure*}

The set of ensembles with different values for the signal cross
section is used to assess bias in the cross section measurement, that
is, the difference between the input cross section and the mean of the
distribution of measured cross sections. For each multivariate
analysis, the bias is estimated by applying the entire analysis chain
to the ensembles of pseudodatasets that each have a different value
for the single top quark cross section.  Straight-line fits of the
average of the measured cross sections versus the input cross section
for the three multivariate analyses are shown in
Fig.~\ref{fig:calibration}. From this measurement, we conclude that
the bias in all three analyses is small.  Moreover, when compared with
the variances of the ensemble distributions of measured values, the
biases are negligible. We thus perform no correction to the expected
or measured cross section values.

%---------------------------------------------------------------------
%---------------------------------------------------------------------
\vspace{-0.15in}
\section{Cross-Check Studies}
\label{sec:crosschecks}
\vspace{-0.05in}

In order to check the background model, we apply the multivariate
discriminants to two background-dominated samples defined by the
following criteria: (i)~2~jets, 1~$b$~tag, and
$H_T(\ell,{\met},{\mathrm{alljets}}) < 175$~GeV for a ``$W$+jets''
sample; and (ii)~4~jets, 1~$b$~tag, and
$H_T(\ell,{\met},{\mathrm{alljets}}) > 300$~GeV for a ``{\ttbar}''
sample. The first sample is mostly $W$+jets and almost no \ttbar,
while the second is mostly {\ttbar} and almost no $W$+jets.

\begin{figure*}[!h!tbp]
\includegraphics[width=0.30\textwidth]
{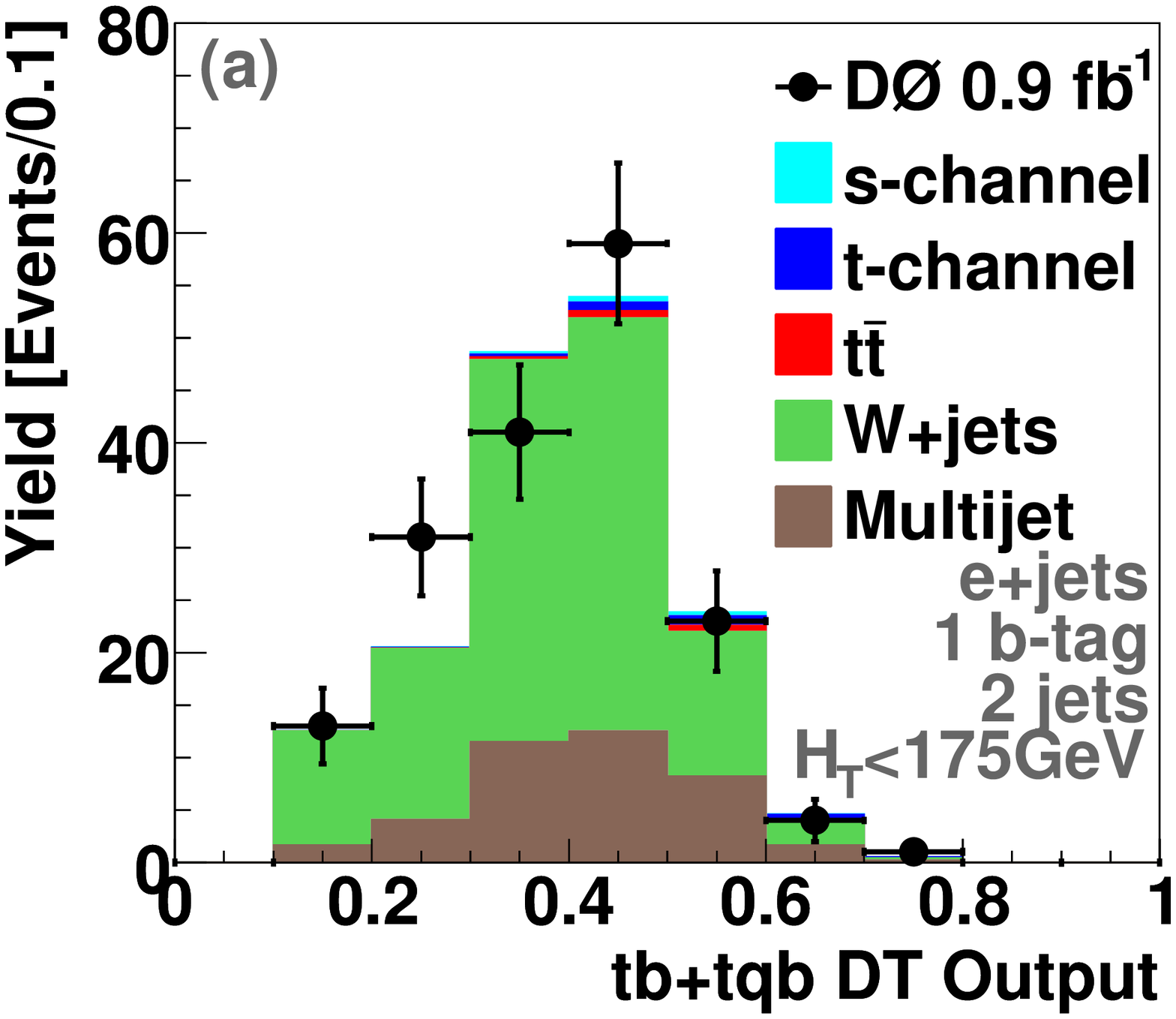}
\includegraphics[width=0.30\textwidth]
{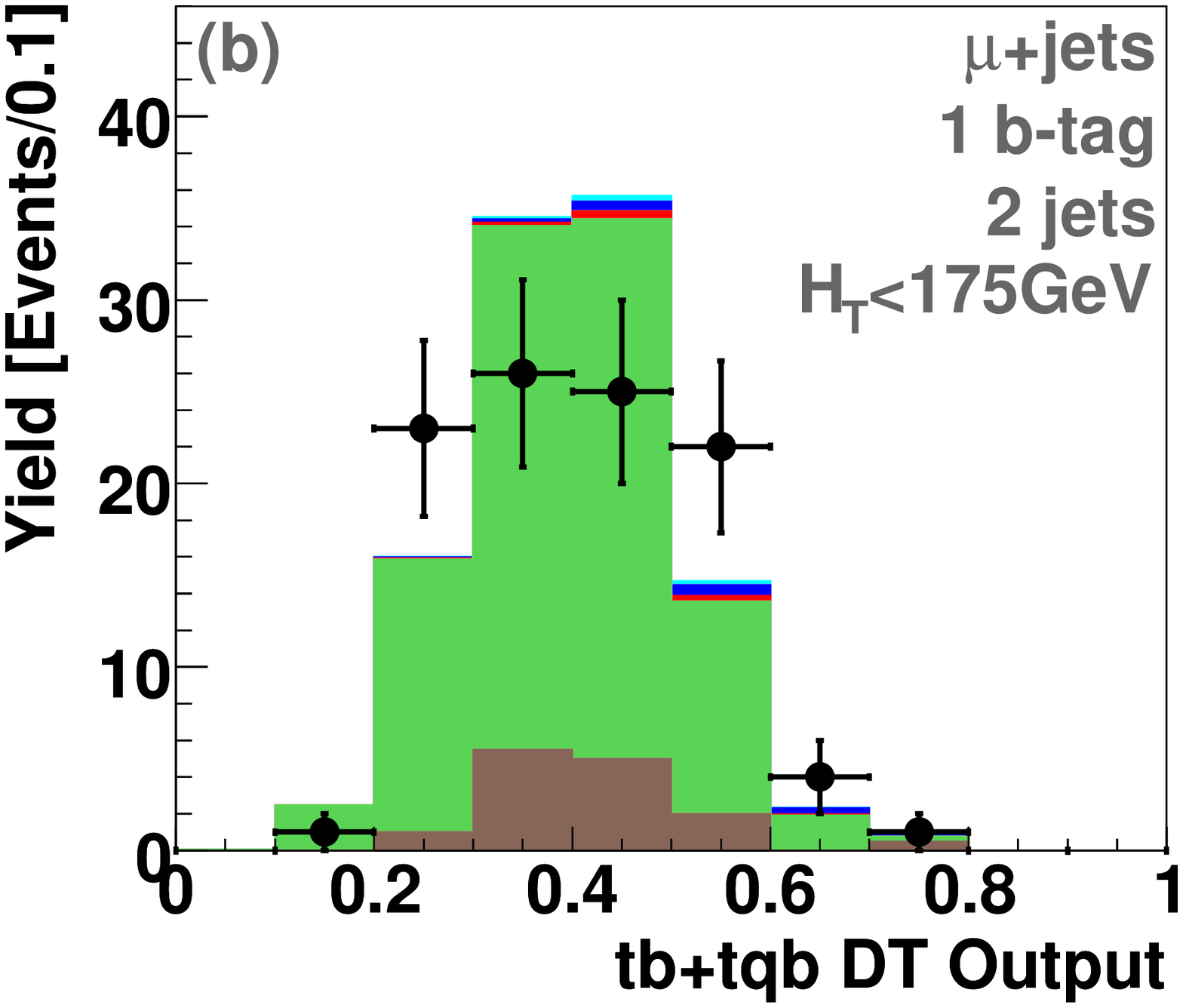}

\includegraphics[width=0.30\textwidth]
{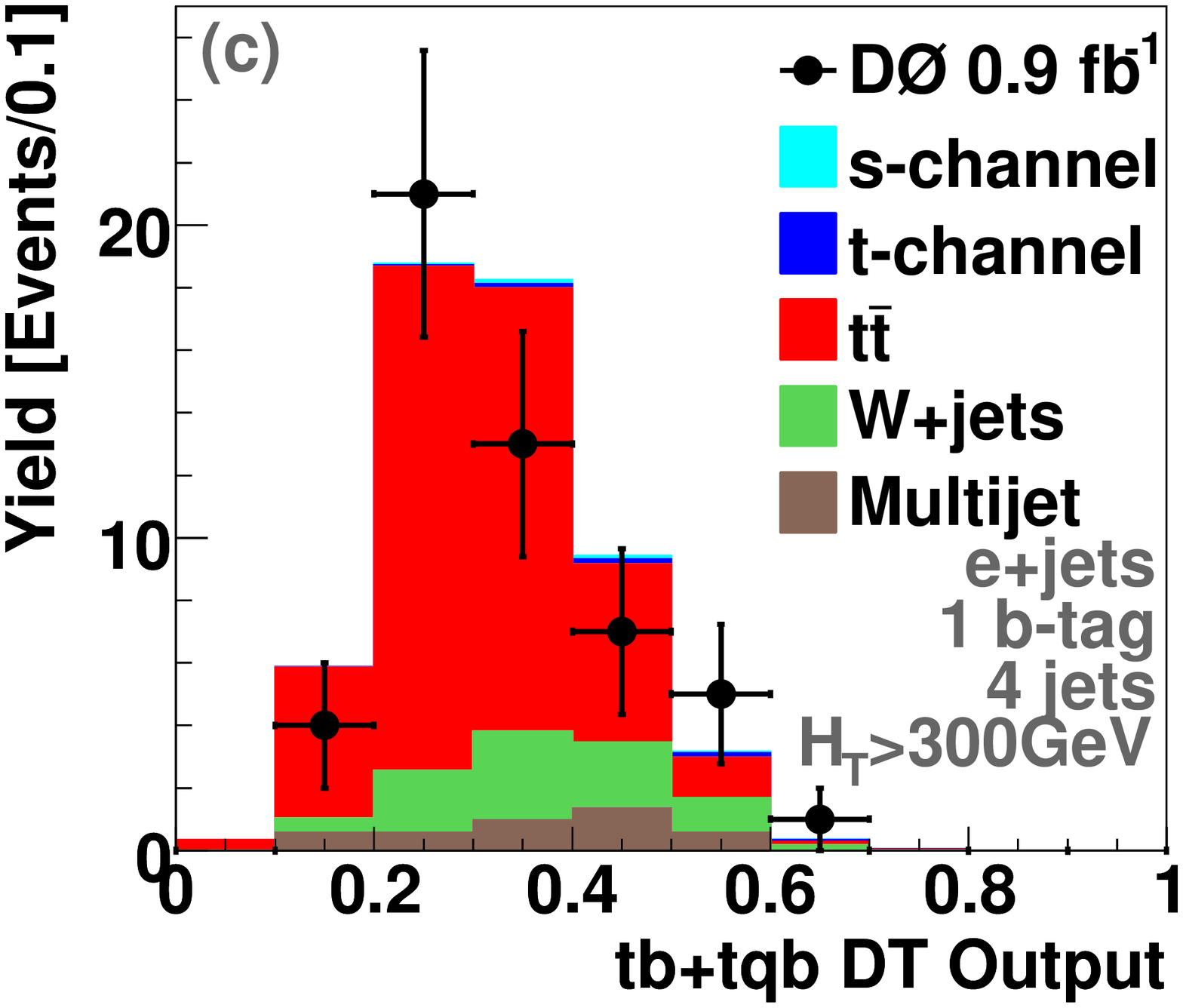}
\includegraphics[width=0.30\textwidth]
{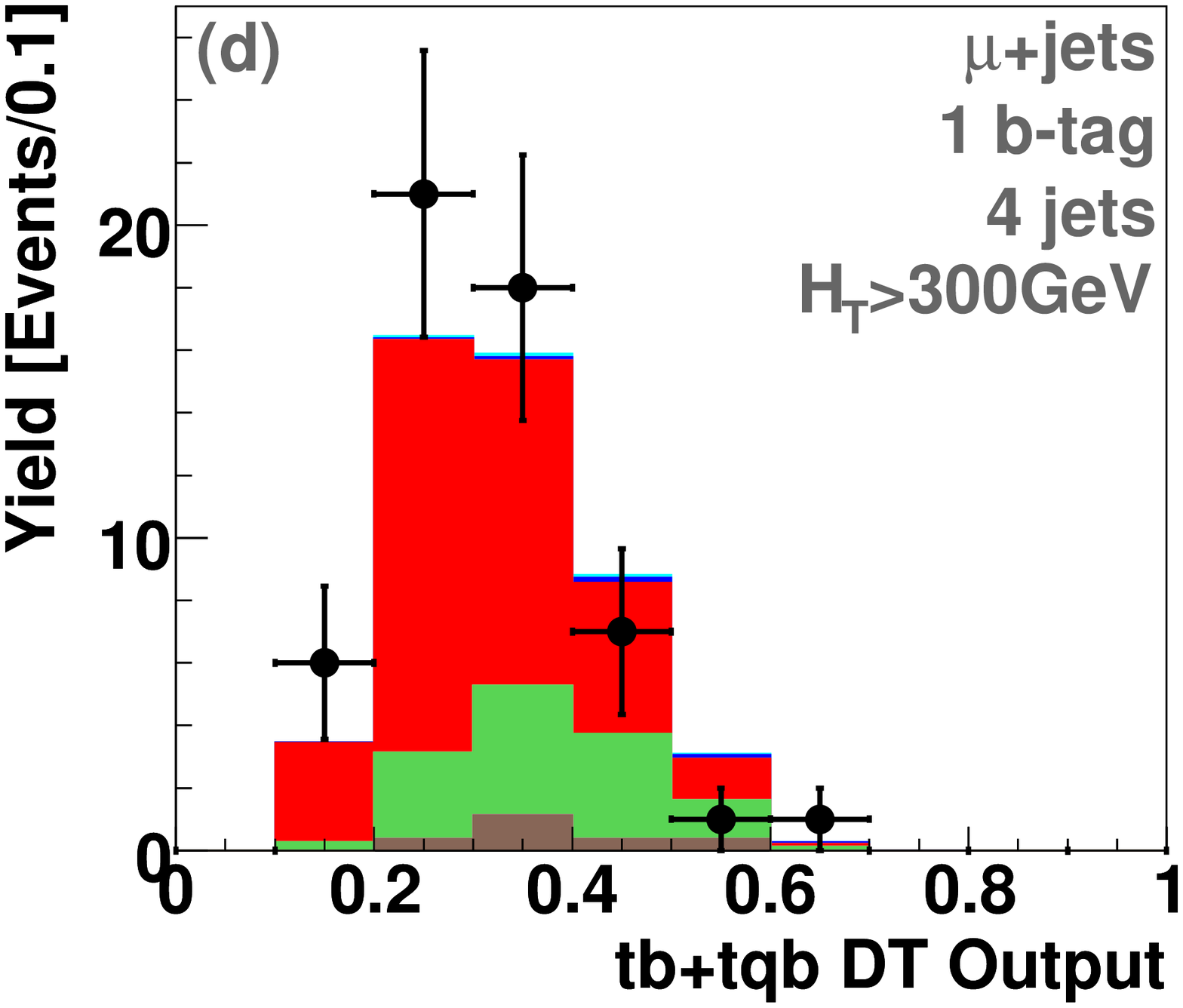}
\vspace{-0.1in}
\caption[dt-crosschecks]{DT outputs from the $W$+jets (upper row) and
{\ttbar} (lower row) cross-check samples for $e$+jets events (left
column) and $\mu$+jets events (right column).}
\label{fig:DT-wjets-ttbar-crosscheck}
\end{figure*}

\begin{figure*}[!h!tbp]
\includegraphics[width=0.30\textwidth]
{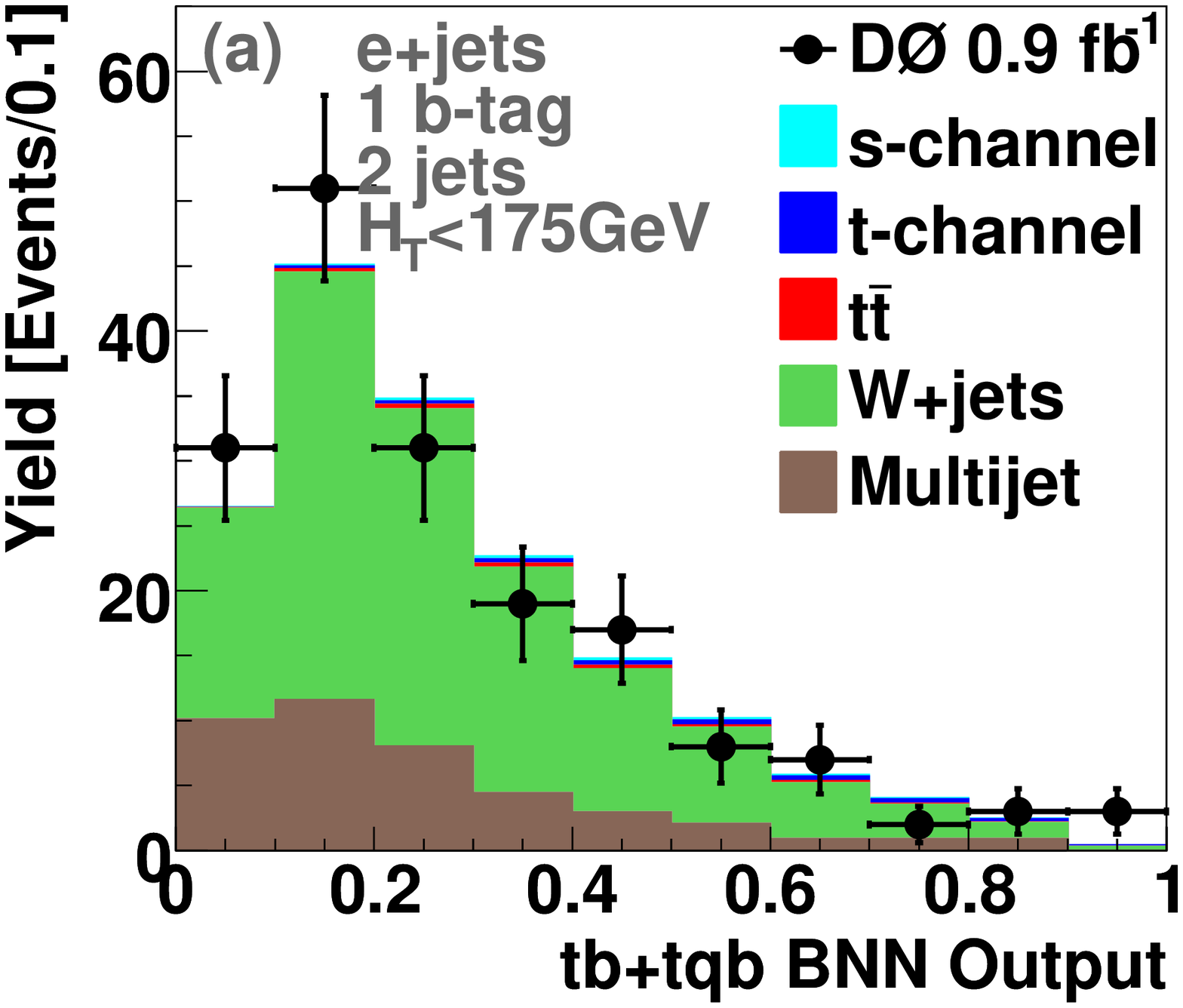}
\includegraphics[width=0.30\textwidth]
{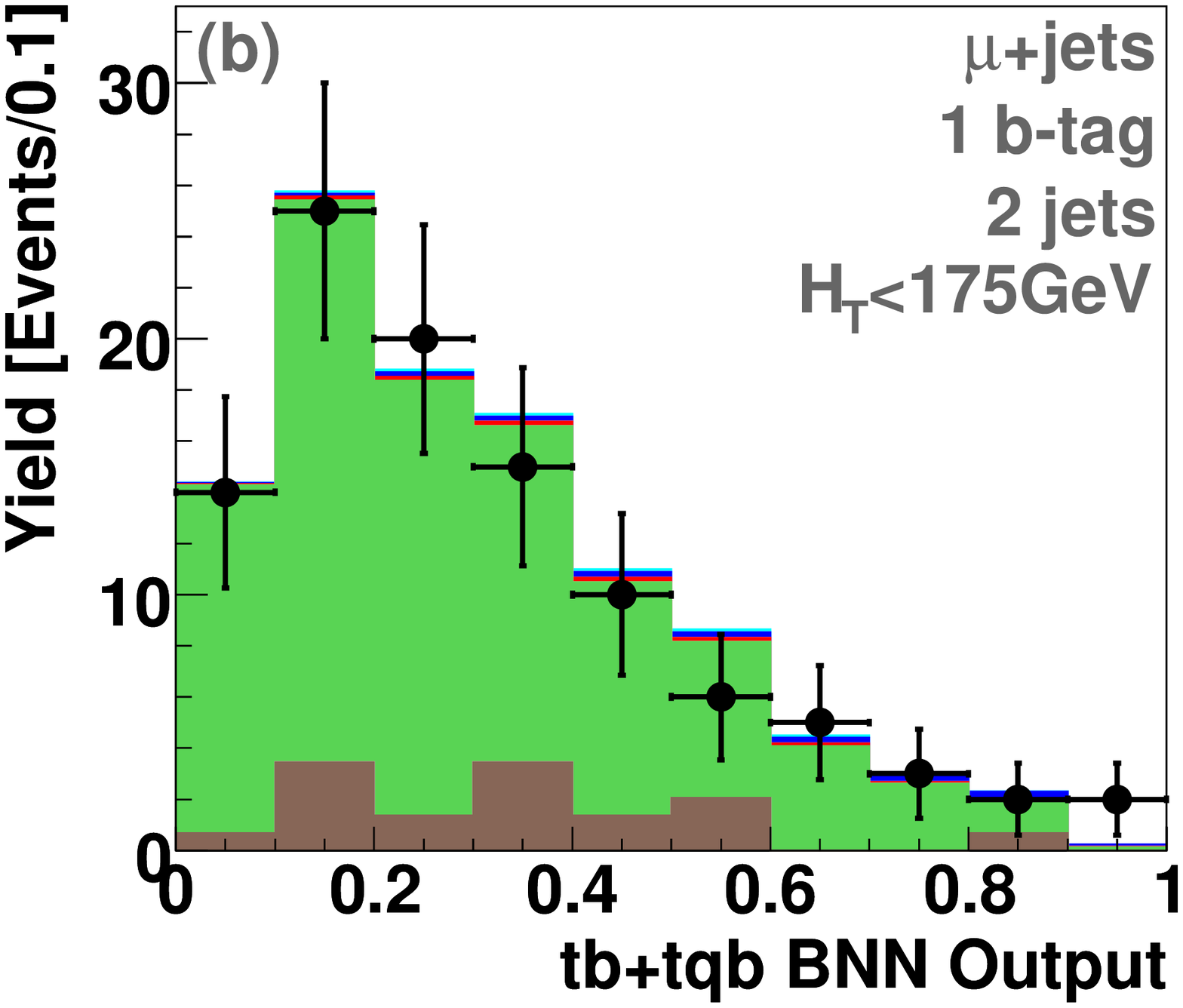}

\includegraphics[width=0.30\textwidth]
{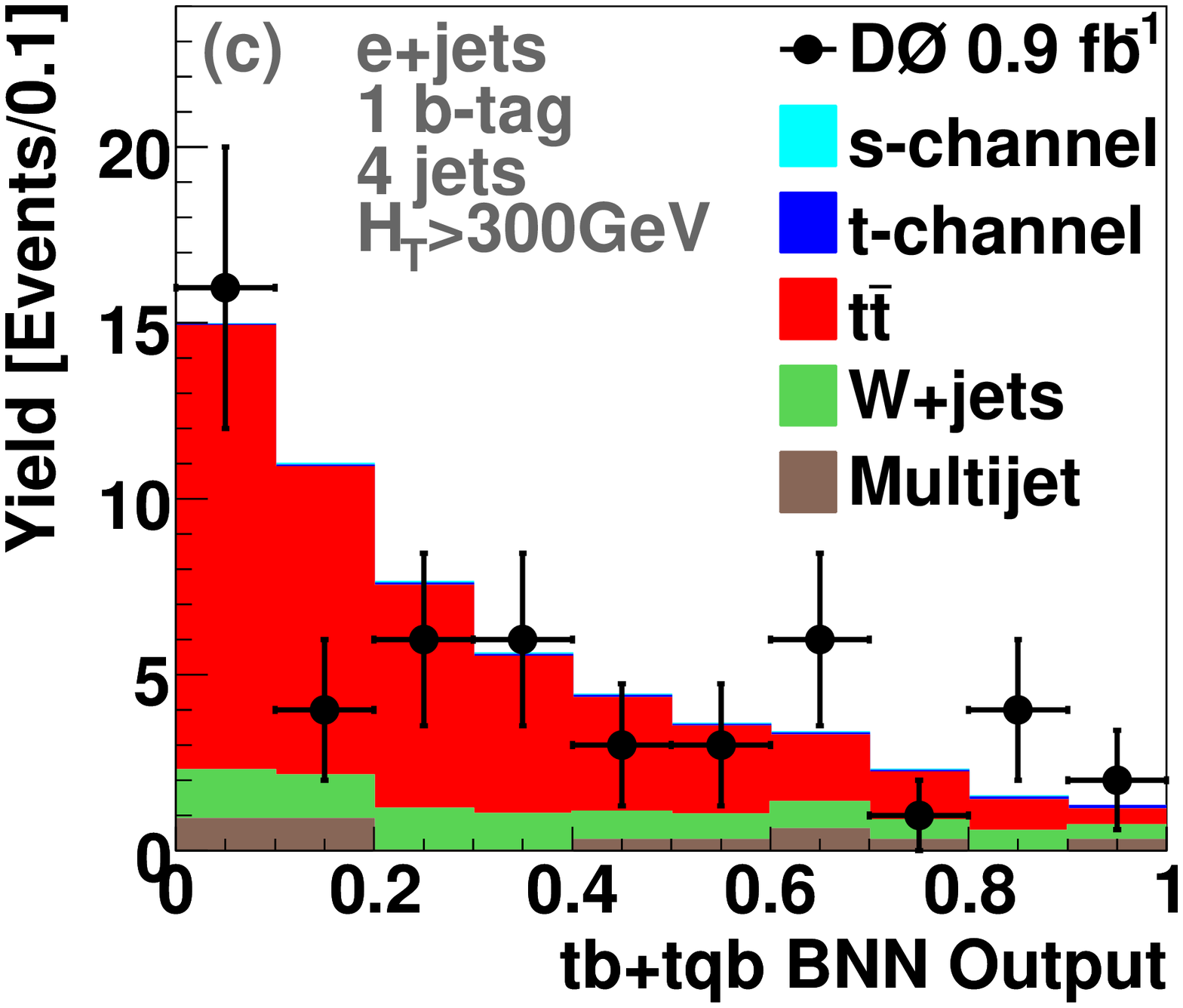}
\includegraphics[width=0.30\textwidth]
{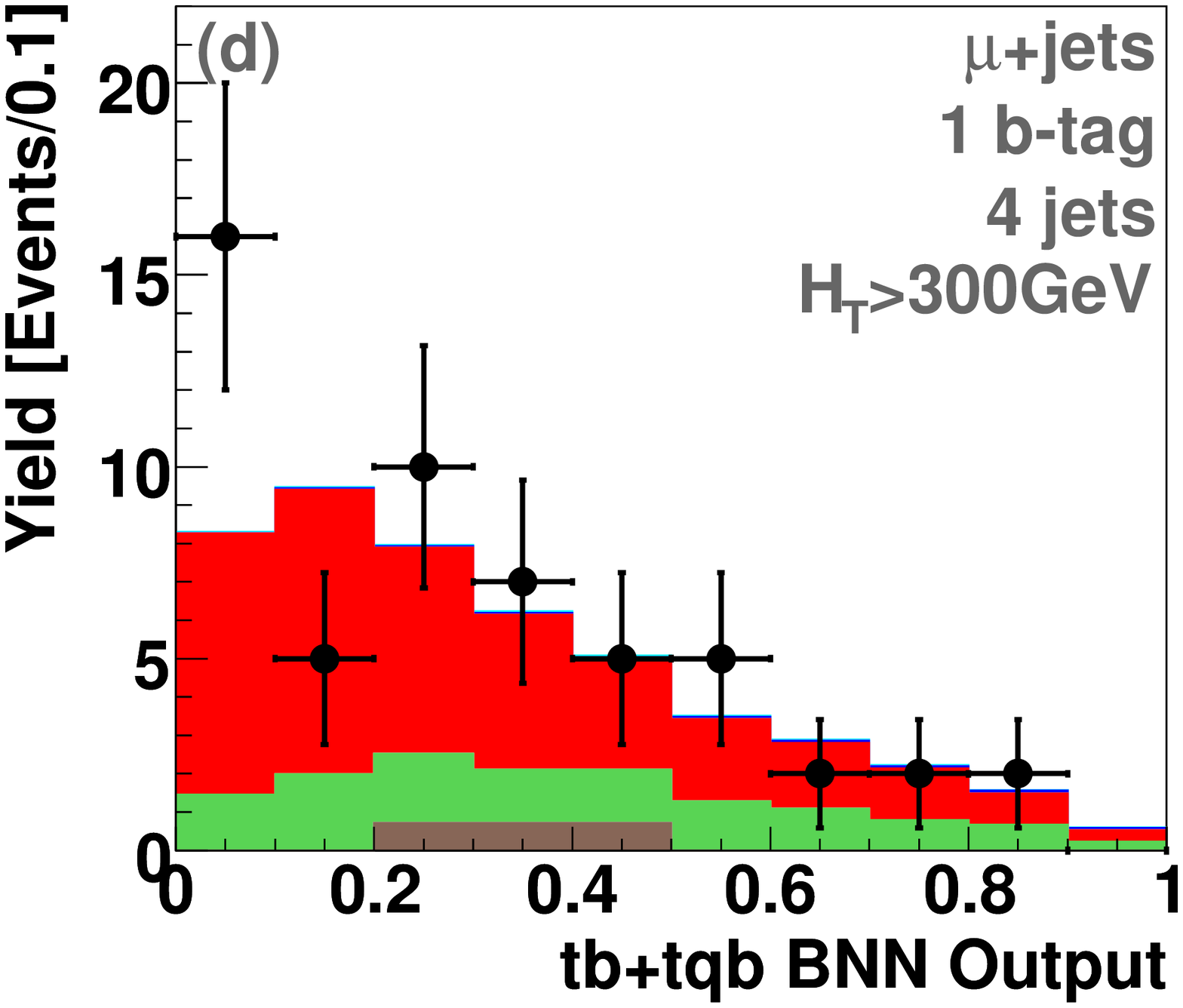}
\vspace{-0.1in}
\caption[bnn-crosschecks]{BNN outputs from $W$+jets (upper row) and
{\ttbar} (lower row) cross-check samples for $e$+jets events (left
column) and $\mu$+jets events (right column).}
\label{fig:BNN-wjets-ttbar-crosscheck}
\end{figure*}

The $tb$+$tqb$ decision tree output distributions for these
cross-check samples are shown in
Fig.~\ref{fig:DT-wjets-ttbar-crosscheck} and the corresponding
Bayesian neural network output distributions are shown in
Fig.~\ref{fig:BNN-wjets-ttbar-crosscheck}. From these data-background
comparisons, we conclude that there is no obvious bias in our
measurement. The background model describes the data within
uncertainties.

\begin{figure*}[!h!tbp]
\includegraphics[width=0.30\textwidth]
{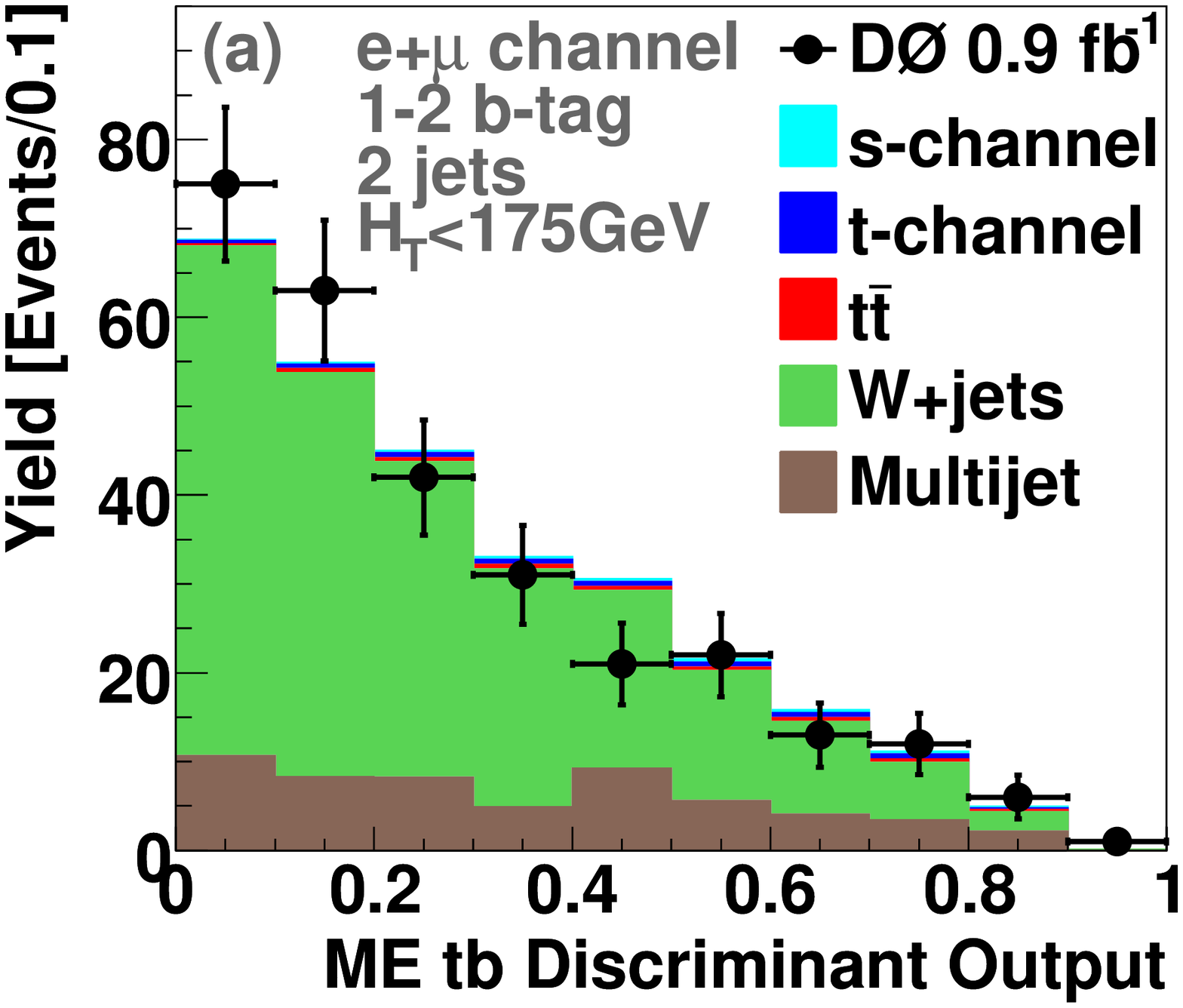}
\includegraphics[width=0.30\textwidth]
{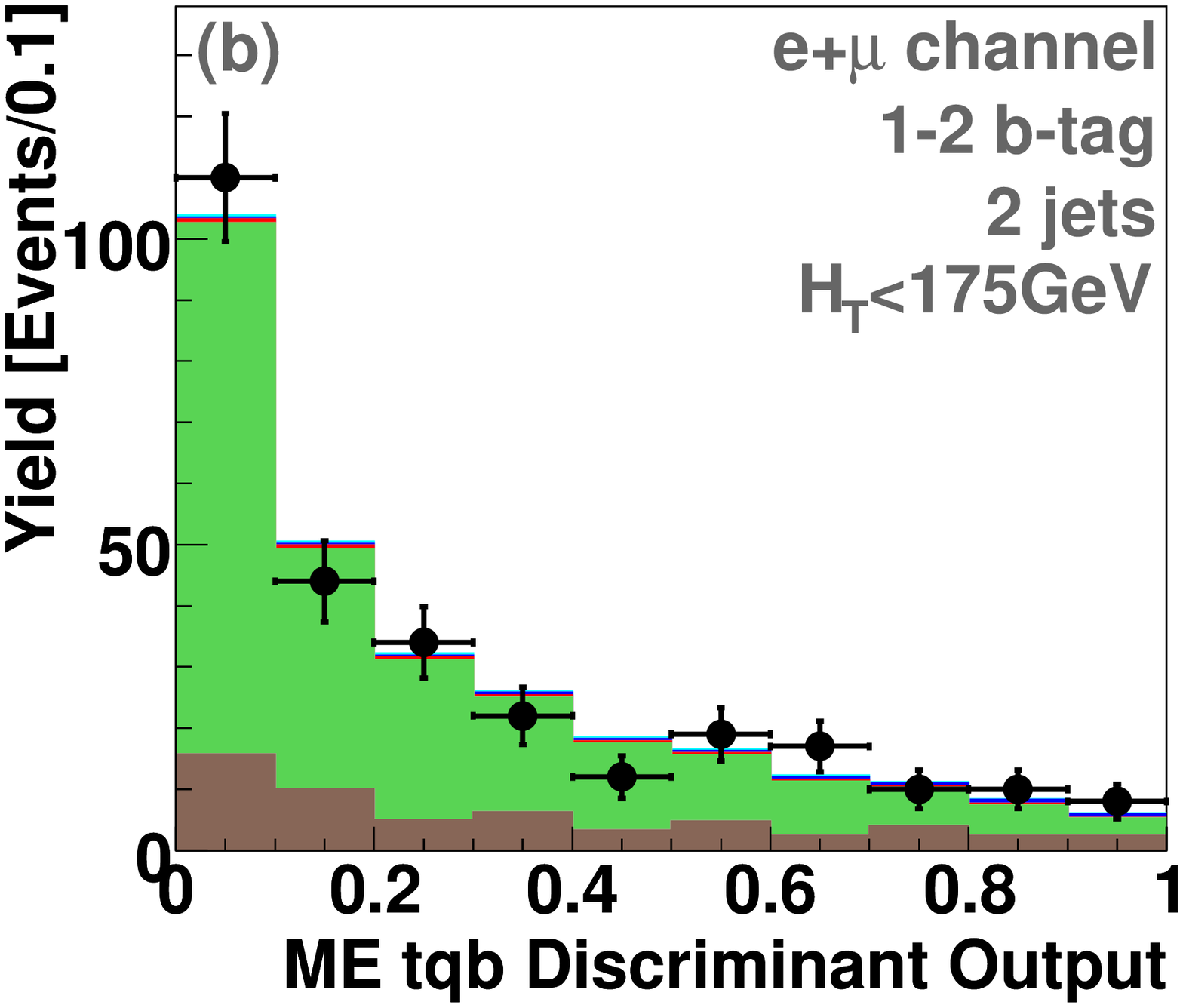}

\includegraphics[width=0.30\textwidth]
{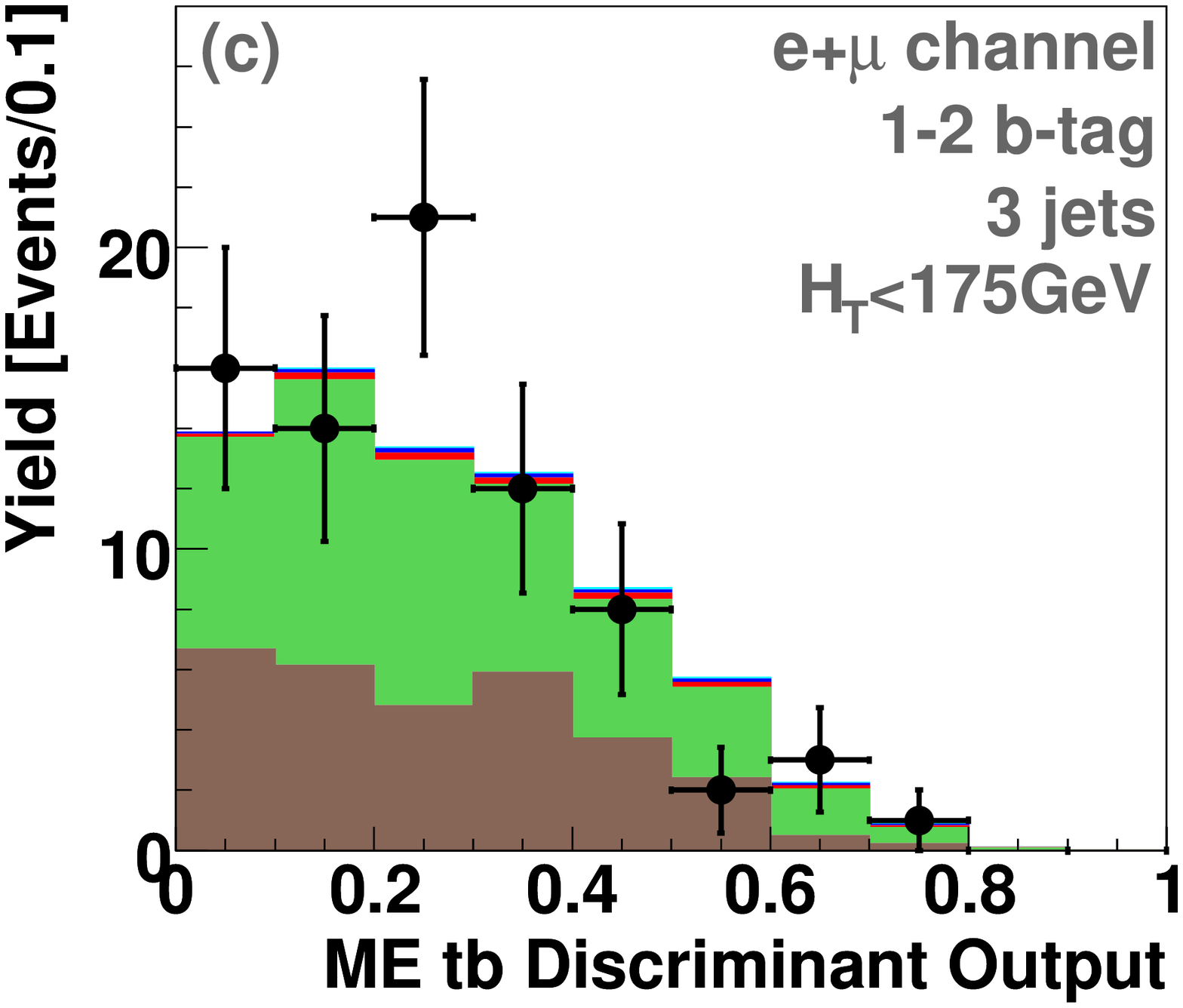}
\includegraphics[width=0.30\textwidth]
{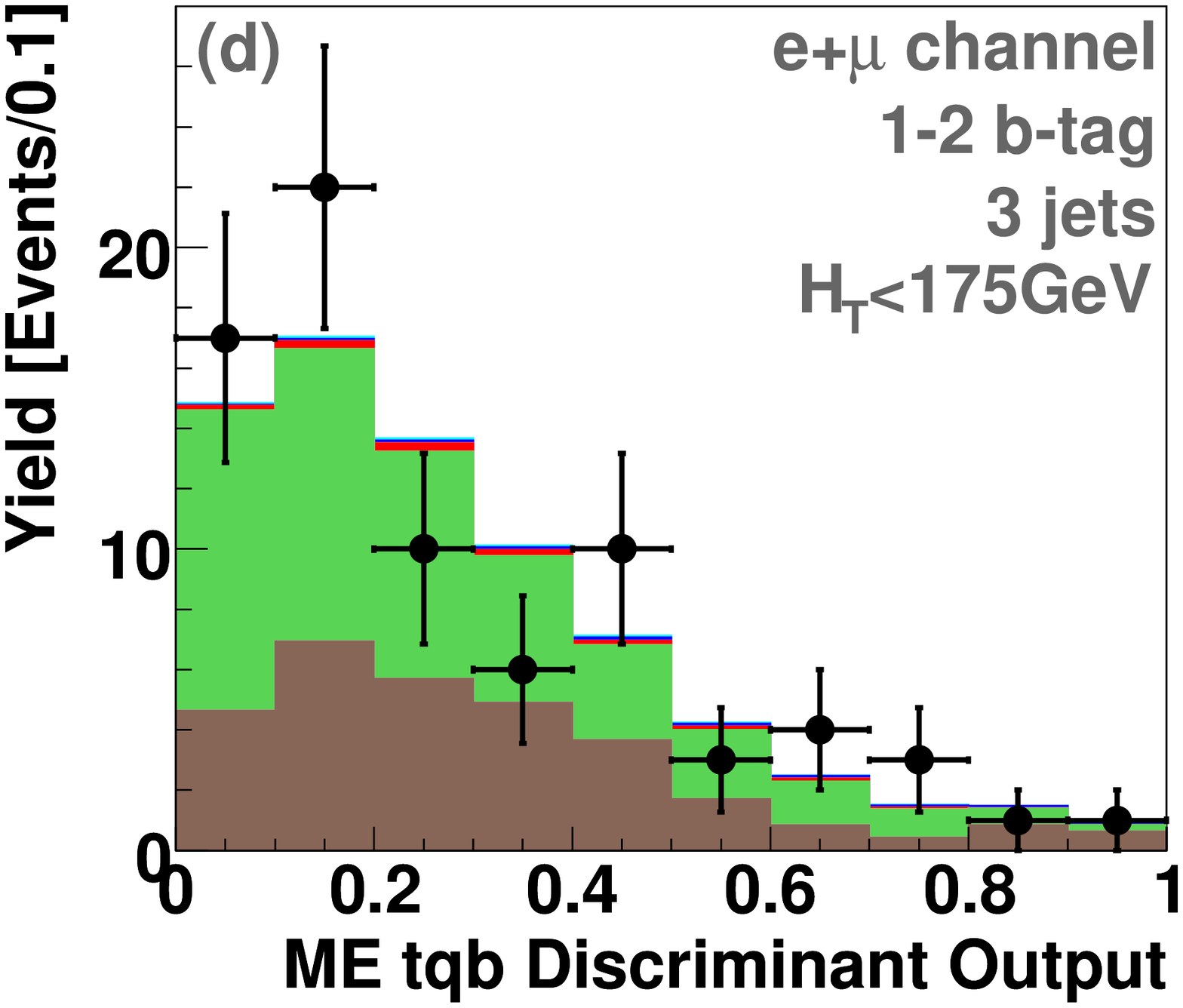}
\vspace{-0.1in}
\caption[me-crosschecks-lowHT]{$H_T < 175$~GeV cross-check plots in
two-jet (upper row) and three-jet (lower row) events for the s-channel
ME discriminant (left column) and the t-channel ME discriminant (right
column). The plots have electrons and muons, one and two $b$~tags
combined.}
\label{fig:ME-lowHT-crosscheck}
\end{figure*}

\begin{figure*}[tbp]
\includegraphics[width=0.30\textwidth]
{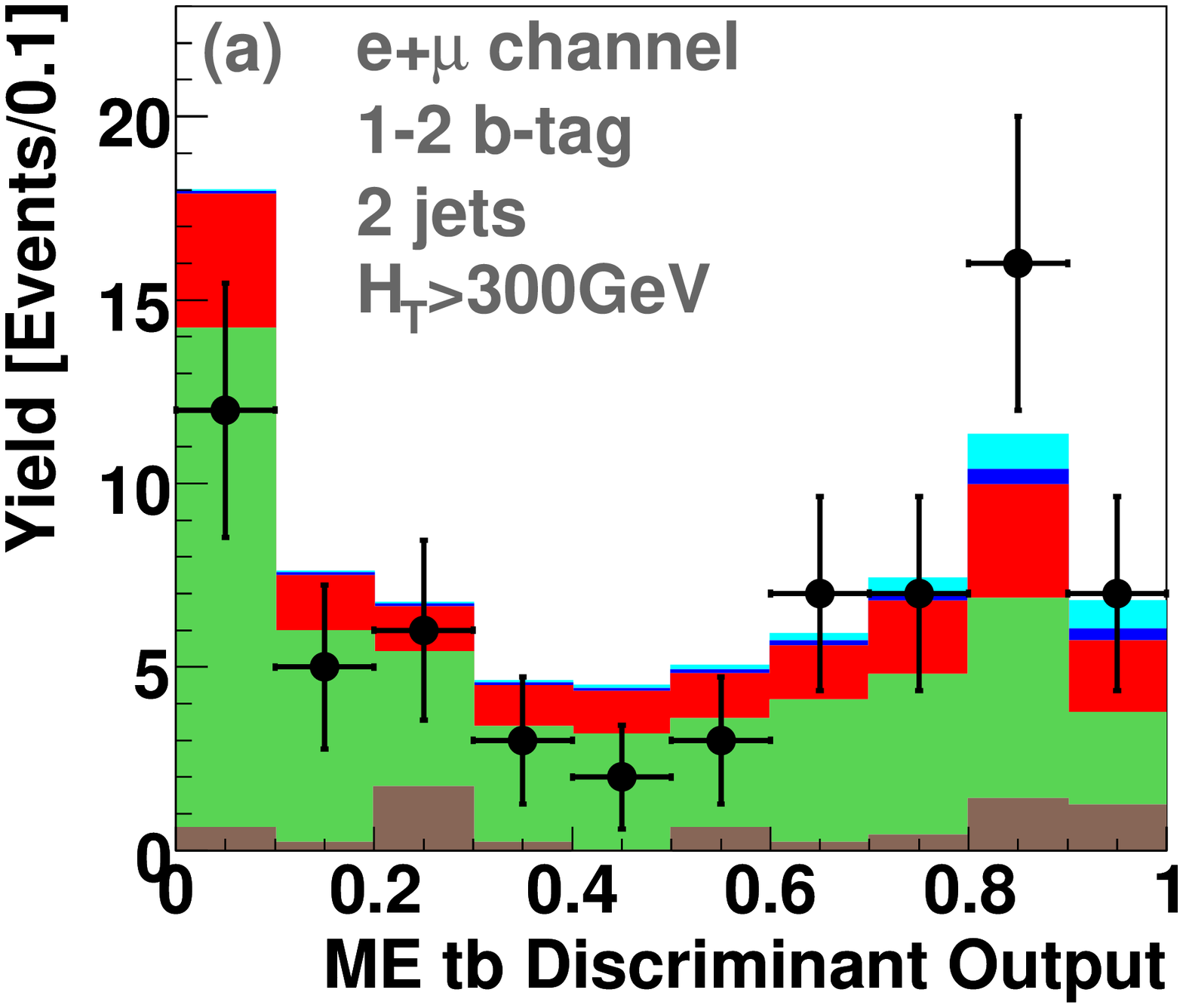}
\includegraphics[width=0.30\textwidth]
{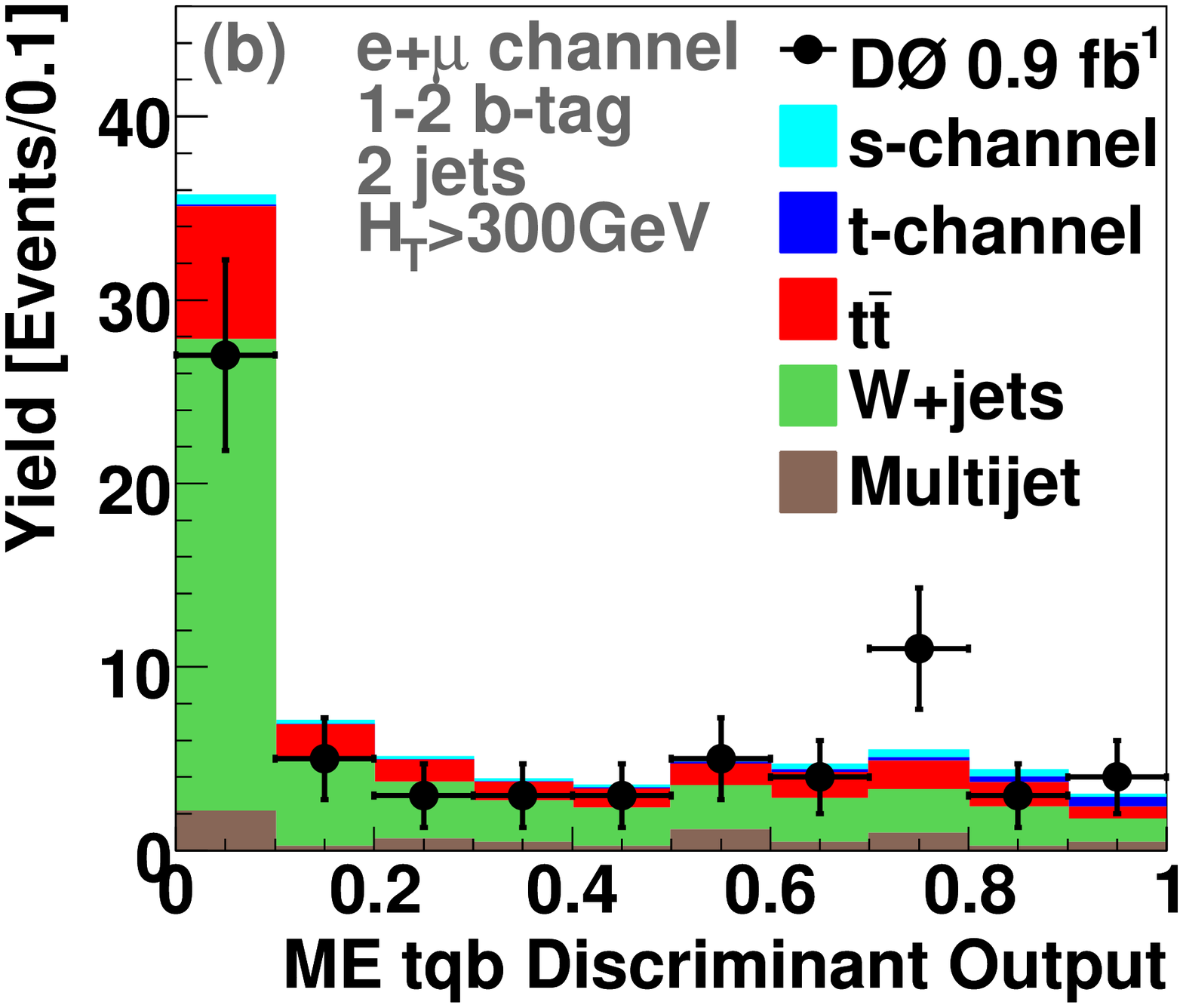}

\includegraphics[width=0.30\textwidth]
{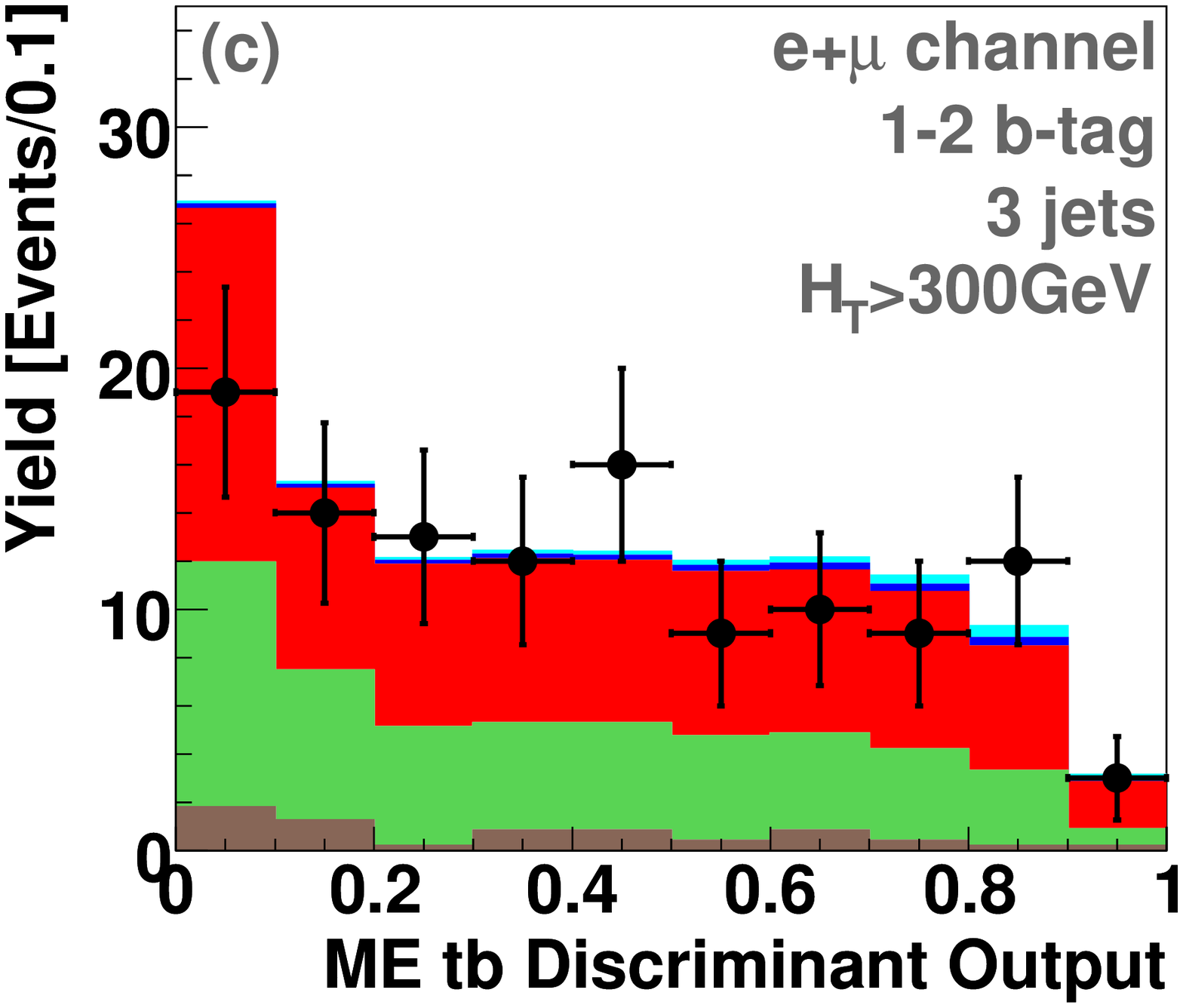}
\includegraphics[width=0.30\textwidth]
{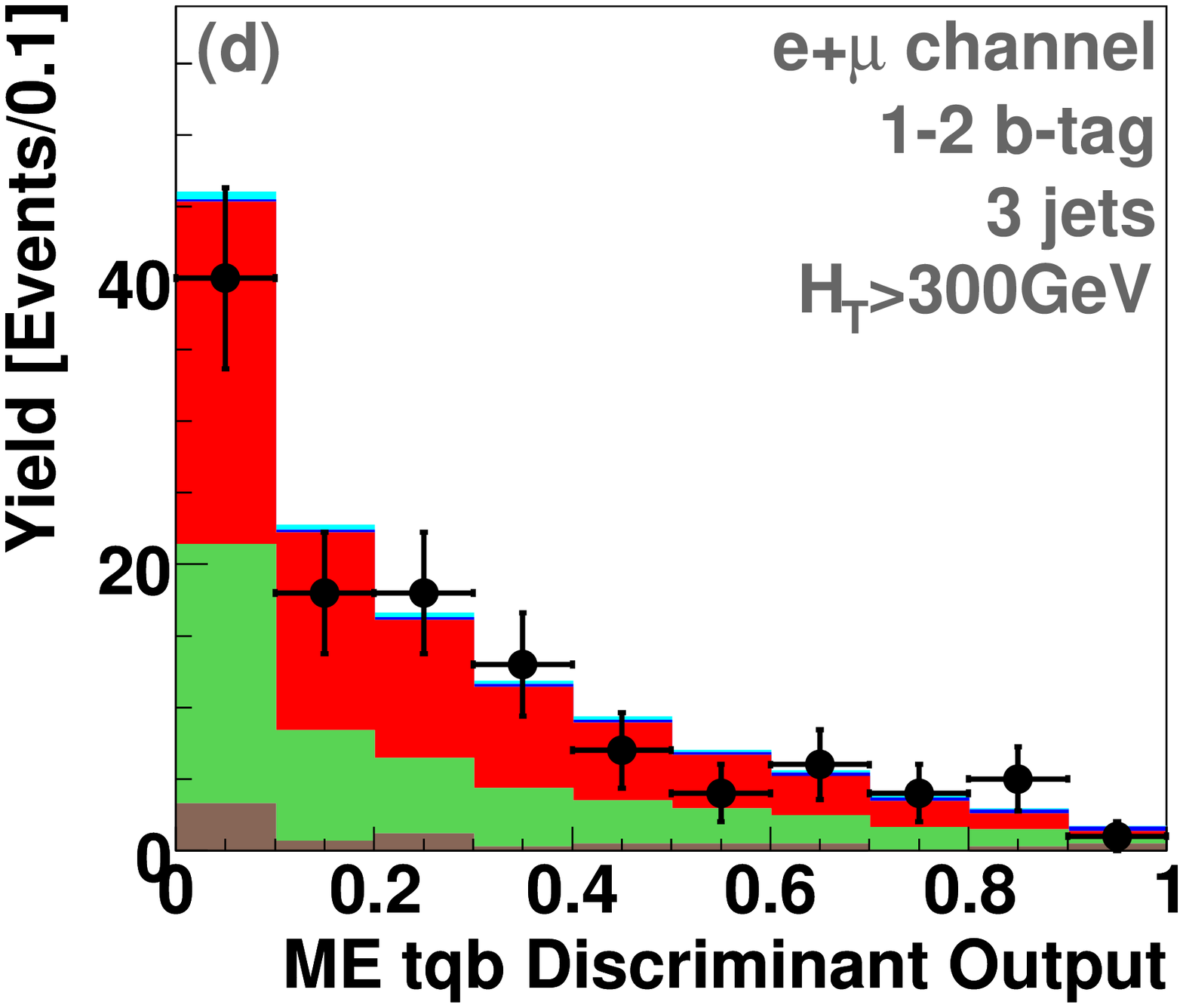}
\caption[ME-crosschecks-highHT]{$H_T > 300$~GeV cross-check plots in
two-jet (upper row) and three-jet (lower row) events for the s-channel
ME discriminant (left column) and the t-channel ME discriminant (right
column). The plots have electrons and muons, one and two $b$~tags
combined.} 
\label{fig:ME-highHT-crosscheck}
\end{figure*}

The matrix element analysis does not use four-jet events, so the
cross-check samples are defined to have $H_T < 175$~GeV or $H_T >
300$~GeV for any number of jets. Figure~\ref{fig:ME-lowHT-crosscheck}
shows the s- and t-channel discriminant outputs for two-jet and
three-jet events for the $H_T < 175$~GeV cross-check samples. The
plots have the electron and muon channels and the one and two $b$-tag
channels combined for increased statistics.
Figure~\ref{fig:ME-highHT-crosscheck} shows the same for the $H_T >
300$~GeV samples.

\clearpage
%---------------------------------------------------------------------
%---------------------------------------------------------------------
\section{Cross Section Measurements}
\label{sec:cross_section_measurements}

We use a Bayesian approach~\cite{bayes-xsec-bertram,bayes-xsec-jaynes}
to extract the cross section $\sigma({\ppbar}{\rargap}tb+X,tqb+X)$
from the observed binned discriminant distributions. In principle, the
binning of data should be avoided because information is lost. In
practice, however, an unbinned likelihood function is invariably
approximate because of the need to fit smooth functions to the
distributions of the unbinned data. Consequently, the uncertainty in
the fits induces an uncertainty in the likelihood function that grows
linearly with the number of events. Without study, it is not clear
whether an unbinned, but approximate, likelihood function will yield
superior results to those obtained from a binned but exact one. Since
we have not yet studied the matter, we choose to bin the data and
avail ourselves of an exact likelihood function.

\subsection{Bayesian Analysis}
\label{bayesian-xsec}

For a given bin, the likelihood to observe count $D$, if the mean
count is $d$, is given by the Poisson distribution
\begin{equation}
\label{eq:poisson}
 L(D|d) = \frac{e^{-d} \, d^D}{\Gamma(D+1)} \, , 
\end{equation}
where $\Gamma$ is the gamma function. (We write the Poisson
distribution in this form to permit the use of noninteger counts in
the calculation of expected results. For observed results, the counts
are of course integers.) The mean count $d$ is the sum of the
predicted contributions from the signal and background sources
\begin{equation}
\label{eq:MeanCount}
 d = \alpha \, \mathcal{L} \, \sigma + \sum_{i=1}^N b_i
    \equiv a \sigma + \sum_{i=1}^N b_i \, ,
\end{equation}
where $\alpha$ is the signal acceptance, $\mathcal{L}$ the integrated
luminosity, $\sigma$ the single top quark production cross section,
$b_i$ the mean count (that is, yield) for background source $i$, $N$
the number of background sources, and $a \equiv \alpha \,
\mathcal{L}$ is the effective luminosity for the signal. For analyses
in which the signal comprises s- and t-channel simulated events, the
latter are combined in the ratio predicted by the standard model.
(Without this assumption, the probability of count $D$ would depend on
the s- and t-channel cross sections $\sigma_{\rm s}$ and $\sigma_{\rm
t}$ explicitly.)

For a distribution of observed counts, the single-bin likelihood is
replaced by a product of likelihoods
\begin{equation}
\label{eq:channellLikelihood}
 L({\bf D}|{\bf d}) \equiv L({\bf D}|\sigma, {\bf a}, {\bf b})
 = \prod_{i=1}^{M} \, L(D_{i}|d_{i}) \, ,
\end{equation}
where ${\bf D}$ and ${\bf d}$ represent vectors of the observed and
mean counts, and ${\bf a}$ and ${\bf b}$ are vectors of effective
luminosity and background yields. The product is over $M$
statistically independent bins: either all bins of a given lepton
flavor, $b$-tag multiplicity, or jet multiplicity, or all bins of a
combination of these channels.

From Bayes' theorem, we can compute the posterior probability density
of the parameters, $p(\sigma, {\bf a}, {\bf b}| {\bf D})$, which is
then integrated with respect to the parameters ${\bf a}$ and ${\bf b}$
to obtain the posterior density for the single top quark production
cross section, given the observed distribution of counts ${\bf D}$,
\begin{equation}
\label{eq:posterior}
 p(\sigma | {\bf D}) = \frac{1}{\mathcal{N}}
   \int \! \! \int L({\bf D} | \sigma, {\bf a}, {\bf b})
   \pi(\sigma, {\bf a}, {\bf b}) \, d{\bf a} \, d{\bf b}\, .
\end{equation}
Here, $\mathcal{N}$ is an overall normalization obtained from the
requirement $\int { p(\sigma|{\bf D})}d\sigma = 1$, where the
integration is performed numerically up to an upper bound $\sigma_{\rm
max}$ when the value of the posterior is sufficiently close to
zero. In this analysis, varying $\sigma_{\rm max}$ from 30 to
150~pb has negligible effect on the result.

The function $\pi(\sigma, {\bf a}, {\bf b})$ is the prior probability
density, which encodes our knowledge of the parameters $\sigma$, ${\bf
a}$, and ${\bf b}$. Since our knowledge of the cross section $\sigma$
does not inform our prior knowledge of ${\bf a}$ and ${\bf b}$, we may
write the prior density as
\begin{equation}
\label{eq:prior}
 \pi(\sigma, {\bf a}, {\bf b}) = \pi({\bf a}, {\bf b}) \,\pi(\sigma) \,.
\end{equation}
The prior density for the cross section is taken to be a nonnegative
flat prior, $\pi(\sigma) = 1/\sigma_{\rm max}$ for $\sigma \ge 0$, and
$\pi(\sigma) = 0$ otherwise. We make this choice because it is simple
to implement and yields acceptable results in ensemble studies (see
Sec.~\ref{sec:ensembles}). The posterior probability density for the
signal cross section is therefore
\begin{equation}
\label{eq:finalPosterior}
 p(\sigma | {\bf D})
 = \frac{1}{\mathcal{N}\sigma_{\rm max}}
 \int \! \! \int L({\bf D} | \sigma, {\bf a}, {\bf b})
 \pi({\bf a}, {\bf b}) \, d{\bf a} \, d{\bf b} \, . 
\end{equation}
We take the mode of $p(\sigma | {\bf D})$ as our measure of the cross
section, and the $68\%$ interval about the mode as our measure of the
uncertainty with which the cross section is measured. We have verified
that these intervals, although Bayesian, have approximately 68\%
coverage probability and can therefore be interpreted as approximate
frequentist intervals if desired.

The integral in Eq.~\ref{eq:finalPosterior} is done numerically using
Monte Carlo importance sampling. We generate a large number $K$ of
points $({\bf a}_k, {\bf b}_k)$ randomly sampled from the prior
density $\pi({\bf a},{\bf b})$ and estimate the posterior density
using
\begin{eqnarray}
\label{eq:MCIntegration}
 p(\sigma|{\bf D})
 & \propto & \int \! \! \int L({\bf D} | \sigma, {\bf a}, {\bf b})
 \pi({\bf a}, {\bf b}) \, d{\bf a} \, d{\bf b} , \nonumber \\
 & \approx & \frac{1}{K} \sum_{k=1}^K \,
 L({\bf D}|\sigma, {\bf a}_k, {\bf b}_k) \, .
\end{eqnarray}

In the presence of two signals, we use the same procedure and
calculate the posterior probability density according to
Eq.~\ref{eq:MCIntegration}, replacing $\sigma$ by
$\sigma_{tb},\sigma_{tqb}$ everywhere. We also replace the term
$a\sigma$ in Eq.~\ref{eq:MeanCount} by $a_{tb}\sigma_{tb} +
a_{tqb}\sigma_{tqb}$, where $a_{tb}$ and $a_{tqb}$ are the effective
luminosities of the $tb$ and $tqb$ signals, and $\sigma_{tb}$ and
$\sigma_{tqb}$ are their cross sections. The prior density for the
cross section $\pi(\sigma)$ in Eq.~\ref{eq:MeanCount} becomes
$\pi(\sigma_{tb},\sigma_{tqb})=1/(\sigma_{tb,{\rm
max}}+\sigma_{tqb,{\rm max}})$ if both $\sigma_{tb}$ and
$\sigma_{tqb}$ are $\ge 0$, and $\pi(\sigma_{tb},\sigma_{tqb})=0$
otherwise. With these two replacements, the posterior probability
density becomes a two-dimensional distribution as a function of the
two cross sections.

\subsection{Prior Density}
\label{sec:priordensity}

The prior density $\pi({\bf a}, {\bf b})$ encodes our knowledge of the
effective signal luminosities and the background yields (see
Sec.~\ref{systematics}). The associated uncertainties fall into two
classes: those that affect the overall normalization only, such as the
integrated luminosity measurement, and those that also affect the
shapes of the discriminant distributions, which are the jet energy
scale and $b$-tag modeling.

The normalization effects are modeled by sampling the effective signal
luminosities {\bf a} and the background yields {\bf b} from a
multivariate Gaussian, with the means set to the estimated yields and
the covariance matrix computed from the associated uncertainties. The
covariance matrix quantifies the correlations of the systematic
uncertainties across different sources of signal and background.

The shape effects are modeled by changing, one at a time, the jet
energy scale and $b$-tag probabilities by plus or minus one standard
deviation with respect to their nominal values. Therefore, for a given
systematic effect, we create three model distributions: the nominal
one, and those resulting from the plus and minus shifts. For each bin,
Gaussian fluctuations, with standard deviation defined by the plus and
minus shifts in bin yield, are generated about the nominal yield, and
added linearly to the nominal yields generated from the sampling of
the normalization-only systematic effects. Since effects such as a
change in jet energy scale affect all bins coherently, we assume
100$\%$ correlation across all bins and sources. This is done by
sampling from a zero mean, unit variance Gaussian and using the same
variate to generate the fluctuations in all bins.

\subsection{Bayes Ratio}
\label{bayes-ratio}

Given two well-defined hypotheses $H_0$ and $H_1$ (e.g., the
background-only and the signal+background hypotheses), it is natural
in a Bayesian context to consider a Bayes factor $B_{10}$,
\begin{eqnarray}
\label{eq:bayesfactor}
 B_{10} & = & \frac{L({\bf D}|H_1)}{L({\bf D}|H_0)}, \nonumber \\
        & = & \frac{\int L({\bf D}|\sigma) \, \pi(\sigma) \, d\sigma}
                   {L({\bf D}|\sigma=0)},
\end{eqnarray}
as a way to quantify the significance of hypothesis $H_1$ relative to
$H_0$. Here,
\begin{equation}
 L({\bf D} | \sigma)
 = \int \! \! \int L({\bf D} | \sigma, {\bf a}, {\bf b}) \,
   \pi({\bf a}, {\bf b}) \, d{\bf a} \, d{\bf b}
\end{equation}
is the marginal (or integrated) likelihood and $\pi(\sigma)$ is the
cross section prior density, which could be taken as a Gaussian about
the standard model predicted value.

Another possible use of a Bayes factor is as an objective function to
be maximized in the optimization of analyses; the optimal analysis
would be the one with the largest expected Bayes factor. These
considerations motivate a quantity akin to a Bayes factor that is
somewhat easier to calculate, which we have dubbed a Bayes ratio,
defined by
\begin{equation}
\label{eq:bayesratio}
 {\rm Bayes~ratio~} = 
  \frac{p(\hat{\sigma}|{\bf D})}{p(\sigma=0|{\bf D})},
\end{equation}
where $\hat{\sigma}$ is the mode of the posterior density. The three
analyses are optimized using the expected Bayes ratio, which is
computed by setting the distribution {\bf D} to the expected one.

%---------------------------------------------------------------------
%---------------------------------------------------------------------
\section{Results}
\label{sec:results}

\subsection{Expected Sensitivity}
\label{sec:expected-sensitivity}

Before making a measurement using data, it is useful to calculate the
expected sensitivity of these analyses. Furthermore, this expected
sensitivity is used to optimize the choice of parameters in the
analyses. For each case under consideration we calculate an expected
Bayes ratio as defined in Sec.~\ref{bayes-ratio}. The highest Bayes
ratio corresponds to the optimal parameter choice.

Table~\ref{tab:expbayesratios} shows the expected Bayes ratio for each
possible combination of analysis channels in the DT analysis. It can
be seen from the numbers in the table that combining the two single
top quark signals (i.e., searching for $tb$+$tqb$ together) results in
the best expected sensitivity. The single-tag two-jet channel
contributes the most to this sensitivity, as expected from the high
signal acceptance and reasonable signal-to-background ratio, but the
addition of the other channels does improve the result; including the
poorer ones does not degrade it. While the result from this table
refers specifically to the DT analysis, the conclusions hold for all
three multivariate techniques. Therefore, from this point onward, the
2--4 jets 1--2 tags result, using electrons and muons in the
$tb$+$tqb$ channel will be considered as default (2--3 jets for the ME
analysis).

\clearpage

\begin{table*}[!h!tbp]
\begin{center}
\begin{minipage}{5in}
\caption[expbayesratios]{Expected Bayes ratios from the decision tree
analysis, including systematic uncertainties, for many combinations of
analysis channels. The best values from all channels combined are
shown in bold type.}
\label{tab:expbayesratios}
\begin{ruledtabular}
\begin{tabular}{l||cc|cc|ccc|c}
\multicolumn{9}{c}{\hspace{0.5in}\underline{Expected Bayes Ratios}}\vspace{0.05in}\\
& \multicolumn{2}{c|}{1--2tags, 2--4jets}& \multicolumn{2}{c|}{$e+\mu$, 2--4jets}
& \multicolumn{3}{c|}{$e+\mu$, 1--2tags}& All \\
                 &  $e$-chan & $\mu$-chan& 1 tag & 2 tags& 2 jets& 3 jets& 4 jets&channels\\
\hline
~$tb$           &  1.1  &  1.1  &  1.1  &  1.1  &  1.2  &  1.0  &  1.0  & {\bf 1.2} \\
~$tqb$          &  2.5  &  1.8  &  4.5  &  1.1  &  3.1  &  1.5  &  1.1  & {\bf 4.7} \\
~$tb$+$tqb$~~   &  3.2  &  2.3  &  6.7  &  1.3  &  4.8  &  1.5  &  1.1  & {\bf 8.0}
\end{tabular}
\end{ruledtabular}
\end{minipage}
\end{center}
\end{table*}

\subsection{Expected Cross Sections}
\label{sec:expected-cross-sections}

We measure the expected cross sections for the various channels by
setting the number of data events in each channel equal to the
(noninteger) expected number of background events plus the expected
number of signal events (using the SM cross section of 2.86~pb at
$m_{\rm top} = 175$~GeV), and obtain the following results:
$$
\begin{array}{llll}
 \sigma^{\rm exp}\left({\ppbar}{\rargap}tb+X,~tqb+X\right)
 & = & 2.7 ^{+1.6}_{-1.4}~{\rm pb} & {\rm (DT)}  \\
 & = & 2.7 ^{+1.5}_{-1.5}~{\rm pb} & {\rm (BNN)} \\
 & = & 2.8 ^{+1.6}_{-1.4}~{\rm pb} & {\rm (ME)}.
\end{array}
$$

The expected cross sections agree with the input cross section. The
small deviation, less than $10\%$, is from the nonsymmetric nature of
several of the systematic uncertainties, in particular the jet energy
scale and $b$~tagging. This effect is also observed in the
pseudodatasets. 

The linearity of the methods to measure the appropriate signal cross section was
discussed in Sec.~\ref{sec:ensembles} and Fig.~\ref{fig:calibration}, and no
calibration is necessary based on those results.

\subsection{Measured Cross Sections}
\label{sec:measured-cross-sections}

\begin{figure*}[!h!tbp]
\includegraphics[width=0.32\textwidth]
{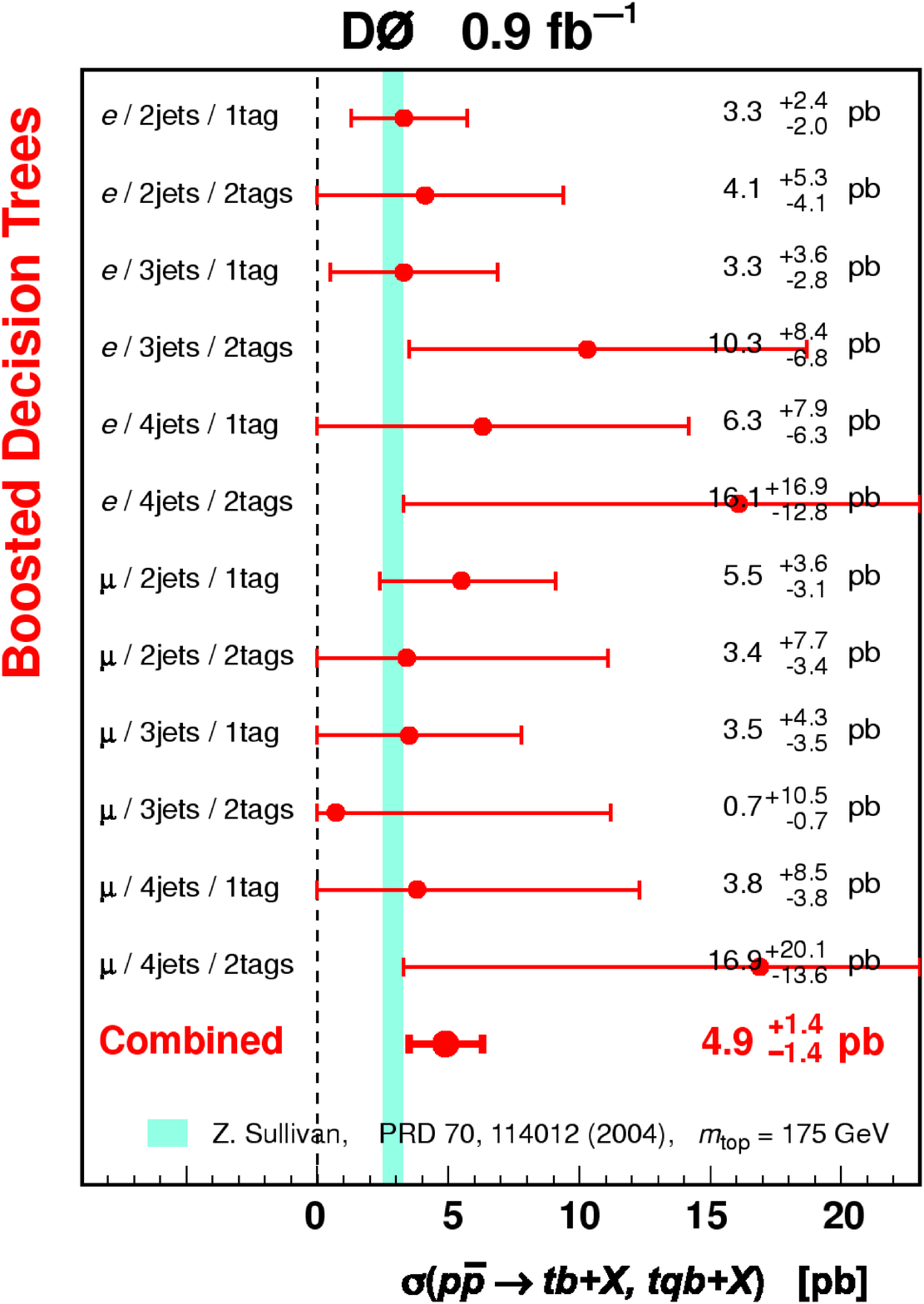} 
\includegraphics[width=0.32\textwidth]
{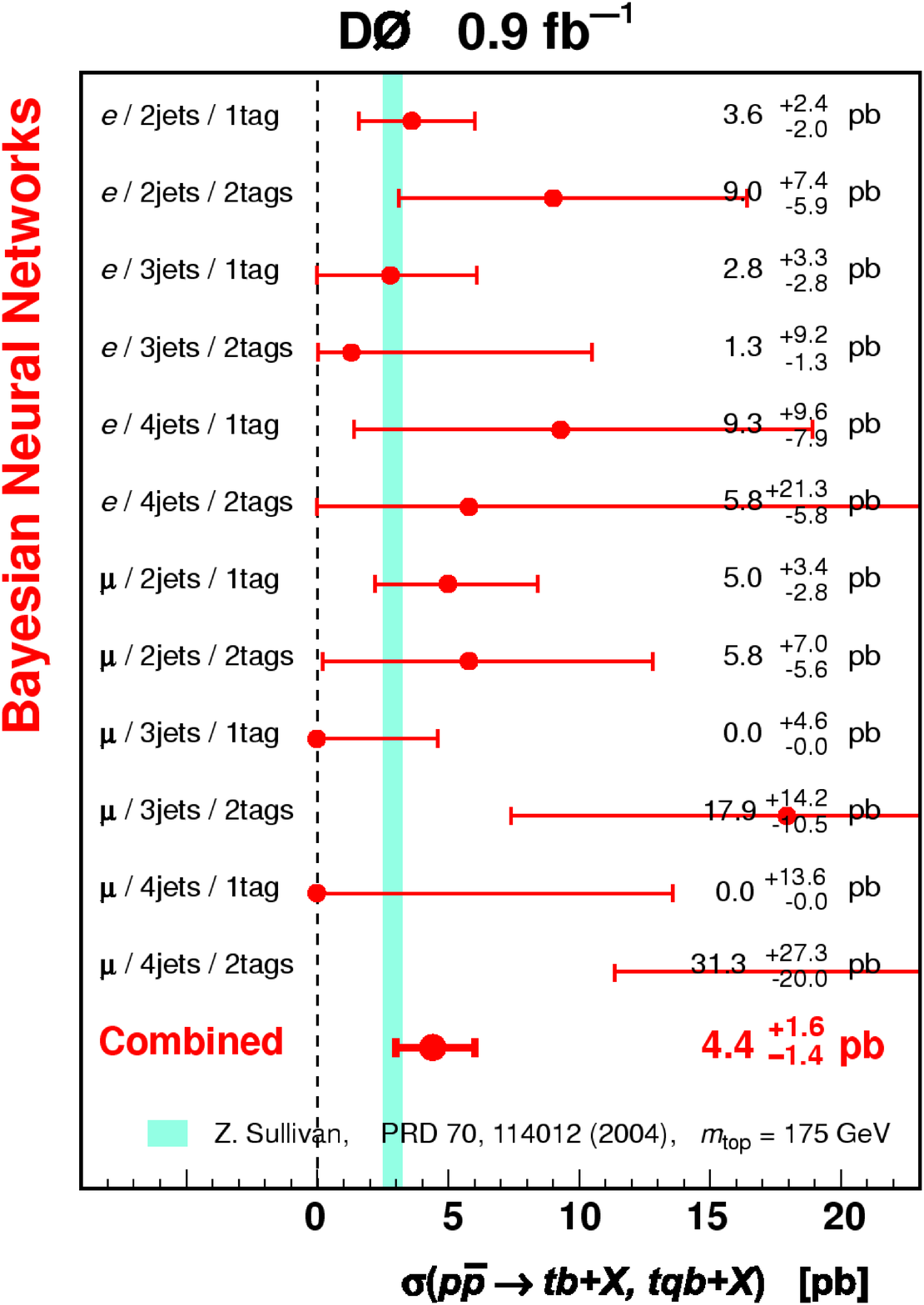} 
\includegraphics[width=0.32\textwidth]
{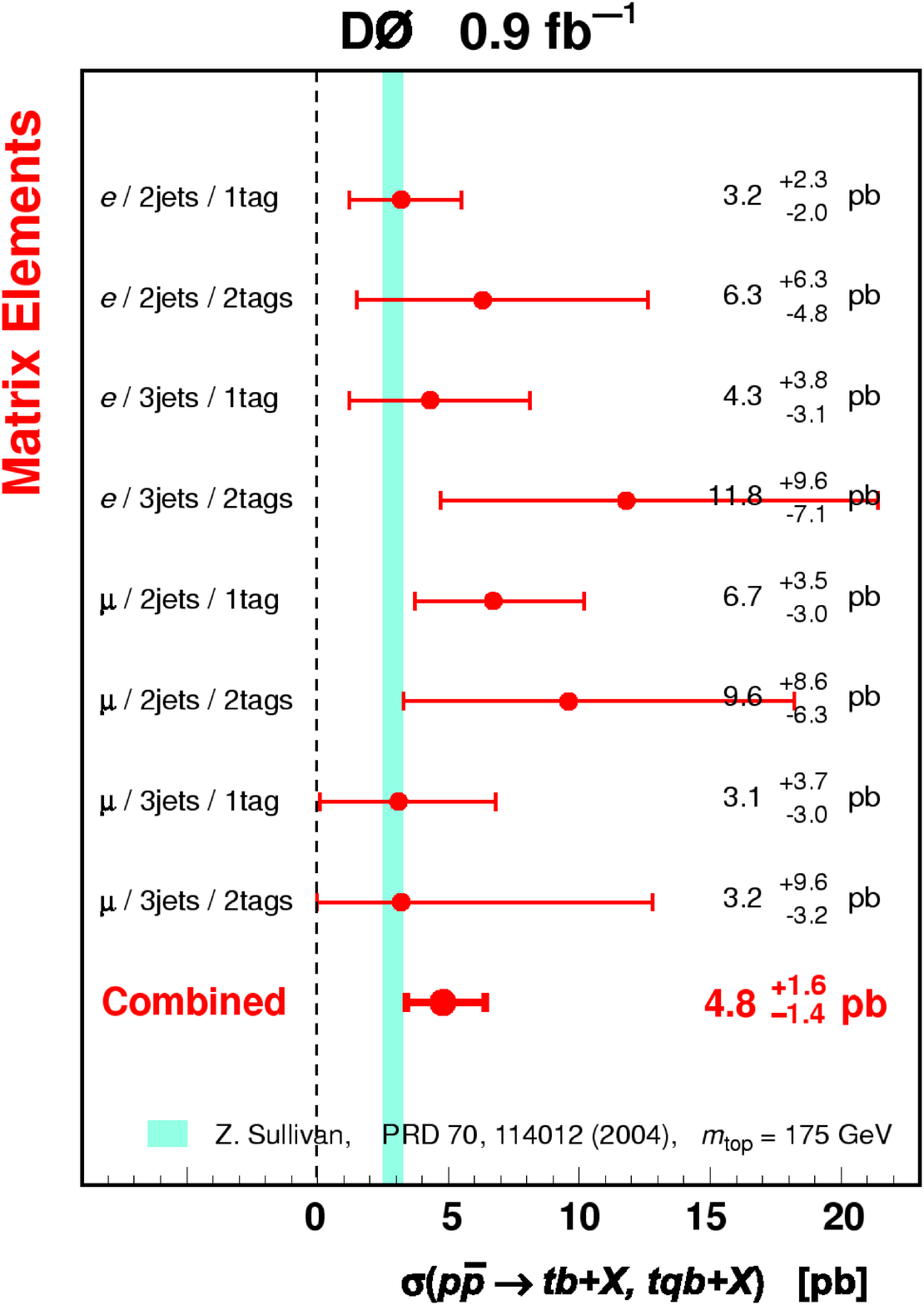} 
\vspace{-0.1in}
\caption[crosssec-channels]{Summaries of the cross section
measurements using data from each multivariate technique. The left
plot is DT, the middle one is BNN, and the right plot is ME.}
\label{fig:crosssec-channels}
\end{figure*}

The cross sections measured using data with the three multivariate
techniques are shown in Fig.~\ref{fig:crosssec-channels} where each
measurement represents an independent subset of the data, for example,
the 2-jet sample with 1 $b$~tag in the electron channel.

\begin{figure*}[!h!tbp]
\includegraphics[width=0.38\textwidth]{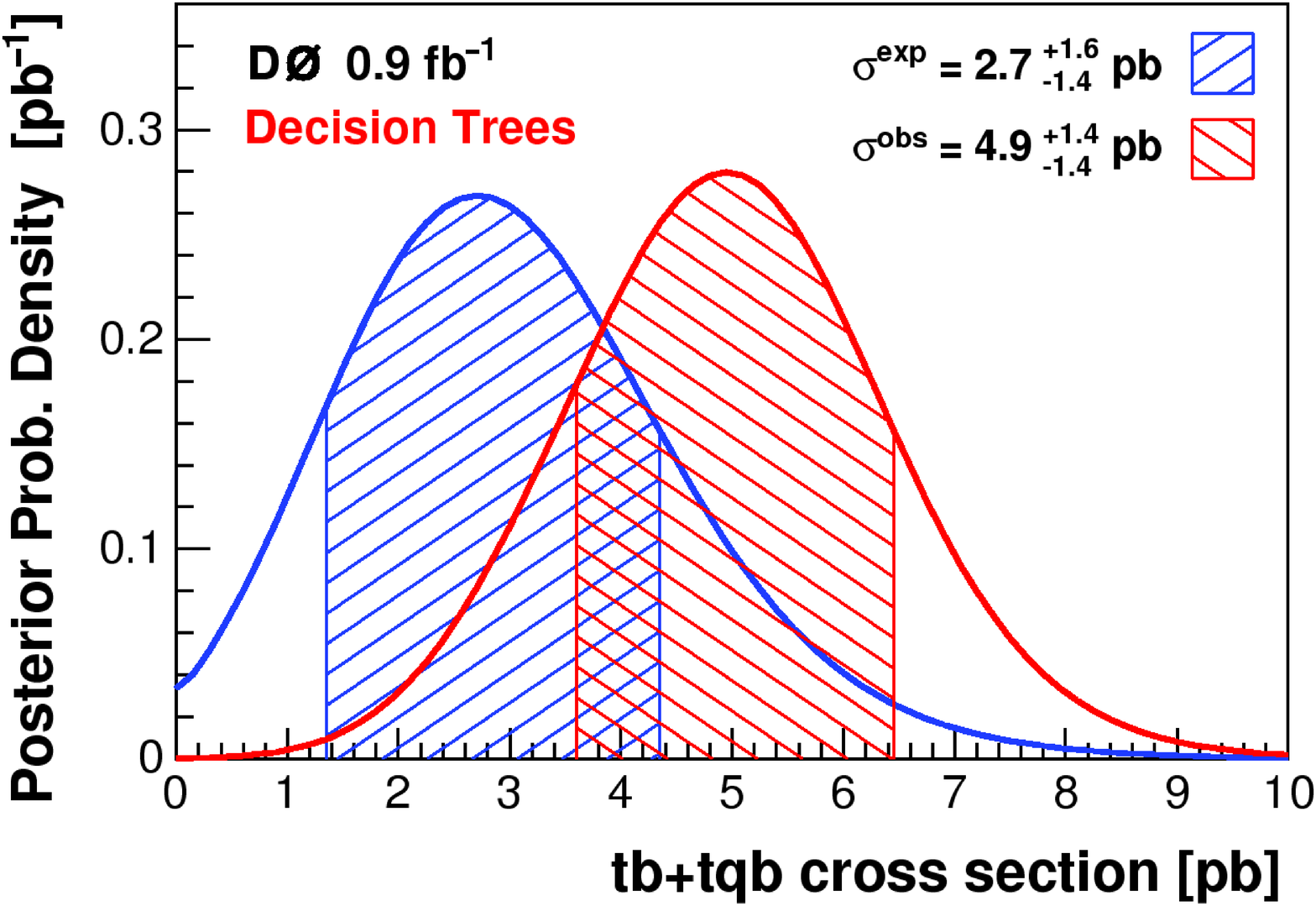}
\includegraphics[width=0.38\textwidth]{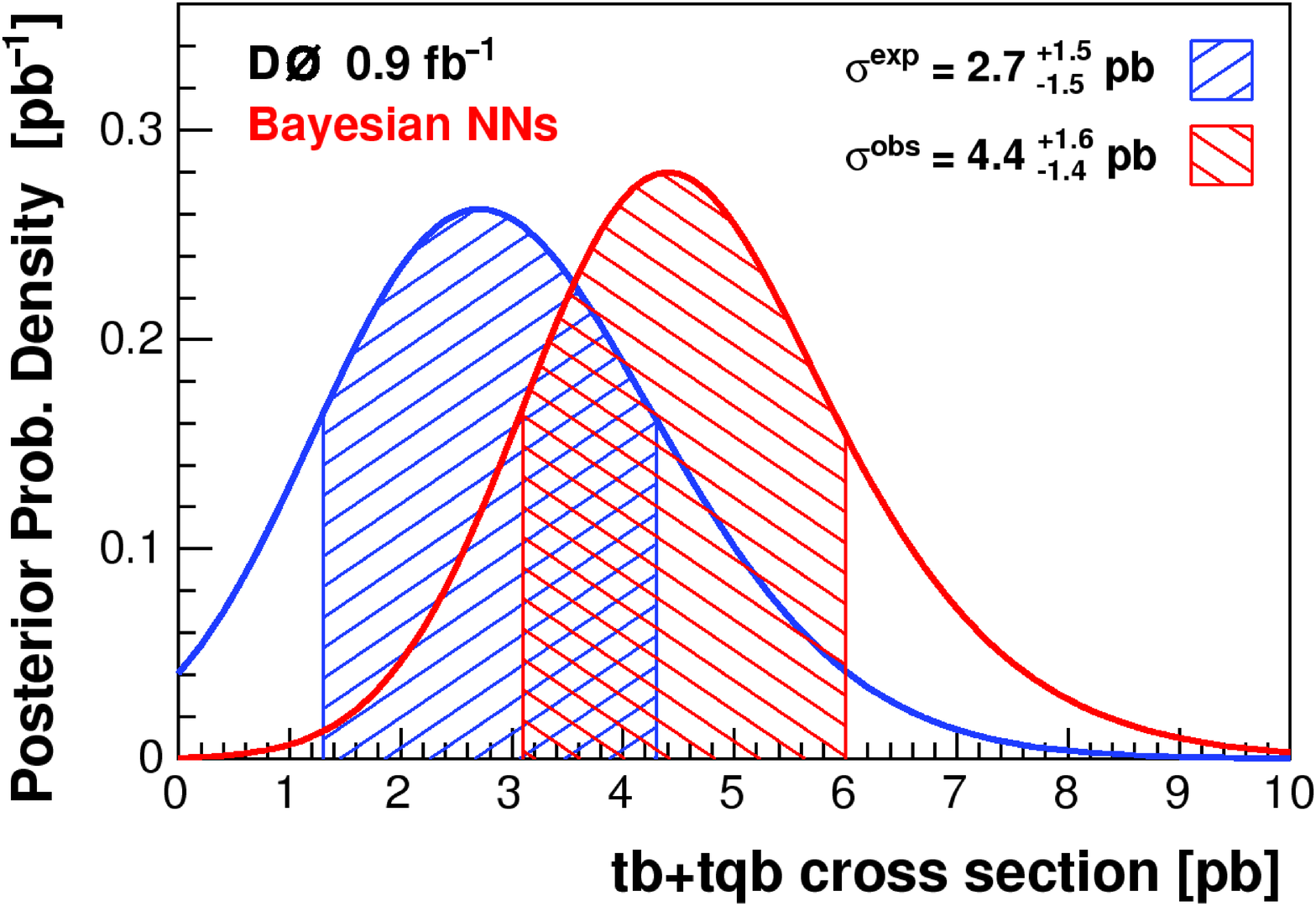}
\includegraphics[width=0.38\textwidth]{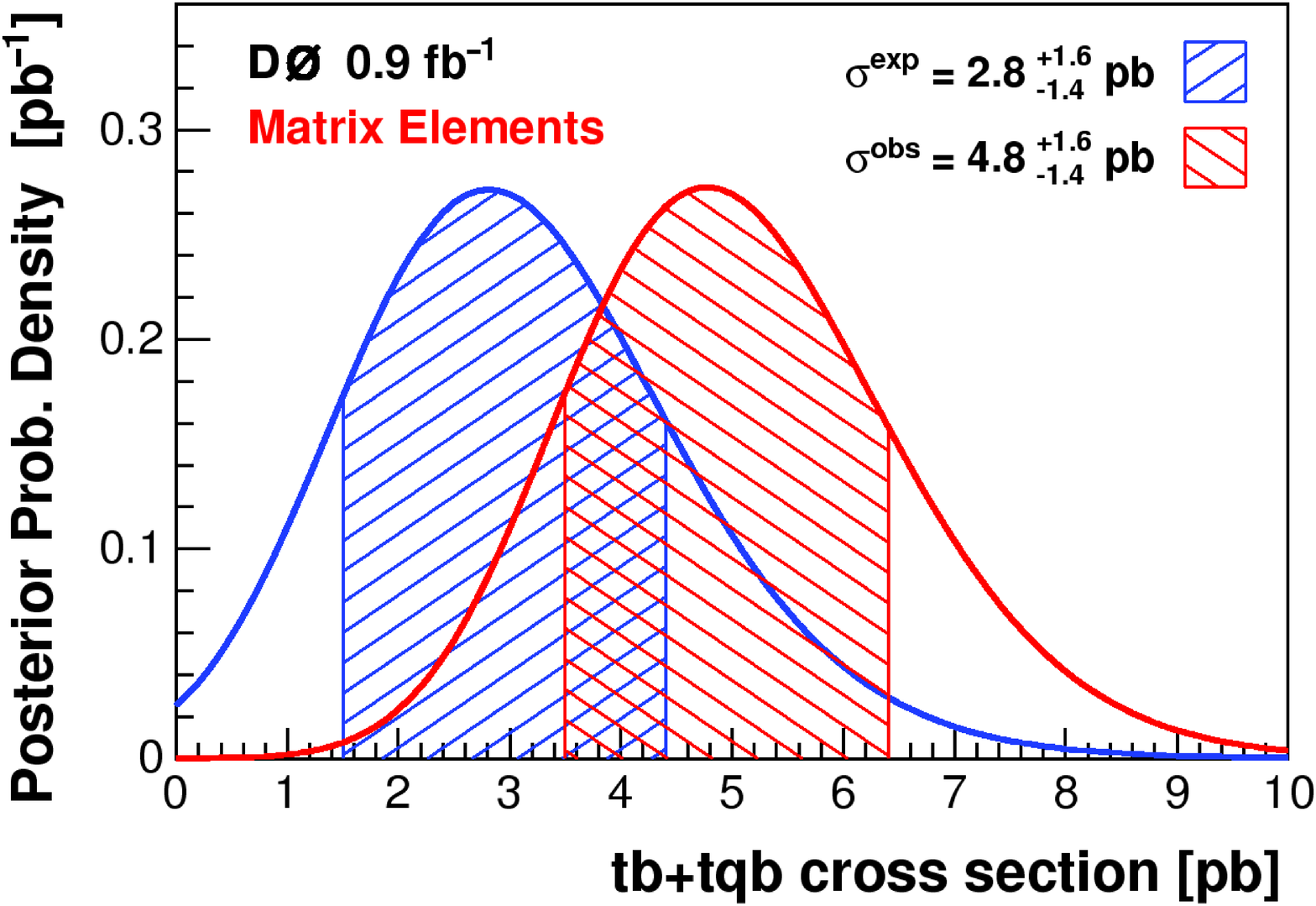}
\vspace{-0.1in}
\caption[finalposteriors]{Expected SM and observed Bayesian posterior
density distributions for the DT, BNN and ME analyses. The shaded
regions indicate one standard deviation above and below the peak
positions.}
\label{fig:finalposteriors}
\end{figure*}

The full combination of available channels (the most sensitive case)
yields the Bayesian posterior density functions shown in
Fig.~\ref{fig:finalposteriors} and cross sections of:
$$
\begin{array}{llll}
 \sigma^{\rm obs}\left({\ppbar}{\rargap}tb+X,~tqb+X\right)
 & = & 4.9 ^{+1.4}_{-1.4}~{\rm pb} & {\rm (DT)}  \\
 & = & 4.4 ^{+1.6}_{-1.4}~{\rm pb} & {\rm (BNN)} \\
 & = & 4.8 ^{+1.6}_{-1.4}~{\rm pb} & {\rm (ME)}.
\end{array}
$$

\begin{figure*}[!h!tbp]
\includegraphics[width=0.32\textwidth]
{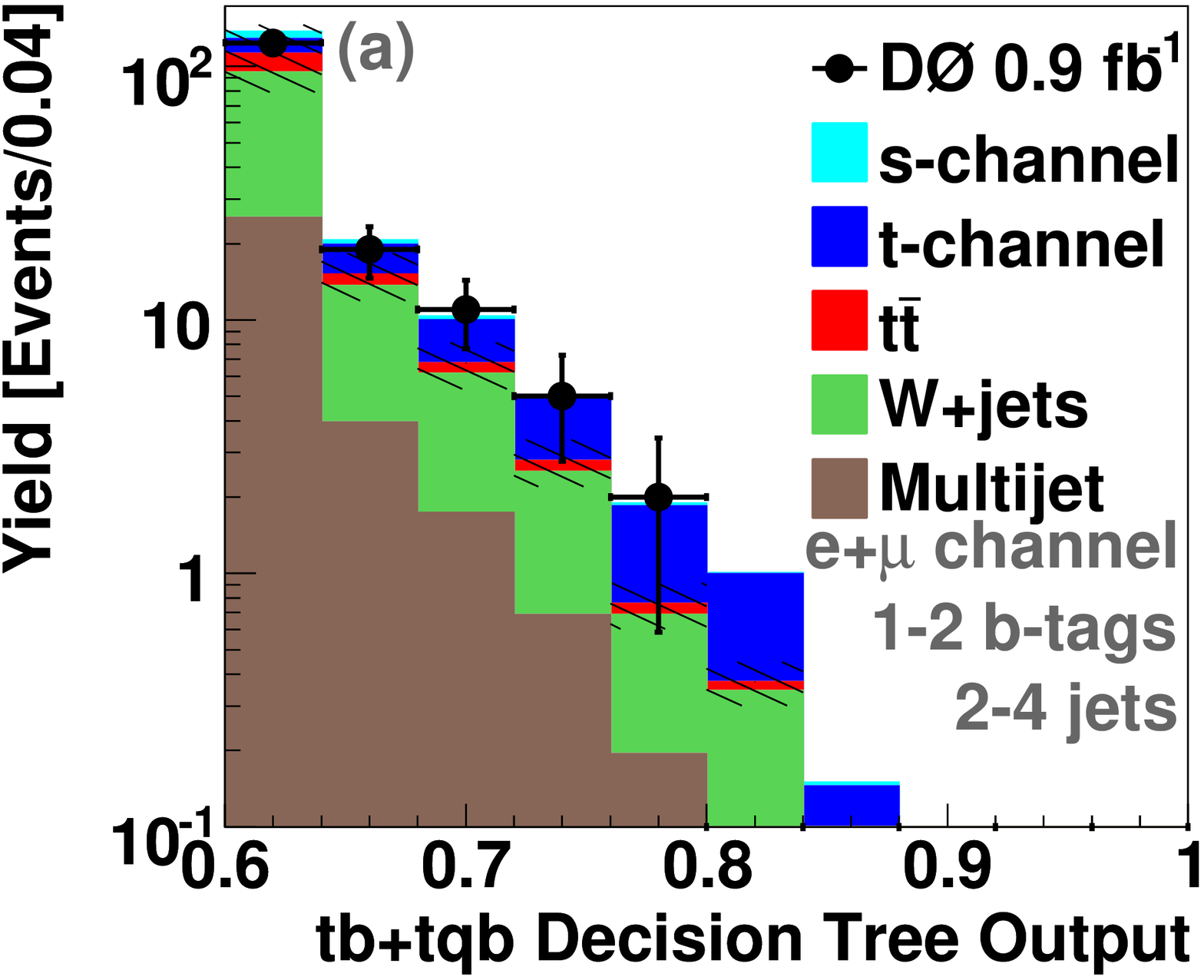}
\includegraphics[width=0.32\textwidth]
{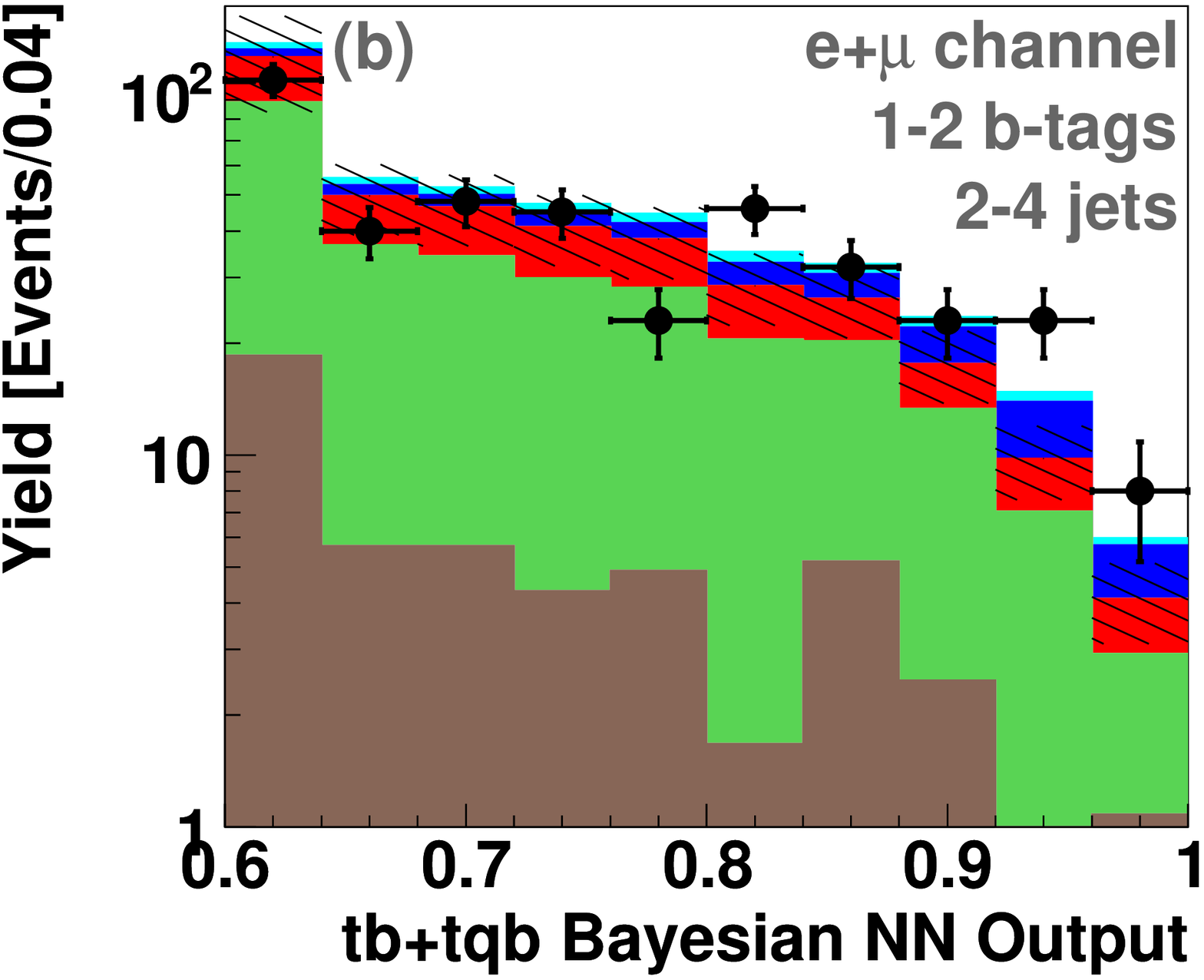}

\includegraphics[width=0.32\textwidth]
{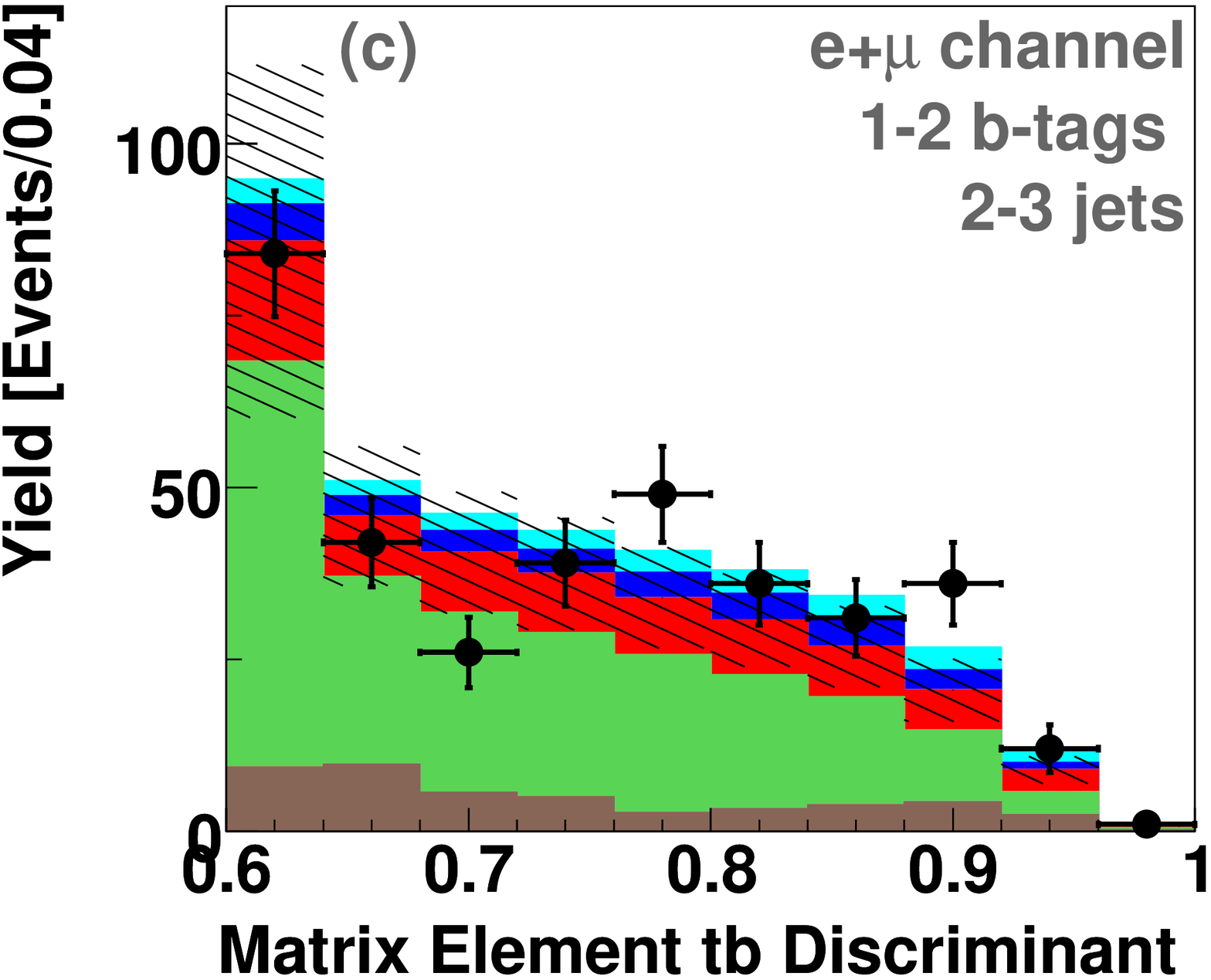}
\includegraphics[width=0.32\textwidth]
{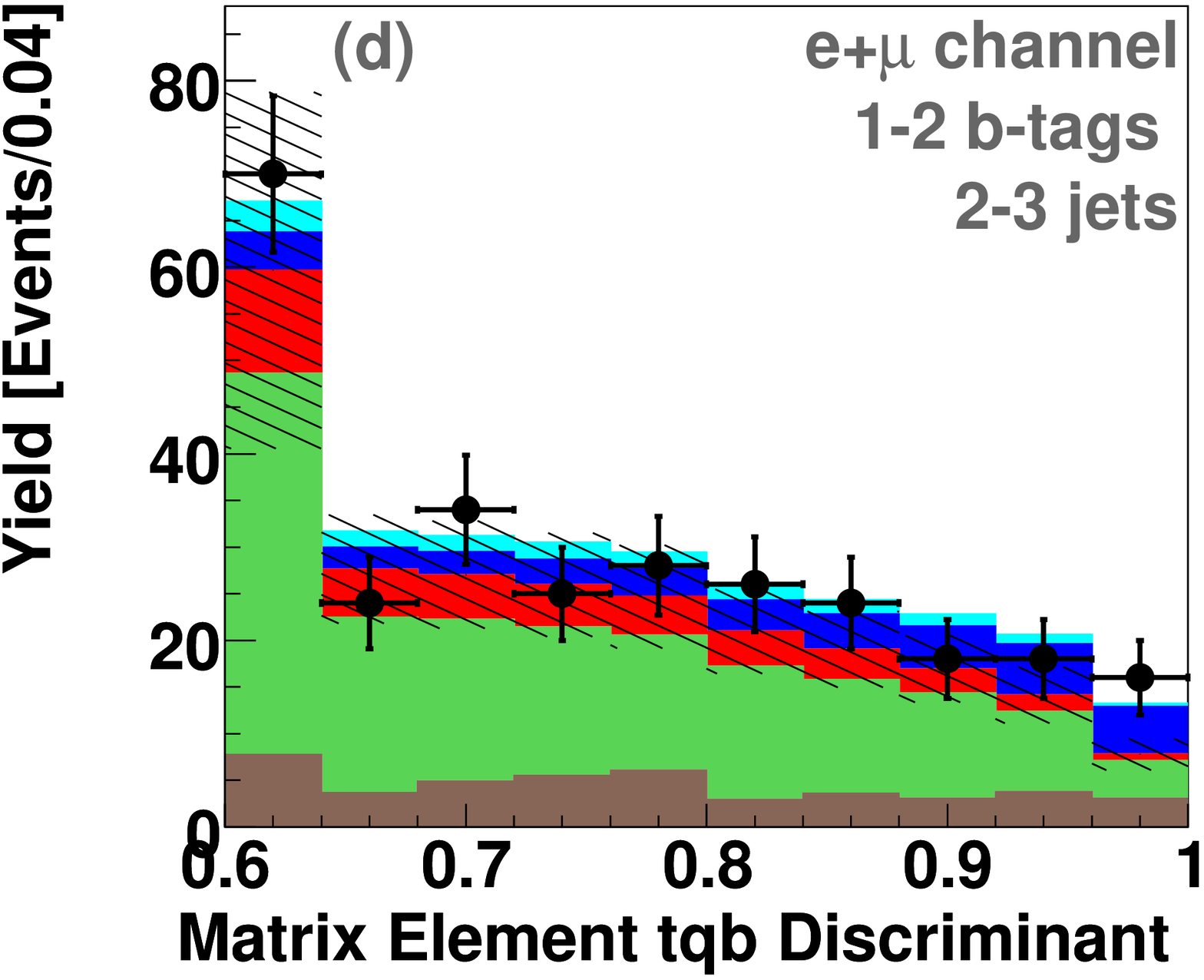}
\vspace{-0.1in}
\caption[disc-zooms]{Zooms of the high-discriminant output regions of
the three multivariate discriminants: (a)~DT, (b)~BNN, (c)~ME
s-channel, and (d)~ME t-channel discriminants. The signal component is
normalized to the cross section measured from data in each case. The
hatched bands show the 1~$\sigma$ uncertainty on the background.}
\label{fig:disc-zooms}
\end{figure*}

Figure~\ref{fig:disc-zooms} shows the high-discriminant regions for
each of the multivariate methods, with the signal component normalized
to the cross section measured from data.  Clearly, a model including a
signal contribution fits the data better than does a background-only
model.

\begin{figure*}[!h!tbp]
\includegraphics[width=0.32\textwidth]
{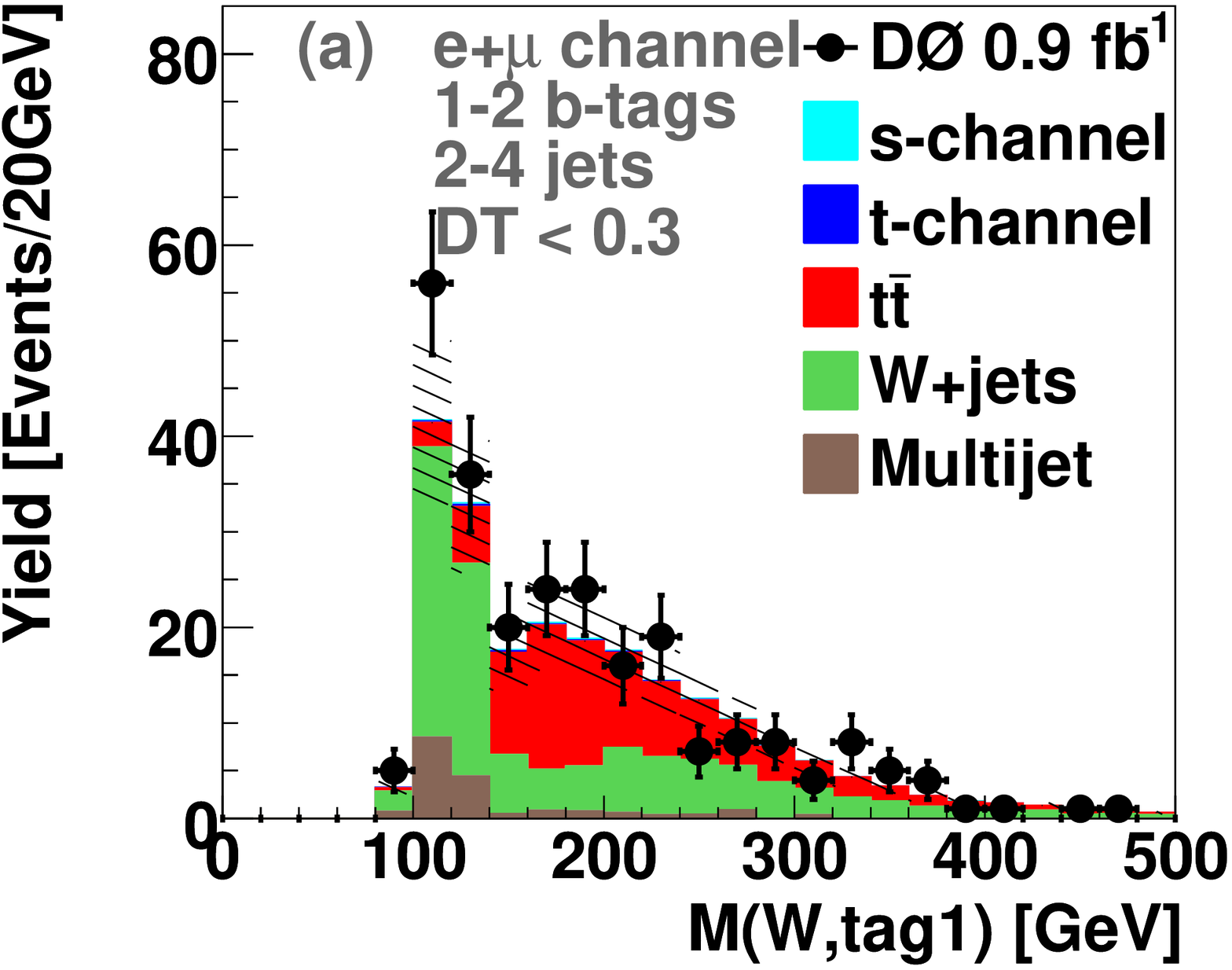}
\includegraphics[width=0.32\textwidth]
{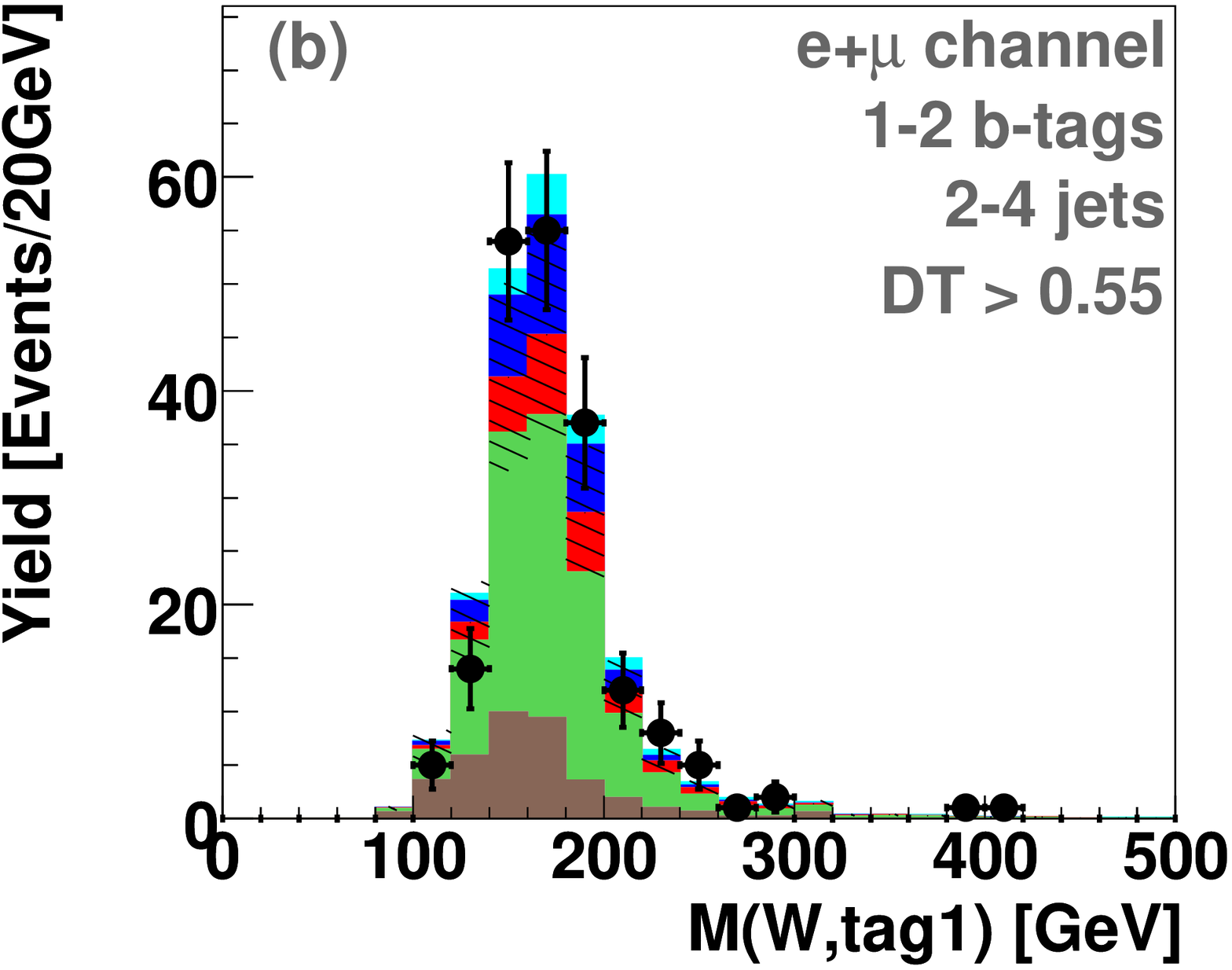}
\includegraphics[width=0.32\textwidth]
{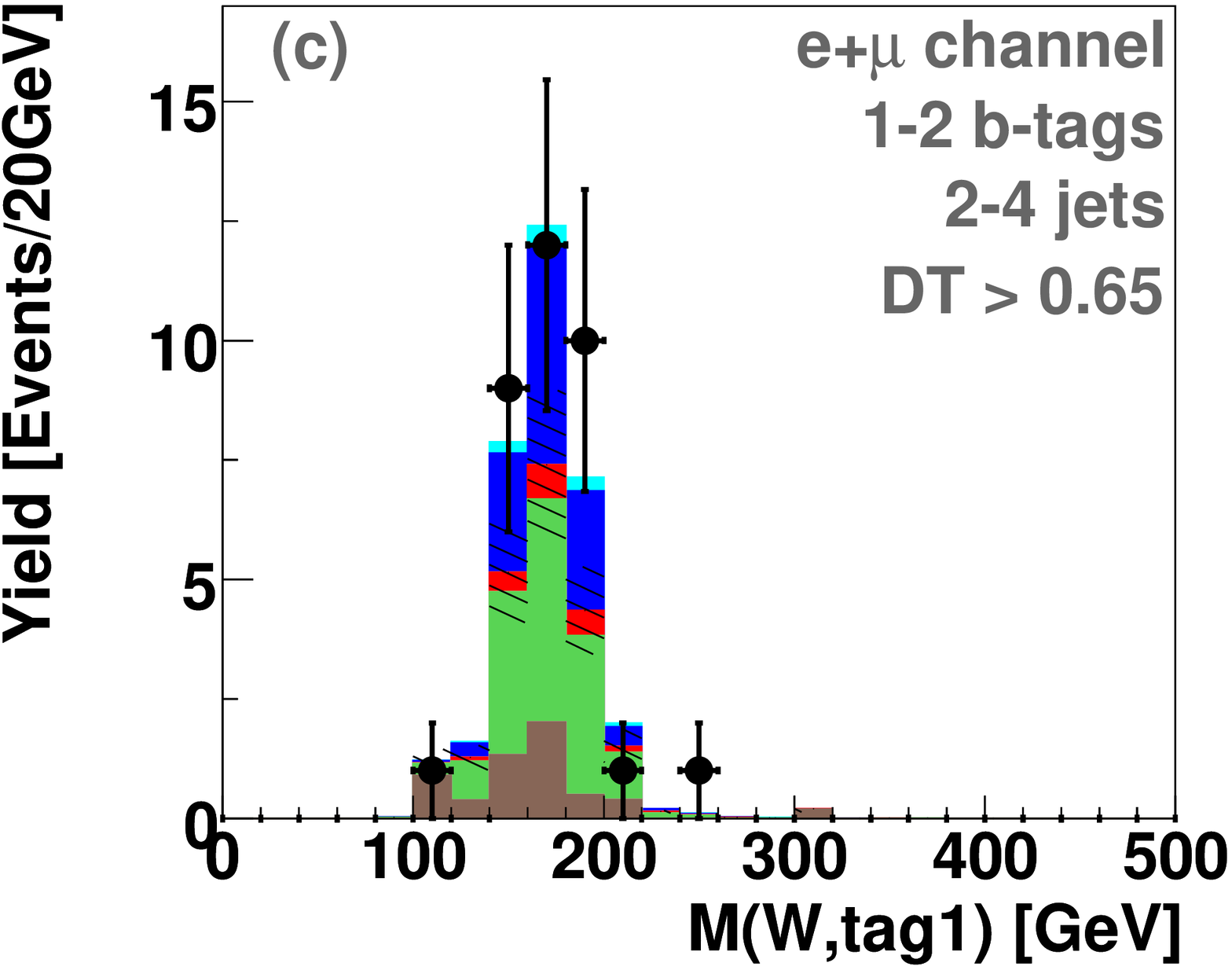}
\includegraphics[width=0.32\textwidth]
{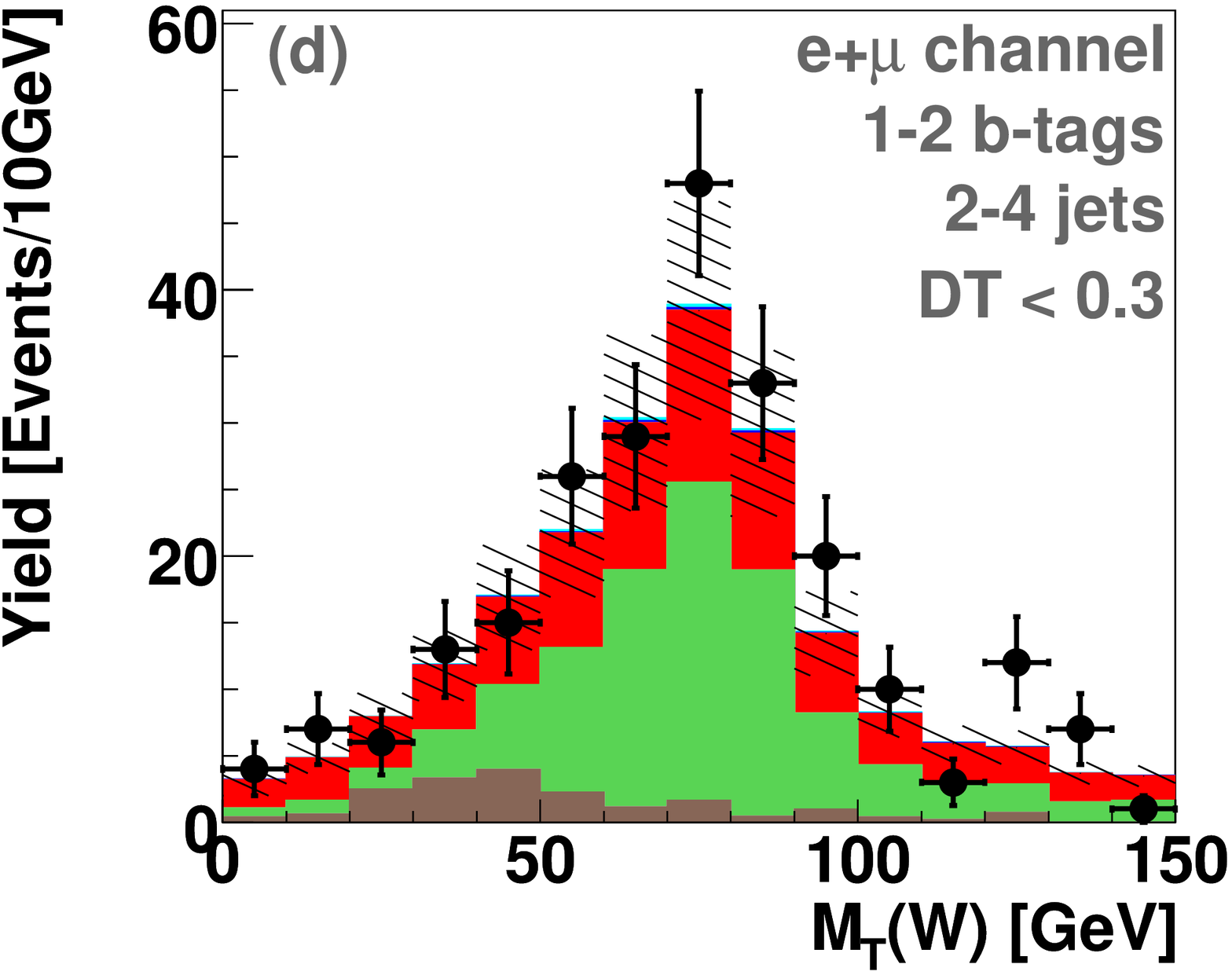}
\includegraphics[width=0.32\textwidth]
{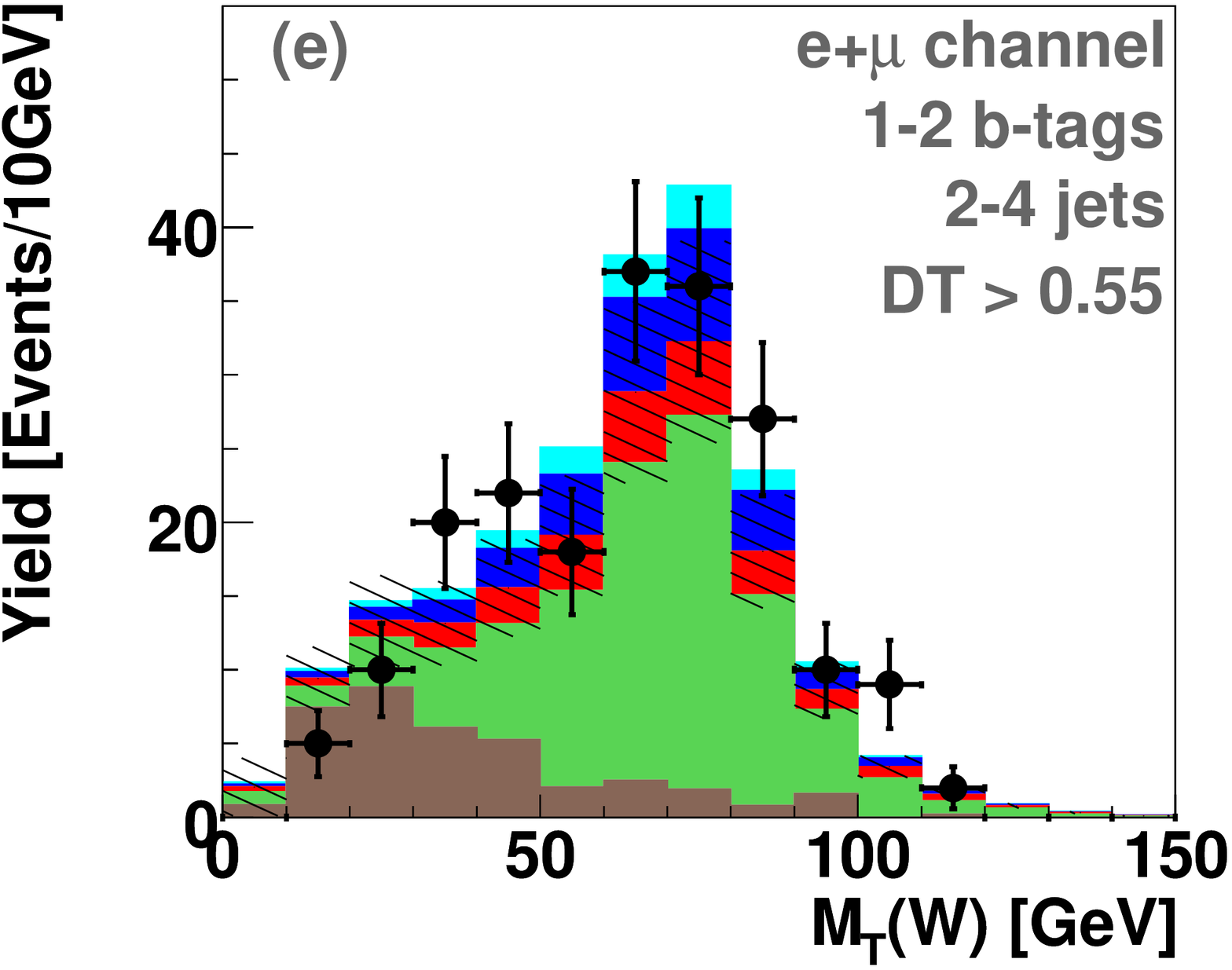}
\includegraphics[width=0.32\textwidth]
{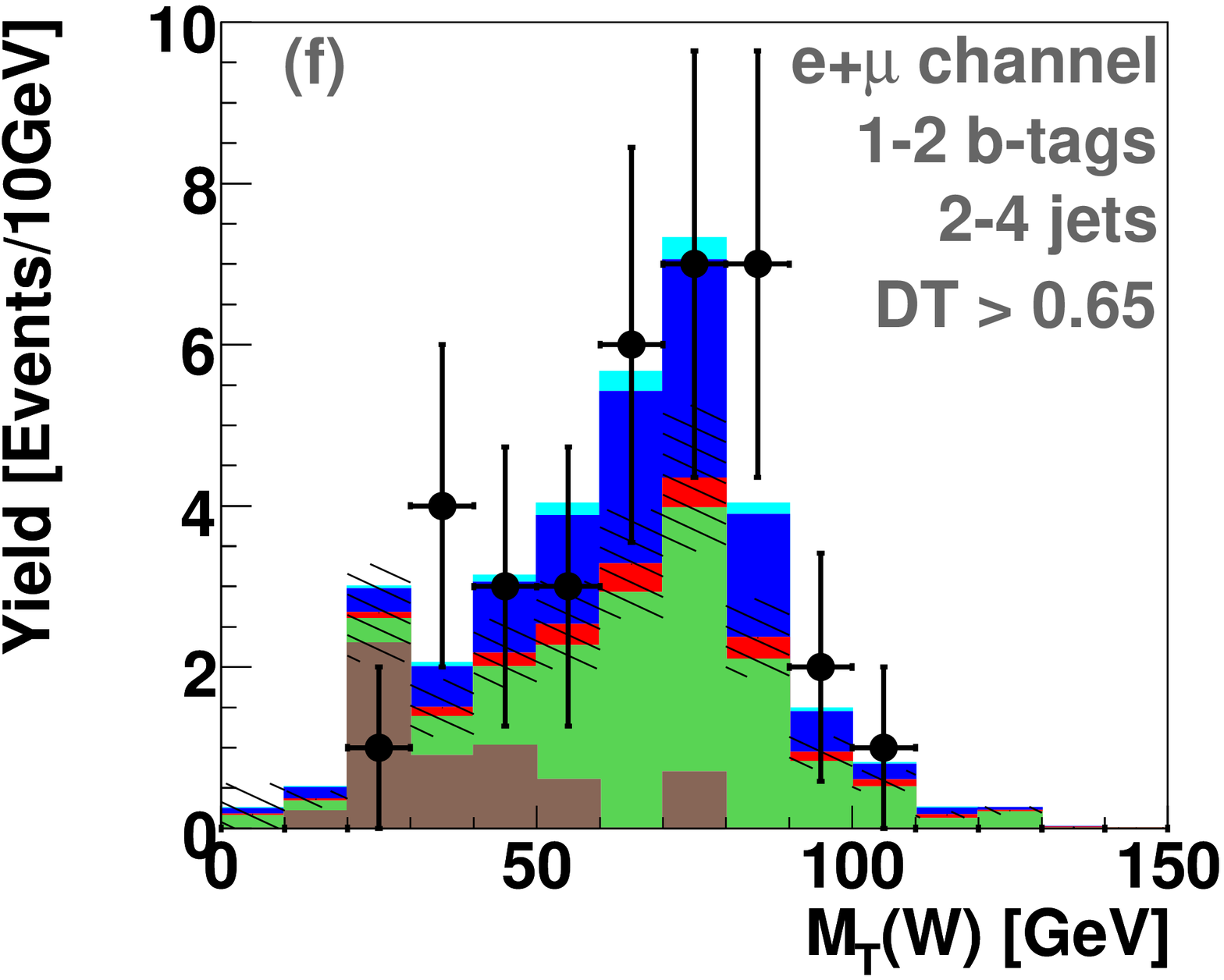}
\includegraphics[width=0.32\textwidth]
{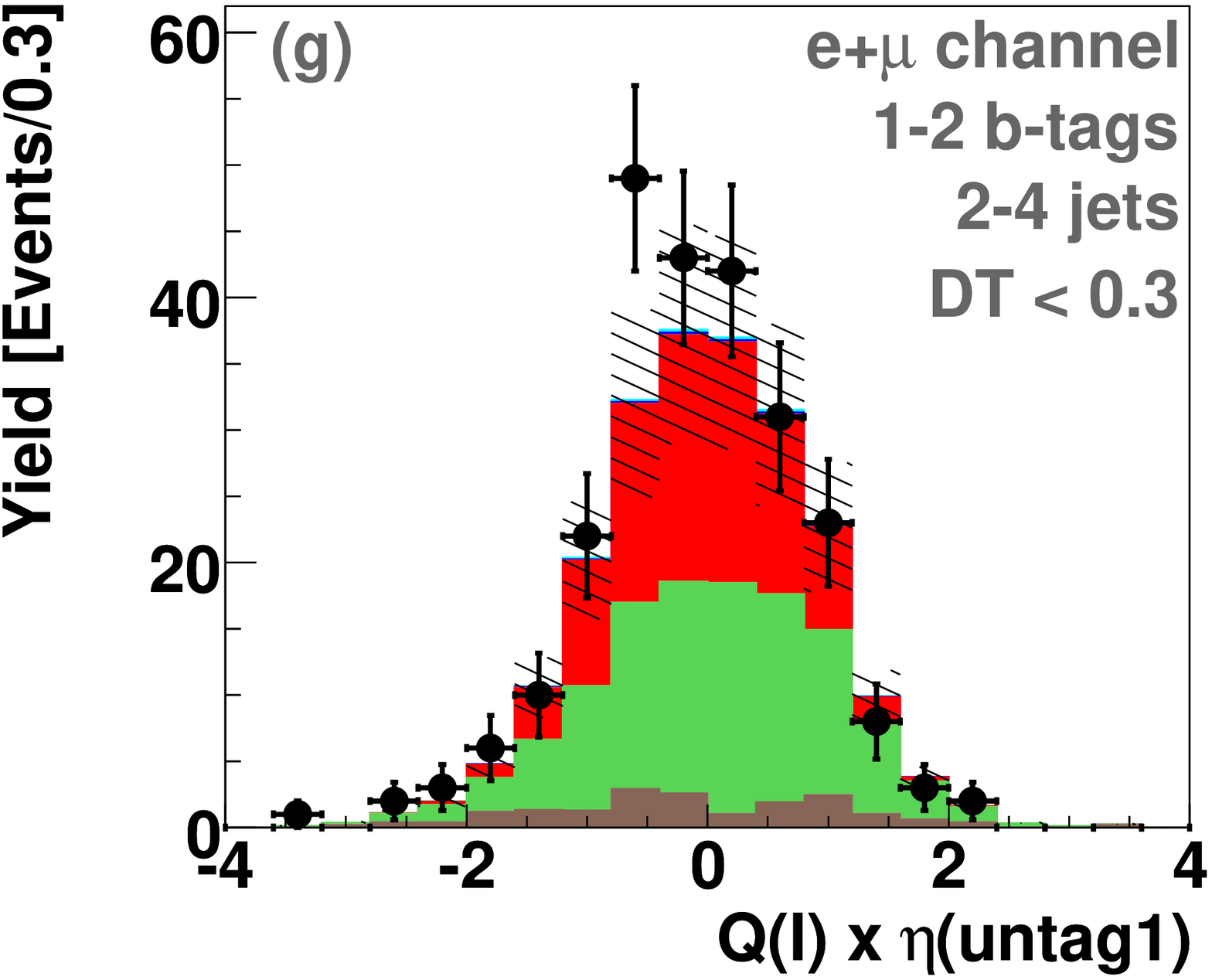}
\includegraphics[width=0.32\textwidth]
{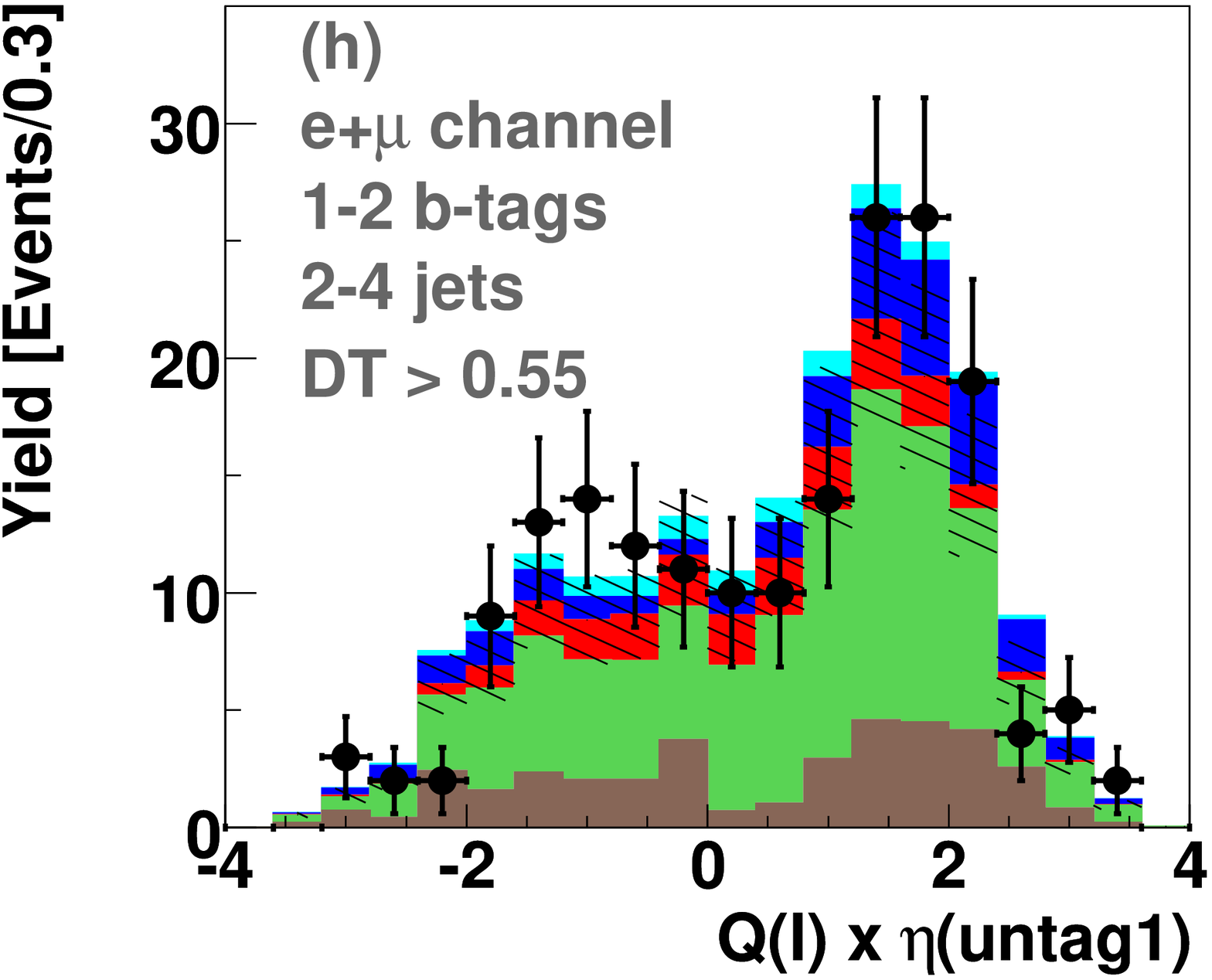}
\includegraphics[width=0.32\textwidth]
{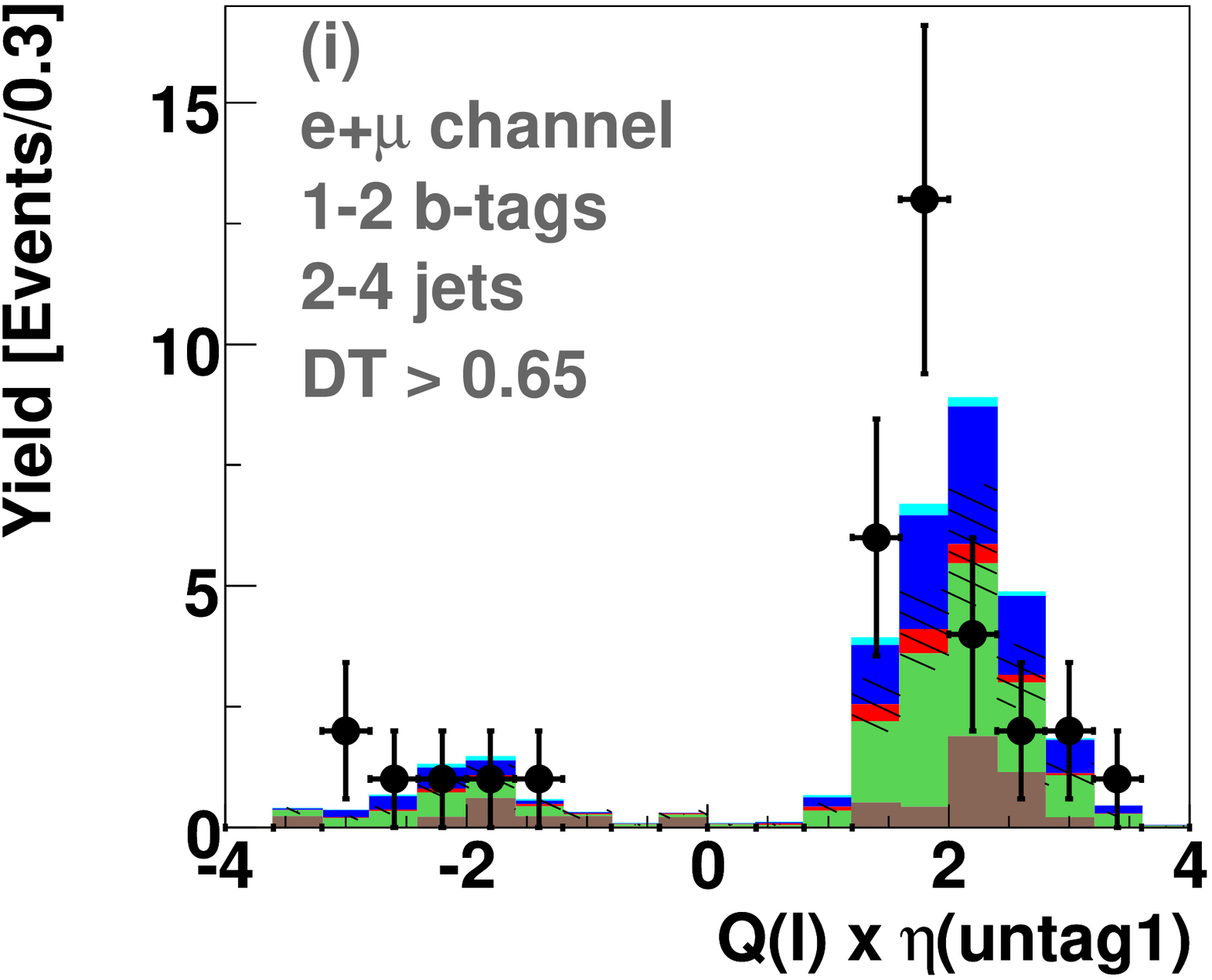}
\vspace{-0.1in}
\caption[eventchars]{The $b$-tagged top quark mass (top row),
$W$~boson transverse mass (second row) and
$Q$(lepton)$\times$$\eta$(untag1) (third row) for the $tb$+$tqb$
analysis with a low ($<0.3$, left column), high ($>0.55$, middle
column), and very high ($>0.65$, right column) DT output, for lepton
flavor ($e$,$\mu$), number of $b$-tagged jets (1,2), and jet
multiplicity (2,3,4) combined. Hatched areas represent the systematic
and statistical uncertainties on the background model. The signal
cross section is the measured one (4.9~pb).}
\label{fig:eventchars}
\end{figure*}

To further illustrate the excess of data events over background in the
high-discriminant region, Fig.~\ref{fig:eventchars} shows three
variables that are inputs to the DT analysis: invariant mass of
lepton+$b$-tagged jet+neutrino, $W$ transverse mass, and so-called
``$Q \times \eta$'' (lepton charge times $\eta$ of the leading
untagged jet). They are each shown for low discriminant output, high
output, and very high output. The excess of data over a
background-only model clearly increases as the discriminant cut is
increased.

The DT analysis has also measured the s- and t-channel cross sections
separately. The cross sections are found to be:
$$
\begin{array}{lll}
 \sigma^{\rm obs}\left({\ppbar}{\rargap}tb+X\right)
 & = & 1.0\pm 0.9~{\rm pb} \\
 \sigma^{\rm obs}\left({\ppbar}{\rargap}tqb+X\right)
 & = & 4.2^{+1.8}_{-1.4}~{\rm pb}.
\end{array}
$$
These measurements each assume the standard model value of the single
top quark cross sections not being measured, since the s-channel
measurement considers the t-channel process as a background and vice
versa.

\begin{figure}[!h!tbp]
\includegraphics[width=0.40\textwidth]{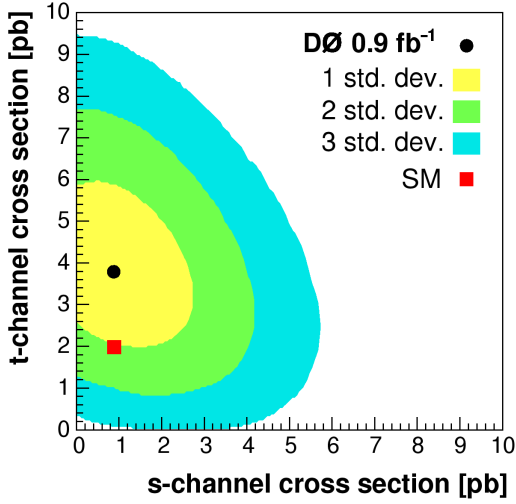}
\caption[2dpost]{Posterior probability density as a function of
$\sigma_{tb}$ and $\sigma_{tqb}$, when both cross sections are
allowed to float in the fit of the $tb+tqb$ DT analysis. Shown are the
contours of equal probability density corresponding to one, two, and
three standard deviations and the location of the most probable value,
together with the SM expectation.}
\label{fig:2dpost}
\end{figure}

We can remove the constraint of the standard model ratio and form the
posterior probability density as a function of the $tb$ and $tqb$
cross sections. This model-independent posterior is shown in
Fig.~\ref{fig:2dpost} for the DT analysis, using the $tb$+$tqb$
discriminant. The most probable value corresponds to cross sections of
$\sigma(tb)=0.9$~pb and $\sigma(tqb)=3.8$~pb. Also shown are the one,
two, and three standard deviation contours. While this result favors a
higher value for the $t$-channel contribution than the SM expectation,
the difference is not statistically significant. Several models of new
physics that are also consistent with this result are shown in
Ref.~\cite{beyond-SM-tait}.

\clearpage

%---------------------------------------------------------------------
%---------------------------------------------------------------------
\section{Combination of Results}
\label{sec:combination} 

Since each multivariate analysis uses the same dataset to measure the
single top quark cross section, their results are highly
correlated. However, because the correlation is rather less than
100\%, one can still gain some additional sensitivity by combining
the results. We combine the three cross section measurements,
$\sigma_i$ ($i$~=~DT, BNN, ME) using the best linear unbiased estimate
(BLUE) method~\cite{BLUE-lyons,BLUE-barlow,BLUE-cowan}; that is, we
take as the new estimate of the cross section the weighted sum
\begin{equation}
 \sigma = \sum_i w_i \, \sigma_i,
\end{equation}
with $\sum_i w_i = 1$, and with the weights chosen so as to minimize
the variance
\begin{equation}
\label{eq:varcor}
 {\rm Var}(y) =
 \sum_i \sum_j \, w_i \, w_j  \, {\rm Cov}(\sigma_i, \sigma_j),
\end{equation}
where ${\mathrm{Cov}}\left( \sigma_i, \sigma_j \right) 
\equiv \langle \sigma_i \sigma_j \rangle - \langle \sigma_i \rangle
\langle \sigma_j \rangle$ are the matrix elements of the covariance
matrix of the measurements. The variance is minimized when
\begin{equation}
\label{eq:wtcor}
 w_i = \frac{\sum_j {\rm Cov}^{-1}(\sigma_i, \sigma_j)}
            {\sum_i \sum_j {\rm Cov}^{-1}(\sigma_i, \sigma_j)},
\end{equation}
where ${\rm Cov}^{-1}(\sigma_i, \sigma_j)$ denotes the matrix elements
of the inverse of the covariance matrix. In order to estimate the
correlation matrix, each analysis is run on the same ensemble of
pseudodatasets, specifically, the SM ensemble with systematics, which
comprises 1,900 pseudodatasets common to all three analyses. To
estimate the $p$-value of the combined result, the analyses are run on
72,000 pseudodatasets of the background-only ensemble.

\subsection{Weights, Coverage Probability, and Combined Measurement}

\begin{figure}[!h!tbp]
\includegraphics[width=0.38\textwidth]{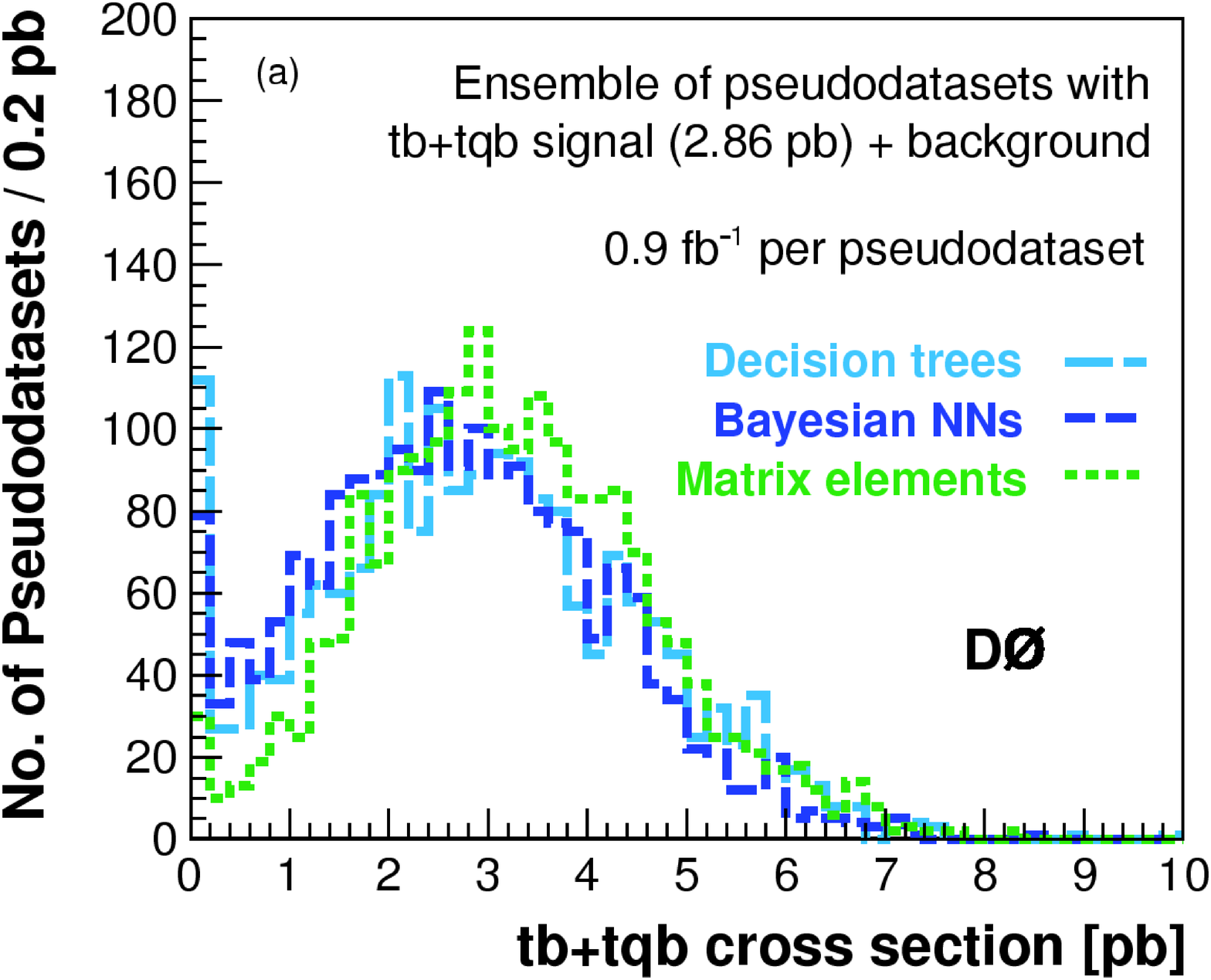}
\hspace{0.2in}
\includegraphics[width=0.38\textwidth]{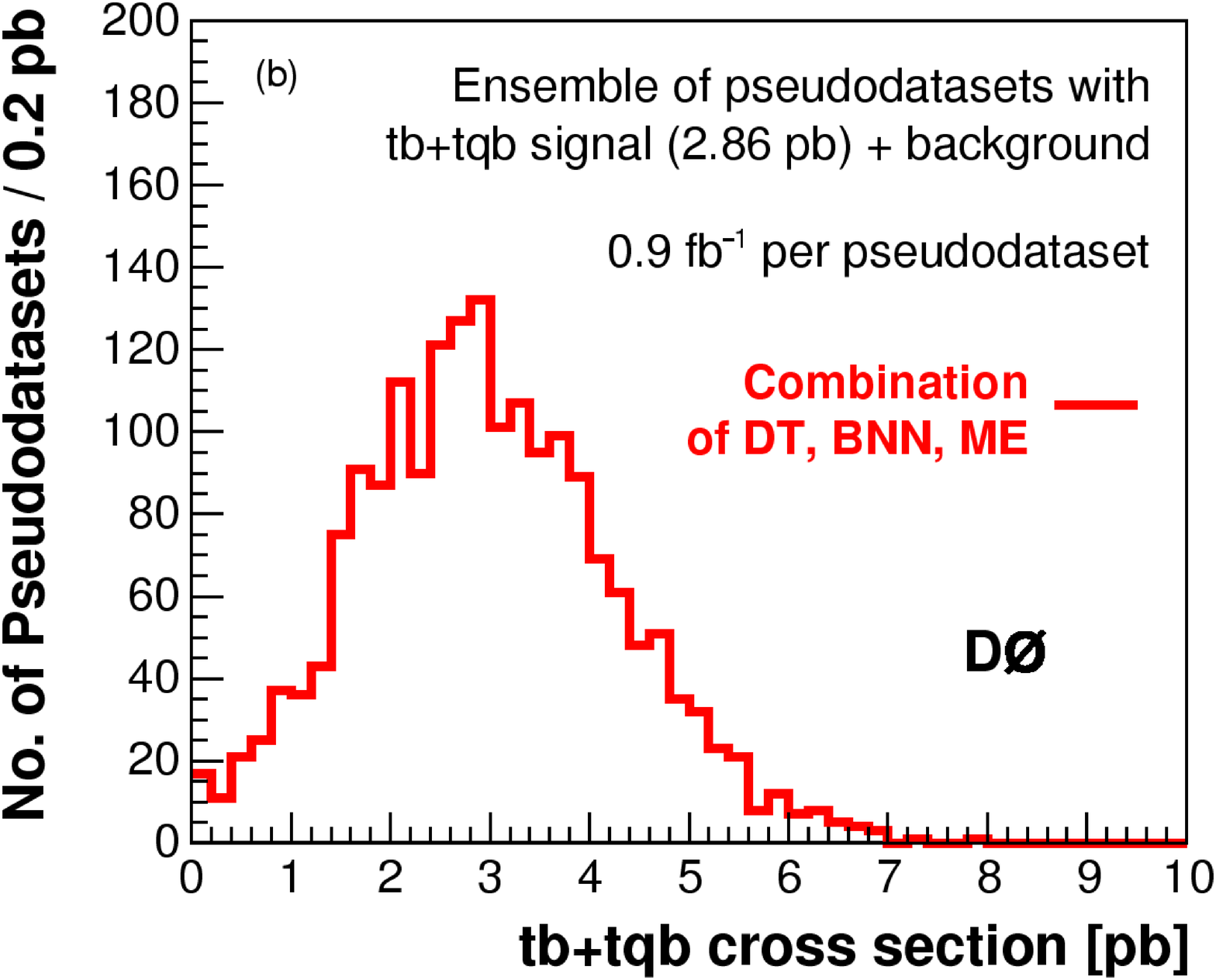}
\vspace{-0.1in}
\caption[measured-xsec-comb]{Distributions of the measured cross
sections from (a) the individual analyses, and (b) the combined
analysis, using the SM ensemble with systematics.}
\label{fig:com_fig1}
\end{figure}

We use the SM ensemble with systematics to determine the weights $w_i$
and to check the coverage probability of the confidence intervals
calculated as described in Sec.~\ref{bayesian-xsec}. The cross section
measurements from this ensemble are shown in Fig.~\ref{fig:com_fig1}
for the individual and combined analyses. The mean and square root of
the variance obtained from these distributions give the following
results:
$$
\begin{array}{ll}
 \sigma^{\rm SM{\mbox{\small -}}ens}
        ({\ppbar}{\rargap}tb+X, & ~tqb+X) \\
 & = 2.9 \pm 1.6~{\rm pb~~(DT)}  \\
 & = 2.7 \pm 1.5~{\rm pb~~(BNN)} \\
 & = 3.2 \pm 1.4~{\rm pb~~(ME)}  \\
 & = 3.0 \pm 1.3~{\rm pb~~(Combined).}  \\
\end{array}
$$

The weights $w_i$ for the three analyses are found to be
\begin{myitemize}
\item $w_{\rm DT}$  = 0.127,
\item $w_{\rm BNN}$ = 0.386,
\item $w_{\rm ME}$  = 0.488.
\end{myitemize}

The correlation matrix is
\begin{equation}
\begin{array}{ccc}
 & & \!\!\!\!\!\!\!\!\!\!\!\!\!\!\rotatebox{45}{DT}~~\rotatebox{45}{BNN}
 ~~\rotatebox{45}{ME} \\
 {\mathrm{Correlation~matrix}} & = & \left(
 \begin{array}{ccc}
   1   & 0.66 & 0.64 \\
  0.66 &  1   & 0.59 \\
  0.64 & 0.59 &  1   
 \end{array}
 \right)
 \begin{array}{l}
  \rm{DT}  \\
  \rm{BNN} \\ 
  \rm{ME} 
 \end{array},
\end{array}
\end{equation}
and the one-standard-deviation coverage probability of the (Bayesian)
confidence interval is 0.67.

\begin{figure}[!h!tbp]
\includegraphics[width=0.465\textwidth]
{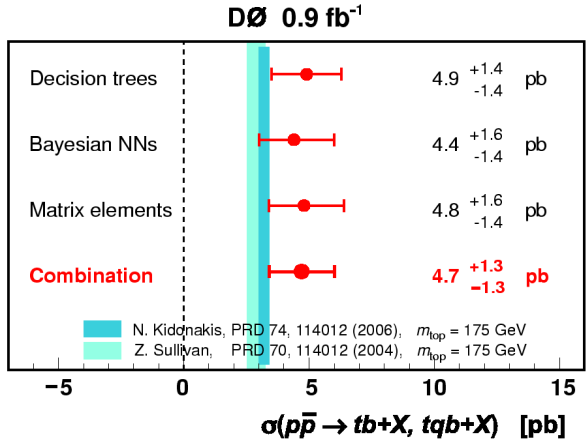}
\vspace{-0.1in}
\caption[results-summary]{The measured single top quark cross sections
from the individual analyses and their combination.}
\label{fig:com_fig2}
\end{figure}

The result from combining the DT, BNN, and DT measurements for the
single top quark cross section is
$$
\begin{array}{ll}
 \sigma^{\rm obs}({\ppbar}{\rargap}tb+X, & ~tqb+X) \\
 & = 4.7 \pm 1.3~{\rm pb~~(Combined)},
\end{array}
$$
using the values listed at the beginning of
Sec.~\ref{sec:measured-cross-sections}. Figure~\ref{fig:com_fig2}
summarizes the measurements of the $tb$+$tqb$ cross section from the
individual analyses as well as the combination.

\subsection{Measurement Significance}

\begin{figure}[!h!tbp]
\includegraphics[width=0.38\textwidth]{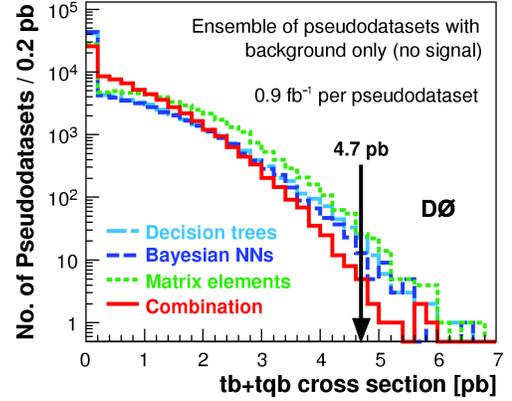}
\vspace{-0.1in}
\caption[meas-xsec-ensemble]{Distributions of the cross sections
measured from data by the three analyses and their combination, using
the background-only ensemble. The arrow shows the combined cross
section measurement, 4.7~pb.}
\label{fig:com_fig3}
\end{figure}

Having determined the combined result for the single top quark cross
section, we can now determine the signal significance corresponding to
this measurement. Distributions of the results from all the analyses
are shown in Fig.~\ref{fig:com_fig3}.

The expected $p$-value (and the associated significance in
Gaussian-like standard deviations) is obtained by counting how many
background-only pseudodatasets yield a measured cross section
greater than the SM value of 2.86~pb. The result is $1.1\%$ or 2.3
standard deviations, as shown in Table~\ref{com_tab2}.

\begin{table}[!h!tbp]
\begin{center}
\caption[com_table2]{The expected $tb$+$tqb$ cross sections,
$p$-values, and significances for the individual and combined
analyses, using the SM value of 2.86~pb for the single top quark
production cross section as the reference point in
Fig.~\ref{fig:com_fig3}.}
\label{com_tab2}
\begin{ruledtabular}
\begin{tabular}{l||ccc}
\multicolumn{4}{c}
{\hspace{0.5in}\underline{Expected Results}}\vspace{0.05in} \\
           &   Expected    & Expected  & Expected     \\
           & cross section & $p$-value & significance \\
Analysis   &    [pb]       &           & (std. dev.)  \\
\hline
 DT        &    2.7        &   0.018   &     2.1      \\
 BNN       &    2.7        &   0.016   &     2.2      \\
 ME        &    2.8        &   0.031   &     1.9      \\
 Combined~~&    2.8        &   0.011   &     2.3
\end{tabular}
\end{ruledtabular}
\end{center}
\end{table}

The observed $p$-value is similarly calculated by counting how many
background-only pseudodatasets result in a cross section above the
value of 4.7~pb measured from data. The result is $0.014\%$ or 3.6
standard deviations. The observed cross sections, $p$-values, and
significances from all the analyses are summarized in
Table~\ref{com_tab3}.

\begin{table}[!h!tbp]
\begin{center}
\caption[table3]{The cross sections measured from data, $p$-values,
and significances for the individual and combined analyses, the latter
two obtained using the background-only ensemble.}
\label{com_tab3}
\begin{ruledtabular}
\begin{tabular}{l||ccc}
\multicolumn{4}{c}
{\hspace{0.5in}\underline{Observed Results}}\vspace{0.05in}\\
           &    Measured   & Measured  &   Measured   \\
           & cross section & $p$-value & significance \\
Analysis   &    [pb]       &           & (std. dev.)  \\
\hline
 DT        &  4.9 &  0.00037  &  3.4  \\
 BNN       &  4.4 &  0.00083  &  3.1  \\
 ME        &  4.8 &  0.00082  &  3.2  \\
 Combined~~&  4.7 &  0.00014  &  3.6
\end{tabular}
\end{ruledtabular}
\end{center}
\end{table}

Finally, using the SM ensemble with systematics, we quantify the
compatibility of our result with the SM expectation by counting how
many pseudodatasets result in a cross section with the observed value
or higher for each of the analyses. The probabilities for the
different analyses are $10\%$ for the DT analysis, $13\%$ for the ME
analysis, $13\%$ for the BNN analysis, and $10\%$ for the combined
analysis.

\subsection{Discriminant Comparison}
\label{sec:powercurves}

\begin{figure}[!h!tbp]
\includegraphics[width=0.45\textwidth]{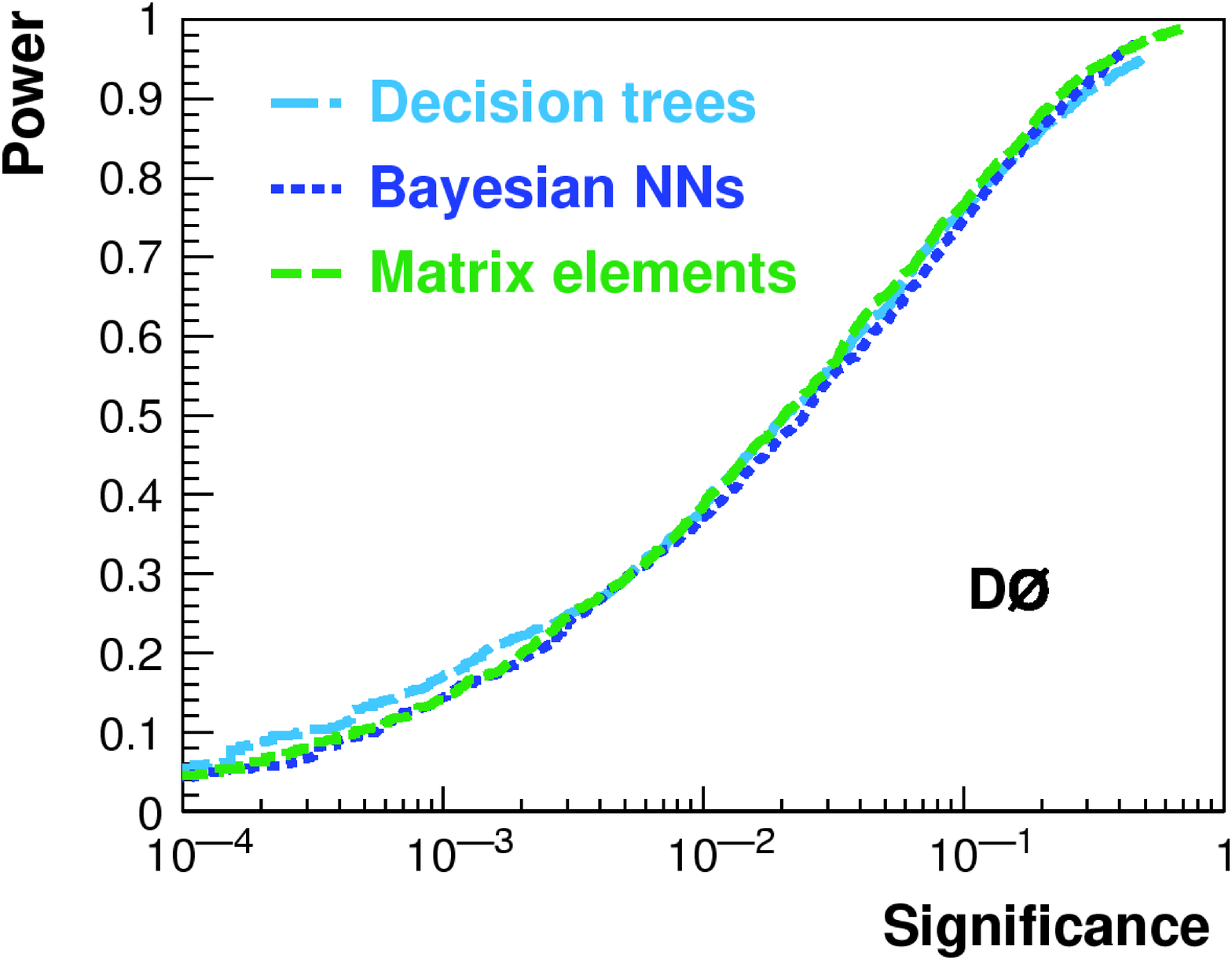}
\vspace{-0.1in}
\caption[powercurve]{The $p$-value computed from the
SM-signal+background ensemble versus the $p$-value from the
background-only ensemble for reference cross sections varying
monotonically from 0--10~pb. For a given significance, that is, the
probability to reject the background-only hypothesis if true, the
power is the probability to accept the signal+background hypothesis if
it is true. For a given significance, one wants the power to be
a large as possible.}
\label{powercurve}
\end{figure}

In order to compare the expected performance of the three multivariate
techniques, it is instructive to compute a power curve for each method using the
two hypotheses $H_1$ = SM-signal+background and $H_0$ = background only. The
power curve in Fig.~\ref{powercurve} is a plot of the probability to accept
hypothesis $H_1$, if it is true, versus the significance level, that is, the
probability to reject hypothesis $H_0$, if it is true. Figure~\ref{powercurve}
shows that all three methods exhibit comparable performance.

%---------------------------------------------------------------------
%---------------------------------------------------------------------
\section{$\mathbf{|V_{tb}|}$ Measurement}

Within the SM with three generations of quarks, the charged-current
interactions of the top quark are of the type $V$--$A$, and involve a
$W$ boson and a down-type quark $q$ ($q=d,s,b$):
\begin{equation}
\label{couplingsm}
 \Gamma_{Wtq}^\mu
 = -\frac{g}{\sqrt{2}}V_{tq}f_1^L\bar{u}(p_b)\gamma^\mu P_L u(p_t),
\end{equation}
where $|V_{tq}|$ is one of the elements of the 3$\times$3 unitary CKM
matrix~\cite{ckm-matrix-c,ckm-matrix-km}, $f_1^L=1$ in the SM, and
$P_{L}=(1-\gamma_5)/2$ is the left-handed ($-$) projection operator.
Under the assumption of three generations and a unitary CKM matrix,
the $|V_{tq}|$ elements are severely
constrained~\cite{particle-data-book}:
\begin{equation}
\begin{array}{lll}
 |V_{td}| & = & (8.14^{+0.32}_{-0.64}) \times 10^{-3} \\
 |V_{ts}| & = & (41.61^{+0.12}_{-0.78})\times 10^{-3} \\
 |V_{tb}| & = & 0.999100^{+0.000034}_{-0.000004}.
\end{array}
\end{equation}

In several extensions of the SM involving, for example, a fourth
generation of quarks or an additional heavy quark singlet that mixes
with the top quark, the 3$\times$3 CKM matrix is no longer required to
be unitary, and $|V_{tb}|$ can be significantly smaller than
unity~\cite{Vtb-not-one-alwall}.

This paper describes in detail the first direct measurement of
$|V_{tb}|$, based on the single top quark production cross section
measurement using decision trees~\cite{run2-d0-prl-evidence}. The
$|V_{tb}|$ measurement is a relatively straightforward extension of
the cross section measurement using the same dataset and analysis
infrastructure, since the cross section for single top quark
production is directly proportional to $|V_{tb}|^2$. This measurement
of $|V_{tb}|$ makes no assumptions on the number of generations or
unitarity of the CKM matrix. However, some assumptions are made in the
generation of our signal MC samples and the extraction of $|V_{tb}|$
from the cross section measurement. In particular, we assume the
following: (i)~there are only SM sources of single top quark
production; (ii)~top quarks decay to $Wb$; and (iii)~the $Wtb$
interaction is $CP$-conserving and of $V$--$A$ type. We discuss these
assumptions in more detail here.

First, we assume that the only production mechanism for single top
quarks involves an interaction with a $W$~boson. Therefore,
extensions of the SM where single top quark events can be produced,
for example, via flavor-changing neutral current
interactions~\cite{run2-d0-fcnc} or heavy scalar or vector boson
exchange~\cite{run2-d0-wprime}, are not considered here.

The second assumption is that $|V_{td}|^2+|V_{ts}|^2 \ll
|V_{tb}|^2$. In other words, we assume $|V_{ts}|$ and $|V_{td}|$ are
negligible compared to $|V_{tb}|$, without making any assumption on
the magnitude of $|V_{tb}|$. This is reasonable given the measurements
of
\begin{equation}
 R = \frac{|V_{tb}|^2}{|V_{td}|^2+|V_{ts}|^2+|V_{tb}|^2},
\end{equation}
by the CDF~\cite{run2-cdf-R} and D0~\cite{run2-d0-R} collaborations,
obtained by comparing the rates of $t\bar{t}$ events with zero, one
and two $b$-tagged jets. For instance, D0's measurement results in
$|V_{td}|^2+|V_{ts}|^2 = (-0.03^{+0.18}_{-0.16}) |V_{tb}|^2$. The
requirement that $|V_{td}|^2+|V_{ts}|^2 \ll |V_{tb}|^2$ implies that
$B(t{\rar}Wb)\simeq 100\%$ and that single top quark production is
completely dominated by the $Wtb$ interaction. This assumption is made
explicitly when measuring the combined $tb$+$tqb$ cross section when
assuming the SM ratio of
$\sigma_{tb}/\sigma_{tqb}$~\cite{Vtb-not-one-alwall}, as well as in
the generation of single top quark and ${\ttbar}$ simulated samples.

Finally, we assume that the $Wtb$ vertex is charge-parity ($CP$)
conserving and of the $V$--$A$ type as given in Eq.~\ref{couplingsm},
but it is allowed to have an anomalous strength $f_1^L$. We do not
allow for right-handed or tensor couplings that may occur in the most
general $Wtb$ vertex~\cite{top-CP-kane,top-CP-whisnant}. The
simulated samples can still be used under the assumption of an
anomalous $f_1^L$ coupling: the {\ttbar} cross section and kinematics,
as well as the $tb$ and $tqb$ kinematics are completely unaffected. An
anomalous value for $f_1^L$ would only rescale the single top quark
cross section, allowing it to be larger or smaller than the SM
prediction, even under the assumption of $|V_{tb}| = 1$. Therefore,
strictly speaking, we are measuring the strength of the $V$--$A$
coupling, i.e., $|V_{tb}f_1^L|$, which is allowed to be $> 1$.
Limiting our measurement to the $[0,1]$ range implies the additional
assumption that $f_1^L=1$.

\subsection{Statistical Analysis}
\label{sec:vtbanalysis}

This measurement uses exactly the same machinery as used to obtain the
single top quark cross section posterior. Following standard
convention for parameters that multiply the cross section, we choose a
prior that is nonnegative and flat in {\vtbsq}, which means it is flat
in the cross section. However, in one of the two cases presented
below, we restrict the prior to the SM allowed region [0,1].

\subsection{Systematic Uncertainties}
\label{sec:vtbsystematics}

In order to extract {\vtb} from the measured cross section, additional
theoretical uncertainties~\cite{singletop-xsec-sullivan} need to be
considered. These uncertainties are applied separately to the $tb$ and
$tqb$ samples in order to take the correlations into account
properly. They are listed in Table~\ref{tab:sys}. The uncertainty on
the top quark mass of 5.1~GeV~\cite{topmass-d0-me-1} is used when
estimating the $\ttbar$ cross section uncertainty and the $tb$ and
$tqb$ cross section uncertainties.

\begin{table}[!h!tbp]
\begin{minipage}{2.75in}
\caption[vtb-systematics]{Systematic uncertainties on the
cross section factor required to extract {\vtb}.}
\label{tab:sys}
\begin{ruledtabular}
\begin{tabular}{l|cc}
\multicolumn{3}{c}{\hspace{0.2in}
{\underline{Additional $|V_{tb}|$ Uncertainties}}} \vspace{0.05in} \\
                      &  $tb$ & $tqb$  \\
\hline
Top quark mass        &  8.5\%  &  13.0\%  \\
Factorization scale   &  4.0\%  &   5.5\%  \\ 
Parton distributions~~&  4.5\%  &  10.0\%  \\
$\alpha_s$            &  1.4\%  &   0.01\%
\end{tabular}
\end{ruledtabular}
\end{minipage}
\end{table}

\subsection{$\mathbf{|V_{tb}|}$ Result}
\label{sec:vtbresult}

\begin{figure}[!h!tbp]
\includegraphics[width=0.35\textwidth]{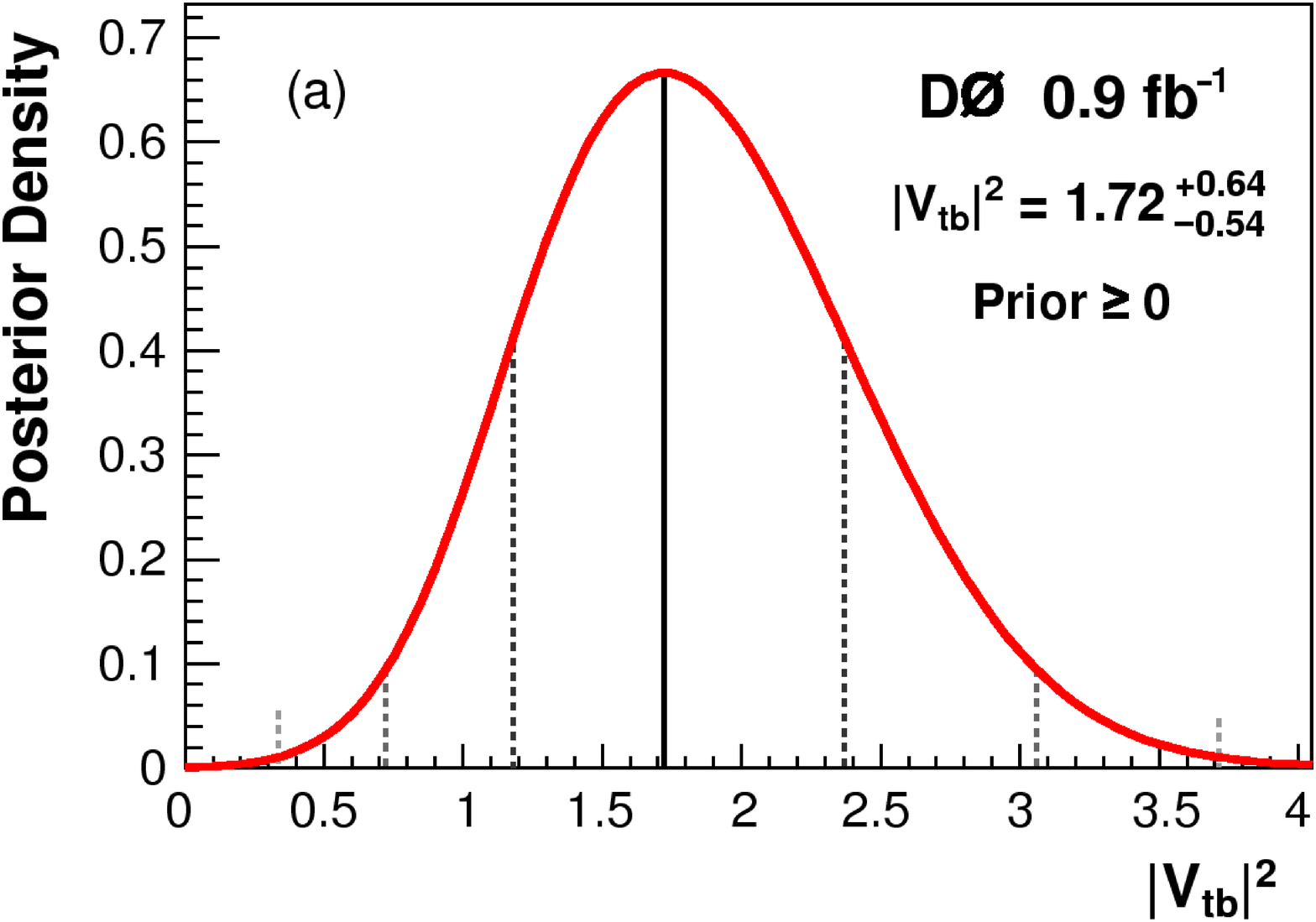}
\includegraphics[width=0.35\textwidth]{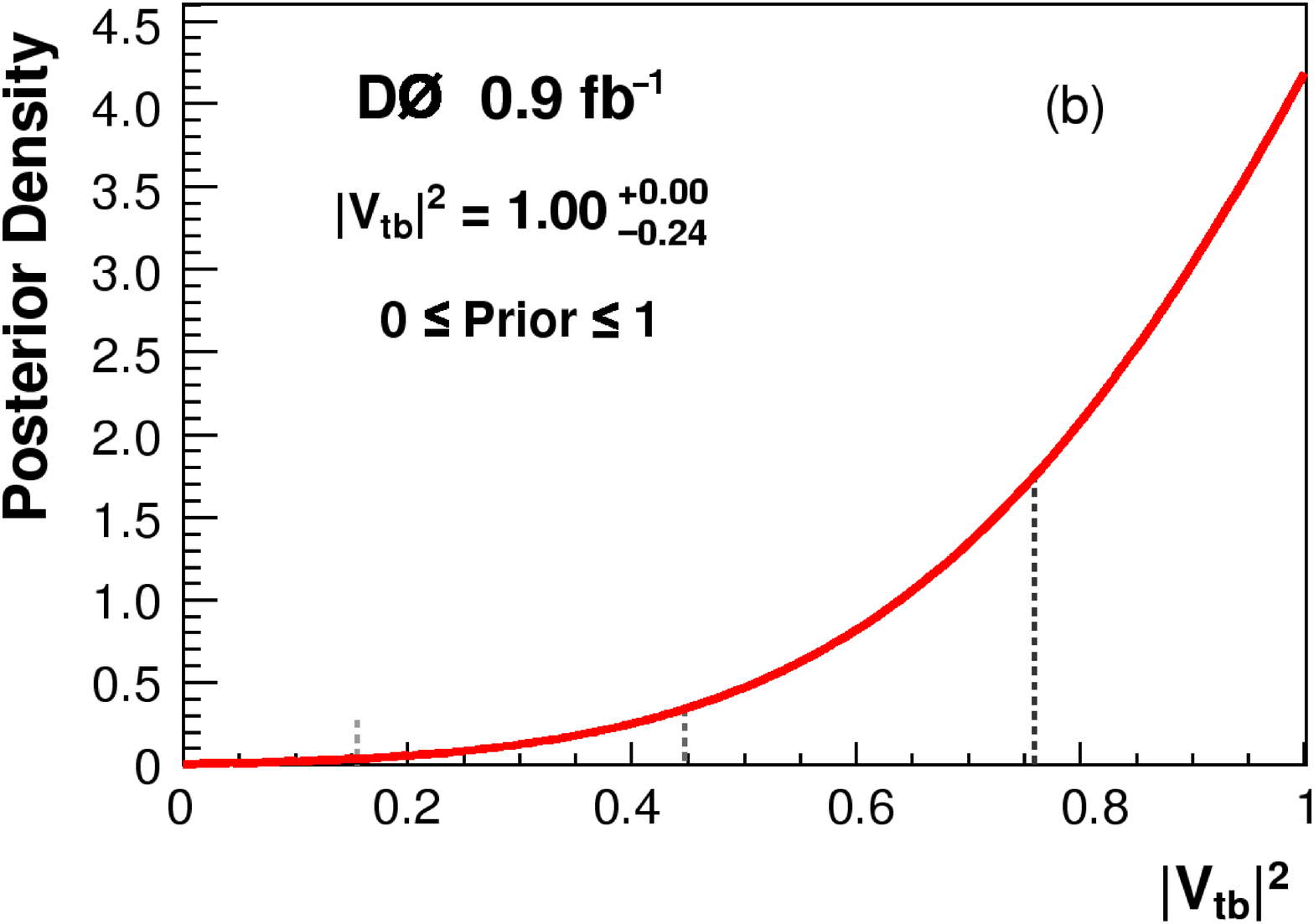}
\vspace{-0.1in}
\caption[vtbposteriors]{The posterior density distributions for
$|V_{tb}|^2$ for (a) a nonnegative flat prior, and (b) a flat prior
restricted to the region [0,1] and assuming $f_1^L = 1$. The dashed
lines show the positions of the one, two, and three standard deviation
distances away from the peak of each curve.}
\label{fig:vtb-posteriors}
\end{figure}

The measurement for the CKM matrix element is obtained from the most
probable value of {\vtbsq}, given by $|V_{tb}| = \sqrt{|V_{tb}|^2}$,
and the uncertainty is computed as $\Delta|V_{tb}| =
\Delta|V_{tb}|^2/2|V_{tb}|$. We have used the decision tree result to
derive a posterior for $|V_{tb}|$. The posterior without the prior
restricted to be only nonnegative gives $|V_{tb}f_1^L|^2 =
1.72^{+0.64}_{-0.54}$, which results in
$$
|V_{tb}f_1^L| = 1.31^{+0.25}_{-0.21}.
$$
The posterior with the prior restricted to the [0,1] region gives
$|V_{tb}|^2 = 1.00^{+0.00}_{-0.24}$, which results in
$$
|V_{tb}| = 1.00^{+0.00}_{-0.12}.
$$
The corresponding $95\%$ C.L. lower limit on {\vtbsq} is 0.46,
corresponding to a lower limit of
$$
|V_{tb}| > 0.68.
$$
The posterior densities for $|V_{tb}|^2$ for each choice of prior are
shown in Fig.~\ref{fig:vtb-posteriors}.

%---------------------------------------------------------------------
%---------------------------------------------------------------------
\section{Summary}

Using approximately 0.9~fb$^{-1}$ of D0 data, we have performed an
analysis of events with a single isolated lepton (electron or muon),
missing transverse energy, and 2--4 jets (1 or 2 of them
$b$~tagged). Using three different multivariate techniques, decision
trees, Bayesian neural networks, and matrix elements, we have searched
for single top quark events from the s-channel ($tb$) and t-channel
($tqb$) processes combined. We measure the cross section to be
$$
 \sigma\left({\ppbar}{\rargap}tb+X,~tqb+X\right)
 = 4.7 \pm 1.3~{\rm pb}.
$$
This corresponds to an excess of 3.6 Gaussian-equivalent standard
deviation significance and constitutes the first evidence of a single
top quark signal. Ensemble tests have shown this result to be
compatible with the standard model cross section with $10\%$
probability.

The decision tree cross section result has been used to extract the
first direct measurement of the CKM matrix element $|V_{tb}|$. This
result does not assume three-generation unitarity of the matrix. The
model independent measurement is
$$
 |V_{tb}f_1^L| = 1.31^{+0.25}_{-0.21},
$$
where $f_1^L$ is a generic left-handed vector coupling.  If we
constrain the value of $|V_{tb}|$ to the standard model region
(i.e., $|V_{tb}| \le 1$ and $f_1^L=1$), then at $95\%$ C.L.,
$|V_{tb}|$ has been measured to be
$$
0.68 < |V_{tb}| \le 1.
$$
\vspace{0.75in}

% acknowledgement_paragraph_r2.tex                   2/19/08
%
We thank the staffs at Fermilab and collaborating institutions, and
acknowledge support from the
DOE and NSF (USA);
CEA and CNRS/IN2P3 (France);
FASI, Rosatom and RFBR (Russia);
CNPq, FAPERJ, FAPESP and FUNDUNESP (Brazil);
DAE and DST (India);
Colciencias (Colombia);
CONACyT (Mexico);
KRF and KOSEF (Korea);
CONICET and UBACyT (Argentina);
FOM (The Netherlands);
STFC (United Kingdom);
MSMT and GACR (Czech Republic);
CRC Program, CFI, NSERC and WestGrid Project (Canada);
BMBF and DFG (Germany);
SFI (Ireland);
The Swedish Research Council (Sweden);
CAS and CNSF (China);
and the
Alexander von Humboldt Foundation.
%
%---------------------------------------------------------------------
%---------------------------------------------------------------------


\begin{thebibliography}{199}
\bibitem{eta}
Pseudorapidity is defined as $\eta = -\ln[\tan(\theta/2)]$, where
$\theta$ is the polar angle with the origin at the primary vertex.
The detector pseudorapidity, or $\deta$, measures $\theta$ as the
polar angle with origin at the center of the detector.
\end{thebibliography}
\end{document}